\documentclass[]{iopart}

\usepackage{graphicx}
\usepackage{graphicx,epsf, epsfig, amssymb}%
\usepackage{bm}
\usepackage{longtable}                 
\usepackage{verbatim}
\usepackage{yfonts}
\usepackage{psfrag}
\usepackage{url}
\makeatletter
\def\url@leostyle{%
  \@ifundefined{selectfont}{\def\UrlFont{\sf}}{\def\UrlFont{\small\ttfamily}}}
\makeatother
\def\lappreq{\! \stackrel{\scriptscriptstyle <}{\scriptscriptstyle
\sim}\!}
\def\gappreq{\!\stackrel{\scriptscriptstyle >}{\scriptscriptstyle \sim}\!}
\def\be{\begin{equation}}
\def\ee{\end{equation}}
\def\beq{\begin{eqnarray}}
\def\eeq{\end{eqnarray}}
\def\f{\frac}
\def\nn{\nonumber}
   
\def\f{\frac}
\def\om{\omega_{lmn}}

\def\ph{\phi_{lmn}}
\def\ta{\tau_{lmn}}

\def\Slm{S_{lmn}}

\newcommand{\qn}{{\textswab{q}}}
\newcommand{\wn}{{\textswab{w}}}
\newcommand{\hj}[1]{\vert \mbox{\boldmath{j}}_{#1}        \vert}

\newcommand{\hjf}  {\vert \mbox{\boldmath{j}}_{\rm fin}   \vert}

\usepackage{fancyhdr}

\begin{document}

\pagestyle{fancy}
\lhead{Quasinormal modes of black holes and black branes}
\chead{}
\rhead{\thepage}
\lfoot{}
\cfoot{}
\rfoot{}

\begin{center}
\topical{Quasinormal modes of black holes and black branes}
\end{center}

\author{Emanuele Berti$^{1,2}$, Vitor Cardoso$^{1,3}$, Andrei O. Starinets$^{4}$}

\address{$^1$~Department of Physics and Astronomy, The University of
  Mississippi, University, MS 38677-1848, USA}

\address{$^2$ Theoretical Astrophysics 130-33, California Institute of
  Technology, Pasadena, CA 91125, USA}

\address{$^3$~Centro Multidisciplinar de Astrof\'{\i}sica - CENTRA,
  Departamento de F\'{\i}sica, Instituto Superior T\'ecnico, Av. Rovisco Pais
  1, 1049-001 Lisboa, Portugal}

\address{$^4$~Rudolf Peierls Centre for Theoretical Physics, 
Department of Physics, University of Oxford, 1 Keble Road, Oxford, OX1 3NP,
United Kingdom}

\ead{berti@phy.olemiss.edu, vitor.cardoso@ist.utl.pt,
  andrei.starinets@physics.ox.ac.uk}

\begin{abstract}
Quasinormal modes are eigenmodes of dissipative systems. Perturbations of
classical gravitational backgrounds involving black holes or branes naturally
lead to quasinormal modes. The analysis and classification of the quasinormal
spectra requires solving non-Hermitian eigenvalue problems for the associated
linear differential equations.  Within the recently developed gauge-gravity
duality, these modes serve as an important tool for determining the
near-equilibrium properties of strongly coupled quantum field theories, in
particular their transport coefficients, such as viscosity, conductivity and
diffusion constants.  In astrophysics, the detection of quasinormal modes in
gravitational wave experiments would allow precise measurements of the mass
and spin of black holes as well as new tests of general relativity.  This
review is meant as an introduction to the subject, with a focus on the recent
developments in the field.
\end{abstract}


{\tableofcontents}
\section{Introduction}


{\it ``The mathematical perfectness of the black holes of Nature is
[...] revealed at every level by some strangeness
in the proportion in conformity of the parts to one another and to the whole.''}
\hspace{.5cm} 
S. Chandrasekhar, ``The Mathematical Theory of Black Holes''
\vskip 2mm

Characteristic modes of vibration are persistent in everything around us.
They make up the familiar sound of various musical instruments but they are
also an important research topic in such diverse areas as seismology,
asteroseismology, molecular structure and spectroscopy, atmospheric science
and civil engineering. All of these disciplines are concerned with the
structure and composition of the vibrating object, and with how this
information is encoded in its characteristic vibration modes: to use a famous
phrase, the goal of studying characteristic modes is to ``hear the shape of a
drum'' \cite{Kac:1966xd}.  This is a review on the characteristic oscillations
of black holes (BHs) and black branes (BHs with plane-symmetric horizon),
called quasinormal modes (QNMs). We will survey the theory behind them, the
information they carry about the properties of these fascinating objects, and
their connections with other branches of physics.

Unlike most idealized macroscopic physical systems, perturbed BH spacetimes
are intrinsically dissipative due to the presence of an event horizon.  This
precludes a standard normal-mode analysis because the system is not
time-symmetric and the associated boundary value problem is non-Hermitian.  In
general, QNMs have complex frequencies, the imaginary part being associated
with the decay timescale of the perturbation. The corresponding eigenfunctions
are usually not normalizable, and, in general, they do not form a complete set
(see Refs.~\cite{RevModPhys.70.1545,Nollert:1998ys} for more extensive
discussions). Almost any real-world physical system is dissipative, so one
might reasonably expect QNMs to be ubiquitous in physics. QNMs are indeed
useful in the treatment of many dissipative systems, e.g. in the context of
atmospheric science and leaky resonant cavities.

Two excellent reviews on BH QNMs \cite{Kokkotas:1999bd,Nollert:1999ji} were
written in 1999. However, much has happened in the last decade that is not
covered by these reviews. The recent developments have brought BH oscillations
under the spotlight again.  We refer, in particular, to the role of QNMs in
gravitational wave astronomy and their applications in the gauge-gravity
duality.
This work will focus on a critical review of the new developments, providing
our own perspective on the most important and active lines of research in the
field.

After a general introduction to QNMs in the framework of BH perturbation
theory, we will describe methods to obtain QNMs numerically, as well as some
important analytic solutions for special spacetimes. Then we will review the
QNM spectrum of BHs in asymptotically flat spacetimes, asymptotically
(anti-)de Sitter (henceforth AdS or dS) spacetimes and other spacetimes of
interest. After this general overview we will discuss what we regard as the
most active areas in QNM research. Schematically, we will group recent
developments in QNM research into three main branches:

\noindent {\bf (i) AdS/CFT and holography.}  In 1997-98, a powerful new
technique known as the AdS/CFT correspondence or, more generally, the
gauge-string duality was discovered and rapidly developed
\cite{Maldacena:1997re}.  The new method (often referred to as holographic
correspondence) provides an effective description of a non-perturbative,
strongly coupled regime of certain gauge theories in terms of
higher-dimensional classical gravity.  In particular, equilibrium and
non-equilibrium properties of strongly coupled thermal gauge theories are
related to the physics of higher-dimensional BHs and black branes and their
fluctuations.  Quasinormal spectra of the dual gravitational backgrounds give
the location (in momentum space) of the poles of the retarded correlators in
the gauge theory, supplying important information about the theory's
quasiparticle spectra and transport (kinetic) coefficients.  Studies of QNMs
in the holographic context became a standard tool in considering the
near-equilibrium behavior of gauge theory plasmas with a dual gravity
description.  Among other things, they revealed the existence of a
universality of the particular gravitational frequency of generic black branes
(related on the gauge theory side to the universality of the viscosity-entropy
ratio in the regime of infinitely strong coupling), as well as intriguing
connections between the dynamics of BH horizons and hydrodynamics
\cite{Son:2007vk}.  The duality also offers a new perspective on notoriously
difficult problems, such as the BH information loss paradox, the nature of BH
singularities and quantum gravity.  Holographic approaches to these problems
often involve QNMs.  This active area of research is reviewed in Section
\ref{sec:holography}.

\noindent {\bf (ii) QNMs of astrophysical black holes and gravitational wave
  astronomy.}  The beginning of LIGO's first science run (S1) in 2002 and the
achievement of design sensitivity in 2005 marked the beginning of an era in
science where BHs and other compact objects should play a prominent
observational role. While electromagnetic observations are already providing
us with strong evidence of the astrophysical reality of BHs
\cite{Narayan:2005ie}, gravitational wave observations will incontrovertibly
show if these compact objects are indeed rotating (Kerr) BHs, as predicted by
Einstein's theory of gravity. BH QNMs can be used to infer their mass and
angular momentum \cite{Echeverria:1989hg} and to test the no-hair theorem of
general relativity \cite{Berti:2005ys,Berti:2007zu}.  Dedicated ringdown
searches in interferometric gravitational wave detector data are ongoing
\cite{Creighton:1999pm,Tsunesada:2005fe}.  The progress on the experimental
side was accompanied by a breakthrough in the numerical simulation of
gravitational wave sources. Long-term stable numerical evolutions of BH
binaries have been achieved after 4 decades of efforts
\cite{Pretorius:2005gq,Campanelli:2005dd,Baker:2005vv}, confirming that
ringdown plays an important role in the dynamics of the merged system.
These developments are reviewed in Section \ref{sec:astro}.

\noindent {\bf (iii) Other developments.} In 1998, Hod suggested that
highly-damped QNMs could bridge the gap between classical and quantum gravity
\cite{Hod:1998vk}. The following years witnessed a rush to compute and
understand this family of highly damped modes. The interest in this subject
has by now faded substantially but, at the very least, Hod's proposal has
contributed to a deeper analytical and numerical understanding of QNM
frequencies in many different spacetimes, and it has highlighted certain
general properties characterizing some classes of BH solutions. These ideas
and other recent developments (including a proposed connection between QNMs
and BH phase transitions, the QNMs of analogue BHs, the stability of naked
singularities and its relation with the so-called algebraically special modes)
are reviewed in Section \ref{sec:otherdev}.

The present work is mostly intended to make the reader familiar with the new
developments by summarizing the vast (and sometimes confusing) bibliography on
the subject.  We tried to keep the review as self-contained as possible, while
avoiding to duplicate (as far as possible to preserve logical consistency)
material that is treated more extensively in other reviews on the topic, such
as
Refs.~\cite{Kokkotas:1999bd,Nollert:1999ji,RevModPhys.70.1545,Ferrari:2007dd}.
A detailed understanding of BH QNMs and their applications requires some
specialized technical background. QNM research has recently expanded to
encompass a very wide range of topics: a partial list includes analogue
gravity, alternative theories of gravity, higher-dimensional spacetimes,
applications to numerical relativity simulations, explorations of the
gauge-gravity duality, the stability analysis of naked singularities and
ringdown searches in LIGO. Because of space limitations we cannot discuss all
of this material in detail, and we refer the reader to other reviews. Topics
that are treated in more detail elsewhere include: (1) a general overview of
gravitational radiation \cite{Cutler:2002me,Sathyaprakash:2009xs} and its
multipolar decomposition \cite{Thorne:1980ru}; (2) BH perturbation theory
\cite{MTB,Nakamura:1987zz,Mino:1997bx,Frolov:1998wf,Sasaki:2003xr,Nagar:2005ea};
(3) the issue of quantifying QNM excitation in different physical scenarios
(see e.g.~\cite{Berti:2006hb} for an introduction pre-dating the numerical
relativity breakthroughs of 2005, and \cite{BertiCampanelli} for a more
updated overview of the field); (4) tests of general relativity and of the
no-hair theorem that either do not make use of ringdown
\cite{AmaroSeoane:2007aw}, or do not resort to gravitational wave observations
at all \cite{Barcelo:2005fc,Psaltis:2008bb,Will:2005va}; (5) BH solutions in
higher dimensions \cite{Emparan:2008eg}; (6) many aspects of the gauge-gravity
duality
\cite{Son:2007vk,Aharony:1999ti,Gubser:2009md,Hartnoll:2009sz,Herzog:2009xv}.
The reviews listed above provide more in-depth looks at different aspects of
QNM research, but we tried to provide concise introductions to all of these
topics while (hopefully) keeping the presentation clear and accessible.


Chandrasekhar's fascination with the mathematics of BHs was due to their
simplicity. BHs in four-dimensional, asymptotically flat spacetime must belong
to the Kerr-Newman family, which is fully specified by only three parameters:
mass, charge and angular momentum (see e.g. Ref.~\cite{Heusler:1998ua}, or
Carter's contribution to Ref.~\cite{Hawking:1979ig}).  One expresses this by
saying that BHs have no hair (or more precisely, that they have three
hairs). A consequence of the no-hair theorem is that all perturbations in the
vicinities of a BH must decay to one and the same final state, i.e.  that all
hairs (except three) must be lost. Perturbative and numerical calculations
show that the hair loss proceeds, dynamically, via quasinormal ringing. The
gravitational wave signal from a perturbed BH can in general be divided in
three parts: (i) A prompt response at early times, that depends strongly on
the initial conditions and is the counterpart to light-cone propagation; (ii)
An exponentially decaying ``ringdown'' phase at intermediate times, where QNMs
dominate the signal, which depends entirely on the {\it final} BH's
parameters; (iii) A late-time tail, usually a power-law falloff of the field
\cite{Ching:1995tj}. Mathematically, each of these stages arises from
different contributions to the relevant Green's function (see Section
\ref{inversioncontour}). QNM frequencies depend only on the BH's parameters,
while their amplitudes depend on the source exciting the oscillations.

Numerical and analytical analysis of processes involving BHs confirm these
expectations. QNMs were observed for the first time in numerical simulations
of the scattering of Gaussian wavepackets by Schwarzschild BHs in 1970, soon
after the BH concept itself was introduced and popularized by John
Wheeler. Vishveshwara \cite{vish} noticed that the waveform at late times
consists of a damped sinusoid, with ringing frequency almost independent of
the Gaussian's parameters. Ringdown was observed again in the linearized
approximation to the problem of a test mass falling from infinity into a
Schwarzschild BH \cite{Davis:1971gg}. By now, decades of experience have shown
that any event involving BH dynamics is likely to end in this same
characteristic way: the gravitational wave amplitude will die off as a
superposition of damped sinusoids.

\begin{figure*}[ht]
\begin{center}
\begin{tabular}{cc}
\epsfig{file=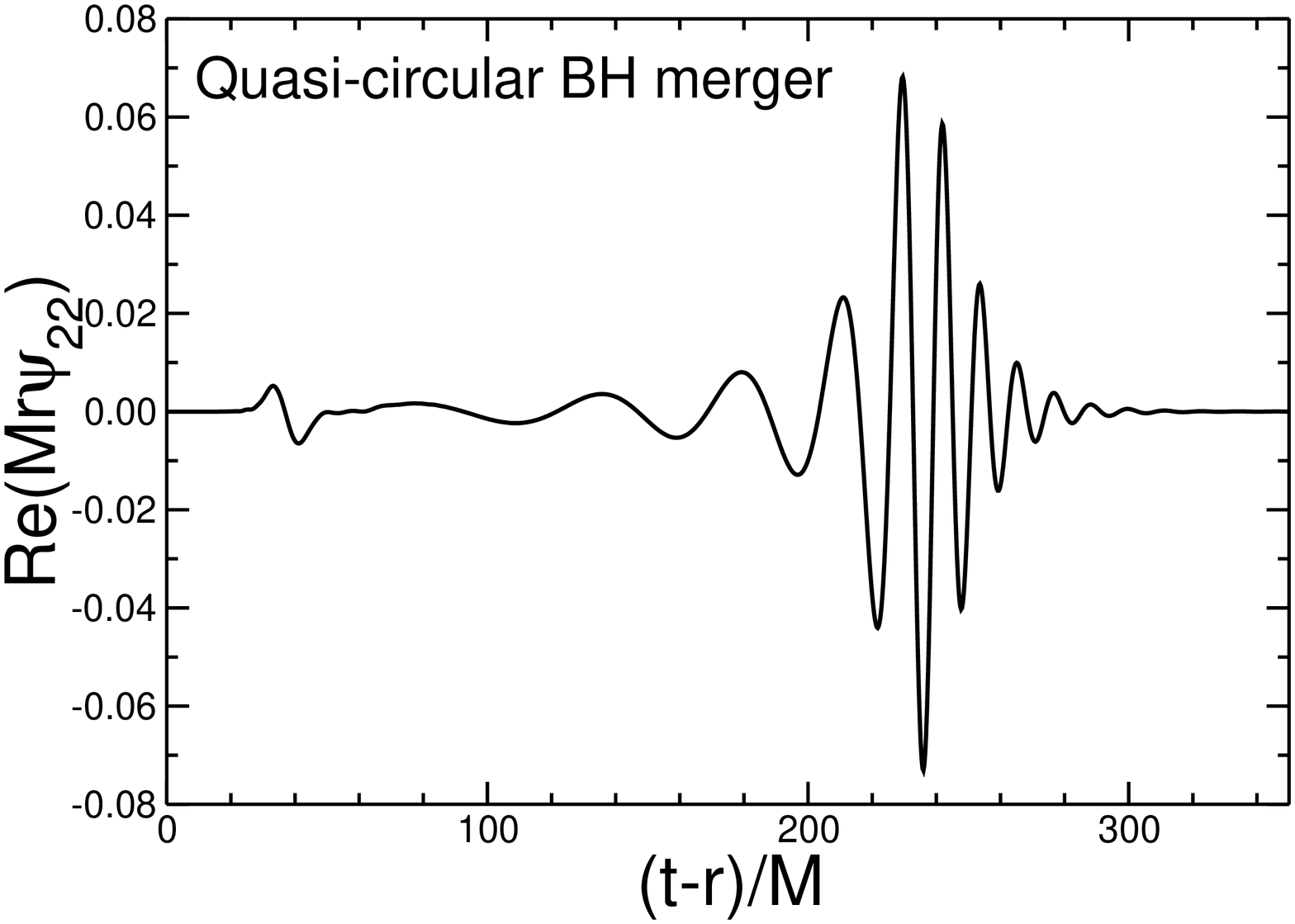,width=5cm,angle=-90} &
\epsfig{file=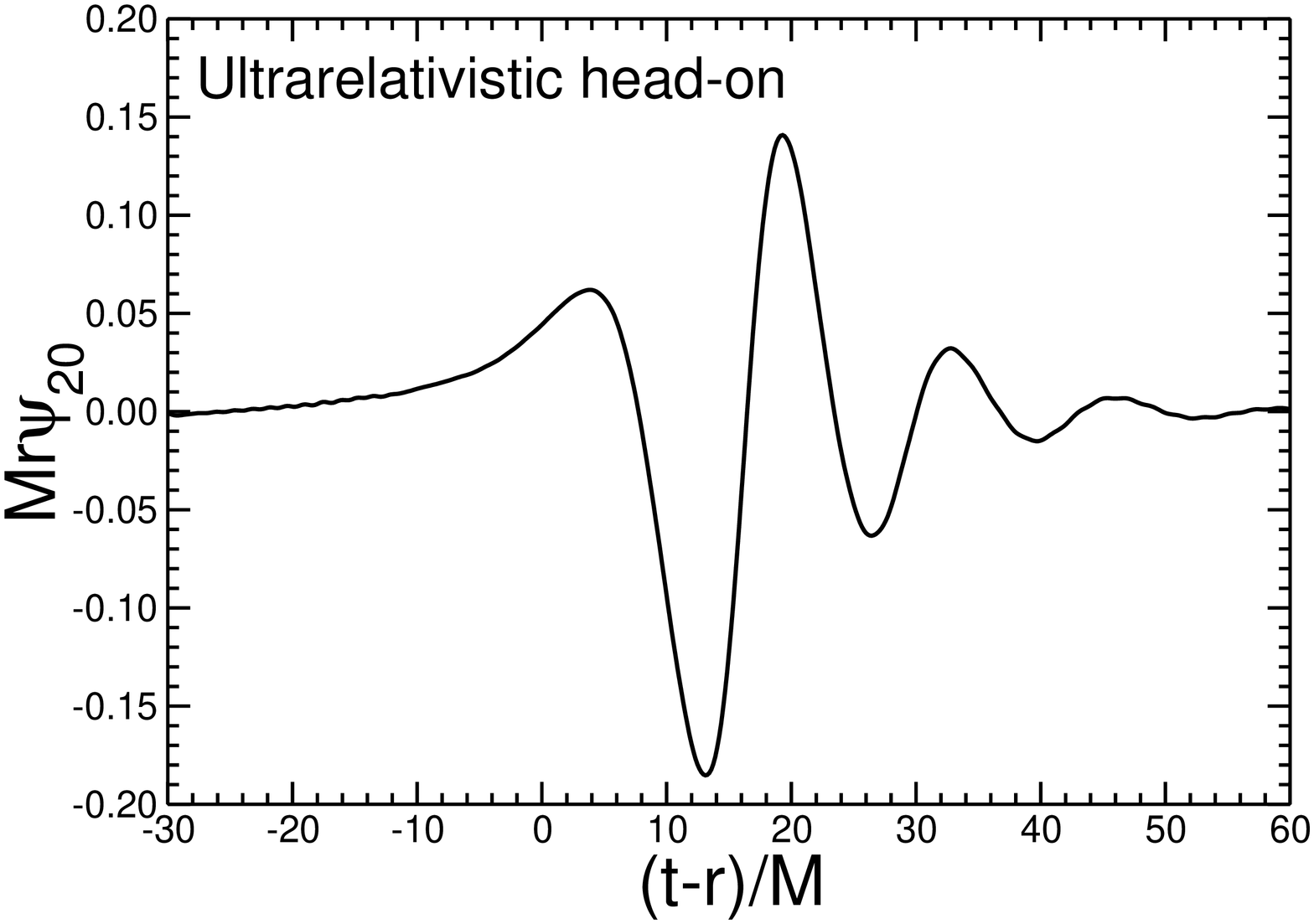,width=5cm,angle=-90}\\
\epsfig{file=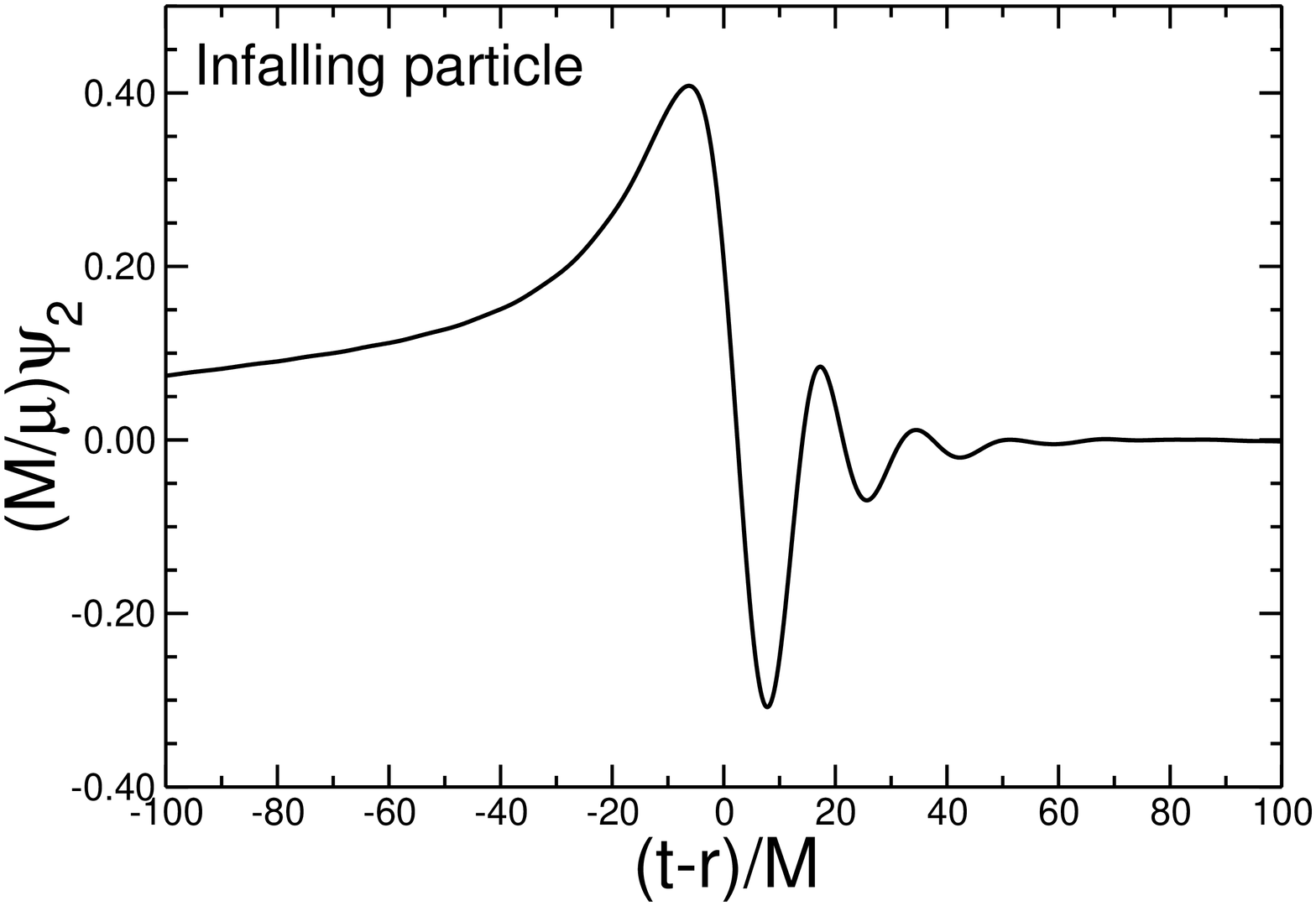,width=5cm,angle=-90} &
\epsfig{file=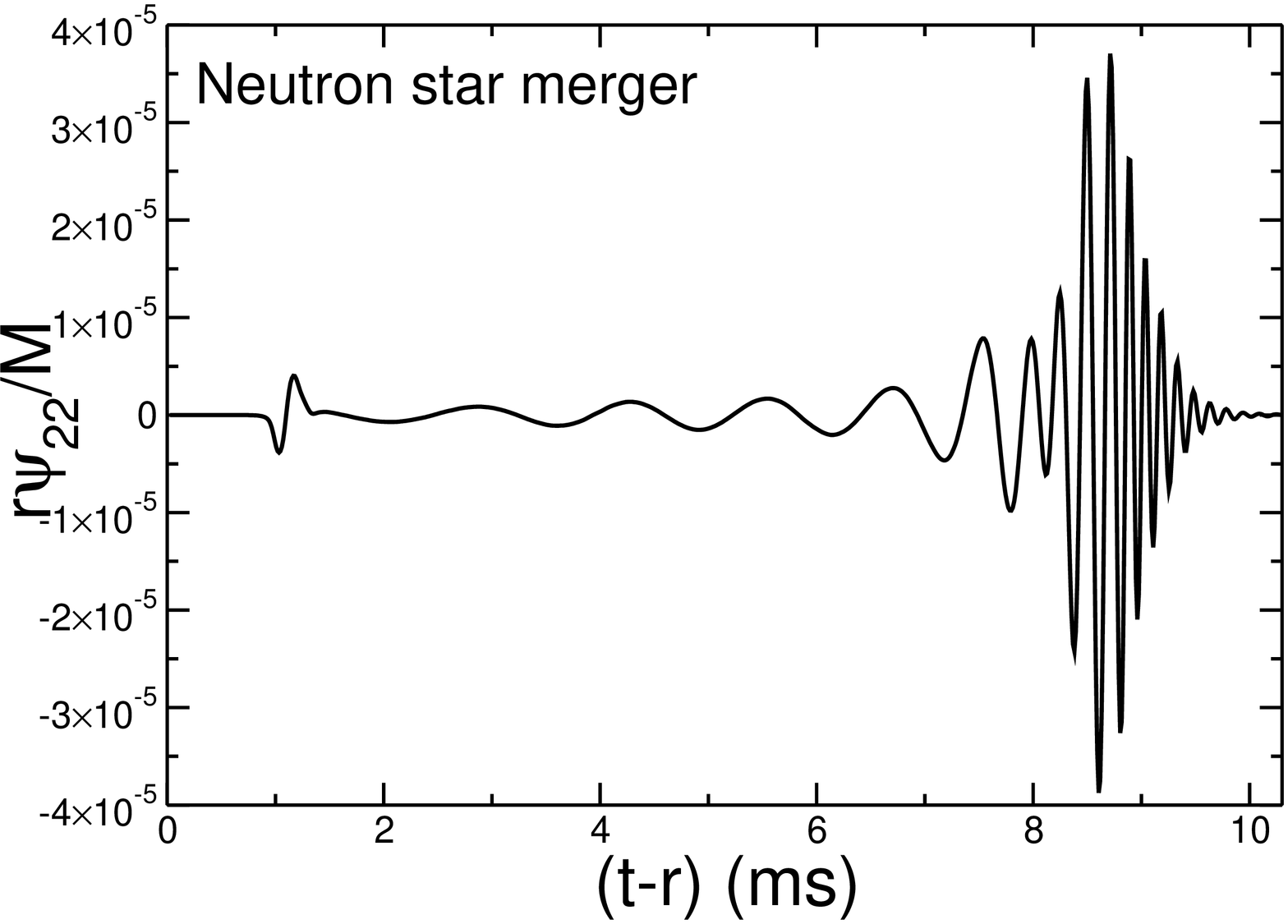,width=5cm,angle=-90}
\end{tabular}
\end{center}
\caption{Four different physical processes leading to substantial quasinormal
  ringing (see text for details). With the exception of the infalling-particle
  case (where $M$ is the BH mass, $\mu$ the particle's mass and $\psi_2$ the
  Zerilli wavefunction), $\psi_{22}$ is the $l=m=2$ multipolar component of
  the Weyl scalar $\Psi_4$, $M$ denotes the total mass of the system and $r$
  the extraction radius (see
  e.g. Ref.~\cite{Berti:2007fi}).} \label{fig:qnmexcitation}
\end{figure*}
Figure \ref{fig:qnmexcitation} shows four different processes involving BH
dynamics. In all of them, quasinormal ringing is clearly visible. The
upper-left panel (adapted from Ref.~\cite{Berti:2007fi}) is the signal from
two equal-mass BHs initially on quasi-circular orbits, inspiralling towards
each other due to the energy loss induced by gravitational wave emission,
merging and forming a single final BH \cite{Pretorius:2005gq}. The upper-right
panel of Fig.~\ref{fig:qnmexcitation} shows gravitational waveforms from
numerical simulations of two equal-mass BHs, colliding head-on with $v/c=0.94$
in the center-of-mass frame: as the center-of-mass energy grows (i.e., as the
speed of the colliding BHs tends to the speed of light) the waveform is more
and more strongly ringdown-dominated \cite{Sperhake:2008ga}. The bottom-left
panel shows the gravitational waveform (or more precisely, the dominant, $l=2$
multipole of the Zerilli function) produced by a test particle of mass $\mu$
falling from rest into a Schwarzschild BH \cite{Davis:1971gg}: the shape of
the initial precursor depends on the details of the infall, but the subsequent
burst of radiation and the final ringdown are universal features. The
bottom-right panel (reproduced from Ref.~\cite{Baiotti:2008ra}) shows the
waves emitted by two massive neutron stars (NSs) with a polytropic equation of
state, inspiralling and eventually collapsing to form a single BH.

QNM frequencies for gravitational perturbations of Schwarzschild and Kerr BHs
have been computed by many authors. Rather than listing numerical tables of
well-known results, we have set up a web page providing tabulated values of
the frequencies and fitting coefficients for the QNMs that are most relevant
in gravitational wave astronomy \cite{rdweb}. On this web page, we also
provide {\it Mathematica} notebooks to compute QNMs of Kerr and asymptotically
AdS BHs \cite{rdweb}.

\subsection{Milestones }

QNM research has a fifty-year-long history. We find it helpful to provide the
reader with a ``roadmap'' in the form of a chronological list of papers that,
in our opinion, have been instrumental to shape the evolution of the
field. Our summary is necessarily biased and incomplete, and we apologize in
advance for the inevitable omissions. A more complete set of references can be
found in the rest of this review.

\noindent $\bullet$ 1957 -- Regge and Wheeler \cite{Regge:1957rw} analyze a
special class of gravitational perturbations of the Schwarzschild
geometry. This effectively marks the birth of BH perturbation theory a decade
before the birth of the BH concept itself. The ``one-way membrane'' nature of
the horizon is not yet fully understood, and the boundary conditions of the
problem are not under control.

\noindent $\bullet$ 1961 -- Newman and Penrose \cite{Newman:1961qr} develop a
formalism to study gravitational radiation using spin coefficients. 

\noindent $\bullet$ 1963 -- Kerr \cite{Kerr:1963ud} discovers the mathematical
solution of Einstein's field equations describing rotating BHs. In the same
year, Schmidt identifies the first quasar (``quasi-stellar radio
source''). Quasars (compact objects with luminosity $\sim 10^{12}$ that of our
sun, located at cosmological distance \cite{Schmidt:1963}) are now believed to
be supermassive BHs (SMBHs), described by the Kerr solution.

\noindent $\bullet$ 1964 -- The UHURU orbiting X-ray observatory makes the
first surveys of the X-ray sky discovering over 300 X-ray ``stars'', most of
which turn out to be due to matter accreting onto compact objects. One of
these X-ray sources, Cygnus X-1, is soon accepted as the first plausible
stellar-mass BH candidate (see e.g.~\cite{Bolton:1972}).

\noindent $\bullet$ 1967 -- Wheeler \cite{Ruffini:1971,wheeler} coins the term
``black hole'' (see the April 2009 issue of {\em Physics Today}, and
Ref.~\cite{Wheeler:1998vs} for a fascinating, first-person historical
account).

\noindent $\bullet$ 1970 -- Zerilli \cite{Zerilli:1970se,Zerilli:1971wd}
extends the Regge-Wheeler analysis to general perturbations of a Schwarzschild
BH. He shows that the perturbation equations can be reduced to a pair of
Schr\"odinger-like equations, and applies the formalism to study the
gravitational radiation emitted by infalling test particles.

\noindent $\bullet$ 1970 -- Vishveshwara \cite{vish} studies numerically the
scattering of gravitational waves by a Schwarzschild BH: at late times the
waveform consists of damped sinusoids (now called ``ringdown waves'').

\noindent $\bullet$ 1971 -- Press \cite{pressringdown} identifies ringdown
waves as the free oscillation modes of the BH.

\noindent $\bullet$ 1971 -- Davis {\it et al.} \cite{Davis:1971gg} carry out
the first quantitative calculation of gravitational radiation emission within
BH perturbation theory, considering a particle falling radially into a
Schwarzschild BH. Quasinormal ringing is excited when the particle crosses the
maximum of the potential barrier of the Zerilli equation, which is located at
$r\simeq 3M$ (i.e., close to the unstable circular orbit corresponding to the
``light ring'').

\noindent $\bullet$ 1972 -- Goebel \cite{goebel} points out that the
characteristic modes of BHs are essentially gravitational waves in spiral
orbits close to the light ring.

\noindent $\bullet$ 1973 -- Teukolsky \cite{Teukolsky:1972my} decouples and
separates the equations for perturbations in the Kerr geometry using the
Newman-Penrose formalism \cite{Newman:1961qr}.

\noindent $\bullet$ 1974 -- Moncrief \cite{Moncrief:1974am} introduces a
gauge-invariant perturbation formalism.

\noindent $\bullet$ 1975 -- Chandrasekhar and Detweiler
\cite{Chandrasekhar:1975qn} compute numerically some weakly damped
characteristic frequencies. They prove that the Regge-Wheeler and Zerilli
potentials have the same spectra.

\noindent $\bullet$ 1978 -- Cunningham, Price and Moncrief
\cite{Cunningham1,Cunningham2,Cunningham3} study radiation from relativistic
stars collapsing to BHs using perturbative methods. QNM ringing is excited.

\noindent $\bullet$ 1979 -- Gerlach and Sengupta give a comprehensive and
elegant mathematical foundation for gauge-invariant perturbation theory
\cite{Gerlach:1979rw,Gerlach:1980tx}.

\noindent $\bullet$ 1983 -- Chandrasekhar's monograph \cite{MTB} summarizes the
state of the art in BH perturbation theory, elucidating connections between
different formalisms.

\noindent $\bullet$ 1983 -- York \cite{York:1983zb} attempts to relate the QNM
spectrum to Hawking radiation. To our knowledge, this is the first attempt to
connect the (purely classical) QNMs with quantum gravity.

\noindent $\bullet$ 1983 -- Mashhoon \cite{Mashhoon:1985} suggests to use WKB
techniques to compute QNMs. Ferrari and Mashhoon \cite{Ferrari:1984zz}
analytically compute QNMs using their connection with bound states of the {\em
  inverted} BH effective potentials.

\noindent $\bullet$ 1985 -- Stark and Piran \cite{Stark:1985da} extract
gravitational waves from a simulation of rotating collapse to a BH in
numerical relativity. QNM excitation is observed, as confirmed by more recent
work \cite{Baiotti:2006wm}.

\noindent $\bullet$ 1985 -- Confirming the validity of Goebel's arguments
\cite{goebel}, Mashhoon \cite{mashhoon} regards QNMs as waves orbiting around
the unstable photon orbit and slowly leaking out, and estimates analytically
some QNM frequencies in Kerr-Newman backgrounds.

\noindent $\bullet$ 1985 -- Schutz and Will \cite{Schutz:1985km} develop a WKB
approach to compute BH QNMs.

\noindent $\bullet$ 1985 -- Leaver \cite{Leaver:1985ax,leJMP,Leaver:1986gd}
provides the most accurate method to date to compute BH QNMs using continued
fraction representations of the relevant wavefunctions, and discusses their
excitation using Green's function techniques.

\noindent $\bullet$ 1986 -- McClintock and Remillard \cite{McClintock:1986}
show that the X-ray nova A0620-00 contains a compact object of mass almost
certainly larger than $3M_\odot$, paving the way for the identification of
many more stellar-mass BH candidates. 

\noindent $\bullet$ 1989 -- Echeverria \cite{Echeverria:1989hg} estimates the
accuracy with which one can estimate the mass and angular momentum of a BH
from QNM observations. The formalism is later improved by Finn
\cite{Finn:1992wt} and substantially refined in Ref.~\cite{Berti:2005ys},
where ringdown-based tests of the no-hair theorem of general relativity are
shown to be possible. An Appendix of Ref.~\cite{Berti:2005ys} provides QNM
tables to be used in data analysis and in the interpretation of numerical
simulations; these data are now available online \cite{rdweb}.

\noindent $\bullet$ 1992 -- Nollert and Schmidt \cite{nollertschmidt} use
Laplace transforms to compute QNMs. Fr\"oman {\it et al.}
\cite{Froeman:1992gp} first introduce phase-integral techniques in the context
of BH physics.

\noindent $\bullet$ 1993 -- Anninos {\it et al.} \cite{Anninos:1993zj} first
succeed in simulating the head-on collision of two BHs, and observe QNM
ringing of the final BH.

\noindent $\bullet$ 1993 -- Bachelot and Motet-Bachelot \cite{Bachelot:1993dp}
show that a potential with compact support does not cause power-law tails in
the evolution of Cauchy data.  Subsequently Ching {\it et al.}
\cite{Ching:1994bd,Ching:1995tj} generalize this result to potentials falling
off faster than exponentially.

\noindent $\bullet$ 1996 -- Gleiser {\it et al.}  \cite{Gleiser:1995gx} extend
the perturbative formalism to second order and use it to estimate radiation
from colliding BHs employing the so-called ``close limit'' approximation,
quantifying the limits of validity of linear perturbation theory
\cite{Gleiser:1996yc}.

\noindent $\bullet$ 1997 -- Maldacena \cite{Maldacena:1997re} formulates the
AdS/CFT (Anti-de Sitter/Conformal Field Theory) duality conjecture. Shortly
afterward, the papers by Gubser, Klebanov, Polyakov \cite{Gubser:1998bc} and
Witten \cite{Witten:1998qj} establish a concrete quantitative recipe for the
duality. The AdS/CFT era begins.

\noindent $\bullet$ 1998 -- The AdS/CFT correspondence is generalized to
non-conformal theories in a variety of approaches (see \cite{Aharony:1999ti}
for a review). The terms ``gauge-string duality'', ``gauge-gravity duality''
and ``holography'' appear, referring to these generalized settings.

\noindent $\bullet$ 1998 -- Flanagan and Hughes \cite{Flanagan:1997sx} show
that, under reasonable assumptions and depending on the mass range, the
signal-to-noise ratio for ringdown waves is potentially larger than the
signal-to-noise ratio for inspiral waves in both Earth-based detectors (such
as LIGO) and planned space-based detectors (such as LISA).

\noindent $\bullet$ 1998 -- Hod \cite{Hod:1998vk} uses earlier numerical
results by Nollert \cite{Nollert:1993} to conjecture that the real part of
highly-damped QNMs is equal to $T \ln 3$ ($T$ being the Hawking temperature),
a conjecture later proven by Motl \cite{Motl:2002hd} using the continued
fraction method.  Hod also proposes a connection between QNMs and Bekenstein's
ideas on BH area quantization.

\noindent $\bullet$ 1999 -- Creighton \cite{Creighton:1999pm} describes a
search technique for ringdown waveforms in LIGO.

\noindent $\bullet$ 1999 -- Two reviews on QNMs appear: {\it Quasinormal modes
  of stars and black holes}, by Kokkotas and Schmidt \cite{Kokkotas:1999bd}
and Nollert's {\it Quasinormal modes: the characteristic ``sound'' of black
  holes and neutron stars} \cite{Nollert:1999ji}.

\noindent $\bullet$ 1999 -- Horowitz and Hubeny \cite{Horowitz:1999jd} compute
QNMs of BHs in AdS backgrounds of various dimensions and relate them to
relaxation times in the dual CFTs.

\noindent $\bullet$ 2000 -- Shibata and Uryu \cite{Shibata:1999wm} perform the
first general relativistic simulation of the merger of two neutron stars. More
recent simulations confirm that ringdown is excited when the merger leads to
BH formation \cite{Baiotti:2008ra}.

\noindent $\bullet$ 2001 -- Birmingham, Sachs and Solodukhin
\cite{Birmingham:2001pj} point out that QNM frequencies of the
$(2+1)$-dimensional Ba\~nados-Teitelboim-Zanelli (BTZ) BH
\cite{Cardoso:2001hn} coincide with the poles of the retarded correlation
function in the dual $(1+1)$-dimensional CFT.

\noindent $\bullet$ 2002 -- Baker, Campanelli and Lousto \cite{Baker:2001sf}
complete the ``Lazarus'' program to ``resurrect'' early, unstable numerical
simulations of BH binaries and extend them beyond merger using BH perturbation
theory.

\noindent $\bullet$ 2002 -- Dreyer \cite{Dreyer:2002vy} proposes to resolve an
ambiguity in Loop Quantum Gravity using the highly damped QNMs studied by Hod
\cite{Hod:1998vk}.

\noindent $\bullet$ 2002 -- Son and Starinets \cite{Son:2002sd} formulate a
recipe for computing real-time correlation functions in the gauge-gravity
duality.  They use the recipe to prove that, in the gauge-gravity duality, QNM
spectra correspond to poles of the retarded correlation functions.

\noindent $\bullet$ 2002 -- QNMs of black branes are computed
\cite{Starinets:2002br}. The lowest QNM frequencies of black branes in the
appropriate conserved charges channels are naturally interpreted as
hydrodynamic modes of the dual theory \cite{Nunez:2003eq}.

\noindent $\bullet$ 2003 -- Motl and Neitzke \cite{Motl:2003cd} use a monodromy
technique (similar to the phase integral approaches of
Ref.~\cite{Froeman:1992gp}) to compute analytically highly damped BH QNMs.

\noindent $\bullet$ 2003 -- In a series of papers
\cite{Kodama:2003jz,Ishibashi:2003ap,Kodama:2003kk}, Kodama and Ishibashi
extend the Regge-Wheeler-Zerilli formalism to higher dimensions.

\noindent $\bullet$ 2003 -- In one of the rare works on probing quantum aspects
of gravity with gauge theory in the context of the gauge-gravity duality
(usually, the correspondence is used the other way around), Fidkowski {\it et
  al.} \cite{Fidkowski:2003nf} study singularities of BHs by investigating the
spacelike geodesics that join the boundaries of the Penrose diagram. The
complexified geodesics' properties yield the large-mass QNM frequencies
previously found for these BHs. This work is further advanced in
Ref.~\cite{Festuccia:2005pi} and subsequent publications.

\noindent $\bullet$ 2004 -- Following Motl and Neitzke \cite{Motl:2003cd},
Nat\'ario and Schiappa analytically compute and classify asymptotic QNM
frequencies for $d$-dimensional BHs \cite{Natario:2004jd}.

\noindent $\bullet$ 2005 -- The LIGO detector reaches design sensitivity
\cite{:2007kva}.

\noindent $\bullet$ 2005 -- Pretorius \cite{Pretorius:2005gq} achieves the
first long-term stable numerical evolution of a BH binary. Soon afterwards,
other groups independently succeed in evolving merging BH binaries using
different techniques \cite{Campanelli:2005dd,Baker:2005vv}. The waveforms
indicate that ringdown contributes a substantial amount to the radiated
energy.

\noindent $\bullet$ 2005 -- Kovtun and Starinets \cite{Kovtun:2005ev} extend
the QNM technique in the gauge-gravity duality to vector and gravitational
perturbations using gauge-invariant variables for black brane fluctuations. A
classification of the fluctuations corresponding to poles of the stress-energy
tensor and current correlators in a dual theory in arbitrary dimension is
given.  These methods and their subsequent development and application in
\cite{Parnachev:2005hh,Benincasa:2005iv,Buchel:2005cv,Benincasa:2005qc} become
a standard approach in computing transport properties of strongly coupled
theories from dual gravity.

\noindent $\bullet$ 2006--2008 -- An analytic computation of the lowest QNM
frequency in the shear mode gravitational channel of a generic black brane
\cite{Starinets:2008fb} reveals universality, related to the universality of
the shear viscosity to entropy density ratio in dual gauge theories. This and
further developments \cite{Iqbal:2008by,Bhattacharyya:2008jc} also point to a
significance of the QNM spectrum in the context of the BH membrane paradigm
(for a recent review of the membrane paradigm approach, see
\cite{Damour:2008ji}).

\noindent $\bullet$ 2008--2009 -- QNM spectra are computed in applications of
the gauge-gravity duality to condensed matter theory
\cite{Hartnoll:2009sz,Herzog:2009xv}.

\subsection{Notation and conventions}

Unless otherwise and explicitly stated, we use geometrized units where
$G=c=1$, so that energy and time have units of length. We also adopt the
$(-+++)$ convention for the metric.  For reference, the following is a list of
symbols that are used often throughout the text.

\begin{table}[h]
\begin{tabular}{ll}
  $d$ & Total number of spacetime dimensions (we always consider one timelike\\
      & and $d-1$ spatial dimensions).\\
  $L$ & Curvature radius of (A)dS spacetime, related to the negative \\
      & cosmological constant $\Lambda$ in the Einstein equations 
        ($G_{\mu\nu}+\Lambda g_{\mu\nu}=0$) \\
  & through $L^2=\mp(d-2)(d-1)/(2\Lambda)$. The $-$ sign is for AdS, $+$ for
  dS.\\
  $M$ & Mass of the BH spacetime. \\
  $a$ & Kerr rotation parameter: $a = J/M \in [0,M]$.\\
  $r_+$ & Radius of the BH's event horizon in the chosen coordinates.\\
  $\omega$ & Fourier transform variable. The time dependence of any
  field is $\sim e^{-i\omega t}$.  \\
  & For stable spacetimes, ${\rm Im}(\omega)<0$. Also useful is $\wn\equiv
  \omega/2\pi T$.\\
  $\omega_R,\,\omega_I$ & Real and imaginary part of the QNM
  frequencies.\\
  $s$ & Spin of the field.\\
  $l$ & Integer angular number, related to the eigenvalue $A_{lm}=l(l+d-3)$\\ 
  &of scalar spherical harmonics in $d$ dimensions.\\ 
  $n$ & Overtone number, an integer labeling the QNMs by increasing $|{\rm
    Im}(\omega)|$.\\
  & We conventionally start counting from a ``fundamental mode'' with $n=0$.\\
\end{tabular}
\end{table}

\section{\label{sec:bhperts}A black hole perturbation theory primer}

Within general relativity (and various extensions thereof involving
higher-derivative gravity), QNMs naturally appear in the analysis of linear
perturbations of fixed gravitational backgrounds.  The perturbations obey
linear second-order differential equations, whose symmetry properties are
dictated by the symmetries of the background.  In most cases, these symmetries
allow one to separate variables with an appropriate choice of coordinates
reducing the system to a set of linear ordinary differential equations (ODEs)
or a single ODE.  The ODEs are supplemented by boundary conditions, usually
imposed at the BH's horizon and at spatial infinity.  QNMs are the eigenmodes
of this system of equations.  The precise choice of the boundary conditions is
physically motivated, but it is clear that the presence of the horizon, acting
for classical fields as a one-sided membrane, is of crucial importance: it
makes the boundary value problem non-hermitian and the associated eigenvalues
complex.
The methods used to reduce the problem to a single ODE depend on the metric
under consideration; some of them are discussed and compared in
Chandrasekhar's book \cite{MTB}. Given the progress in the field in recent
years and the vast literature on the subject, we will not attempt to describe
these techniques in detail. As a simple example illustrating the main
extensions of the formalism described in \cite{MTB} we discuss field
perturbations in $d$-dimensional, non-rotating geometries. For the interested
reader, Sections \ref{sec:asflat}, \ref{sec:asAdS} and \ref{sec:catalogue}
provide references on other background geometries.

\subsection{Perturbations of the Schwarzschild-anti-de Sitter geometry}

Consider the Einstein-Hilbert gravitational action for a $d-$dimensional
spacetime with cosmological constant $\Lambda$:
\be S=\frac{1}{16\pi G}\, \int d^dx\sqrt{-g}
\left( R - 2 \Lambda\right) + \int  d^dx\sqrt{-g} \, {\cal L}_m\,, 
\label{action}
\ee
where ${\cal L}_m$ is the Lagrangian representing a generic contribution of
the ``matter fields'' (scalar, Maxwell, $p-$form, Dirac and so on) coupled to
gravity. The specific form of ${\cal L}_m$ depends on the particular theory.
The Einstein equations read
\be
G_{\mu\nu} + \Lambda g_{\mu\nu} = 8 \pi G T_{\mu\nu}\,,
\label{field-eqs}
\ee 
where $T_{\mu\nu}$ is the stress-energy tensor associated with ${\cal L}_m$.
Eq.~(\ref{field-eqs}) should be supplemented by the equations of motion for
the matter fields. Together with Eq.~(\ref{field-eqs}), they form a
complicated system of non-linear partial differential equations describing the
evolution of all fields including the metric.  A particular solution of this
system forms a set of {\it background} fields ${g_{\mu\nu}^{BG}, \Phi^{BG}}$,
where $\Phi$ is a cumulative notation for all matter fields present. By
writing $g_{\mu\nu} = g_{\mu\nu}^{BG} + h_{\mu\nu}$, $\Phi = \Phi^{BG} +
\phi$, and linearizing the full system of equations with respect to the
perturbations $h_{\mu\nu}$ and $\phi$, we obtain a set of linear differential
equations satisfied by the perturbations.

Maximally symmetric vacuum ($T_{\mu\nu}^{BG}=0$) solutions to the field
equations are Minkowski, de Sitter (dS) and anti-de Sitter (AdS) spacetimes,
depending on the value of the cosmological constant (zero, positive or
negative, respectively).  Generic solutions of Eq.~(\ref{field-eqs}) are
asymptotically flat, dS or AdS.  We will be mostly interested in
asymptotically flat or AdS spacetimes. AdS spacetimes of various dimension
arise as a natural groundstate of supergravity theories and as the
near-horizon geometry of extremal BHs and $p$-branes in string theory, and
therefore they play an important role in the AdS/CFT correspondence
\cite{Aharony:1999ti,Klebanov:2000me,Maldacena:2003nj,Duff:1999rk,Cotsakis:2000bb}.

BHs in asymptotically AdS spacetimes form a class of solutions interesting
from a theoretical point of view and central for the gauge-gravity duality at
finite temperature.  Their relation to dual field theories is discussed in
Section \ref{sec:holography}.  In addition to the simplest Schwarzschild-AdS
(SAdS) BH, one finds BHs with toroidal, cylindrical or planar topology
\cite{Lemos:1994fn,Lemos:1994xp,Lemos:1995cm,Mann:1996gj,Vanzo:1997gw,
  Birmingham:1998nr} as well as the Kerr-Newman-AdS family
\cite{Carter:1968ks}.  The standard BH perturbation theory \cite{MTB} is
easily extended to asymptotically AdS spacetimes
\cite{Kodama:2003jz,Mellor:1989ac,Cardoso:2001bb,Cardoso:2001vs}.  For
illustration we consider the non-rotating, uncharged $d$-dimensional SAdS (or
SAdS$_d$) BH with line element
\be ds^{2}= -fdt^{2}+f^{-1}dr^{2}+r^{2}d\Omega_{d-2}^2\,,
\label{lineelementads} \ee
where $f(r) = 1 + r^2/L^2 - r_0^{d-3}/r^{d-3}$, $d\Omega_{d-2}^2$ is the
metric of the $(d-2)$-sphere, and the AdS curvature radius squared $L^2$ is
related to the cosmological constant by $L^2 = -(d-2)(d-1)/2\Lambda$. The
parameter $r_0$ is proportional to the mass $M$ of the spacetime:
$M=(d-2)A_{d-2}r_0^{d-3}/16\pi$,
where $A_{d-2} = 2 \pi^{(d-1)/2}/\Gamma{\left[(d-1)/2\right]}$.  The
well-known Schwarzschild geometry corresponds to $L\to \infty$.
\subsubsection*{Scalar field perturbations}
Let us focus, for a start, on scalar perturbations in vacuum. The
action for a complex scalar field with a conformal coupling is given by 
$S_m \equiv \int d^dx\sqrt{-g}\, {\cal L}_m$, where
\be
 {\cal L}_m = - \left(\partial_{\mu}\Phi\right)^\dagger\partial^{\mu}\Phi -
\frac{d-2}{4(d-1)} \gamma \, R \, \Phi^\dagger \Phi 
- m^2  \Phi^\dagger \Phi\,.
\ee
For $\gamma =1$, $m=0$ the action is invariant under the conformal
transformations $g_{\mu\nu}\rightarrow
\Omega^2g_{\mu\nu}\,,\Phi\rightarrow\Omega^{1-d/2}\Phi$, and for $\gamma=0$,
$m=0$ one recovers the usual minimally coupled massless scalar. The equations
of motion satisfied by the fields $g_{\mu\nu}$ and (massless) $\Phi$ are
\be \nabla_{\mu}\nabla^{\mu}\Phi=\frac{d-2}{4(d-1)}\gamma R
\Phi\,,\label{kg}\qquad G_{\mu\nu}+ \Lambda\,
g_{\mu\nu}=8 \pi G \, T_{\mu\nu} \,,\ee
where $T_{\mu\nu}$ is quadratic in $\Phi$.  Considering perturbations of the
fields, $g_{\mu\nu} = g_{\mu\nu}^{BG} + h_{\mu\nu}$ and $\Phi = \Phi^{BG} +
\phi$ with $\Phi^{BG}=0$, we observe that the linearized equations of motion
for $h_{\mu\nu}$ and $\phi$ decouple, and thus the metric fluctuations
$h_{\mu\nu}$ can be consistently set to zero.  The background metric satisfies
$G_{\mu\nu}^{BG} +\Lambda g_{\mu\nu}^{BG}=0$.  We choose $g_{\mu\nu}^{BG}$ to
be the SAdS$_d$ metric (\ref{lineelementads}). The scalar fluctuation
satisfies the equation
\be \frac{1}{\sqrt{-g_{\mbox{\tiny BG}}}}\partial_{\mu}
\Biggl( \sqrt{-g_{\mbox{\tiny BG}}}\,  g^{\mu\nu}_{{\mbox{\tiny BG}}}\, \partial_{\nu}\phi\Biggr) = 
\frac{d(d-2)\gamma}{4L^2}\, \phi\,. \label{kg2} \ee
The time-independence and the spherical symmetry of the metric imply the
decomposition
\be \phi(t,r,\theta \,)=\sum _{lm} e^{-i\omega t}
\frac{\Psi_{s=0}(r)}{r^{(d-2)/2}}Y_{lm}(\theta)\,,
\label{separationscalar} \ee
where $Y_{lm}(\theta)$ denotes the $d-$dimensional scalar spherical
harmonics, satisfying $\Delta_{\Omega_{d-2}}Y_{lm}=-l(l+d-3)Y_{lm}$, with
$\Delta_{\Omega_{d-2}}$ the Laplace-Beltrami operator, and the ``$s=0$'' label
indicates the spin of the field. Here and in the rest of this paper, for
notational simplicity, we usually omit the integral over frequency in the
Fourier transform.  Substituting the decomposition into Eq.~(\ref{kg2}) we get
a radial wave equation for $\Psi_{s=0}(r)$:
\beq
f^2\frac{d^2\Psi_{s=0}}{dr^2}+ff'\frac{d\Psi_{s=0}}{dr}+\left( \omega^2-
V_{s=0}\right)  \Psi_{s=0}=0\,. \label{waveeqscalar} \eeq
We will see shortly that perturbations with other spins satisfy similar
equations.  In the particular case of $s=0$, the radial potential $V_s$ is
given by
\be
V_{s=0}=f\,\left[\frac{l(l+d-3)}{r^2}+\frac{d-2}{4}\left (\frac{(d-4)f}{r^2}+\frac{2f'}{r}+\frac{d\gamma}{L^2}\right)\right]\,.
\ee
Finally, if we define a ``tortoise'' coordinate $r_*$ by the relation
$dr_*/dr=1/f$, Eq.~(\ref{waveeqscalar}) can be written in the form of a
Schr\"{o}dinger equation with the potential $V_s$
\be \label{waveeq}
\frac{d^2 \Psi_{s}}{dr_*^2} + \left(\omega^2-V_{s}\right)\Psi_{s}=0\,.
\ee
Notice that the tortoise coordinate $r_* \rightarrow -\infty$ at the horizon
(i.e. as $r\to r_+$), but its behavior at infinity is strongly dependent on
the cosmological constant: $r_*\rightarrow +\infty$ for asymptotically-flat
spacetimes, and $r_* \rightarrow {\rm constant}$ for the SAdS$_d$ geometry.

\subsubsection*{Electromagnetic, gravitational and half-integer spin perturbations}

Equations for linearized Maxwell field perturbations in curved spacetimes can
be obtained along the lines of the scalar field example above. To separate the
angular dependence we now need {\it vector spherical harmonics}
\cite{Nollert:1999ji,Ruffini,edmonds}.  In $d=4$, electromagnetic
perturbations can be completely characterized by the wave equation
(\ref{waveeq}) with the potential
\be V_{s=1}^{d=4}=f\left\lbrack\frac{l(l+1)}{r^2}\right\rbrack \,.
\label{potentialmaxwell} \ee
A comprehensive treatment of the four-dimensional case can be found in
Ref.~\cite{Ruffini} for the Schwarzschild spacetime, and in
Ref.~\cite{Cardoso:2001bb} for the SAdS geometry. Higher-dimensional
perturbations are discussed in Ref.~\cite{Crispino:2000jx}.

The classification of gravitational perturbations $h_{\mu\nu}(x)$ on a fixed
background $g_{\mu\nu}^{BG}(x)$ is more complicated.  We focus on the SAdS$_4$
geometry.  After a decomposition in tensorial spherical harmonics, the
perturbations fall into two distinct classes: odd (Regge-Wheeler or
vector-type) and even (Zerilli or scalar-type), with parities equal to
$(-1)^{l+1}$ and $(-1)^l$, respectively
\cite{Nagar:2005ea,Thorne:1980ru,Zerilli:1971wd,mathewsharmonics}.
In the Regge-Wheeler gauge
\cite{Kokkotas:1999bd,Nollert:1999ji,Nagar:2005ea,Regge:1957rw,Vishveshwara:1970cc},
the perturbations are written as $h_{\mu \nu}= e^{-i\omega t} \tilde{h}_{\mu
  \nu}$, where for odd parity
\begin{eqnarray}
\tilde{h}_{\mu \nu}= \left[
 \begin{array}{cccc}
 0 & 0 &0 & h_0(r)
\\ 0 & 0 &0 & h_1(r)
\\ 0 & 0 &0 & 0
\\ h_0(r) & h_1(r) &0 &0
\end{array}\right]
\left(\sin\theta\frac{\partial}{\partial\theta}\right)
Y_{l0}(\theta)\,, \label{oddgauge}
\end{eqnarray}
whereas for even parity
\begin{eqnarray}
\tilde{h}_{\mu \nu}= \left[
 \begin{array}{cccc}
 H_0(r) f & H_1(r) &0 & 0
\\ H_1(r) & H_2(r)/f  &0 & 0
\\ 0 & 0 &r^2K(r) & 0
\\ 0 & 0 &0 & r^2K(r)\sin^2\theta
\end{array}\right] \, Y_{l0}(\theta)\,. \label{evengauge}
\end{eqnarray}
The angular dependence of the perturbations is dictated by the structure of
tensorial spherical harmonics
\cite{Nagar:2005ea,Thorne:1980ru,Zerilli:1971wd,mathewsharmonics}.  Inserting
this decomposition into Einstein's equations one gets ten coupled second-order
differential equations that fully describe the perturbations: three equations
for the odd radial variables, and seven for the even variables. The odd
perturbations can be combined in a single Regge-Wheeler or vector-type
gravitational variable $\Psi^-_{s=2}$, and the even perturbations can likewise
be combined in a single Zerilli or scalar-type gravitational wavefunction
$\Psi^+_{s=2}$. The Regge-Wheeler and Zerilli functions ($\Psi^-_{s=2}$ and
$\Psi^+_{s=2}$, respectively) satisfy the Schr\"odinger-like equation
(\ref{waveeq}) with the potentials
\begin{equation}
V_{s=2}^{-}=  f(r)
\left\lbrack\frac{l(l+1)}{r^2}-\frac{6M}{r^3}\right\rbrack\,
\label{vodd}
\end{equation}
and 
\begin{equation}
V_{s=2}^{+}= \frac{2f(r)}{r^3}
\frac{9M^3+3\lambda^2Mr^2+\lambda^2\left(1+\lambda\right)r^3+9M^2\left(\lambda
r+\frac{r^3}{L^2}\right)} {\left(3M+\lambda
r\right)^2} \,. \label{veven}
\end{equation}
The parameters $h_0$ and $h_1$ of the vector-type perturbation are related to
$\Psi^-_{s=2}$ by
\begin{equation}
\Psi^-_{s=2}=  \frac{f(r)}{r} h_1(r)\,,\qquad
h_0=\frac{i}{\omega}\frac{d}{dr_*}\left(r\Psi^-_{s=2}\right)\,.
\label{qodd}
\end{equation}
For the scalar-type gravitational perturbation, the functions $H_1$ and $K$
can be expressed through $\Psi^+_{s=2}$ via
\begin{eqnarray}
K&=& \frac{6M^2+\lambda \left(1+\lambda\right)r^2+3M\left(\lambda
r-\frac{r^3}{L^2}\right)} {r^2\left(3M+\lambda r\right)}
\Psi^+_{s=2}
+\frac{d\Psi^+_{s=2}}{dr_*} \,, \\
H_1&=& \frac{i\omega\left(3M^2+3\lambda Mr-\lambda r^2+3M
\frac{r^3}{L^2}\right)} {r \left(3M+\lambda r\right)
f(r)}\Psi^+_{s=2} -
\frac{i \omega r}{f(r)}\frac{d\Psi^+_{s=2}}{dr_*}\,, \label{4.11}
\end{eqnarray}
where $\lambda\equiv (l-1)(l+2)/2$, and $H_0$ is then obtained from the
algebraic relation
\beq
&&\left [(l-1)(l+2)+\frac{6M}{r}\right ]H_0+\left[i\frac{l(l+1)}{\omega\,r^2}(M+r^3/L^2)-2i\omega\,r\right ]H_1\,\nonumber\\
&-&\left [(l-1)(l+2)+rf'-\frac{4\omega^2r^2+r^2f'^2}{2f}\right ]K=0\,.
\eeq
A complete discussion of Regge-Wheeler or vector-type gravitational
perturbations of the four-dimensional Schwarzschild geometry can be found in
the original papers by Regge and Wheeler \cite{Regge:1957rw} as well as in
Ref.~\cite{Edelstein:1970sk}, where some typos in the original work are
corrected. For Zerilli or scalar-type gravitational perturbations, the
fundamental reference is Zerilli's work \cite{Zerilli:1970se,Zerilli:1971wd};
typos are corrected in Appendix A of Ref.~\cite{Sago:2002fe}. An elegant,
gauge-invariant decomposition of gravitational perturbations of the
Schwarzschild geometry is described by Moncrief \cite{Moncrief:1974am} (see
also \cite{Gerlach:1979rw,Gerlach:1980tx,Sarbach:2001qq,Martel:2005ir}).
These papers are reviewed by Nollert \cite{Nollert:1999ji} and Nagar and
Rezzolla \cite{Nagar:2005ea}. For an alternative treatment, see
Chandrasekhar's book \cite{MTB}. Chandrasekhar's book and papers
\cite{chandrarelation,chandraspecial} use a different notation, exploring
mathematical aspects of the relations between different gravitational
perturbations (see \ref{app:relpotentials}). Extensions to the SAdS$_4$
geometry can be found in Ref.~\cite{Cardoso:2001bb}, while the general
$d-$dimensional case has been explored in a series of papers by Kodama and
Ishibashi \cite{Kodama:2003jz,Ishibashi:2003ap,Kodama:2003kk}.

The case of Dirac fields seems to have been discussed first by Brill and
Wheeler \cite{Brill:1957fx}, with important extensions of the formalism by
Page \cite{Page:1976df}, Unruh \cite{Unruh:1976fm} and Chandrasekhar
\cite{MTB}. For the treatment of Rarita-Schwinger fields, see \cite{castillo}.

To summarize this Section: in four-dimensional Schwarzschild or SAdS
backgrounds, scalar ($m=0$, $\gamma=0,\,s=0$), electromagnetic ($s=\pm 1$) and
Regge-Wheeler or vector-type gravitational ($s=2$) perturbations, can be
described by the master equation (\ref{waveeq}) with the potential
\be
V_{s}=f\,\left[\frac{l(l+1)}{r^2}+(1-s^2)\left (\frac{2M}{r^3}+\frac{4-s^2}{2L^2}\right )\right] \label{vsimple}\,.
\ee
The potentials for the scalar-type gravitational perturbations and
half-integer spin perturbations have forms different from (\ref{vsimple}), see
for instance Refs. \cite{Kodama:2003jz,Cho:2003qe}.  However, the vector-type
(Regge-Wheeler) and scalar-type (Zerilli) potentials have the remarkable
property of being {\em isospectral}, i.e. they possess the same QNM
spectrum. The origin of this isospectrality, first discovered by Chandrasekhar
\cite{MTB}, is reviewed in \ref{app:relpotentials}.

\subsection{Higher-dimensional gravitational perturbations}

The literature on gravitational perturbations can be quite confusing. Naming
conventions were already unclear in 1970, so much so that Zerilli decided to
list equivalent terminologies referring to odd and even tensor spherical
harmonics (cf. Table II of Ref.~\cite{Zerilli:1971wd}). The situation got even
worse since then. Chandrasekhar's book, which is the most complete reference
in the field, established a different terminology: ``odd'' (Regge-Wheeler)
perturbations were called ``axial'' and described by a master variable
$\Psi^-$, while ``even'' (Zerilli) perturbations were renamed ``polar'' and
described by a master variable $\Psi^{+}$. In recent years, Kodama and
Ishibashi \cite{Kodama:2003jz,Ishibashi:2003ap,Kodama:2003kk} extended the
gauge-invariant perturbation framework to higher-dimensional, non-rotating
BHs. In higher dimensions {\it three} master variables are necessary to
completely describe the perturbations \cite{Kodama:2003jz}. Two of them (the
{\it vector-type gravitational perturbations} and the {\it scalar-type
  gravitational perturbations}) reduce to the Regge-Wheeler and Zerilli master
variables in $d=4$. Kodama and Ishibashi refer to the third type of
perturbations, which have no four-dimensional analogue, as {\it tensor-type
  gravitational perturbations}.  In this review we will usually adopt the
Kodama-Ishibashi terminology.

\subsection{\label{sec:Kerr}Weak fields in the Kerr background: the Teukolsky equation}
In four-dimensional asymptotically flat spacetimes, the most general vacuum BH
solution of Einstein's equations is the Kerr metric. In the standard
Boyer-Linquist coordinates, the metric depends on two parameters: the mass $M$
and spin $J=aM$. The spacetime has a Cauchy horizon at
$r=r_-=M-\sqrt{M^2-a^2}$ and an event horizon at $r=r_+=M+\sqrt{M^2-a^2}$. The
separation of variables for a minimally coupled scalar field in the Kerr
background was first reported by Brill {\it et al.} \cite{Brill:1972xj}.

Teukolsky \cite{Teukolsky:1972my,Teukolsky:1973ap} showed that if one works
directly in terms of curvature invariants, the perturbation equations decouple
and separate for all Petrov type-D spacetimes. He derived a master
perturbation equation governing fields of general spin, including the most
interesting gravitational perturbations (see \cite{MTB,breuerbook} for
reviews). Teukolsky's approach is based on the Newman-Penrose
\cite{Newman:1961qr} formalism. In this formalism one introduces a tetrad of
null vectors ${\bf l\,,n\,,m\,,m^*}$ at each point in spacetime, onto which
all tensors are projected. The Newman-Penrose equations are relations linking
the tetrad vectors, the spin coefficients, the Weyl tensor, the Ricci tensor
and the scalar curvature \cite{Newman:1961qr}. The most relevant perturbation
variables, which both vanish in the background spacetime, are the Weyl scalars
${\bf \Psi}_0$ and ${\bf \Psi}_4$, obtained by contracting the Weyl tensor
$C_{\mu\nu\lambda\sigma}$ \cite{waldbook} on the tetrad legs (roughly
speaking, these quantities describe ingoing and outgoing gravitational
radiation):
\beq 
{\bf \Psi}_0&=&-C_{1313}=-C_{\mu\nu\lambda\sigma}l^\mu m^\nu l^\lambda m^\sigma\,,\\
{\bf \Psi}_4 &=& -C_{2424}=-C_{\mu\nu\lambda\sigma}n^\mu m^{*\nu} n^\lambda m^{*\sigma}\,.
\eeq
Two analogous quantities ${\bf \Phi}_0$ and ${\bf \Phi}_2$ describe
electromagnetic perturbations:
\beq
{\bf \Phi}_0&=&F_{\mu\nu}l^{\mu} m^{\nu}\,,\quad 
{\bf \Phi}_2=F_{\mu\nu}m^{*\,\mu}n^{\nu}\,.
\eeq
By Fourier-transforming a spin-$s$ field $\psi(t\,,r\,,\theta\,,\phi)$ and
expanding it in spin-weighted {\it spheroidal} harmonics as follows:
\be 
\psi(t\,,r\,,\theta\,,\phi) =\frac{1}{2\pi}\int e^{-i\omega t}
\sum_{l=|s|}^{\infty}\sum_{m=-l}^{l}
e^{im\phi}\,{}_sS_{lm}(\theta)R_{lm}(r)d\omega\,, 
\label{psiteuk}
\ee
Teukolsky finds separated ODEs for ${}_sS_{lm}$ and $R_{lm}$
\cite{Teukolsky:1972my,Teukolsky:1973ap}:
\beq && {\biggl[}\frac{\partial}{\partial u} (1-u^2)\frac{\partial}{\partial u}{\biggr]}\,{}_sS_{lm}\nonumber\\
&+&{\biggl[}a^2\omega^2u^2-2a\omega s u +s+\, _sA_{lm}
-\frac{(m+s u)^2}{1-u^2}  {\biggr]}\,{}_sS_{lm}=0\,,\nonumber \label{angular}\\
&&\Delta \partial^2_r
R_{lm}+(s+1)(2r-2M)\partial_rR_{lm}+VR_{lm}=0\,. \label{radial}
\eeq
Here $u\equiv \cos\theta$, $\Delta = (r-r_-)(r-r_+)$ and
\beq V&=&2is\omega r -a^2\omega^2-\, _sA_{lm}+\frac{1}{\Delta} {\biggl[}
(r^2+a^2)^2\omega^2- 4M am\omega r +a^2m^2\nn\\
&&+is\left(am(2r-2M)-2M\omega(r^2-a^2)        \right ) {\biggr]} \,.
\eeq
The solutions to the angular equation (\ref{angular}) are known in the
literature as spin-weighted spheroidal harmonics:
${}_sS_{lm}={}_sS_{lm}(a\omega\,,\theta\,,\phi)$.
For $a\omega =0$ the spin-weighted spheroidal harmonics reduce to
spin-weighted {\it spherical} harmonics ${_s}Y_{lm}(\theta\,,\phi)$
\cite{Goldberg:1967sp}. In this case the angular separation constants
$_sA_{lm}$ are known analytically: ${}_sA_{lm}(a=0)=l(l+1)-s(s+1)$.  The
determination of the angular separation constant in more general cases is a
non-trivial problem (see \cite{Berti:2005gp} and references therein).
\begin{table}[ht]
  \centering \caption{\label{tab:teukolskyquantities} Teukolsky wavefunction
    $\psi$, as in (\ref{psiteuk}), for each value $s$ of the
    spin. The spin-coefficient $\rho \equiv -1/(r-ia\cos\theta)$.  The
    quantities $\chi_0$ and $\chi_1$ refer to the components of the neutrino
    wavefunction along dyad legs.}
\begin{tabular}{|c||cccc|}
\hline
$s$& 0& $(+1/2,-1/2)$&$(+1,-1)$&$(+2,-2)$\\
$\psi$&$\Phi$&$(\chi_0,\rho^{-1}\chi_1)$&$({\bf
\Phi}_0,\rho^{-2}{\bf \Phi}_2)$&$({\bf \Psi}_0,\rho^{-4}{\bf \Psi}_4)$\\
\hline \hline
\end{tabular}
\end{table}

The field's spin weight $s$ is equal to $0\,,\pm 1/2\,,\pm 1\,,\pm 2$ for
scalar, Dirac, electromagnetic and gravitational perturbations,
respectively. The Teukolsky master variable $\psi$ is related to the
perturbation fields by the relations listed in Table
\ref{tab:teukolskyquantities} (see also Appendix B of
Ref.~\cite{Teukolsky:1973ap}). Relations between the Regge-Wheeler-Zerilli and
the Teukolsky variables are explored in Ref.~\cite{MTB}. Reconstructing the
metric from the Teukolsky functions is a highly non-trivial problem which is
still not completely solved (see
e.g.~\cite{Chrzanowski:1975wv,Wald:1978vm,Kegeles:1979an,Lousto:2002em,Whiting:2005hr,Yunes:2005ve}).

\section{Defining quasinormal modes}

\subsection{Quasinormal modes as an eigenvalue problem}

In a spherically symmetric background, the study of BH perturbations due to
linearized fields of spin $s$ can be reduced to the study of the differential
equation (\ref{waveeq}). Henceforth, to simplify the notation, we will usually
drop the $s$-subscript in all quantities.  To determine the free modes of
oscillation of a BH, which correspond to ``natural'' solutions of this
unforced ODE, we must impose physically appropriate boundary conditions at the
horizon ($r_*\to -\infty$) and at spatial infinity ($r_*\to \infty$). These
boundary conditions are discussed below.

\subsubsection*{Boundary conditions at the horizon}
%
For most spacetimes of interest the potential $V \rightarrow 0$ as $r_* \to -
\infty$, and in this limit solutions to the wave equation (\ref{waveeq})
behave as
$\Psi \sim e^{-i\omega (t\pm r_*)}$. Classically nothing should leave the
horizon: only ingoing modes (corresponding to a plus sign) should be present,
and therefore
\be \Psi \sim e^{-i\omega (t+r_*)}\,,\quad r_* \rightarrow-
\infty\,~(r\rightarrow r_+)\label{bchorizon} \,.\ee
This boundary condition at the horizon can also be seen to follow from
regularity requirements. For non-extremal spacetimes, the tortoise coordinate
tends to
\be r_*=\int f^{-1}\,dr \sim
\left[f'(r_+)\right]^{-1}\log{(r-r_+)}\,,\quad r\sim r_+ \,,\ee
with $f'(r_+)>0$.
Near the horizon, outgoing modes behave as
\be e^{-i\omega (t-r_*)}=e^{-i\omega v}e^{2i\omega r_*}\sim
e^{-i\omega v}(r-r_+)^{2i\omega/f'(r_+)}\,,\label{outedd} \ee
where $v=t+r_*$.  Now Eq.~(\ref{outedd}) shows that unless $2i\omega/f'(r_+)$
is a positive integer the outgoing modes cannot be smooth, i.e. of class
${\cal C}^{\infty}$, and they must be discarded.  An elegant discussion of the
correct boundary conditions at the horizon of rotating BHs can be found in
Appendix B of Ref.~\cite{Bardeen:1972fi}.
\subsubsection*{Boundary conditions at spatial infinity: asymptotically flat spacetimes}

For asymptotically flat spacetimes, the metric at spatial infinity tends to
the Minkowski metric. From Eq.~(\ref{vsimple}) with $L\to \infty$ we see that
the potential is zero at infinity. By requiring
\be 
\Psi \sim e^{-i\omega (t-r_*)},\,\quad r\rightarrow \infty 
\label{bcinfinityflat}\,, 
\ee
we discard unphysical waves ``entering the spacetime from infinity''. 

The main difference between QNM problems and other prototypical physical
problems involving small oscillations, such as the vibrating string, is that
the system is now dissipative: waves can escape either to infinity or into the
BH. For this reason an expansion in normal modes is not possible
\cite{Kokkotas:1999bd,Nollert:1999ji,Leaver:1986gd,nollertschmidt}. There is a
discrete infinity of QNMs, defined as eigenfunctions satisfying the above
boundary conditions. The corresponding eigenfrequencies $\omega_{\rm QNM}$
have both a real and an imaginary part, the latter giving the (inverse)
damping time of the mode. One usually sorts the QNM frequencies by the
magnitude of their imaginary part, and labels them by an integer $n$ called
the overtone number. The fundamental mode $n=0$ is the least damped mode, and
being very long-lived it usually dominates the ringdown waveform.

A seemingly pathological behavior occurs when one imposes the boundary
conditions (\ref{bchorizon}) and (\ref{bcinfinityflat}). When the mode
amplitude decays in time, the characteristic frequency $\omega_{\rm QNM}$ must
have a {\it negative} imaginary component. Then, the amplitude near infinity
($r_*\rightarrow +\infty$) must blow up. So it is in general impossible to
represent regular initial data on the spacetime as a sum of QNMs. QNMs should
be thought of as {\it quasistationary} states which cannot have existed for
all times: they decay exponentially with time, and are excited only at a
particular instant in time (see \cite{Andersson:1996cm} for an alternative
viewpoint on ``dynamic'' QNM excitation). In more formal terms, QNMs do not
form a complete set of wavefunctions \cite{Nollert:1999ji}.

\subsubsection*{Boundary conditions at spatial infinity: asymptotically
  anti-de Sitter spacetimes}

When the cosmological constant does not vanish, by inspection of
Eq.~(\ref{waveeq}) we see that
\be
\Psi_{s=0} \sim Ar^{-2}+Br\,,\quad \Psi_{s=1,2} \sim A/r+B\,,\quad r\rightarrow \infty\,.
\ee
Regular scalar field perturbations should have $B=0$, corresponding to
Dirichlet boundary conditions at infinity.  The case for electromagnetic and
gravitational perturbations is less clear: there is no {\it a priori}
compelling reason for a specific boundary condition. A popular choice
implements Dirichlet boundary conditions for the Regge-Wheeler and Zerilli
variables \cite{Cardoso:2001bb}, but other boundary conditions were
investigated, e.g., in Ref.~\cite{Moss:2001ga}. A discussion of preferred
boundary conditions in the context of the AdS/CFT correspondence can be found
in Refs.~\cite{Michalogiorgakis:2006jc,Bakas:2008gz,Bakas:2008zg} (see also
Section \ref{dualqnms}).
\subsection{\label{inversioncontour}Quasinormal modes as poles in the Green's function}

\begin{figure*}[ht]
\begin{center}
\epsfig{file=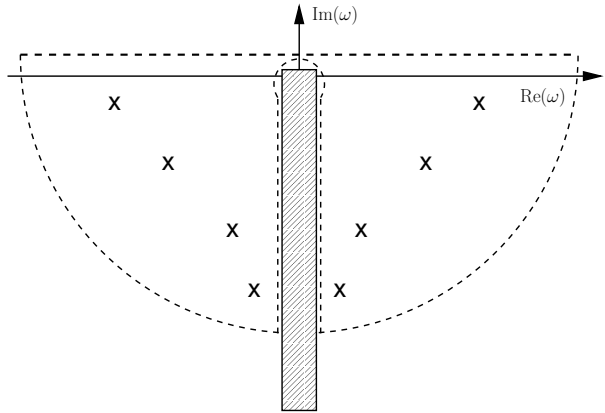,width=8cm,angle=0}
\caption{Integration contour for Eq.~(\ref{originalu}). The hatched area is
  the branch cut and crosses mark zeros of the Wronskian $W$ (the QNM
  frequencies).} \label{fig:contour2}
\end{center}
\end{figure*}

The QNM contribution to the BH response to a generic perturbation can be
identified formally by considering the Green's function solution to an
inhomogeneous wave equation
\cite{Leaver:1986gd,nollertschmidt,Andersson:1995zk,Andersson:1996cm,Berti:2006wq}.
Consider the Laplace transform of the field, ${\cal L}\Psi(t,r) \equiv
\Psi(\omega\,,r)= \int_{t_0}^{\infty} \Psi(t,r)e^{i\omega t} dt$, which is
well defined if $\omega_I\geq c$ (the usual Laplace variable is $s=-i\omega$;
we use $\omega$ for notational consistency with previous works).
The problem of computing the gravitational waveform produced when a BH is
perturbed by some material source (such as a particle of mass $m\ll M$ falling
into the BH) can be reduced to a wave equation of the form (\ref{waveeq}) with
a source term:
\be
\frac{d^2\Psi}{dr_*^2}+\left (\omega^2-V\right)\Psi=I(\omega\,,r)\,.
\label{waveeqsources}
\ee
We can solve this equation by the standard Green's function technique
\cite{morsefeshbach} (see
Refs.~\cite{Leaver:1986gd,nollertschmidt,Andersson:1995zk,Andersson:1996cm,Berti:2006wq}
for applications in this context), focusing for definiteness on asymptotically
flat spacetimes.  Take two independent solutions of the homogeneous equation:
one has the correct behavior at the horizon,
\beq & & \lim_{r \to r_+} \Psi_{r_+}\sim e^{-i\omega r_*}\,,
\label{asrplus}\\
& & \lim_{r \to \infty }\Psi_{r_+}\sim A_{\rm
in}(\omega)e^{-i\omega r_*}+A_{\rm out}(\omega)e^{i\omega r_*}\,,
\eeq
and the second independent solution $\Psi_{\infty_+} \sim e^{i\omega r_*}$ for
large $r$. The Wronskian of these two wavefunctions is $W=2i\omega A_{\rm
  in}$, and we can express the general solution as \cite{Leaver:1986gd}
\be
\Psi(\omega\,,r)=\Psi_{\infty_+}\int_{-\infty}^{r_*}\frac{I(\omega\,,r)\Psi_{r_+}}{2i\omega
A_{\rm in}}dr_*'
+\Psi_{r_+}\int_{r_*}^{\infty}\frac{I(\omega\,,r)\Psi_{\infty_+}}{2i\omega
A_{\rm in}}dr_*' \,. \ee
The time-domain response is obtained by inversion of the Laplace transform:
\be \Psi(t,r)=\frac{1}{2\pi}\int_{-\infty +ic}^{\infty+ic}
\Psi(\omega\,, r)e^{-i\omega t} d\omega\,.\label{originalu} \ee
The frequency integral can be performed by the integration contour shown in
Fig.~\ref{fig:contour2}.
There are in general three different contributions to the integral. The
integral along the large quarter-circles is the flat-space analogue of the
prompt response, i.e. waves propagating directly from the source to the
observer at the speed of light. Depending on the asymptotic structure of the
potential, there is usually a branch point at $\omega=0$ \cite{Ching:1995tj}.
To prevent it from lying inside the integration contour, we place a branch cut
along the negative imaginary-$\omega$ axis and split the half circle at
$|\omega|\rightarrow \infty$ into two quarter circles
\cite{nollertthesis,Jensen:1985in,Ching:1994bd}. The branch-cut contribution
gives rise to late-time tails \cite{price,Leaver:1986gd,Ching:1994bd}:
physically, these tails are due to backscattering off the background curvature
\cite{price}, and therefore they depend on the asymptotics of the
spacetime. Tails are absent for certain backgrounds, such as the SAdS
\cite{Horowitz:1999jd} or the Nariai spacetime \cite{Casals:2009zh}.
The third contribution comes from a sum-over-residues at the poles in the
complex frequency plane, which are the zeros of $A_{\rm in}$.  These poles
correspond to perturbations satisfying {\it both} in-going wave conditions at
the horizon and out-going wave conditions at infinity, so (by the very
definition of QNMs) they represent the QNM contribution to the
response. Although we used the asymptotic behavior of the solutions of the
wave equation, this discussion can be trivially generalized to any spacetime.

Far from the source, the QNM contribution in asymptotically flat spacetimes
can be written as \cite{Andersson:1995zk,Berti:2006wq}
\be
\Psi(t,r) =
-{\rm Re} \left[ \sum_{n}C_n e^{-i\omega_n(t-r_*)} \right]\,,
\label{init2} 
\ee
where the sum is over all poles in the complex plane.  The $C_n$'s are called
{\it quasinormal excitation coefficients} and they quantify the QNM content of
the waveform. They are related to initial-data independent quantities, called
the {\it quasinormal excitation factors} (QNEFs) and denoted by $B_n$, as
follows:
\be
\label{excfactors} 
C_n=B_n \int_{-\infty}^{\infty} \frac{
I(\omega\,,r)\Psi_{r_+}}{A_{\rm out}}dr_*' \,,\quad
B_n= 
\left. \frac{A_{\rm out}}{2\omega} \left (\frac{dA_{\rm
in}}{d\omega}\right)^{-1} \right|_{\omega=\omega_{n}} \,.
\ee
In general the QNM frequencies $\omega_n$, the $B_n$'s and the $C_n$'s depend
on $l$, $m$ and on the spin of the perturbing field $s$, but to simplify the
notation we will omit this dependence whenever there is no risk of confusion.

The QNEFs play an important role in BH perturbation theory: they depend only
on the background geometry, and (when supplemented by specific initial data)
they allow the determination of the QNM content of a signal, i.e. of the
$C_n$'s. This has been known for over two decades, but relatively little
effort has gone into understanding how these modes are excited by physically
relevant perturbations. The QNEFs have long been known for scalar,
electromagnetic and gravitational perturbations of Schwarzschild BHs
\cite{Leaver:1986gd,Sun:1988tz,Andersson:1995zk,Andersson:1996cm}, and they
have recently been computed for general perturbations of Kerr BHs
\cite{Berti:2006wq}. In \cite{Berti:2006wq} it was also shown that for large
overtone numbers ($n \rightarrow \infty$) $B_n \propto n^{-1}$ for all
perturbing fields in a large class of non-rotating spacetimes.  The excitation
factors $C_n$ have mainly been computed for the simple case where the initial
data are Gaussian pulses of radiation
\cite{Andersson:1995zk,Berti:2006wq}. The only work we are aware of studying
the excitation factors $C_n$ for a point particle falling into a Schwarzschild
BH is Leaver's classic paper \cite{Leaver:1986gd}.

Besides the theoretical interest of quantifying QNM excitation by generic
initial data in the framework of perturbation theory, excitation factors and
excitation coefficients have useful applications in gravitational wave data
analysis.
First of all, a formal QNM expansion of the BH response can simplify the
calculation of the self-force acting on small bodies orbiting around BHs. It
was shown recently, using as a model problem the Nariai spacetime, that a QNM
expansion of the Green's function can be used for a matched expansion of the
``quasi-local'' and ``distant-past'' contributions to the self force
\cite{Casals:2009zh}.
The problem of quantifying QNM excitation is of paramount importance to search
for inspiralling compact binaries in gravitational wave detector data. All
attempts to match an effective-one-body description of inspiralling binaries
to numerical relativity simulations found that the inclusion of several
overtones in the ringdown waveform is a crucial ingredient to improve
agreement with the numerics \cite{Buonanno:2009qa,Damour:2009kr}.  So far the
matching of the inspiral and ringdown waveforms has been performed by {\em ad
  hoc} procedures. For example, the amplitudes and phases of the $C_n$'s have
been fixed by requiring continuity of the waveform on a grid of points, or
``comb'' \cite{Damour:2007xr}. These matching procedures have their own
phenomenological interest, but a self-consistent estimation of the excitation
coefficients within perturbation theory is needed to improve our physical
understanding of the inspiral-ringdown transition.

Finally, we point out that recent investigations have addressed the issue of
mode excitation in the gauge/gravity duality
\cite{Amado:2008ji,Amado:2007yr}. There it was found, for example, that the
residue of the diffusion and shear mode decays at small wavelength, so these
modes effectively cease to exist.

\section{\label{methodsqnm}Computing quasinormal modes}

To determine the QNMs and compute their frequencies we must solve the
eigenvalue problem represented by the wave equation (\ref{waveeq}), with
boundary conditions specified by Eq.~(\ref{bchorizon}) at the horizon and
Eq.~(\ref{bcinfinityflat}) at infinity. There is no universal prescription to
compute QNMs.  In this section we discuss various methods to obtain such a
solution, pointing out that different methods are better suited to different
spacetimes.

For a start we consider the exceptional cases where an exact, analytical
solution to the wave equation can be found.  In some spacetimes the potential
appearing in the wave equation can be shown to reduce to the P\"oschl-Teller
potential \cite{poschlteller}, for which an exact QNM calculation is possible
\cite{Ferrari:1984zz}. These spacetimes and their QNM spectra are reviewed in
Section~\ref{sec:exactsol}. In the general case, QNM calculations require
approximations or numerical methods
\footnote{Fiziev \cite{Fiziev:2005ki,Fiziev:2009kh} actually showed that the
  Regge-Wheeler and Teukolsky equations can be solved analytically in terms of
  confluent Heun's functions. These solutions allow a high-accuracy
  determination of the Schwarzschild quasinormal frequencies. The calculation
  of Heun's functions is quite contrived, but it already proved useful to
  address interesting open problems in perturbation theory
  \cite{Borissov:2009bj,Fiziev:2009ud}.}.
Some of these (including WKB approximations, monodromy methods, series
solutions in asymptotically AdS backgrounds and Leaver's continued fraction
method) are reviewed in Sections~\ref{sec:wkb}-\ref{sec:Leaver}.

\subsection{\label{sec:exactsol}Exact solutions}

In general exact solutions to the wave equation are hard to find, and they
must be computed numerically. There are a few noteworthy exceptions, some of
which we summarize here.

We begin our review of exact solutions by sketching the analytical derivation
of the QNMs of the P\"oschl-Teller potential. Many of the difficulties in
computing QNMs in BH spacetimes arise from the slow decay of the potential as
$r\to \infty$, which is due (mathematically) to the presence of a branch cut
and gives rise (physically) to backscattering of gravitational waves off the
gravitational potential and to late-time tails.  Ferrari and Mashhoon realized
that these difficulties can be removed and exact solutions can be found if one
considers instead a potential that decays exponentially as $r\to \infty$,
while recovering the other essential features of the Schwarzschild
potential. Such a potential is the P\"oschl-Teller potential
\cite{Ferrari:1984zz}. After reviewing QNM solutions for the P\"oschl-Teller
potential we briefly review the modes of pure AdS and dS spacetimes. Then we
show that perturbations of the near-extreme Schwarzschild de-Sitter (SdS) and
of the Nariai spacetime reduce to a wave equation with a P\"oschl-Teller
potential, so they can be solved analytically.  We also discuss two
asymptotically AdS BH spacetimes which have improved our understanding of the
role of QNMs in the AdS/CFT correspondence: the $(2+1)$-dimensional BTZ BH
\cite{Banados:1992wn} and the $d-$dimensional topological (massless) BH
\cite{Lemos:1994fn,Lemos:1994xp,Lemos:1995cm,Mann:1996gj,Vanzo:1997gw,Birmingham:1998nr}.
The BTZ BH is the first BH spacetime for which an exact, analytic expression
for the QNMs has been derived \cite{Cardoso:2001hn}, and it offers interesting
insights into the validity of the AdS/CFT correspondence
\cite{Birmingham:2001pj,Sachs:2008gt,Sachs:2008yi}.

\subsubsection*{\label{sec:PT}The P\"oschl-Teller potential}

In this section we analytically compute the QNMs of the P\"oschl-Teller
potential \cite{poschlteller,Ferrari:1984zz}, which will serve as a prototype
for several BH spacetimes to be discussed below. Consider the equation
\be \frac{\partial^{2} \Psi}{\partial r_*^2} +
\left\lbrack\omega^2-\frac{V_0}{\cosh^2{\alpha(
r_*-\bar{r}_*)}}\right\rbrack \Psi=0 \,. \label{wavePoschl} \ee
The quantity $\bar{r}_*$ is the point $r_*$ at which the potential attains a
maximum, i.e. $dV/dr_*(\bar{r}_*)=0$, and $V_0$ is the value of the
P\"oschl-Teller potential at that point: $V_0=V(\bar{r}_*)$. The quantity
$\alpha$ is related to the second derivative of the potential at
$r_*=\bar{r}_*$, $\alpha^2\equiv -(2V_0)^{-1} d^2V/dr_*^2(\bar{r}_*)$.
The solutions of Eq.~(\ref{wavePoschl}) that satisfy both boundary conditions
(\ref{bchorizon}) and (\ref{bcinfinityflat}) are the QNMs
\cite{Ferrari:1984zz}. To find these solutions we define a new independent
variable
$\xi=\left[1+e^{-2\alpha(r_*-\bar{r}_*)}\right]^{-1}$ and rewrite
Eq.~(\ref{wavePoschl}) as
\be \xi^2(1-\xi)^2\frac{\partial^{2} \Psi}{\partial
\xi^2}- \xi(1-\xi) (2\xi-1)\frac{\partial \Psi}{\partial
\xi} + \left\lbrack \frac{\omega^2}{4\alpha^2 }-\frac{V_0}{\alpha^2 }\xi (1-\xi)\right\rbrack \Psi=0
\,. \ee
Near spatial infinity $1-\xi \sim e^{-2\alpha(r_*-\bar{r}_*)}$, and near the
horizon $\xi \sim e^{2\alpha(r_*-\bar{r}_*)}$.
If we define $a=\left[\alpha+\sqrt{\alpha^2-4V_0}-2i\omega\right]/(2\alpha)
\,, b=\left[\alpha-\sqrt{\alpha^2-4V_0}-2i\omega\right]/(2\alpha)\,, c=1-
i\,\omega/\alpha $ and set $\Psi=\left (\xi(1-\xi)\right
)^{-i\omega/(2\alpha)}y$, we get a standard hypergeometric equation for $y$
\cite{Abramowitz:1970as}:
\be \xi(1\!-\!\xi)\partial_\xi^2 y+[c-(a+b+1)\xi]\partial_\xi y-ab
y=0\,, \ee
and therefore
\begin{eqnarray}
\Psi &=& A\, \xi^{i\,\omega/(2\alpha)}(1-\xi)^{-i\omega/(2\alpha)}
F(a-c+1,b-c+1,2-c,\xi)\nonumber \\
&+&B\,\left (\xi(1-\xi)\right )^{-i\omega/(2\alpha)}F(a,b,c,\xi) \,.
 \label{hypergeometric solution}
\end{eqnarray}
Recalling that $F(a_1,a_2,a_3,0)=1$ and that $\xi^{i\,\omega/(2\alpha)} \sim
e^{i\omega r_*}$ near the horizon, we see that the first term represents,
according to our conventions, an outgoing wave at the horizon, while the
second term represents an ingoing wave. QNM boundary conditions require
$A=0$. To investigate the behavior at infinity, one uses the $z \rightarrow
1-z$ transformation law for the hypergeometric function
\cite{Abramowitz:1970as}:
\begin{eqnarray}
& \hspace{-2cm} F(a,b,c,z)= (1\!-\!z)^{c-a-b}
\frac{\Gamma(c)\Gamma(a+b-c)}{\Gamma(a)\Gamma(b)}
 \,F(c\!-\!a,c\!-\!b,c\!-\!a\!-\!b\!+\!1,1\!-\!z) & \nonumber \\
&  \hspace{-0.2cm}+
\frac{\Gamma(c)\Gamma(c-a-b)}{\Gamma(c-a)\Gamma(c-b)}
 \,F(a,b,-c\!+\!a\!+\!b\!+\!1,1\!-\!z)\,.
 \label{transformation law}
\end{eqnarray}
The boundary condition at infinity implies either $1/\Gamma(a)=0$ or
$1/\Gamma(b)=0$, which are satisfied whenever
\be 
\omega=\pm \sqrt{V_0-\alpha^2/4}-i\alpha(2n+1)/2\,,\quad n=0,\,1,\,2,\,....
\ee
where $n$ is the overtone index.

A popular approximation scheme to compute BH QNMs consists in replacing the
true potential in a given spacetime by the P\"oschl-Teller potential.  This
approximation works well for the low-lying modes of the Schwarzschild
geometry. It predicts $M\omega=0.1148-0.1148i,\,0.3785-0.0905i$ for the
fundamental $l=s=0,2$ perturbations, respectively \cite{Ferrari:1984zz}. This
can be compared to the numerical result
\cite{Leaver:1985ax,Berti:2005ys,rdweb} $M
\omega=0.1105-0.1049i,\,0.3737-0.0890i$.  In the eikonal limit ($l\rightarrow
\infty$) the P\"oschl-Teller approximation yields the correct solution: it
predicts the behavior $3\sqrt{3}M\omega=\pm (l+1/2)-i(n+1/2)$, in agreement
with WKB-based calculations \cite{pressringdown,Iyer:1986np,Iyer:1986nq} (see
also \cite{barretozworski}).

The P\"oschl-Teller approximation provides a solution which is more and more
accurate for near-extremal SdS BHs, since the event horizon and the
cosmological horizon coalesce in the extremal limit
\cite{Moss:2001ga,Cardoso:2003sw,LopezOrtega:2007sr}.
\subsubsection*{Normal modes of the anti-de Sitter spacetime}

A physically interesting analytical solution concerns the QNMs of pure AdS
spacetime, which can be obtained by setting $r_0=0$ in the metric
(\ref{lineelementads}). In this case, QNMs are really {\em normal} modes of
the spacetime, and have been computed for scalar field perturbations by
Burgess and L\"{u}tken \cite{Burgess:1984ti}. They satisfy
\be L\,\omega=2n+d+l-1\,,\quad n=0,\,1,\,2,...\quad
s=0\,,\ee
where $l$ is related to the eigenvalue of spherical harmonics in $d$
dimensions by $A_{lm}=l(l+d-3)$ \cite{Berti:2005gp}. Normal modes of
electromagnetic perturbations in $d=4$ were shown to be the same as normal
modes for gravitational perturbations (see Appendix in
Ref.~\cite{Cardoso:2003cj}); for general $d$, they can be computed from the
potentials for wave propagation derived by Kodama and Ishibashi
\cite{Kodama:2003kk}. Ref.~\cite{Natario:2004jd} studies the normal modes of
gravitational perturbations.  Tensor gravitational perturbations obey the same
equation as $s=0$ fields, and therefore their normal modes are
\be L\,\omega=2n+d+l-1\,,\quad n=0,\,1,\,2,...\quad
s=2\,\,{\rm tensor-type}\,.\ee
Vector-type gravitational perturbations have normal modes with the following
characteristic frequencies
\be L\,\omega=2n+d+l-2\,,\quad n=0,\,1,\,2,...\quad s=2 \,\,
{\rm vector-type}\,.\label{normaladsd4vector}\ee
Finally, scalar-type gravitational perturbations have a somewhat surprising
behavior. For $d=4$ they are given by Eq.~(\ref{normaladsd4vector}). For $d=5$
they have a continuous spectrum, and for $d>5$ one finds \cite{Natario:2004jd}
\be L\,\omega=2n+d+l-3\,,\quad n=0,\,1,\,2,...\quad
s=2\,\,{\rm scalar-type}\,.\ee
%

\subsubsection*{Quasinormal modes of the de Sitter spacetime}

The dS spacetime is an extensively studied solution of the Einstein field
equations, most of the early investigations being motivated by cosmological
considerations. It satisfies Eq.~(\ref{lineelementads}) with $r_0=0$ and $f(r)
= 1 - r^2/L^2$.  Nat\'ario and Schiappa found that no QNM solutions are
allowed in even-dimensional dS space \cite{Natario:2004jd}. For scalar fields
and tensor-type gravitational perturbations in odd-dimensional dS backgrounds
the QNM frequencies are purely imaginary, and given by
\be L\,\omega=-i\left (2n+d+l\right )\,,\quad
n=0,\,1,\,2,...{\rm tensor-type}\,.\ee
For the other types of gravitational perturbations Ref.~\cite{Natario:2004jd}
finds
\beq 
L\,\omega&=&-i\left (2n+d+l+1\right )\,,\quad
n=0,\,1,\,2,...{\rm vector-type}\,,\\
L\,\omega&=&-i\left (2n+d+l+2\right )\,,\quad
n=0,\,1,\,2,...{\rm scalar-type}\,.
\eeq
Notice the striking similarity with the pure AdS results when one replaces
$L\rightarrow iL$. Fields of other spins were considered in
Refs.~\cite{LopezOrtega:2006my,LopezOrtega:2007sr}.

\subsubsection*{Nearly-extreme SdS and the Nariai spacetime}
%
The metric for $d$-dimensional Schwarzschild de-Sitter (SdS$_d$) BHs can be
obtained from Eq.~(\ref{lineelementads}) by the replacement $L\to iL$, i.e.,
$f(r) = 1 - r^2/L^2 - r_0^{d-3}/r^{d-3}$.  The corresponding spacetime has two
horizons: an event horizon at $r=r_+$ and a cosmological horizon at
$r=r_c$. It was observed in Ref.~\cite{Cardoso:2004uz} that for $r_c/r_+-1\ll
1$ perturbations of this spacetime satisfy a wave equation with a
P\"oschl-Teller potential. In particular, setting $k_b \equiv
(r_c-r_+)/2r_c^2$, for near-extreme SdS$_d$ one finds $M/r_+^{d-3}\sim
1/(d-1)\,,\,r_+^2/L^2\sim (d-3)/(d-1)$ \cite{Cardoso:2004uz,Molina:2003ff}.
QNM frequencies for scalar field and tensor-like gravitational perturbations
are then given by \cite{Cardoso:2004uz,Molina:2003ff}
\be 
\f{\omega}{k}=\sqrt{l(l+d-3)-\f{1}{4}}\,-i\left
(n+\f{1}{2}\right )\,,
\quad {\rm tensor-type}
\,,\ee
where $k=(d-3)(r_c-r_+)/(2r_+^2)$ is the surface gravity of the
BH. For gravitational perturbations the result is
\beq
\f{\omega}{k}&=&\sqrt{(l-1)(l+d-2)-\f{1}{4}}\,-i\left
(n+\f{1}{2}\right )\,,
\quad {\rm vector-type}\,,\\
\nn
\f{\omega}{k}&=&\sqrt{(l-1)(l+d-2)-d+\f{15}{4}}\,-i\left
(n+\f{1}{2}\right )\,,
\quad {\rm scalar-type}\,. 
\eeq

Through an appropriate limiting procedure \cite{Cardoso:2004uz,Vanzo:2004fy},
the nearly-extreme SdS$_d$ geometry can yield a spacetime with a different
topology, the Nariai spacetime
\cite{Cardoso:2004uz,Nariai1,Nariai2,Caldarelli:2000wc}, of the form
\be ds^2=-\left (-r^2/L^2+1\right )dt^2 +\left
(-r^2/L^2+1\right )^{-1}
dr^2+r^2d\Omega_{d-2}^2\,.
\ee
This manifold has topology $dS_2\,\times \,S_{d-2}$ and two horizons (with the
same surface gravity). The QNM frequencies are the same as for nearly-extreme
SdS$_d$, if $k$ is replaced by the surface gravity of each horizon
\cite{Vanzo:2004fy}.

\subsubsection*{The BTZ black hole}

Ichinose and Satoh \cite{Ichinose:1994rg} were the first to realize that the
wave equation in the ($2+1$)-dimensional, asymptotically AdS BTZ BH
\cite{Banados:1992wn} can be solved in terms of hypergeometric functions. An
analytical solution for the QNMs of this BH was first found in
Ref.~\cite{Cardoso:2001hn}.  The non-rotating BTZ BH metric is given by
\cite{Banados:1992wn}
\be ds^{2}= \left (r^2/L^2-M\right )dt^{2}-
 \left (r^2/L^2-M \right )^{-1}dr^{2} - r^2d{\phi}^2
\,, \label{metricbtz} \ee
where $M$ is the BH mass and $r_+ =M^{1/2}L$ is the horizon radius. In $2+1$
dimensions gravity is ``trivial'': the full curvature tensor is completely
determined by the local matter distribution and the cosmological constant. In
particular, in vacuum the curvature tensor $R_{\mu\nu\lambda\rho}=\Lambda\left
(g_{\mu \lambda}g_{\nu\rho}-g_{\mu\rho}g_{\nu\lambda}\right )$ and
$R=6\Lambda$. Curvature effects produced by matter do not propagate through
the spacetime. There are no dynamical degrees of freedom, and no gravitational
waves \cite{Brown:1988am,Cardoso:2002pa}. Therefore, we will focus on scalar
fields and assume an angular dependence of the form $e^{im\phi}$.  Scalar QNM
frequencies are given by
\be \omega L=\pm m-2i(n+1)r_+/L\,,
\ee
with $m$ the azimuthal number, and $n$ the overtone number
\cite{Cardoso:2001hn}. This result has been generalized by Birmingham, Sachs
and Solodukhin \cite{Birmingham:2001pj} to the rotating BTZ BH. For general
massive scalar perturbations with mass parameter $\mu$ they find
\be \omega L=\pm m-i \lbrack
2n+(1+\sqrt{\mu^2+1})\rbrack\,(r_+-r_-)/L\,.
\label{BTZQNMs}
\ee
The BTZ background has provided a first quantitative test of the AdS/CFT
correspondence: the QNM frequencies (\ref{BTZQNMs}) match the poles of the
retarded correlation function of the corresponding perturbations in the dual
CFT \cite{Birmingham:2001pj}.  Recently the QNMs of BTZ BHs were shown to be
Breit-Wigner-type resonances generated by surface waves supported by the
boundary at infinity, which acts as a photon sphere \cite{Decanini:2009dn}.
This interpretation is highly reminiscent of work in asymptotically flat
spacetimes, interpreting QNMs as null particles slowly leaking out of circular
null geodesics (see
Refs.~\cite{goebel,mashhoon,Ferrari:1984zz,stewart,Decanini:2002ha,
  Cardoso:2008bp,Decanini:2009mu} and Section~\ref{sec:wkb} below).

An alternative to Einstein's gravity in three dimensions is the so-called
``topologically massive gravity'', obtained by adding a Chern-Simons term to
the action \cite{Deser:1981wh,Deser:1982vy}. Topologically massive gravity
allows for dynamics, i.e. gravitational waves.  The QNMs of BTZ BHs in this
theory have recently been computed, providing yet another confirmation of the
AdS/CFT correspondence \cite{Sachs:2008gt,Sachs:2008yi}.

\subsubsection*{Massless topological black holes}

It is also possible to obtain exact solutions in a restricted set of
higher-dimensional BH spacetimes. These asymptotically AdS solutions are known
as topological BHs. The horizon is an Einstein space of positive, zero, or
negative curvature
\cite{Lemos:1994fn,Lemos:1994xp,Lemos:1995cm,Mann:1996gj,Vanzo:1997gw,Birmingham:1998nr}.
In the negative-curvature case there is a massless BH playing a role quite
similar to the BTZ BH in three dimensions. Consider the exterior region of the
massless topological BH \cite{Aros:2002te}
\[
ds^{2}=-\left(r^{2}/L^{2}-1\right)
dt^{2}+\left(r^{2}/L^{2}-1\right)^{-1}dr^{2}
+r^{2}d\sigma ^{2}\;.
\]
This is a manifold of negative constant curvature with an event horizon at
$r=L$. Here $d\sigma^{2}$ stands for the line element of a
$(d-2)$-dimensional surface $\Sigma _{d-2}$ of negative constant curvature.

The wave equation for a massive scalar field with non-minimal coupling can be
solved by the ansatz $\Phi=\Psi(r)e^{-i\omega t}Y$, where $Y$ is a harmonic
function of finite norm with eigenvalue $-Q=-\left(d-3\right)^{2}/4-\xi
^{2}$. The parameter $\xi $ is generically restricted, assuming only discrete
values if $\Sigma _{d-2}$ is a closed surface. If the effective mass
$m_{\textrm{{\tiny{eff}}}}^{2}=\mu^{2}-\gamma d(d-2)/4L^{2}$ (with $\mu$ the
mass of the field and $\gamma$ the conformal coupling factor) satisfies the
bound $m_{\textrm{\tiny{eff}}}^{2}L^{2}\geq -[(d-1)/2]^{2}$, one set of QNM
frequencies is given by \cite{Aros:2002te}
\be \omega \,L=\pm \xi -i\left(2n+1+\sqrt{(d-1)^{2}/4+m_{\textrm{{\tiny{eff}}}}^{2}L^{2}}\right) \,,\quad n=0,1,2...\;. \label{omegamas} \ee

If the mass and the coupling constant $\gamma $ satisfy the relations
$\sqrt{(d-1)^{2}/4+m_{\textrm{{\tiny{eff}}}}^{2}L^{2}}= \f{d-1}{2}-
\frac{\gamma}{2}\frac{d-2}{d-1-\gamma (d-2)}$ and
$ -\left( \frac{d-1}{2}\right) ^{2}<m_{\textrm{{\tiny{eff}}}}^{2}L^{2}<1-\left( \frac{d-1}{2}\right) ^{2}$,
there is another set of modes for which the QNM frequencies are given by
\cite{Aros:2002te}
\be \omega \,L=\pm \xi -i\left( 2n+1-\sqrt{(d-1)^{2}/4+m_{ \textrm{{\tiny{eff}}}}^{2}L^{2}}\right) \;. \label{omegamenos} \ee
This computation represents the first exact analytic determination of QNMs in
four and higher dimensions. The generalization to other fields (and in
particular to gravitational perturbations) can be found in
Ref.~\cite{Birmingham:2006zx}.

\subsection{\label{sec:wkb}The WKB approximation}

Normal modes of vibration of an object usually have a simple interpretation in
terms of waves traveling across or around the object. For example, the Earth's
free modes of oscillation were highly excited and measured for the first time
in the 1960 earthquake in Chile \cite{seismicbook}.  These (roughly) one-hour
long periodic oscillations correspond to waves traveling around the globe, and
carry information about the Earth's interior.  Just like the Earth's free
modes of oscillation, BH QNMs can be thought of as waves traveling around the
BH \cite{pressringdown,goebel,Ferrari:1984zz,mashhoon,stewart,Decanini:2002ha,
  Cardoso:2008bp,Decanini:2009mu}.  More precisely, QNMs can be interpreted as waves
trapped at the unstable circular null geodesic (also known as the light-ring)
and slowly leaking out. The instability timescale of the geodesic is the decay
timescale of the QNM, and the oscillation frequency $\omega\sim c/r_{\rm LR}$,
with $c$ the speed of light and $r_{\rm LR}$ the light-ring radius
\cite{Cardoso:2008bp}.

This intuitive picture, first proposed by Goebel \cite{goebel}, is related to
a more rigorous WKB approximation developed by Mashhoon \cite{Mashhoon:1985}
and by Schutz and Will \cite{Schutz:1985km} (see also
\cite{pressringdown,Ferrari:1984zz,mashhoon,stewart,Decanini:2002ha,
Cardoso:2008bp,Decanini:2009mu}).
Their derivation and results closely parallel the
Bohr-Sommerfeld quantization rule from quantum mechanics. The procedure
involves relating two WKB solutions across a ``matching region'' whose limits
are the classical turning points, where $\omega^2=V(r)$. The technique works
best when the classical turning points are close, i.e. when $\omega^2 \sim
V_{\rm max}$, where $V_{\rm max}$ is the peak of the potential.  Under these
assumptions we can expand in a Taylor series around the extremum of the
potential $\bar{r}_*$:
\be Q\equiv \omega^2-V \sim
Q_0+Q_0''(r_*-\bar{r}_*)^2/2\,,\quad Q_0''\equiv
d^2Q/dr_*^2\,.\label{parabolawkb}\ee
In this region, the wave equation $\frac{d^2\Psi}{dr_*^2}+Q\Psi=0$
can be approximated by
\be \frac{d^2\Psi}{dr_*^2}+\left
[Q_0+\frac{1}{2}Q_0''(r_*-\bar{r}_*)^2\right ]\Psi=0 \,.\ee
This equation has an exact solution in terms of parabolic cylinder
functions \cite{Abramowitz:1970as,benderorszag}:
\be
\Psi = AD_{\nu}(z)
+ BD_{-\nu-1}(iz) \,,\quad z\equiv (2Q_0'')^{\frac{1}{4}}e^{i\frac{\pi}{4}}(r_*-\bar{r}_*)\,,\ee
with $\nu=-iQ_0/\sqrt{2Q_0''}-1/2$. Using the asymptotic behavior of cylinder
functions \cite{Abramowitz:1970as,benderorszag} and demanding only outgoing
waves at spatial infinity we get, near the horizon,
\be \Psi \sim  A e^{-i\pi
\nu}z^{\nu}e^{-\frac{z^2}{2}} -
i
\sqrt{2\pi}A
\left[\Gamma(-\nu)\right]^{-1}
e^{5i\pi/4}z^{-\nu-1}e^{\frac{z^2}{2}}
\,.\ee
QNM boundary conditions imply that the outgoing term, proportional to
$e^{z^2/2}$, should be absent, so $1/\Gamma(-\nu)=0$, or
$\nu=n(=0\,,1\,,2\,,...)$. As we anticipated, the leading-order WKB
approximation yields a ``Bohr-Sommerfeld quantization rule'' defining the QNM
frequencies:
\be 
Q_0/\sqrt{2Q_0''}=i(n+1/2)\,,\quad n=0\,,1\,,2\,,...
\ee

\begin{figure*}[htb]
\begin{center}
\begin{tabular}{cc}
\epsfig{file=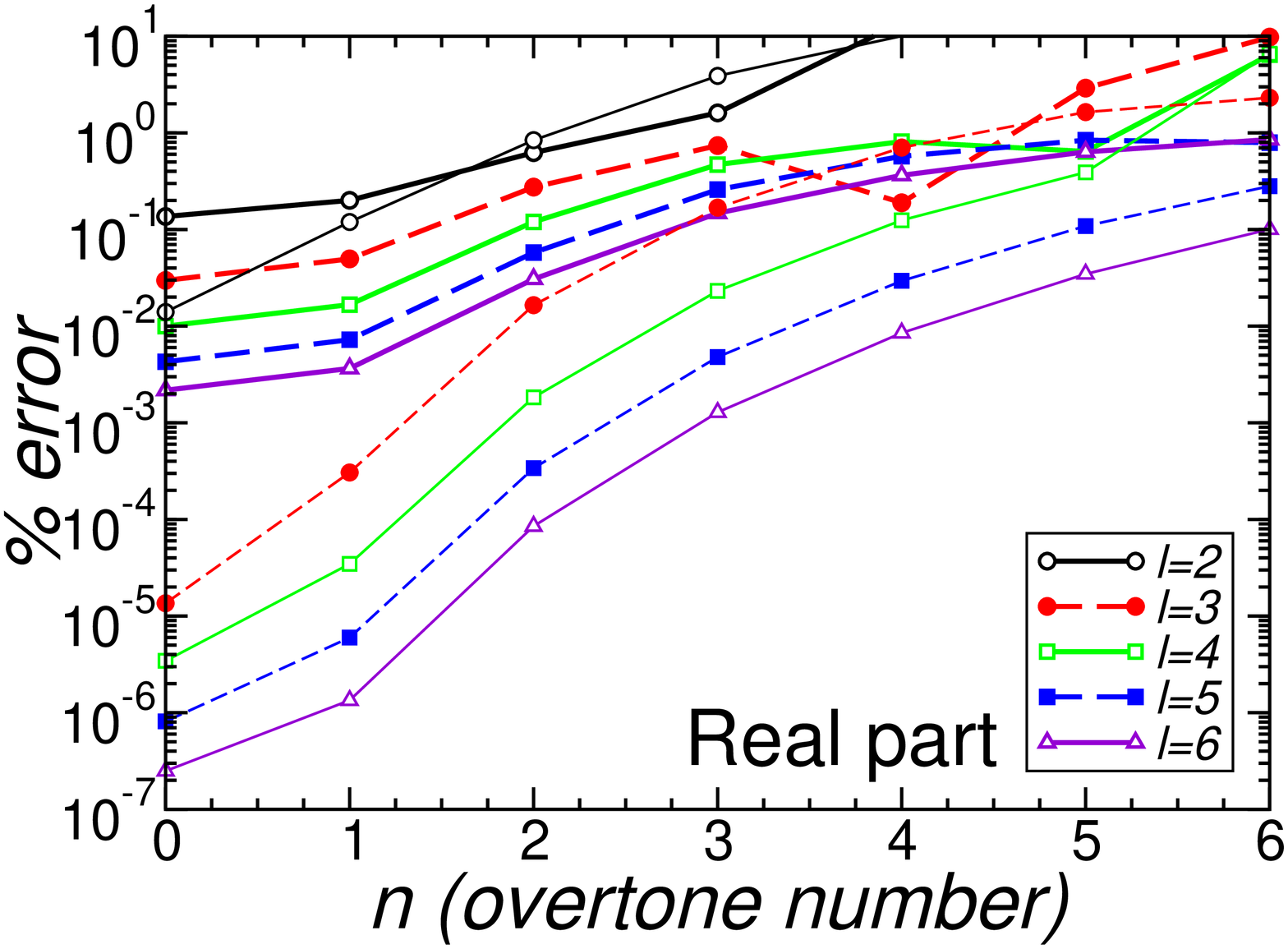,width=5.2cm,angle=-90}
\epsfig{file=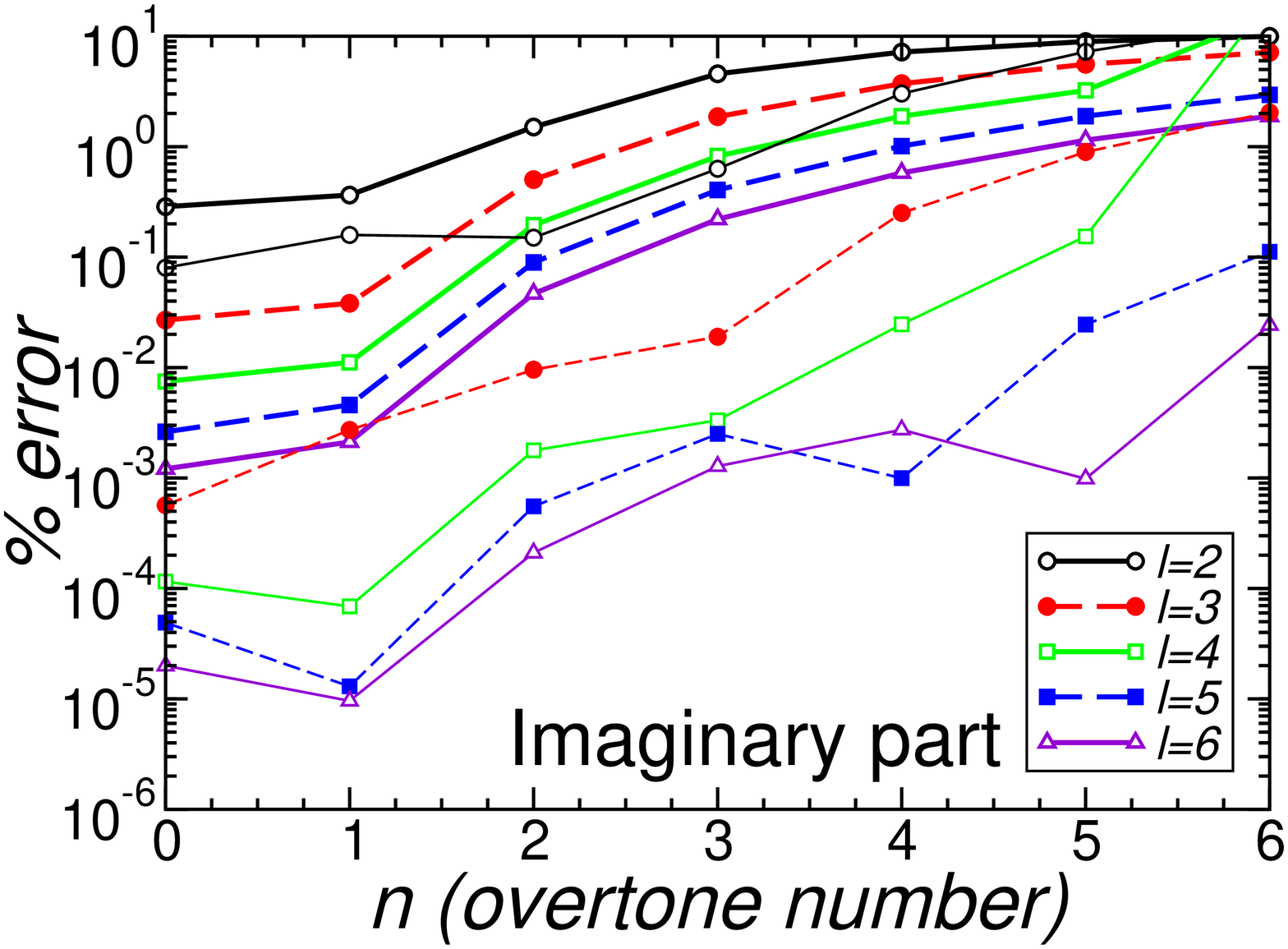,width=5.2cm,angle=-90}
\end{tabular}
\caption{\label{fig:wkbcomparison} Percentage errors for the real (left) and
  imaginary part (right) of the QNM frequencies as predicted from WKB
  calculations. Thick lines: third-order WKB approximation; thin lines:
  sixth-order WKB approximation.}
\end{center}
\end{figure*}

Higher-order corrections to Eq.~(\ref{parabolawkb}) have been computed
\cite{pressringdown}. Iyer and Will \cite{Iyer:1986np,Iyer:1986nq} carried out
a third-order WKB expansion, and more recently Konoplya \cite{Konoplya:2003ii}
pushed the expansion up to sixth order. There is no rigorous proof of
convergence, but the results do improve for higher WKB orders. This is shown
in Fig.~\ref{fig:wkbcomparison}, where we compare numerical results for the
QNMs of Schwarzschild BHs from Leaver's continued fraction method (to be
discussed in Section~\ref{sec:qnmschw} below) against third- and sixth-order
WKB predictions.  The WKB approximation works best for low overtones,
i.e. modes with a small imaginary part, and in the eikonal limit of large $l$
(which corresponds to large quality factors, or large
$\omega_R/\omega_I$). The method assumes that the potential has a single
extremum, which is the case for most (but not all) BH potentials: see
Ref.~\cite{Ishibashi:2003ap} for interesting counterexamples.

Ref.~\cite{DolanPRL} introduces a new, WKB-inspired asymptotic expansion of
QNM frequencies and eigenfunctions in powers of the angular momentum parameter
$l+1/2$. Their asymptotic expansion technique is easily iterated to high
orders, and it seems to provide very accurate results in spherically-symmetric
spacetimes.  The asymptotic expansion also provides physical insight into the
nature of QNMs, nicely connecting the geometrical understanding of QNMs as
perturbations of unstable null geodesics with the singularity structure of the
Green function.

\subsection{\label{monodromy}Monodromy technique for highly-damped modes}

A powerful variant of the WKB approximation, which is particularly useful in
the highly-damped limit $\omega_I\to \infty$, is the so-called monodromy
technique \cite{Motl:2003cd}. The basic idea is related to Stokes' phenomenon
in the theory of asymptotic expansions (see e.g.~\cite{Froeman:1965} for an
excellent introduction to the topic). As shown by Andersson and Howls
\cite{Andersson:2003fh}, the monodromy technique is a simple variant of the
phase-integral approach, whose application to BH physics dates back to the
work by Fr\"oman {\it et al.}  \cite{Froeman:1992gp}.

\begin{figure*}[ht]
\begin{center}
\begin{tabular}{c}
\epsfig{file=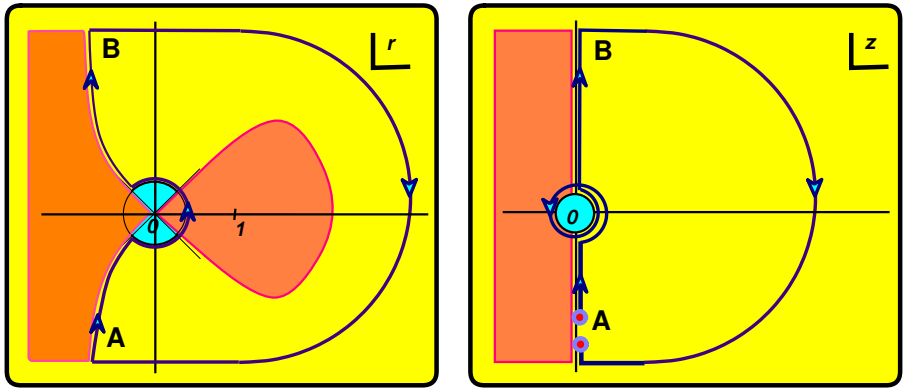,width=12cm,angle=0}
\end{tabular}
\end{center}
\caption{\label{fig:monodromy2} Left panel: contour for calculation of the QNM
  frequencies in the complex-$r$ plane. The different regions are separated by
  the associated Stokes lines. Not shown in the plot are branch cuts from
  $r=-\infty$ to the origin and from $r=1$ to point A. Right panel:
  integration contour in the complex-$z$ plane, with $z\equiv r_*/2M-i\pi$.
  For more details see Ref.~\cite{Motl:2003cd}.}
\end{figure*}

Let us consider the wave equation (\ref{waveeq}) for the Schwarzschild
geometry, but allowing $r$ and $r_*$ to be complex variables.  In the
complex-$r$ plane, solutions to Eq.~(\ref{waveeq}) may be multivalued around
the singular points ($r=0$ and $r=2M$). To deal with the singular points we
introduce branch cuts emanating from $r=0$ and $r=2M$.  The relation $r_*(r)$
is also multi-valued: in the Schwarzschild geometry
$r_*(r)=r+2M\log{\left(r/2M-1\right )}$, and we choose the branch such that
$\log(-1)=i\pi$. We can now define a variable $z\equiv r_*/2M-i\pi$ which
tends to zero as $r\to 0$. The Stokes lines are defined as the lines for which
${\rm Re}(r_*)=0$ \cite{Froeman:1965}, and they are shown in
Fig.~\ref{fig:monodromy2} (near the singular point $r=0$, the Stokes lines
form an angle $\pi/4$).
The idea now is to equate the monodromy as computed in two different ways. The
first computation takes the contour in Fig.~\ref{fig:monodromy2}, starting and
ending at point A and following the Stokes lines, joined at infinity by the
large semi-circles shown in Fig.~\ref{fig:monodromy2}.  If we start from A
with a plane wave $e^{i\omega z}$, we can extrapolate the behavior to near the
singularity $z=0$. This is because for large imaginary $\omega$ the term
$\omega^2$ is much larger than the other terms in the potential. Near the
origin ($r=0$) the solutions to the wave equation can be written as
\be \Psi(z)=A_+\sqrt{4\pi\,M\omega z}J_{\frac{s}{2}}(2M\omega
z)+A_-\sqrt{4\pi\,M\omega z}J_{-\frac{s}{2}}(2M\omega z) \,,\ee
where $A_{\pm}$ are constants and $J_{s/2}$ is a Bessel function. From the
asymptotics of Bessel functions for large $\omega z$ one has
\be \Psi(z)\sim e^{i2M\omega z} \lbrack
A_+e^{-i\alpha_+}+A_-e^{-i\alpha_-}\rbrack+ e^{-i2M\omega z} \lbrack
A_+e^{i\alpha_+}+A_- e^{i\alpha_-}\rbrack \,,\label{asym1}\ee
with $\alpha_{\pm}=\pi(1\pm s)/4$. The second term must vanish because of the
boundary conditions. Continuing along the contour, we make a $3/2\pi$ turn
around $r=0$. In terms of the $z$ coordinate, this means a $3\pi$
rotation. Using the representation for the Bessel functions
\be J_{\pm s/2}(\eta)=\eta^{\pm
s/2}\sum_{n=0}^{+\infty}\frac{(-1)^n \eta^{2n}}{2^{2n\pm
s/2}\Gamma(\pm s/2+n+1)n!}\,, \ee
one can see that $J_{\pm s/2}(e^{3i\pi}2M\omega z)=e^{\pm i 3\pi s/2}J_{\pm
s/2}(2M\omega z)$. Therefore
\be \Psi \sim e^{i2M\omega z} \lbrack
A_+e^{7i\alpha_+}+A_-e^{7i\alpha_-}\rbrack+e^{-i2M\omega z} \lbrack
A_+e^{5i\alpha_+}+A_- e^{5i\alpha_-}\rbrack \,.  \ee
The $e^{-i2M\omega z}$ term is exponentially small, since we are in the region
where ${\rm Re}(z)\gg0$ and $\omega_I\ll 0$.  Using Eq.~(\ref{asym1}) we get
for the coefficient of $e^{i\omega z}$ the monodromy
\be
\frac{A_+e^{7i\alpha_+}+A_-e^{7i\alpha_-}}{A_+e^{-i\alpha_+}+A_-e^{-i\alpha_-}
}\,.\label{asym2}\ee
Requiring that the second term in Eq.~(\ref{asym1}) vanishes,
Eq.~(\ref{asym2}) yields $-\left (1+2\cos(\pi s)\right )$ for the
monodromy. On the other hand, the only singularity inside the contour is at
$r=1$, giving a factor $e^{-4i(2M\omega)}$ and leading to the condition
\be 8\pi M\omega=\pm \log3-i\pi(2n+1)\, \ee
for the highly damped QNM frequencies of Schwarzschild BHs with $s=0\,,2$
($s=1$ is predicted to have a vanishing real part). This formula is in
agreement with numerical studies by Nollert \cite{Nollertlargen} for the
Schwarzschild spacetime. The procedure has been generalized to several other
backgrounds, including Reissner-Nordstr\"om (RN) \cite{Motl:2003cd}, Kerr
\cite{Keshet:2007be,Kao:2008sv}, SAdS \cite{Cardoso:2004up} and other BH
spacetimes \cite{Natario:2004jd}, always finding excellent agreement with
numerical calculations \cite{Berti:2003zu,Berti:2004um,Cardoso:2003cj}.

The application of the monodromy method requires extreme care in identifying
the appropriate Stokes lines and integration contours.  Nat\'ario and Schiappa
\cite{Natario:2004jd} point out some instances of inappropriate applications
of the technique.  A more rigorous mathematical treatment of these methods is
highly desirable.

\subsection{Asymptotically anti-de Sitter black holes: a series solution}

With the exception of the monodromy technique, the methods discussed so far
work for asymptotically flat or (with some minor modifications) dS
spacetimes. For asymptotically AdS spacetimes, the perturbation equations
exhibit regular singularities both at the horizon and at spatial infinity.
Local solutions in the vicinity of the regular singular points are represented
by {\it convergent} Frobenius series \cite{arfken}. In many cases, the radius
of convergence of the series is equal to or larger than the interval of
interest. In this case, a simple numerical procedure (implemented e.g. by
Horowitz and Hubeny in \cite{Horowitz:1999jd}) is possible.
Consider, for example, a scalar field $\Psi$ in the SAdS$_4$ spacetime,
satisfying the wave equation (\ref{waveeq}). Defining $\Phi$ for a generic
wavefunction as $\Phi=e^{i\omega r_*} \Psi$, we find
\be f(r)\frac{d^{2} \Phi}{d r^{2}}+ \left\lbrack
f'-2i\omega \right\rbrack \frac{d \Phi}{d r}
-\frac{V}{f}\Phi=0\,,\label{waveads}\ee
where $f'\equiv df/dr$. To find the frequencies $\omega$ that satisfy the
boundary conditions we first note that Eq.~(\ref{waveads}) has only regular
singularities in the range of interest. To deal with the point at infinity,
we first change the independent variable $r$ to $x=1/r$ and get
\be 
\label{froben}
(x-x_+)s(x)\frac{d^{2} \Phi}{d x^{2}}+
t(x)\frac{d \Phi}{d x}+\frac{u(x)}{x-x_+}\Phi=0\,, \ee
where $x_+=1/r_+\equiv h$, $s= -fx^4/(x-x_+)$, $t=- \left\lbrack 2fx^3-f'x^2-2i\omega x^2
\right\rbrack$, $u=V(x-x_+)/f$.
Eq.~(\ref{froben}) admits a local (near $x=h$) Frobenius solution of the form
\be \Phi(x)= (x-h)^{\alpha}\sum_{n=0}^{\infty} a_{n} (x-h)^n \,,
\label{frobenius} 
\ee
where $a_{n}(\omega)$ is a function of the frequency.  The radius of
convergence of the series is limited by the distance to the next nearest
singular point of the equation.  The index $\alpha$ is determined by imposing
the boundary condition at the horizon.  Writing
\be 
\label{serfrob}
s(x)=\sum_{i=0}^\infty s_i(x-h)^i\,,\,\,
t(x)=\sum_{i=0}^\infty t_i(x-h)^i\,,\,\,
u(x)=\sum_{i=0}^\infty u_i(x-h)^i\,, 
\ee
and substituting Eqs.~(\ref{frobenius}) and (\ref{serfrob}) in
Eq.~(\ref{waveads}) we get an indicial equation for $\alpha$:
$\alpha(\alpha-1)s_0+\alpha t_0+u_0=0$. We also have
$s_0=2h^2\kappa\,,t_0=2h^2(\kappa-i\omega)$ and $u_0=0$, where the surface
gravity is $\kappa=f'(r_+)/2$.  Therefore, either $\alpha=0$ or
$\alpha=i\omega/\kappa$. One can check that the second option corresponds to
outgoing waves at the horizon, so we choose $\alpha=0$.  Inserting the
decomposition (\ref{frobenius}) with $\alpha=0$ in Eq.~(\ref{waveads}) and
comparing powers of $(x-h)$ we get a recurrence relation for the $a_n$'s:
\be a_n=-\frac{1}{n(n-1)s_0+nt_0}\sum_{i=0}^{n-1}\left
[i(i-1)s_{n-i}+it_{n-i}+u_{n-i} \right ]a_i\,. \ee
Since the differential equation is linear, $a_0$ is just an arbitrary
normalization constant. Using the boundary condition $\Phi=0$ at infinity
($x=0$) we finally obtain a simple relation determining the QNM frequencies:
\begin{equation}
\sum_{n=0}^{\infty} a_{n}(-h)^n=0 \,. \label{numerico}
\end{equation}
In practice one computes the $N$-th partial sum of the series
($\ref{numerico}$) and finds the roots $\omega$ of the resulting polynomial
expression, checking for convergence by comparison with (say) the roots
obtained from the $N+1$-th partial sum. The method can be easily implemented:
see e.g. the publicly available {\it Mathematica} notebook \cite{rdweb} that
was used in \cite{Cardoso:2001bb} to compute low-lying modes of the SAdS$_4$
geometry. This notebook can be easily modified to deal with other geometries.

Some mathematical aspects of the series solution method as applied to QNMs in
asymptotically AdS backgrounds are discussed in Ref.~\cite{Starinets:2002br}.

\subsection{\label{sec:resonance}Asymptotically anti-de Sitter black holes: the resonance method}

The series solution described in the previous subsection converges very slowly
for small SAdS BHs.  Fortunately, in this regime there is a simple
alternative: the resonance method \cite{Berti:2009wx}.  This method requires a
numerical integration of the wave equation, but unlike the original numerical
procedure by Chandrasekhar and Detweiler \cite{Chandrasekhar:1975qn}, one only
needs to search for roots on the real-$\omega$ line.

It is well known that quasi-bound states manifest themselves as poles in the
scattering matrix, and as Breit-Wigner resonances in the scattering
amplitude. Chandrasekhar and Ferrari made use of the form of the scattering
cross section near these resonances in their study of gravitational wave
scattering by ultra-compact stars
\cite{ChandrasekharFerrari,Chandrasekhar:1992ey}.  For quasi-bound, trapped
modes of ultra-compact stars, the asymptotic wave amplitude at spatial
infinity $\Psi \sim \alpha \cos \omega r+\beta \sin \omega r$ has a
Breit-Wigner-type behavior close to the resonance:
\be
\alpha^2+\beta^2\approx{\rm constant}\,
\left[(\omega-\omega_R)^2+\omega_I^2\right]\,,
\label{breitwigner}
\ee
where $\omega_I^{-1}$ is the lifetime of the quasi-bound state and
$\omega_R^2$ its characteristic ``energy''.  The example of ultra-compact
stars shows that the search for weakly damped QNMs corresponding to
quasi-bound states ($\omega=\omega_R-i\omega_I$ with $\omega_I\ll \omega_R$)
is extremely simplified. One locates the resonances by looking for minima of
$\alpha^2+\beta^2$ on the $\omega=\omega_R$ line, and the corresponding
damping time $\omega_I$ can then be obtained by a fit to a parabola around the
minimum \cite{ChandrasekharFerrari,Chandrasekhar:1992ey}.

It was shown in \cite{Berti:2009wx} that these ideas can be used very
successfully in BH spacetimes, if one integrates in from spatial infinity to
the BH horizon (i.e., the appropriately re-defined quantities
$\alpha^2+\beta^2$ are taken close to the horizon). The resonance method is
extremely simple and accurate, especially for small SAdS BHs
\cite{Berti:2009wx} (see also Section \ref{sec:sads}). It is presently unclear
whether it can be applied successfully to study QNMs in other BH spacetimes.

\subsection{\label{sec:Leaver}The continued fraction method}

Applications of the continued fractions to eigenvalue problems have an
interesting history dating back to Jaff\'e's 1933 paper on the spectrum of the
hydrogen molecule ion, or perhaps to even earlier times (some details and
relevant references can be found in \cite{Starinets:2002br}).

The method was applied to gravitational problems by Leaver, leading to what is
possibly the most successful algorithm to date to compute QNM frequencies
\cite{Leaver:1985ax,leJMP,Leaver:1986gd}.  Leaver's approach is based on the
observation that the Teukolsky equation is a special case of a general class
of spheroidal wave equations that appear in the calculation of the electronic
spectra of the hydrogen molecule ion \cite{leJMP}.  These equations can be
solved through a three-term recurrence relation, and the boundary conditions
lead to a continued fraction equation characterizing the QNMs.  Originally
devised for the Schwarzschild and Kerr geometries
\cite{Nollertlargen,Onozawa:1996ux,Berti:2003jh,Berti:2004um}, the method has
also been applied to RN BHs \cite{LeaverRN}, Kerr-Newman BHs
\cite{Berti:2005eb}, higher-dimensional BHs
\cite{Cardoso:2003vt,Cardoso:2003qd,Cardoso:2004cj,Morisawa:2004fs}, SdS BHs
\cite{Yoshida:2003zz,Konoplya:2004uk} and acoustic BHs \cite{Cardoso:2004fi}
(see Section~\ref{ref:acoustic} below), among others. We illustrate the main
ideas by considering the Kerr case. Start with the following series solution
for the angular eigenfunctions defined in Eq.~(\ref{angular}):
\begin{equation}\label{leavers}
{}_sS_{lm}(u)=e^{a\omega
u}\left(1+u\right)^{k_-}\left(1-u\right)^{k_+} \sum_{p=0}^\infty
a_p(1+u)^p\,,
\end{equation}
where $k_{\pm}\equiv |m\pm s|/2$. The expansion coefficients $a_p$ are
obtained from the three-term recurrence relation
\be\label{recur} 
\alpha_0 a_{1}+\beta_0 a_{0}=0\,,\quad \alpha_p
a_{p+1}+\beta_p a_{p}+\gamma_p a_{p-1}=0\,, \quad p=1,2\dots \ee
where the constants $\alpha_p,\,\beta_p,\,\gamma_p$ are given in the original
work \cite{Leaver:1985ax}.
Given a complex argument $\omega$, the separation constant ${}_sA_{lm}$ can be
obtained solving numerically the continued fraction $\beta_0-
{\frac{\alpha_0\gamma_1}{\beta_1-}} {\alpha_1\gamma_2\over\beta_2-}
{\alpha_2\gamma_3\over\beta_3-} ...=0$ or any of its inversions
\cite{Leaver:1985ax}.

A solution $R_{r_+}$ of the radial equation (\ref{radial}) should satisfy the
appropriate boundary conditions:
\label{teuk-norm} \beq & & \lim_{r \to r_+}R_{r_+}\sim
(r_+-r_-)^{-1-s+i\omega+i\sigma_+} e^{i\omega
r_+}(r-r_+)^{-s-i\sigma_+}\,, \label{Rrp-norm}
\\
& & \lim_{r \to \infty }R_{r_+}\sim A_{\rm
out}^T(\omega)r^{-1-2s+i\omega}e^{i\omega r}\,. \label{psirmainf}
\eeq
In these relations $\sigma_+=\left(\omega r_+-am\right)/b$, and
$b=\sqrt{1-4a^2}$. A convenient series solution close to the horizon can be
found by methods due to Jaff\'e (see \cite{Leaver:1985ax}):
\be R_{r_+}=e^{i\omega r} (r-r_-)^{-1-s+i\omega+i\sigma_+}
(r-r_+)^{-s-i\sigma_+} \sum_{n=0}^{\infty} a^r_n\left
(\frac{r-r_+}{r-r_-} \right )^n\,.\label{Rrmaisdet} \ee
The coefficients $a_n^r$ can be obtained from a recurrence relation similar to
Eq.~(\ref{recur}). The continued fraction method is very powerful at computing
overtones because the $n$-th overtone turns out to be the most stable
numerical root of the $n$-th inversion of the radial continued fraction (which
in principle should be completely equivalent to any other inversion)
\cite{Leaver:1985ax}. This observation makes it (relatively) easy to
numerically compute Kerr overtones up to $n\sim 50$, and Schwarzschild
overtones up to $n\sim 100$; refinements of the technique to compute even
higher overtones will be discussed in Section \ref{highdamp} below.  Given the
QNM frequencies and the corresponding angular eigenvalues ${}_sA_{lm}$ it is a
simple matter to compute $R_{r_+}$ for moderate values of $r$ (the convergence
of the expansion gets worse for large values of $r$). We provide an
implementation of the method to compute the QNM frequencies, as well as the
radial and angular wavefunctions, in a publicly available {\it Mathematica}
notebook \cite{rdweb}.
On the same web page we also provide numerical data for the QNM frequencies of
the first 8 Kerr gravitational modes ($n=0,\dots,7$) with $2\leq l\leq 7$, as
computed by an independent {\it Fortran} code which will be made available
upon request \cite{Berti:2004md}.
\begin{figure}[h]
\begin{center}
\epsfig{file=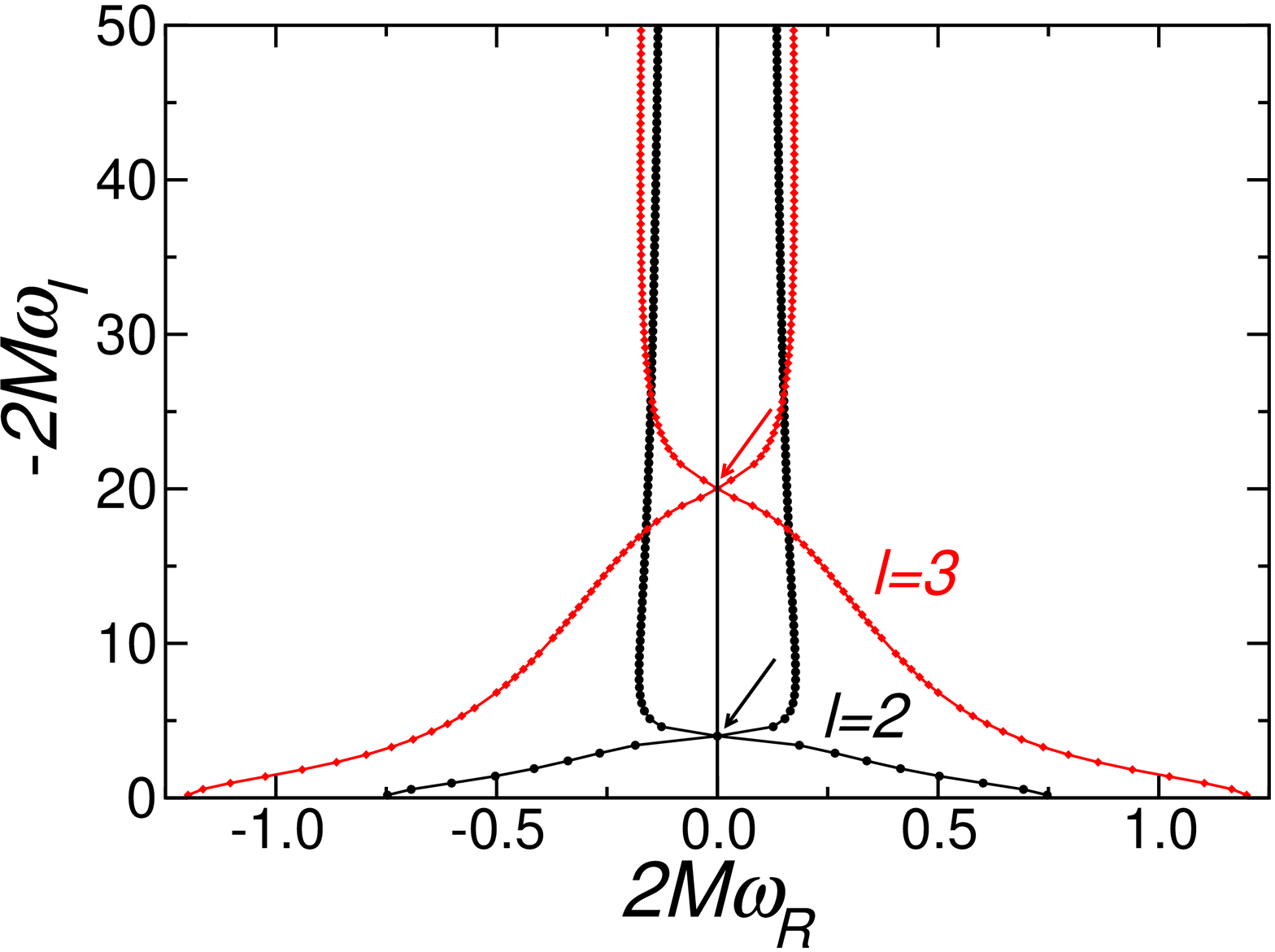,width=8.4cm,angle=-90}
\epsfig{file=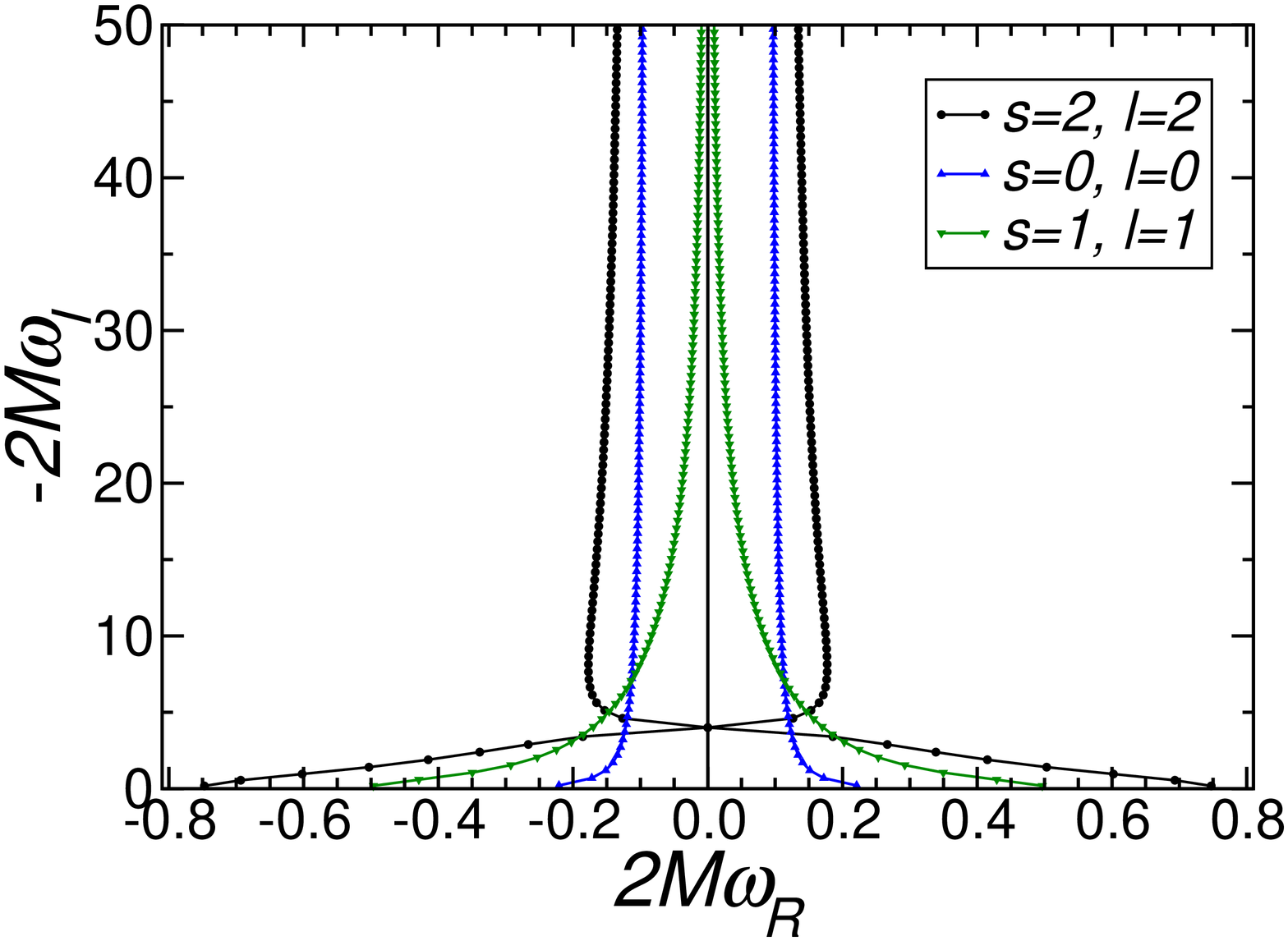,width=8.4cm,angle=-90}
\caption{Top: QNM frequencies for gravitational perturbations with $l=2$
  (black circles) and $l=3$ (red diamonds). In both cases we mark by an arrow
  the algebraically special mode, given analytically by Eq.~(\ref{AlgSp}); a
  more extensive discussion of this mode is given in
  \ref{app:relpotentials}. Notice that as the imaginary part of the frequency
  tends to infinity the real part tends to a finite, {\it $l$-independent}
  limit. Bottom: comparison of the $l=|s|$ QNM frequencies for scalar,
  electromagnetic and gravitational perturbations.}
\label{fig:schw}
\end{center}
\end{figure}

\section{\label{sec:asflat}The spectrum of asymptotically flat black holes}

In this section we briefly review the structure of the QNM spectrum of BHs
belonging to the Kerr-Newman family. More details can be found in
Ref.~\cite{Berti:2004md}.  We focus our discussion on gravitational
perturbations, and refer to relevant works on other kinds of perturbations in
the appropriate sub-sections.

\subsection{\label{sec:qnmschw}Schwarzschild}

For Schwarzschild BHs, scalar-type and vector-type gravitational perturbations
can be shown to give rise to the same QNM spectrum. This remarkable result,
due to Chandrasekhar and collaborators
\cite{MTB,Chandrasekhar:1975qn,chandrarelation}, is reviewed in
~\ref{app:relpotentials} below.

The vector-type potential is simpler than the scalar-type potential, so it is
customary to compute QNMs for vector-type perturbations. We computed QNM
frequencies using the continued fraction technique (see Section
\ref{sec:Leaver}) as improved by Nollert \cite{Nollertlargen}. Schwarzschild
QNM frequencies with $l=2$ and $l=3$ are shown in Fig.~\ref{fig:schw}. The
data are also available online \cite{rdweb}.

It is apparent from Fig.~\ref{fig:schw} that the gravitational QNMs of
Schwarzschild BHs are naturally divided in two categories. A mode whose
frequency is (almost) purely imaginary separates the lower QNM branch from the
upper branch.  This mode is very close to (and may actually coincide with) the
so-called ``algebraically special mode'' \cite{chandraspecial}, discussed in
\ref{app:relpotentials}. It is located at
\be\label{AlgSp}
M \omega_l \approx \pm i (l -1)l (l +1)(l +2)/12\,,
\ee
and it can be taken as roughly marking the onset of the asymptotic
high-damping regime. The algebraically special mode corresponds to an overtone
index $n=9$ when $l=2$, to $n=41$ when $l=3$ and to even larger values of $n$
as $l$ increases. This means that the asymptotic high-damping regime is very
hard to probe numerically as we approach the eikonal limit ($l\gg 1$). The
existence of an almost purely imaginary QNM frequency is unique to
gravitational perturbations: for other known fields, the approach to the
asymptotic regime is monotonic (see the bottom panel of Fig.~\ref{fig:schw}
and Ref.~\cite{Cardoso:2003vt}).  For gravitational perturbations we will,
somewhat arbitrarily, define the weakly damped (highly damped) regime as
corresponding to imaginary parts smaller (larger) than the algebraically
special frequency. For spins of other fields there is no such clear marker
(see Fig.~\ref{fig:schw}), but we will usually call ``weakly damped'' modes
those with $n\lesssim 10$, and ``highly damped'' modes those with $n\gg 10$.

\subsubsection*{The weakly damped modes}

The weakly damped gravitational modes were computed numerically by
Chandrasekhar and Detweiler \cite{Chandrasekhar:1975qn}, Leaver
\cite{Leaver:1985ax} and many others, and they are available online
\cite{rdweb}.  Iyer \cite{Iyer:1986nq} computed the first few modes for
scalar, electromagnetic and gravitational perturbations using the WKB
technique, and compared results against the more accurate continued fraction
method.  In geometrical units, the fundamental $l=|s|$ mode is
$M\omega=0.1105-0.1049i$ for $s=0$ and $M\omega=0.2483-0.0925i$ for $s=1$. For
ringdown detection from astrophysical BHs the most relevant QNM is, in most
situations, the fundamental gravitational mode with $l=2$, with
$M\omega=0.3737-0.0890i$. This mode has oscillation frequency and damping time
given by
\beq\label{QNMscale}
f&=&\omega_R/2\pi=1.207 \left(\frac{10~M_\odot}{M}\right) {\rm kHz}\,,\\
\tau&=&1/|\omega_I|=0.5537 \left(\frac{M}{10~M_\odot}\right) {\rm ms}\,,
\eeq
where $M\sim 10M_\odot$ is a typical value for a stellar-mass BH (see Section
\ref{sec:massspin} below).

Weakly damped QNMs for Schwarzschild perturbations with half-integer spins
have been computed in the WKB approximation by Cho and collaborators
\cite{Cho:2003qe,Cho:2004wj}.

\subsubsection*{The highly damped modes}

Using a variant of Leaver's method, Nollert carried out the first reliable
numerical calculation of highly damped QNM frequencies for gravitational
perturbations \cite{Nollertlargen}. The real part of the QNM frequencies is
well fitted, for large $n$, by a relation of the form
$\omega_R=\omega_{\infty}+\lambda_{s,l}/\sqrt{n}$. The leading-order fitting
coefficient $2M\omega_{\infty}\simeq 0.08742$ is independent of $l$, and it
has the same value for $s=2$ and $s=0$ \cite{Berti:2003zu}. Nollert's
numerical results have been confirmed by various analytical calculations. Motl
\cite{Motl:2002hd} used Leaver's continued fraction conditions to show that in
the limit $n\rightarrow \infty$ the following asymptotic expansion holds:
\be\label{Mresult}
\omega\sim{T \ln 3}-(2n+1)\pi i T+{\cal O}(n^{-1/2})\,,
\ee
where $T=(8\pi M)^{-1}$ is the Hawking temperature.  This conclusion was
confirmed by complex-integration \cite{Motl:2003cd} and phase-integral
techniques \cite{Andersson:2003fh}.

Corrections of order $\sim n^{-1/2}$ to Eq.~(\ref{Mresult}) were first
obtained by Neitzke \cite{Neitzke:2003mz} and Maassen van den Brink
\cite{MaassenvandenBrink:2003as}. Ref.~\cite{Musiri:2003bv} developed a
systematic perturbative approach to determine subdominant corrections as $n\to
\infty$, using a solution of the Regge-Wheeler equation in terms of Bessel
functions. Ignoring contributions of order ${\cal O}(1/n)$, the result for
$s=0$ and $s=2$ can be written as 
\be
\label{eqfs02} 
\frac{\omega}{T} = 
-(2n+1)\pi i + \ln 3 + 
\frac{1+i}{\sqrt{n+1/2}} \ 
\frac{3l(l +1) -(s^2-1)}{18|s^2-1|\sqrt 2 \pi^{3/2}}
\ \Gamma^4 \left(\frac{1}{4}\right).
\ee
The subdominant coefficients are in agreement with fits of numerical results
\cite{Nollertlargen,Berti:2003zu,Cardoso:2003vt}.  For electromagnetic
perturbations, analytic and numerical results suggest that the real part
$\omega_R\to 0$ in the asymptotic limit \cite{Motl:2003cd,Cardoso:2003vt},
which is also apparent from Fig.~\ref{fig:schw}.  The above calculation has
been generalized to massless fermionic perturbations.  By solving the
Teukolsky equation in terms of confluent hypergeometric functions,
Refs.~\cite{Musiri:2006ta,Musiri:2007zz} confirmed the known results for
integer spins, and found in addition that the result for $s=1/2$ and $s=5/2$
can be written as:
\be
\frac{\omega}{T}=
-(2n+1) \pi i +\frac{2(1+i)}{(s^2+23/4)\sqrt{n}}\ 
\left(3l(l+1)+1-s^2\right)\ \Gamma^2\left( \frac{1}{4} \right).
\ee
For $s=3/2$ there is no correction at order ${\cal O}(n^{-1/2})$.  Numerical
studies on $s=1/2$ fields confirm that $\omega_R \to 0$ and the spacing
$\delta \omega_I \to 2\pi T=1/4M$ as $n\to \infty$ \cite{Jing:2005dt}.

In summary, numerical and analytical results for Schwarzschild BHs are in
perfect agreement. As $|\omega_I|\to \infty$, $\omega_R \to T \ln 3$ for
scalar and gravitational oscillation frequencies, and $\omega_R \to 0$ for
perturbations of other spins. The spacing of the imaginary parts for large $n$
is always given by $2\pi T$. By considering the scattering amplitude in the
Born approximation, Padmanabhan showed that this universality in the spacing
arises from the exponential redshift of the wave modes close to the horizon
\cite{Padmanabhan:2003fx}.

\subsubsection*{The eikonal limit}

By their own nature, WKB methods become increasingly accurate for large $l$,
and they can be used to compute the QNM frequencies analytically when $l\gg
1$. Up to order ${\cal O}(l^{-2})$ the result is
\cite{pressringdown,Iyer:1986np,Iyer:1986nq,barretozworski}
\beq M\omega&=&\f{1}{3\sqrt{3}} \left[\left (l+\frac{1}{2}\right )
  -\f{1}{3}\left(\f{5(n+\frac{1}{2})^2}{12}+\f{115}{144}-1+s^2
  \right)\left(\f{1}{l}-\f{1}{2l^2}\right)\right]\nn
\\ &-&i\f{(n+\frac{1}{2})}{3\sqrt{3}}\left[1+\f{1}{9}
  \left(\f{235(n+\frac{1}{2})^2}{432}-\f{1415}{1728}+1-s^2\right)\f{1}{l^2}\right].
\label{eikonal}
\eeq
The convergence of the series gets worse with increasing $n$ (cf. also
Fig.~\ref{fig:wkbcomparison}).  In the eikonal limit the asymptotic behavior
of the potential is not important, so it should not be surprising that a
P\"oschl-Teller approximation of the Schwarzschild potential, discussed in
Section \ref{sec:PT}, yields the correct result (to leading order) for large
$l$.  The eikonal regime of Schwarzschild BHs is related to the properties of
unstable circular null geodesics
\cite{pressringdown,goebel,Ferrari:1984zz,mashhoon,stewart,
  Decanini:2002ha,Cardoso:2008bp,Decanini:2009mu}: the leading order of
Eq.~\ref{eikonal} can be written as $\omega=\Omega_c \, l-i\left (n+1/2 \right
)\lambda$, where $\Omega_c$ is the orbital frequency and $1/\lambda$ is the
instability timescale of the unstable circular null geodesic. Such a
connection can be generalized to any asymptotically flat, static spacetime in
four and higher dimensions \cite{Cardoso:2008bp}.

\subsection{\label{sec:qnmrn}Reissner-Nordstr\"om}

With a few exceptions, BH charge is usually considered astrophysically
negligible. Despite this, the RN metric is of more than academic interest: for
example, charged naked singularities have been proposed as classical models
for elementary particles (see \cite{Berti:2005eb} and references therein).
Handling scalar fields in the background of a charged BH requires only a
straightforward generalization of the uncharged case, resulting in a wave
equation of the form (\ref{waveeq}).  Electromagnetic and gravitational
perturbations are more technically challenging, since they are coupled through
the Einstein-Maxwell equations. It is still possible to reduce the problem to
the study of two wave equations of the general form (\ref{waveeq}) for two
wavefunctions $Z_1^{-},\,Z_2^{-}$ \cite{MTB}:
\beq\label{axialRN}
\left({d^2\over dr_*^2}+\omega^2\right)Z_i^{(-)}=V_i^{(-)}Z_i^{(-)}\,,\\
V_i^{(-)}=\frac{(r^2-2Mr+Q^2)}{r^5}\left[l(l+1)r-q_j
\left(
1+\frac{q_i}{(l-1)(l+2)r}
\right)\right]\,.
\nn
\eeq
Here $q_1+q_2=6M$, $-q_1q_2=4Q^2(l-1)(l+2)$ and $i,j=1,2$ ($i\neq j$). In the
limit when the charge $Q/M$ goes to zero, $Z_1^-,\,Z_2^-$ describe pure
electromagnetic and pure vector-type gravitational perturbations of a
Schwarzschild BH, respectively.  In the limit of maximally charged BHs
($Q/M=1$) the wave equations have a different singularity structure, and
deserve a special treatment \cite{Onozawa:1995vu}.

\begin{figure}[ht]
\begin{center}
\epsfig{file=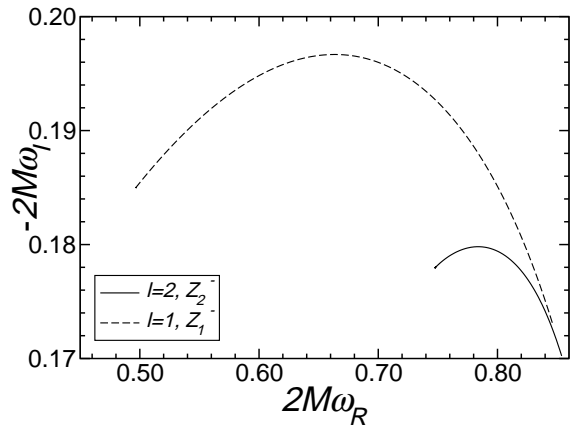,width=6.5cm,angle=-90}
\caption{Trajectory the complex-frequency plane described by the fundamental
  QNM as the charge $Q/M$ is increased. The solid line corresponds to $l=2$
  and $Z_2^-$ (which reduces to vector-type perturbations as $Q/M\to 0$); the
  dashed line corresponds to $l=1$ and $Z_1^-$ (purely electromagnetic
  perturbations in the limit $Q/M\to 0$).  The two modes coalesce in the
  extremal limit ($Q/M\to 1$).}
\label{fig:figrn}
\end{center}
\end{figure}

\subsubsection*{The weakly damped modes}

The behavior of the weakly damped modes of gravitational/electromagnetic
perturbations is exemplified in Fig.~\ref{fig:figrn}. The solid line is the
trajectory described in the complex--frequency plane by the fundamental QNM
with $l=2$ corresponding to $Z_2^-$ (perturbations that reduce to the
vector-type gravitational Schwarzschild case as $Q/M\to 0$). The dashed line
is the trajectory of the fundamental QNM with $l=1$ corresponding to $Z_1^-$
(which reduces to purely electromagnetic perturbations as $Q/M\to 0$). The
Schwarzschild limit corresponds to the bottom left of each curve, and the
trajectories are described clockwise as $Q$ increases. The real part of the
frequency grows monotonically with $Q$, and the imaginary part shows an
extremum.  A striking feature is that modes of $Z_2^-$ with angular index $l$
coalesce with modes of $Z_1^-$ with index $(l-1)$ in the extremal limit. We
will elaborate on the significance of this result later in this section.
In general there are {\em two} algebraically special frequencies for RN BHs
\cite{chandraspecial}, located at
%
\be\label{RNas} \omega_{1,2}=\pm \frac{i}{2}
\frac{(l-1)l(l+1)(l+2)}{3M\mp \sqrt{9M^2+16 Q^2(l-1)(l+2)}}\,.
\ee
Using the numerical procedure described by Chandrasekhar and Detweiler
\cite{Chandrasekhar:1975qn}, Gunter computed the lowest-lying QNMs of
electromagnetic/gravitational perturbations \cite{Gunter:1979}. Extensive
comparisons of Gunter's results against WKB predictions were done by Kokkotas
and Schutz \cite{Kokkotas:1988fm}.  The continued-fraction method can be
generalized to charged BHs with relatively minor modifications
\cite{LeaverRN}, and it was used to compute numerically the weakly and highly
damped QNMs of gravitational/electromagnetic perturbations
\cite{LeaverRN,Andersson1993,Onozawa:1995vu,
  Andersson:1996xw,Berti:2004md,Berti:2003zu}.  Weakly damped modes of massive
charged fields were computed by WKB techniques in Ref.~\cite{Chang:2007zza}.
Dirac QNMs in the Kerr-Newman metric (which includes RN as a special case)
have been computed by continued fractions in Ref.~\cite{Jing:2005pk}. For
generalizations to higher-dimensional charged BHs, see
\cite{Chakrabarti:2008xz} and references therein.

\subsubsection*{The highly damped modes}

Let us focus our attention on the high-damping regime for $Z_2^-$; results for
$Z_1^-$ are similar. For a more detailed discussion of highly-damped modes,
see \cite{Berti:2003zu,Berti:2004md}.
\begin{figure}[ht]
\begin{center}
\begin{tabular}{cc}
\epsfig{file=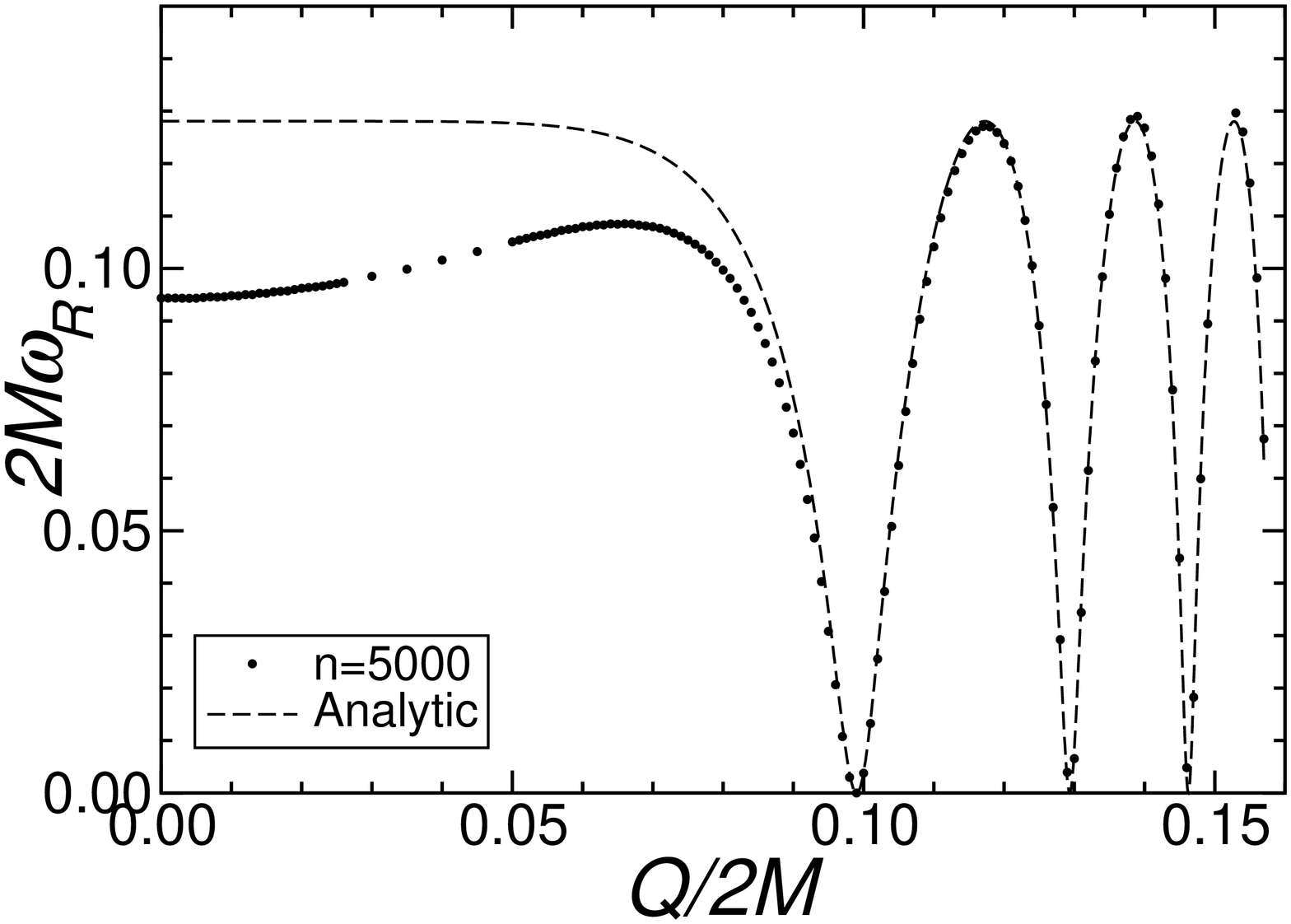,width=5.2cm,angle=-90} &
\epsfig{file=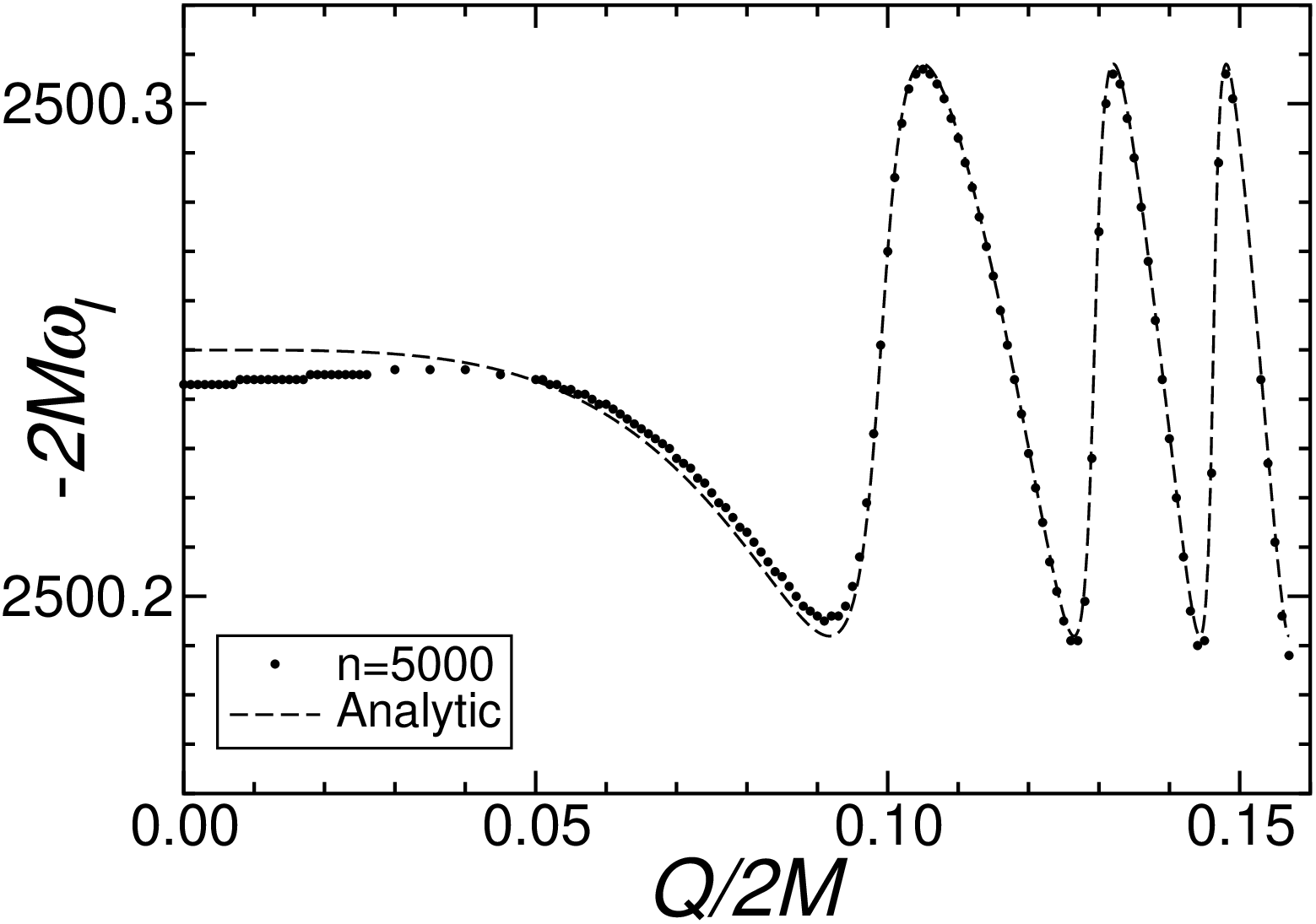,width=5.2cm,angle=-90}
\end{tabular}
\caption{Numerical and analytical predictions for the real and imaginary part
  of the RN QNM frequencies as a function of charge for
  $n=5\times10^3$.}
\label{fig:fig3}
\end{center}
\end{figure}
The modes' behavior is better understood by looking separately at the real and
imaginary parts of their frequencies as functions of charge. Numerical results
are shown in Fig.~\ref{fig:fig3}.  Frequencies and damping times oscillate as
a function of charge whenever $Q$ is larger than some $Q_{\rm crit}$, and
$Q_{\rm crit}/M$ decreases with overtone number. For moderate-to-large values
of $Q/M$, numerical data are in excellent agreement with the analytical
prediction by Motl and Neitzke
\cite{Motl:2003cd,Andersson:2003fh,Natario:2004jd}, valid for scalar and
electromagnetic-gravitational perturbations of a RN BH in the large-damping
limit:
\be\label{MNf}
e^{\omega/T}+2+3e^{(Q^4\,\omega)/(r_+^4\,T)}=0\,,
\quad 
T=\sqrt{M^2-Q^2}/\left[2\pi r_+^2\right]\,,
\ee
where $r_+=M+\sqrt{M^2-Q^2}$ is the radius and $T$ is the Hawking temperature
of the outer horizon. The complex solutions of Eq.~(\ref{MNf}) exactly overlap
with the oscillations observed in Fig.~\ref{fig:fig3} for large enough values
of $Q/M$. This agreement gives support to the asymptotic formula (\ref{MNf}),
while confirming that the numerics are still accurate for large values of $n$
and $Q/M$ \cite{Neitzke:2003mz}.  However, Eq.~(\ref{MNf}) presents us with
some striking puzzles. The predicted asymptotic RN QNM frequencies do not
reduce to the expected Schwarzschild limit of Eq.~(\ref{Mresult}): one finds
instead $\omega_R \rightarrow T \ln 5$ as $Q/M\to 0$
\cite{Motl:2003cd,Neitzke:2003mz}.  Eq.~(\ref{MNf}) should also be taken with
care in the extremal limit ($Q/M\to 1$): in this case the inner and outer
horizons coalesce, the topology of the singularities in the differential
equation changes, and the problem requires a separate analysis
\cite{Natario:2004jd}. Such an analysis shows that the asymptotic oscillation
frequency for extremal RN BHs is {\it not} given by the limit of
Eq.~(\ref{MNf}) as $Q/M\to 1$. Instead, the modes are purely damped
($\omega_R\to 0$).
An interesting classification of the solutions of Eq.~(\ref{MNf}) can be found
in Fig.~3 and Section 4.4 of \cite{Andersson:2003fh}. Besides ``spiralling''
solutions, the equation also admits {\it periodic} solutions when
$\sqrt{1-Q^2/M^2}$ is a rational number, and even {\it pure imaginary}
solutions that may not be QNMs at all. A survey of highly-damped QNM spectra
of charged BHs can be found in Ref.~\cite{Natario:2004jd}. For a complete
account of the asymptotic QNM spectrum of several perturbing fields in the RN
metric, see Ref.~\cite{Cho:2005yc}.

\subsubsection*{The eikonal limit}

The eikonal regime ($l\gg 1$) is well described by a WKB analysis.  The
lowest-order WKB approximation yields 
\be
\omega_R\sim\left (l+1/2\right )\Omega_c\,,\quad \omega_I\sim -1/2\,\Omega_c\,\sqrt{3M/r_0-4Q^2/r_0^2}\,,
\ee
where $r_c$ and $\Omega_c$ are the radius and frequency of the unstable
circular null geodesic, $r_c=\left (3M+\sqrt{9M^2-8Q^2}\right )/2$ and
$\Omega_c=\sqrt{M/r_0^3-Q^2/r_0^4}$
\cite{Kokkotas:1988fm,Andersson:1996xw}. QNMs in this regime can also be
interpreted, from a geometrical-optics point of view, as waves trapped at the
unstable circular null geodesics \cite{mashhoon,Cardoso:2008bp}.

\subsubsection*{\label{sec:rnse}Extremal Reissner-Nordstr\"om}

A direct application of Leaver's method fails in the extremal limit. In this
limit the two horizons coalesce, and the radial wave equation has irregular
singular points at the horizon and at infinity. Onozawa {\it et al.}
\cite{Onozawa:1995vu} managed to reduce the problem to a five-term recurrence
relation.  As anticipated from Fig.~\ref{fig:figrn}, the QNM spectrum for
extremal RN BHs is characterized by an isospectrality between electromagnetic
and gravitational perturbations: modes of $Z_2^-$ with angular index $l$
coalesce with modes of $Z_1^-$ with index $(l-1)$ in the extremal limit. This
has been shown to be a manifestation of supersymmetry
\cite{Onozawa:1996ba,Okamura:1997ic,Kallosh:1997ug}.

The resulting QNM spectrum \cite{Berti:2004md,Berti:2003zu,Onozawa:1995vu}
looks qualitatively similar to the Schwarzschild spectrum of
Fig.~\ref{fig:schw}.  The real part first decreases, approaching the
pure-imaginary axis as the overtone index grows. A QNM seems to be located at
$\omega=(0,-3.0479)$, while Chandrasekhar's formula (\ref{RNas}) predicts a
mode at $\omega=(0,-3)$.  The numerical method used so far becomes unreliable
for $|\omega_I|\gtrsim 10$, and better techniques will be needed to verify
analytical predictions \cite{Motl:2003cd,Natario:2004jd}.

\subsection{\label{sec:qnmkerr}Kerr}

The Kerr QNM spectrum has a rich and complex structure
\cite{Leaver:1985ax,Onozawa:1996ux,Berti:2003jh,Berti:2004md,Detweiler:1980gk}.
The most relevant feature of the spectrum is that rotation acts very much like
an external magnetic field on the energy levels of an atom, causing a {\it
  Zeeman splitting} of QNM frequencies. The determination of the QNM
frequencies is tangled to the solution of an angular equation, the
spin-weighted spheroidal harmonic equation (see Sections \ref{sec:Kerr} and
\ref{sec:Leaver}).  In the general case $a \omega \neq 0$ there are no known
closed-form solutions for the separation constant $A_{lm}$ or for the
spheroidal harmonics. However the spheroidal harmonics satisfy various
symmetry properties \cite{Berti:2005gp,Leaver:1985ax}, namely:

\noindent (i) Eigenvalues for negative and positive $m$ are related:
${}_sA_{lm}={}_sA_{l-m}^*$.

\noindent (ii) Eigenvalues for negative and positive $s$ are related:
$_{-s}A_{lm}={}_sA_{lm}+2s$.

\noindent (iii) If $\omega$ and $_{-s}A_{lm}$ correspond to a solution for
given $(s,~l,~m)$, another solution can be obtained by the following
replacements: $m\to -m$, $\omega \to -\omega^*$, $_{-s}A_{lm}\to\,
_{-s}A_{l-m}^*$.
\noindent
These symmetries are usually a source of some confusion, so we give an
explicit numerical example. For $a=0.6 M$, the methods described in
Section~\ref{sec:Leaver} yield the following eigenvalues for $s=-2$:
\beq
M\omega_{22}&=&0.49404-0.08376i \,,\quad M\omega_{2-2}=-0.49404-0.08376i \,,\\
M\omega_{22}&=&-0.31678-0.08889i \,,\,\, M\omega_{2-2}=0.31678-0.08889i \,.
\eeq

\begin{figure*}[ht]
\begin{center}
\begin{tabular}{cc}
\epsfig{file=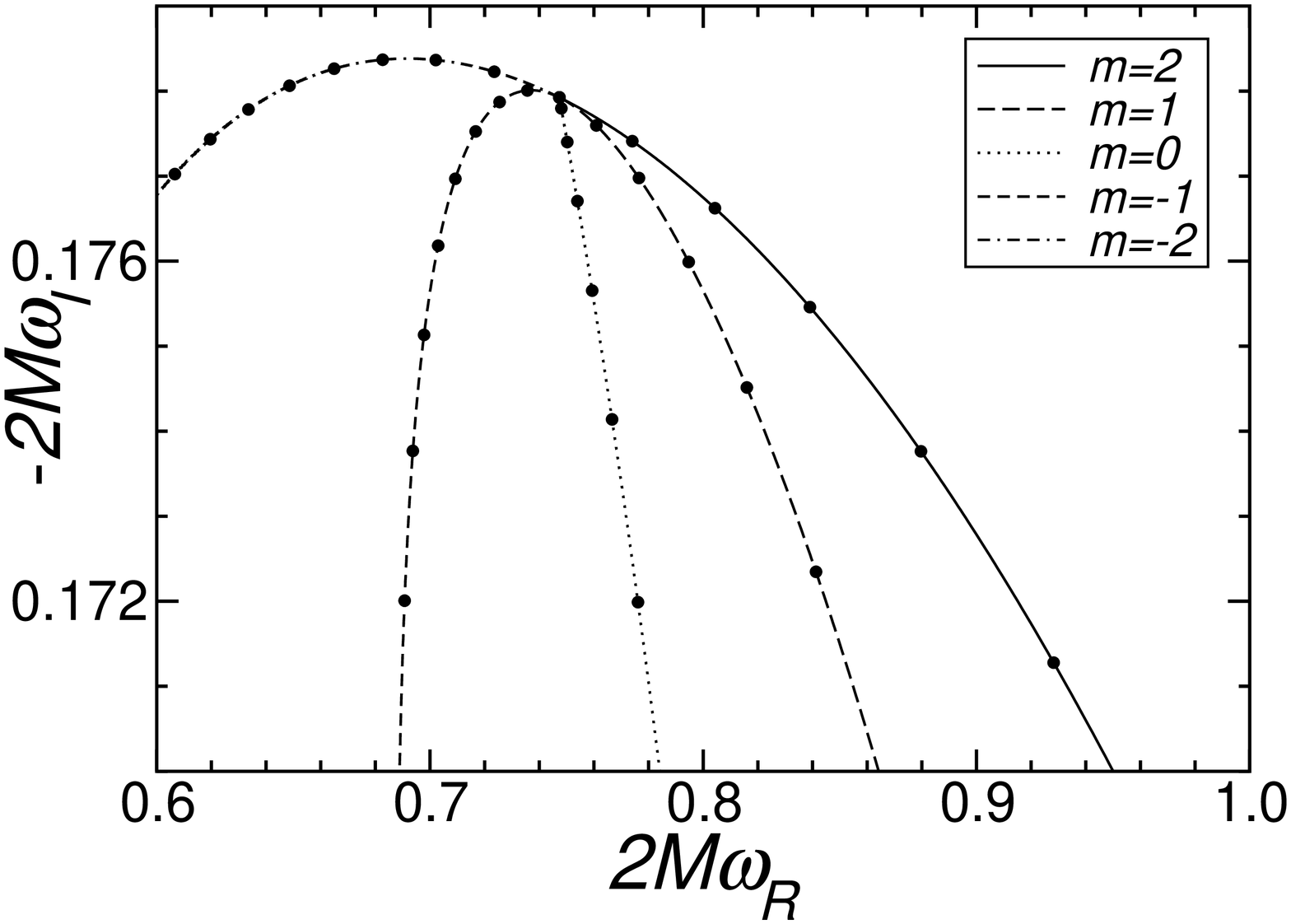,width=5.2cm,angle=-90}
\epsfig{file=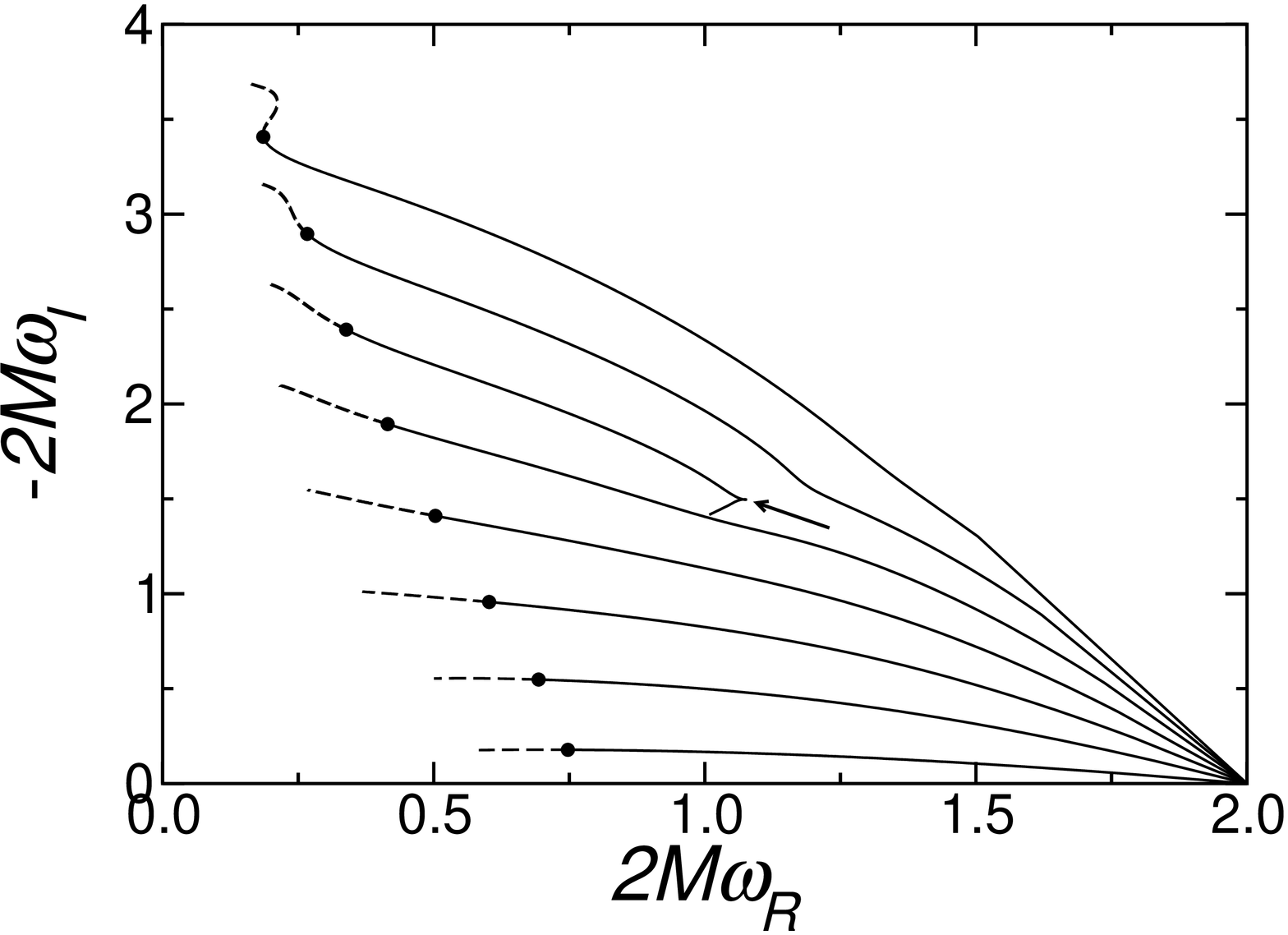,width=5.2cm,angle=-90}
\end{tabular}
\caption{Left: ``Zeeman--like'' splitting of the fundamental gravitational
  mode with $l=2$. We mark by dots the points corresponding to
  $a/M=0.0,0.1,0.2,\dots 1.0$. Right: trajectory of the first eight Kerr QNM
  frequencies with $m=2$ (solid lines) and $m=-2$ (dashed lines). Filled
  circles mark the corresponding mode in the Schwarzschild limit.  An arrow
  indicates the small loop described by the ``exceptional'' QNM with $n=6$,
  that does not tend to the critical frequency for superradiance (see also
  Figs. 3-4 in \cite{Onozawa:1996ux}). The data used to produce this figure
  (and more) are available online \cite{rdweb}. \label{fig:fig5ono}}
\end{center}
\end{figure*}

We illustrate the splitting of the fundamental gravitational QNM with $l=2$ in
the left panel of Figure \ref{fig:fig5ono}. Even though QNMs have both
positive and negative frequencies, it is customary to plot only the
positive-frequency part of the spectrum \cite{Leaver:1985ax}: in view of the
symmetry properties listed above, this yields all the necessary information.
As the rotation parameter $a/M$ increases, the branches with $m=2$ and $m=-2$
move in opposite directions, being tangent to the the branches with $m=1$ and
$m=-1$ in the limit $a/M\to 0$. For low overtone numbers and small values of
$a/M$ the rotation-induced splitting of the modes is roughly proportional to
$m$, as physical intuition would suggest.

\subsubsection*{The weakly damped modes of Kerr black holes}

In the right panel of Figure \ref{fig:fig5ono} we show the first eight
gravitational QNM frequencies with $m=2$ (solid lines) and $m=-2$ (dashed
lines).  A general feature is that almost all modes with $m>0$ cluster at the
critical frequency for superradiance, $2M\omega=m$, as $a/M\to 1$.  This fact
was first observed by Detweiler \cite{Detweiler:1980gk}, and a thorough
examination of the extremal limit can be found in
Refs.~\cite{Glampedakis:2001js,Cardoso:2004hh,Hod:2008zz}.  The mode with
$n=6$ (marked by an arrow) shows a peculiar deviation from the general trend,
illustrating the fact that some positive-$m$ modes do {\it not} tend to this
critical frequency in the extremal limit.

\begin{figure*}[ht]
\begin{center}
\begin{tabular}{cc}
\epsfig{file=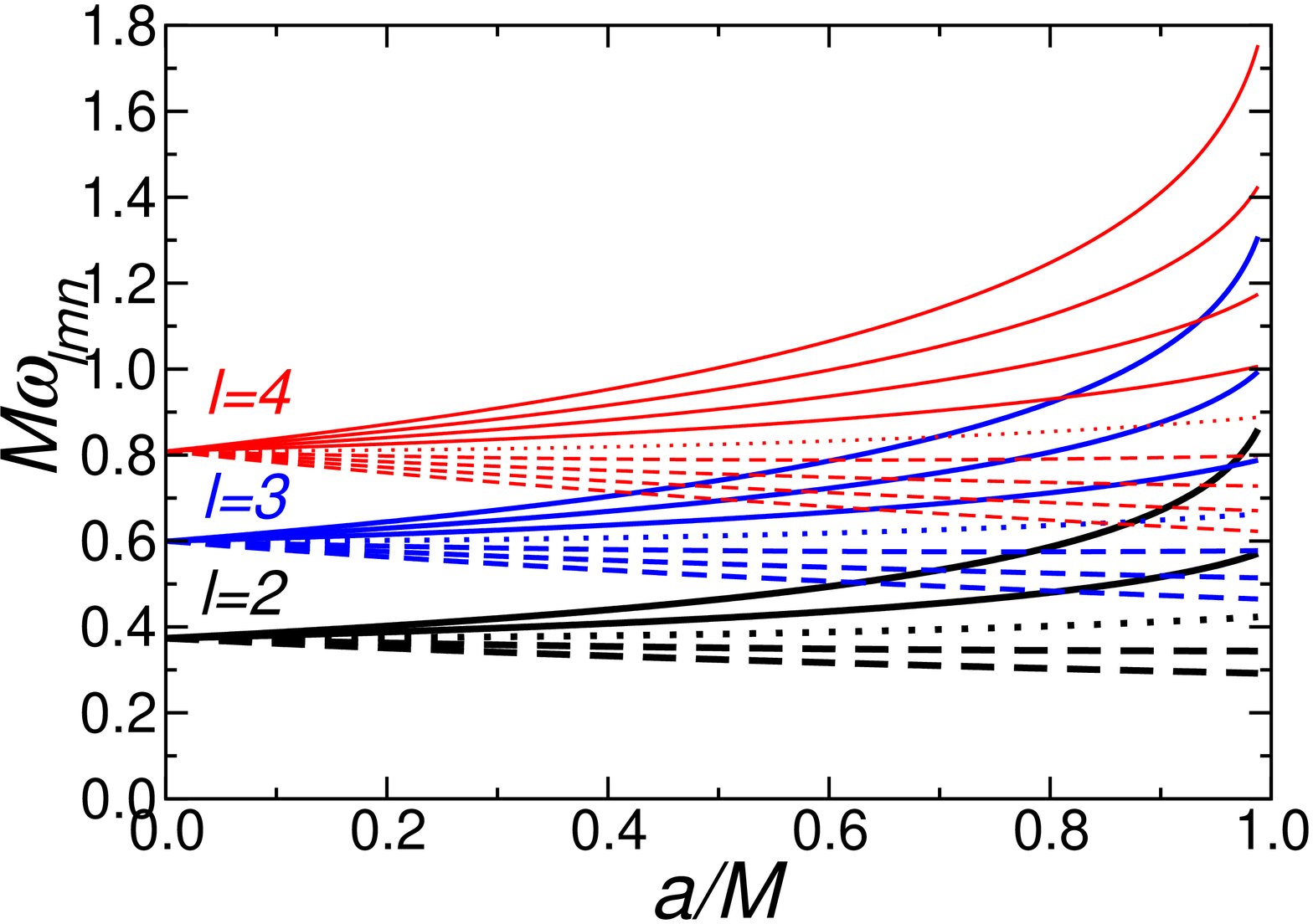,width=5.2cm,angle=-90}&
\epsfig{file=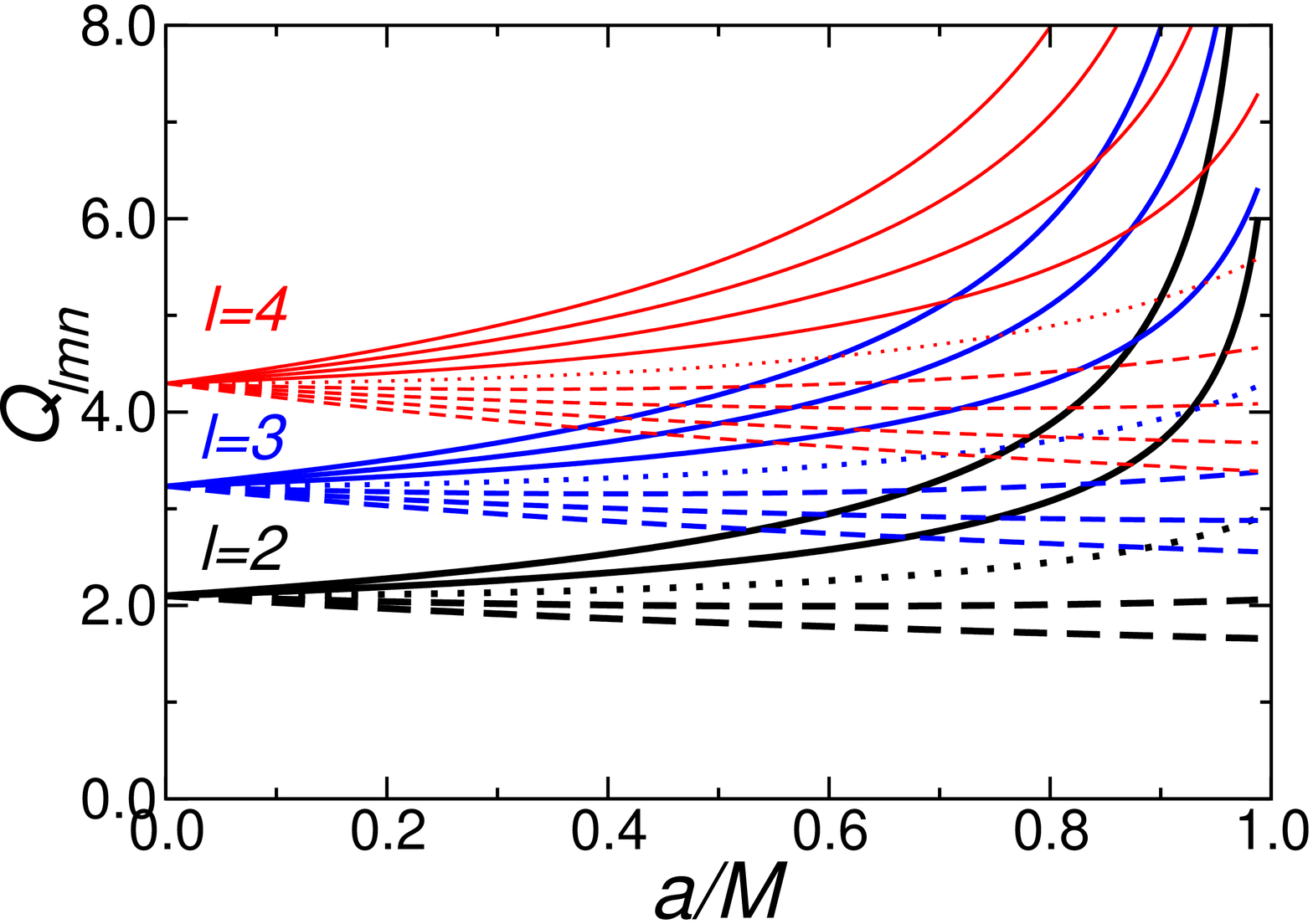,width=5.2cm,angle=-90}\\
\end{tabular}
\caption{Frequencies and quality factors for the fundamental modes with
  $l=2,~3,~4$ and different values of $m$. Solid lines refer to $m = l, .., l$
  (from top to bottom), the dotted line to m = 0, and dashed lines refer to $m
  = -1, ..,-l$ (from top to bottom). Quality factors for the higher overtones
  are lower than the ones we display here.}
\label{fig:fQKerr}
\end{center}
\end{figure*}

For gravitational wave detection we are mostly interested in the frequency and
quality factor of the dominant modes, which determine whether the mode lies in
the sensitive frequency band of a given detector and the number of observable
cycles.  Figure \ref{fig:fQKerr} shows these quantities for QNMs with $l<5$.
Improving on previous results \cite{Echeverria:1989hg,Fryer:2001zw},
Ref.~\cite{Berti:2005ys} presented accurate fits for the first three overtones
with $l=2,3,4$ and all values of $m$, matching the numerical results to within
$5\%$ or better over a range of $a/M \in [0,0.99]$ (see Tables VIII-X in
Ref.~\cite{Berti:2005ys} and the numerical data available online
\cite{rdweb}). For instance, defining ${\hat b}\equiv 1-a/M$, the frequency
$\omega_{lm}=\omega_R$ and quality factor $Q_{lm}\equiv \omega_R/(2\omega_I)$
of the fundamental $l=m=2$ and $l=2\,,m=0$ modes are
\beq
M\omega_{22} &\simeq& 1.5251-1.1568 \,{\hat b}^{0.1292}\,,\quad Q_{22}\simeq 0.700+1.4187\,{\hat b}^{0.4990}\,,\\
M\omega_{20} &\simeq& 0.4437-0.0739\,{\hat b}^{0.3350}\,,\quad Q_{20}\simeq 4.000-1.9550\,{\hat b}^{0.1420}\,,
\eeq
%
\subsubsection*{\label{highdamp}The highly damped modes}

The intermediate- and large-damping regime of the QNM spectrum of Kerr BHs is
even more puzzling than the RN spectrum. The main technical difficulty in
pushing the calculation to higher damping is that Leaver's approach requires
the {\it simultaneous} solution of the radial and angular continued fraction
conditions.  For mode order $n\gtrsim 50$ the method becomes increasingly
unreliable. A way around this ``coupling problem'' is to study the asymptotic
behavior of the angular equation as $|a\omega|\simeq |a \omega_I|\to \infty$
\cite{Berti:2004um,Berti:2005gp}. The leading-order behavior of the separation
constant $_sA_{lm}(a\omega)$ when $\omega\simeq i \omega_I$ and $|a\omega|\to
\infty$ is
\begin{equation}
_{s}A_{lm}=i(2L+1) a\omega+ {\cal O}(1) \,\,,\, |a\omega| \rightarrow \infty \,,
\label{alm}
\end{equation}
with $L=l-|m|$ for $|m| \geq |s|$ and $L=l-|s|$ otherwise
\cite{Berti:2004um,Berti:2005gp}.
By replacing this analytic expansion in the radial continued fraction one
effectively decouples the angular and radial continued fractions, and the
calculation of QNM frequencies with $n\gg 50$ can be performed by solving {\it
  only} the radial continued fraction. It turns out that in the highly damped
limit $\omega_R$ is independent of $(s\,,l)$ and proportional to $m$:
\be
2M\omega_R=m \varpi(a/M)\,. \label{varpikerr}
\ee

\begin{figure*}[ht]
\begin{center}
\begin{tabular}{cc}
\epsfig{file=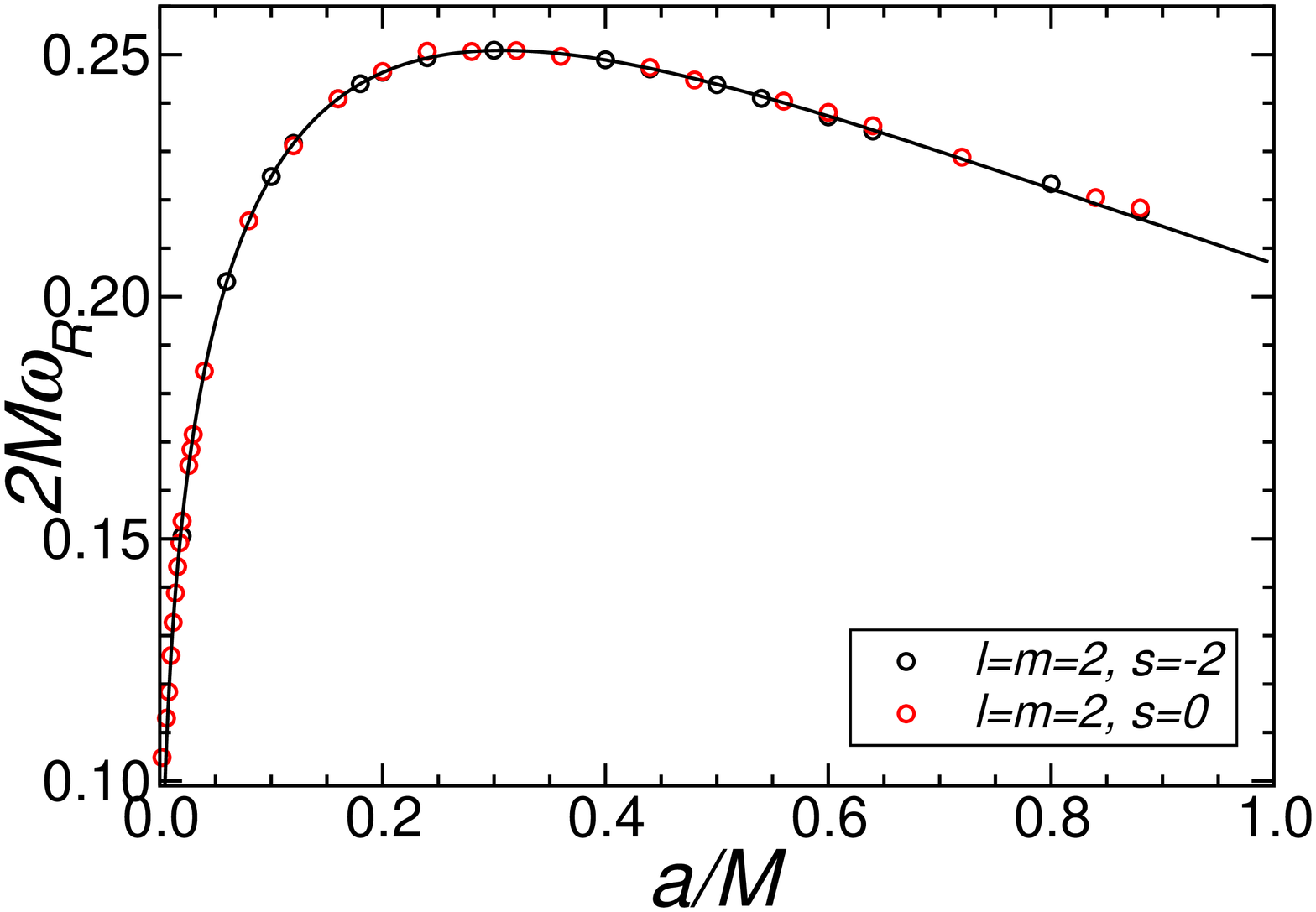,width=5.2cm,angle=-90} &
\epsfig{file=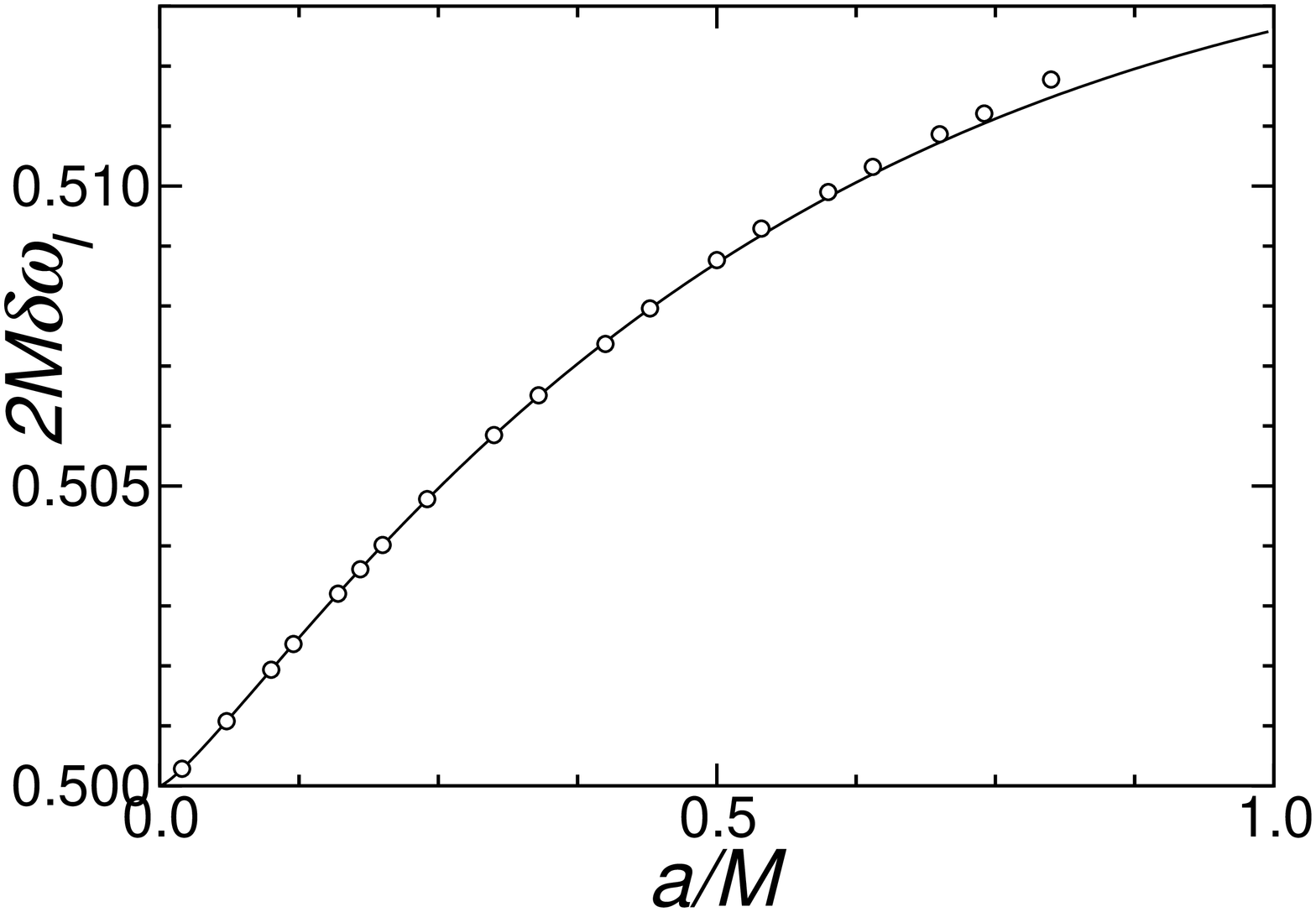,width=5.2cm,angle=-90}
\end{tabular}
\caption{Left: asymptotic real part $2M\omega_R=2\varpi(a/M)$ of the $l=m=2$
  gravitational and scalar QNM frequencies extrapolated from numerical data.
  Points refer to numerical results, the line is the analytical prediction.
  Results are independent of $l$, $s$ and the sign of $m$. Right: same for
  imaginary part. \label{fig:fig13}}
\end{center}
\end{figure*}
Numerical results for $\varpi(a/M)$ are shown in Fig.~\ref{fig:fig13}.  These
numerical results have been confirmed by recent analytical calculations in an
impressive {\it tour de force}
\cite{Hod:2005ha,Keshet:2007nv,Keshet:2007be,Kao:2008sv}.  The final result
can be implicitly expressed as a contour integral, which in turn can be
expressed as a sum of elliptic integrals \cite{Keshet:2007nv,Kao:2008sv}.  The
relevant equations are summarized in \ref{app:highdampinglimitKerr}, and
compared against our own numerics in Fig.~\ref{fig:fig13}.  The agreement is
remarkable, given how involved the numerical and analytical calculations
are. In the Schwarzschild limit ($a/M\to 0$) it can be shown that $\varpi
(a)\simeq 0.30634 \left(a/M\right )^{1/3}$ and $2M \delta \omega_I\simeq 1/2$.
A good fit to the analytical predictions for the real part (accurate to within
$0.8\%$ in the entire range) is
\be
\varpi=0.307\left(a/M\right )^{1/3}-0.308\,a/M+0.156\left(a/M\right )^{2}-0.052\left(a/M\right )^{3}\,.
\ee
In the extremal limit ($a/M \to 1$) the analytical results imply $\varpi
\simeq 0.10341\,,\,\,2M \delta \omega_I\simeq 0.51260$.  For any $a/M$, the
imaginary part $\omega_I$ grows without bound. The spacing between consecutive
modes $\delta \omega_I$ is not simply given by $2\pi T$, but it is a
monotonically increasing function of $a/M$. A power fit in $a/M$ of the
numerical results yields \cite{Berti:2004um}:
\be
2M\delta \omega_I=1/2+0.0219 a/M-0.0089\left(a/M\right)^2\,. 
\label{deltakerr}
\ee
The agreement between the analytical predictions for the mode spacing and the
fit of Eq.~(\ref{deltakerr}) is better than $0.1\%$ in the entire range of
$a/M$. A generalization of these asymptotic results to higher-dimensional
rotating BHs can be found in Ref.~\cite{Kao:2008sv}.

\subsubsection*{The eikonal limit}

The eikonal limit of Kerr QNMs is not yet fully understood. Partial results
concern $l=\pm m$ modes, for which \cite{mashhoon,Berti:2005ys,Cardoso:2008bp}
\be
\omega=\pm m\Omega_{c}-(n+1/2)\left|\Omega_{c}\right
|\sqrt{3M/r_c}\,\delta^{-1}\,,\quad 
(l=|m|\rightarrow \infty)\,,
\ee
where $r_{c}\,,\,\Omega_{c}$ are counter-rotating or co-rotating radius and
orbital frequencies at the unstable circular null geodesics, and
$2\delta=r_{c}(r_{c}-M)/(r_c^2-2Mr_c+a^2)$. Again, this result can be
expressed as $\omega\sim l\Omega_c-i(n+1/2)/\tau$, with $\tau$ the typical
instability timescale of the unstable circular null geodesic
\cite{mashhoon,Berti:2005ys,Cardoso:2008bp}. To our knowledge, a simple
geometrical optics interpretation is still lacking for modes with $l\neq |m|$.

\subsection{Kerr-Newman}

General BH solutions of the Einstein-Maxwell system are described by the
Kerr-Newman metric. For this metric the Klein-Gordon equation and the Dirac
equation are still separable \cite{Page:1976jj}, so QNMs can be computed using
the same general methods that apply to Kerr BHs \cite{Jing:2005pk}. The scalar
spectrum was analyzed by Berti and Kokkotas \cite{Berti:2005eb}, who showed in
particular that the eikonal limit can still be understood in terms of unstable
circular null geodesics.

Unfortunately, studies of the interplay of electromagnetic and gravitational
fields in the Kerr-Newman metric are plagued by a major technical difficulty:
to date, all attempts to decouple the electromagnetic and gravitational
perturbations have failed (see e.g.~Section 111 of Ref.~\cite{MTB}).
Approximations to gravitational/electromagnetic perturbations of the
Kerr-Newman geometry either keep the geometry fixed and perturb the electric
field, or (more interestingly) keep the electric field fixed and perturb the
geometry. This approach should be appropriate for values of the charge $Q$ at
most as large as the perturbations of the spacetime metric. QNMs for
gravitational and electromagnetic perturbations in this approximation scheme
were computed in Refs.~\cite{Kokkotas:1993ef,Berti:2005eb}, but a solution of
the general problem is highly desirable, and it could shed light on many open
problems in BH physics.

\subsection{Higher dimensional Schwarzschild-Tangherlini black holes}
Here we briefly discuss the QNM spectrum of the higher-dimensional analogs of
the Schwarzschild solution, known as Schwarzschild-Tangherlini
BHs. Electromagnetic perturbations of these BHs were considered by Crispino,
Higuchi and Matsas \cite{Crispino:2000jx}. The elegant work by Kodama and
Ishibashi \cite{Kodama:2003jz,Ishibashi:2003ap,Kodama:2003kk} laid the
foundations for the analysis of higher-dimensional RN BHs.

\subsubsection*{Weakly damped modes}
The lowest lying QNMs of a $d-$dimensional Schwarzschild-Tangherlini BH were
computed in a WKB approximation in
Refs.~\cite{Konoplya:2003ii,Konoplya:2003dd,Berti:2003si,Cardoso:2002pa}.
Leaver's method can be generalized to these higher-dimensional BHs
\cite{Cardoso:2003vt,Cardoso:2003qd}, although the number of terms in the
recurrence relation rapidly grows with $d$. For instance, in $d=5$ the
recurrence relation for vector-gravitational and tensor-gravitational
perturbations has four terms, while for scalar-type gravitational QNM
frequencies it has eight terms \cite{Cardoso:2003qd}.  A naive application of
Leaver's method breaks down for $d>9$. For large $d$ more and more
singularities, spaced uniformly on the circle $|r| = r_h$ (where $r_h$ is the
horizon radius) approach the horizon at $r = r_h$. A solution satisfying the
outgoing wave boundary condition at the horizon must be continued to some mid
point, and only then can the continued fraction condition be applied
\cite{Rostworowski:2006bp}. Alternative calculations of weakly damped modes
for scalar and gravitational perturbations in $d$ dimensions make use of time
evolutions \cite{Konoplya:2007jv}.

\subsubsection*{Highly-damped modes} In the large-damping limit, the leading-order
result of the monodromy calculation, Eq.~(\ref{Mresult}), generalizes to the
$d$-dimensional Schwarzschild-Tangherlini metric.  This was first suggested in
Ref.~\cite{Motl:2003cd}, and then it was explicitly shown by Birmingham
\cite{Birmingham:2003rf} (see also \cite{Kunstatter:2002pj}). The perturbative
technique of Ref.~\cite{Musiri:2003bv} has been extended to bosonic fields in
higher-dimensional Schwarzschild-Tangherlini BHs, with the result
\cite{Cardoso:2003vt}
\be  \label{highdampingTangherlini}
\omega=T \ln (1+2\cos\pi j)+(2n+1)\pi i T +k_d\omega_I^{-(d-3)/(d-2)}\,.
\ee
Here $j=0$ for scalar fields and tensor-type gravitational perturbations,
$j=2$ for vector-type gravitational perturbations and
$j=2/(d-2)\,,\,2-2/(d-2)$ for vector- and scalar-type electromagnetic
perturbations, respectively. The coefficient $k_d$ can be determined
analytically for given values of $d$ and $j$ \cite{Cardoso:2003vt}. For
electromagnetic perturbations in $d=5$, Eq.~(\ref{highdampingTangherlini}) is
singular: this either means that there are no asymptotic QNM frequencies at
all (a possibility first suggested in \cite{Motl:2003cd}) or that the
asymptotic frequency is zero. Numerical results support the latter possibility
\cite{Cardoso:2003vt}.

The result above illustrates a general feature of the high-damping regime,
which concerns the damping itself. It was shown by several authors
\cite{Medved:2003rga,Medved:2003pr,Padmanabhan:2003fx} that for spacetimes
with a single horizon, for all values of $l$ one has
\be
\omega_I=-2\pi in /T\,,\quad n\rightarrow \infty\,,
\label{imwqnmgen}
\ee
where $T$ is the Hawking temperature. General results for the oscillation
frequencies were obtained in Ref.~\cite{Das:2004db} for single-horizon
spacetimes (see also \cite{Kettner:2004aw} for BHs in two-dimensional dilaton
gravity).  An elegant unification of these results dealing with a rather
general class of wave equations is given in Ref.~\cite{Daghigh:2005ph} (but
see Ref.~\cite{Daghigh:2006xg} for words of caution on the generality of the
results). A review on the highly damped regime of QNMs for several spacetimes
can be found in Ref.~\cite{Natario:2004jd}.

\subsubsection*{Eikonal limit} For $l \gg 1$, a WKB analysis of the
Schwarzschild-Tangherlini perturbation equations yields
\cite{Konoplya:2003ii,Cardoso:2008bp},
\be
\omega\,r_+=\Omega_c
\left (l+d/2-3/2-i(n+1/2)\sqrt{d-3}\right )\,.
\ee
where $\Omega_c=\sqrt{d-3}/\sqrt{d-1}\left [2/(d-1)\right ]^{\frac{1}{d-3}}$
is the orbital angular velocity at the circular null geodesic
\cite{Cardoso:2008bp}.

\subsubsection*{Infinite-dimensional limit} ($d \rightarrow \infty$) 
To our knowledge, the interesting limit where the dimensionality $d$ tends to
infinity has not been studied in much detail in the literature.  Using a WKB
analysis for scalar fields or gravitational tensor modes we find
\be
\omega \,r_+=d/2-i \kappa\,\sqrt{d/2}\,,\quad
d\rightarrow \infty\,,
\ee
with $\kappa$ a factor of order unity depending on the perturbing field
($\kappa=1,\,1/\sqrt{2}$ for scalar fields and vector gravitational
perturbations, respectively). BHs in higher dimensions are much better
resonators, with a quality factor $Q\equiv \omega_R/[2\omega_I]$ increasing as
$\sqrt{d}$. As far as we know, no numerical studies are available to confirm
this analytical prediction.

It is important to recall that black objects in higher dimensions may have
other topologies beside the spherical one. One family consists of the
so-called black strings, whose gravitational perturbations were studied by
Gregory and Laflamme \cite{Gregory:1993vy,Gregory:1994bj} and Kudoh
\cite{Kudoh:2006bp}. Higher-dimensional ``squashed'' Kaluza-Klein BHs were
considered in \cite{Ishihara:2008re}, and perturbations of brane-localized BHs
were studied, for instance, in
\cite{Kanti:2004nr,Ida:2002ez,Seahra:2004fg}. Apart from specific noteworthy
exceptions \cite{Kunduri:2006qa,Murata:2007gv,Kodama:2009rq,Kodama:2009bf},
gravitational perturbations of higher-dimensional rotating solutions are yet
to be understood. For a review on higher-dimensional BHs, see
Ref.~\cite{Emparan:2008eg}.

\section{\label{sec:asAdS}The spectrum of asymptotically anti-de Sitter black holes}

\subsection{\label{sec:sads}Schwarzschild anti-de Sitter black holes}

\begin{figure*}[ht]
\begin{center}
\epsfig{file=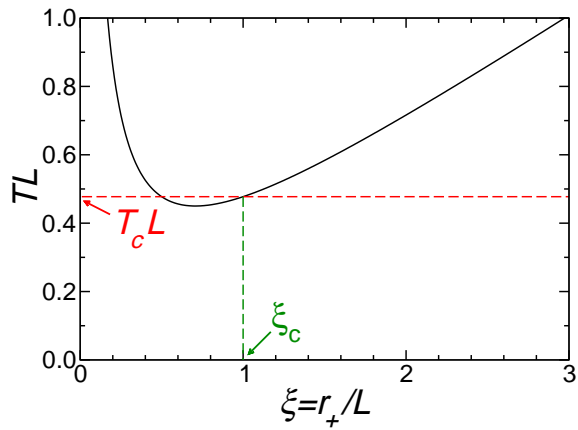,width=6.5cm,angle=-90}
\caption{The (dimensionless) Hawking temperature $TL$ as a function of the
  (dimensionless) horizon radius $r_+/L$ for asymptotically AdS BHs (shown
  here for $d=5$). The horizontal line is the critical temperature of the
  Hawking-Page phase transition.
  \label{fig:hpf}}
\end{center}
\end{figure*}

The metric (\ref{lineelementads}) of an uncharged static BH in an
(asymptotically) AdS spacetime has two parameters, $r_0$ and $L$, related,
respectively, to the mass of the BH and the cosmological constant.  The
horizon radius $r_+$ is the largest root of the equation $f(r)=0$.  The
Hawking temperature associated with the metric (\ref{lineelementads}) is
\be
T = \left[d-3 +(d-1)\xi^2\right]/\left(4\pi \xi L\right)\,,
\ee
where $\xi=r_+/L$. For $d\geq 4$, this function has a characteristic minimum
at $\xi_{\rm min} = \sqrt{(d-3)/(d-1)}$ with $T_{\rm min} =
\sqrt{(d-1)(d-3)}/2\pi L$ (see Fig.~\ref{fig:hpf}).

The specific heat of BHs with $\xi<\xi_{\rm min}$ ($\xi>\xi_{\rm min}$) is
negative (positive). For $T<T_{\rm min}$, no BH solution exists, and the state
with minimal free energy is ``thermal AdS'' (Euclidean AdS$/\mathbb{Z}$).  For
$T>T_{\rm min}$, there are two solutions with the same temperature, ``small''
and ``large'' BHs. Small BHs have negative specific heat and are
thermodynamically unstable.
Large BHs exist in equilibrium with the heat bath. Moreover, for $\xi<1$, the
minimum of the free energy still corresponds to thermal AdS, even for
$T>T_{\rm min}$.  
Thus the stable ground state is given by large BHs with $\xi>1$ (and
$T>T_{c}=(d-2)/2\pi L$). The (first order) transition between thermal AdS and
a large AdS BH is known as the Hawking-Page phase transition
\cite{Hawking:1982dh}.  As shown by Witten \cite{Witten:1998zw}, in the
gauge-gravity duality this transition corresponds to a first order
deconfinement transition in a dual thermal strongly coupled gauge theory on a
sphere of radius $L$.  These considerations can be extended to charged and
rotating AdS BHs (see e.g.~\cite{Chamblin:1999tk,Yamada:2006rx} and references
therein).  The role of small AdS BHs in the gauge-gravity duality is not well
understood (see, however, Ref.~\cite{Asplund:2008xd}).  The limit of extra
large BHs with $r_+/L \to \infty$ in the appropriate scaling leads to black
branes dual to strongly coupled gauge theories in flat space (see Section
\ref{sec:holography}).

BHs in AdS in the context of the gauge-string duality attract considerable
interest because they serve as a good laboratory for studying the most acute
problems of a theory of gravity: the information loss paradox, BH
singularities and some aspects of quantum gravity. Although we will not
venture into this fascinating field in this review, in Section
\ref{sec:holography} we mention some of these topics which involve the use of
QNMs.

Another peculiar feature of asymptotically AdS space is the ``active role''
played by its boundary. In AdS, null geodesics reach the boundary in finite
coordinate time. One thus often refers to an asymptotically AdS space as a
box, having in mind that AdS boundary conditions directly affect the bulk
physics \cite{Wald:1980jn,Ishibashi:2003jd,Ishibashi:2004wx}.  This should be
contrasted with the asymptotically flat case, where the only physically
relevant choice for the boundary conditions of the bulk fluctuations
corresponds to outgoing waves at spatial infinity.  In the gauge-gravity
duality, the choice of the AdS boundary conditions is dictated by a
holographic prescription
\cite{Son:2002sd,Herzog:2002pc,Skenderis:2008dh,Skenderis:2008dg}, see Section
\ref{sec:holography}.

QNMs in asymptotically AdS backgrounds were first considered by Chan and Mann
for a conformally coupled scalar field \cite{Chan:1996yk} (see also
Ref. \cite{Chan:1999sc}).  Subsequent interest was strongly motivated by the
development of the AdS/CFT correspondence.  An interesting albeit somewhat
qualitative discussion of thermalization in AdS/CFT first mentioning QNMs
appeared in Ref.~\cite{KalyanaRama:1999zj} (see also the important work of
Danielsson, Keski-Vakkuri and Kruczenski \cite{Danielsson:1999zt,
  Danielsson:1999fa}).
Horowitz and Hubeny \cite{Horowitz:1999jd} explicitly pointed out the
fundamental link between QNMs of a large AdS BH, which describe the
background's relaxation into a final state, and the dual field theory, where
they describe the approach to thermal equilibrium. Ref.~\cite{Horowitz:1999jd}
computed the QNMs of a minimally coupled massless scalar in intermediate or
large AdS BHs for dimensions $d=4,5,7$.  After this seminal work, a series of
studies in four and higher dimensions led to a deeper understanding of QNMs in
asymptotically AdS backgrounds.

The following is a brief overview of the QNMs of the four and
higher-dimensional SAdS geometries. For more details we refer to the original
works (see \cite{Horowitz:1999jd,Natario:2004jd,Cardoso:2001bb,Cardoso:2001vs,
  Konoplya:2002zu,Cardoso:2003cj,Berti:2003ud,Musiri:2003rv,Musiri:2003rs,
  Cardoso:2004up,Musiri:2005ev,Miranda:2005qx,Friess:2006kw,Govindarajan:2000vq,Zhu:2001vi,Cardoso:2002cf,Wang:2001tk,
  Konoplya:2003dd,Siopsis:2004up,Siopsis:2004as,LopezOrtega:2006vn,Siopsis:2007wn}
and references therein). The results below focus mostly on the large BH regime
$r_+/L\gg 1$ and on Dirichlet boundary conditions, which are more relevant for
holography.  Other boundary conditions were investigated in
Refs.~\cite{Moss:2001ga,Michalogiorgakis:2006jc,Bakas:2008gz,Bakas:2008zg}.

\subsubsection*{The weakly damped modes of large black holes}

The fundamental QNM frequencies for scalar field perturbations were first
computed by Horowitz and Hubeny \cite{Horowitz:1999jd} for intermediate and
large BHs. In Refs.~\cite{Cardoso:2001bb,Cardoso:2003cj} the analysis was
extended to electromagnetic and gravitational perturbations. These works
considered only $d=4$; the $d=5$ case was analyzed in
Ref.~\cite{Friess:2006kw}, and half-integer spins were considered in
Refs.~\cite{Giammatteo:2004wp,Jing:2005ux}. Approximate analytical solutions
have been discussed in Refs.~\cite{Musiri:2003rv,Musiri:2003rs,Musiri:2005ev}
with particular emphasis on the $d=5$ geometry.

\begin{figure*}[ht]
\begin{center}
\epsfig{file=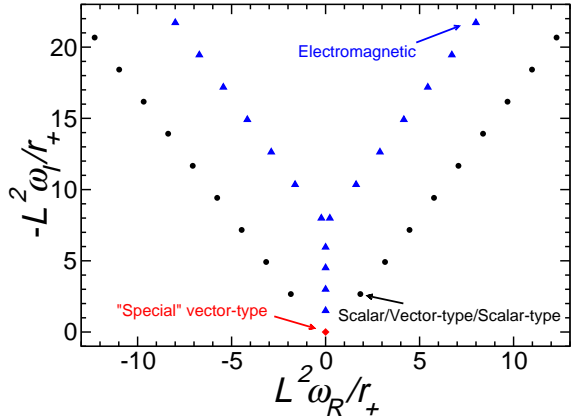,width=6.5cm,angle=-90}
\caption{QNM frequencies for scalar ($s=0$), electromagnetic ($s=1$) and
  gravitational ($s=2$) perturbations of large SAdS BH, computed for
  $r_+/L=100$ and $l=s$. In the large BH regime the frequencies scale with
  $r_+/L$ and are basically independent of $l$ for $l \ll r_+/L$. Furthermore,
  scalar-field, tensor-type and vector-type gravitational perturbations are
  nearly isospectral, except for a special mode belonging to the vector-type
  family and marked by a red diamond. For further details see
  \cite{Cardoso:2003cj,Berti:2003ud}.\label{fig:sads}}
\end{center}
\end{figure*}
The weakly-damped QNM spectrum of a large ($r_+/L=100$) SAdS BH in $d=4$ is
shown in Fig.~\ref{fig:sads}. Usually the QNM frequencies scale with the
horizon radius, $\omega \propto r_+/L^2$
\cite{Horowitz:1999jd,Cardoso:2003cj}, so the modes of {\em any} large BH can
be obtained by rescaling appropriately the numbers in the figure. The
exception to this rule is marked by a red diamond and will be discussed in
more detail below.  The frequencies of different perturbations are very
similar: scalar perturbations and scalar-type gravitational perturbations have
basically the same spectra, while vector-type gravitational perturbations are
displaced by one overtone relative to these two \cite{Cardoso:2003cj}.
The QNM frequencies are practically independent of $l$ for $l \ll r_+/L$; the
large-$l$ limit is discussed below. The QNM spectrum for electromagnetic
perturbations has a peculiar structure
\cite{Cardoso:2001bb,Berti:2003ud,Cardoso:2003cj,Miranda:2008vb,Myung:2008pr}:
the real part of some modes asymptotes to zero when $r_+/L \to \infty$, and for
the dominant mode $\omega_I L^2/r_+ \to -1.5$.

The fundamental QNM for vector-type gravitational perturbations is extremely
long-lived when compared to all other modes of other kinds of
perturbations. This mode has an interesting interpretation in the
gauge/gravity duality, discussed in Section \ref{sec:holography}. The
timescale of this long-lived mode is proportional to $r_+$, and the mode
itself is easily computed numerically by a straightforward application of the
series solution \cite{Cardoso:2001bb,Cardoso:2003cj}. In a general
$d-$dimensional SAdS geometry it is well described by
\be 
r_+\, \omega^{\rm vector-type}_{n=0}=-i(l-1)(l+d-2)/(d-1)\,,\quad
r_+/L\gg1\,. 
\label{longlived}
\ee
This was first observed numerically in
Refs.~\cite{Cardoso:2001bb,Cardoso:2003cj} for $d=4$, and in
Ref.~\cite{Friess:2006kw} for $d=5$. The four-dimensional result was later
confirmed analytically by Miranda and Zanchin \cite{Miranda:2005qx}.
Eq.~(\ref{longlived}) for general $d$ was derived by Siopsis
\cite{Siopsis:2007wn}, and we checked that it agrees with numerical results
for $d=4$ up to $d=8$. In $d=4$ and for $r_+/L\gg l$, corrections to
Eq.~(\ref{longlived}) can be found:
\be 
r_+\, \omega^{\rm vector-type}_{n=0}=
-i(l-1)(l+2)/3-0.0288\,i\,l^4L^2/r_+^2\,.
\label{longlived2}
\ee
We estimate the uncertainty in the $l^4$ term to be about $5\%$.  Analytical
calculations of these corrections have been done in the $r_+ \to \infty$ limit
\cite{Kovtun:2005ev,Natsuume:2007ty,Natsuume:2008iy} and are consistent with
the numerical results above after proper identifications. In this limit the
geometry becomes that of a black brane and the angular wavefunctions $Y_{lm}$
are replaced by $e^{i{\vec p}{\vec x}}$. For large $l,p$ one has the
correspondence $Lp=l$ \cite{Festuccia:2008zx}. In particular, it is found that
in the large $l/L$ limit,
\beq
\omega\,r_+&=&-\frac{il^2}{3}-iL^2l^4\frac{9-9\log3+\sqrt{3}\,\pi}{162r_+^2}\sim-\frac{il^2}{3}-\frac{0.0281L^2l^4}{r_+^2}\,,
\eeq
in quite good agreement with the numerical fits. For similar high-order
analytical corrections in higher-dimensional AdS backgrounds, see
Refs.~\cite{Kovtun:2005ev,Natsuume:2007ty,Natsuume:2008iy}.

If Dirichlet boundary conditions are imposed at infinity, the scalar-type
gravitational sector does not have such a long-lived mode.  It was suggested
that the preferred boundary conditions in the AdS/CFT framework are of Robin
type, and in particular that the perturbations should not deform the metric on
the AdS boundary \cite{Friess:2006kw,Michalogiorgakis:2006jc}.  Using Robin
conditions, a long-lived mode for gravitational perturbations was discovered
in Refs.~\cite{Friess:2006kw,Michalogiorgakis:2006jc,Siopsis:2007wn}.

\subsubsection*{The weakly damped modes of small black holes}

\begin{figure*}[ht]
\begin{center}
\begin{tabular}{cc}
\epsfig{file=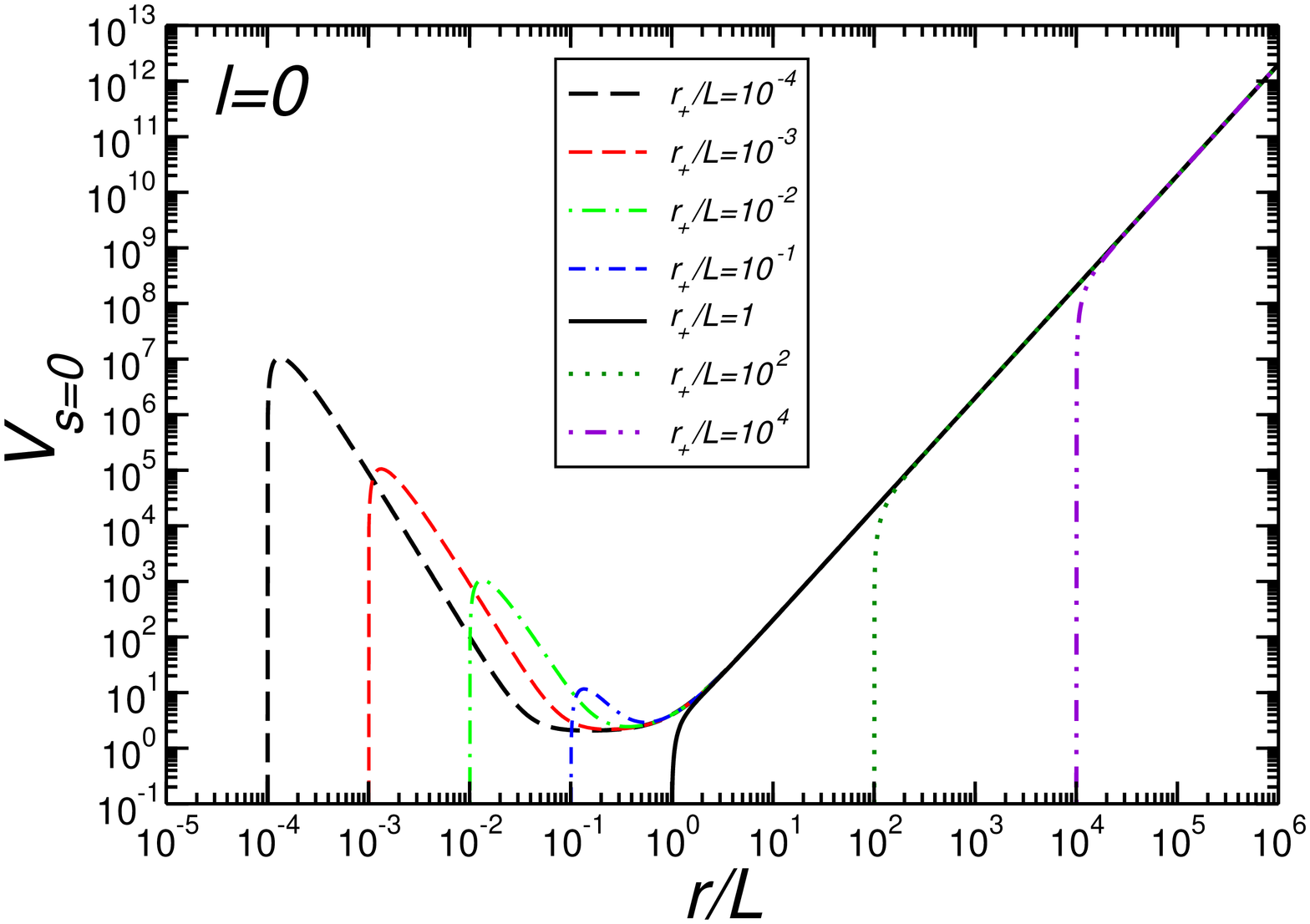,width=5.2cm,angle=-90} &
\epsfig{file=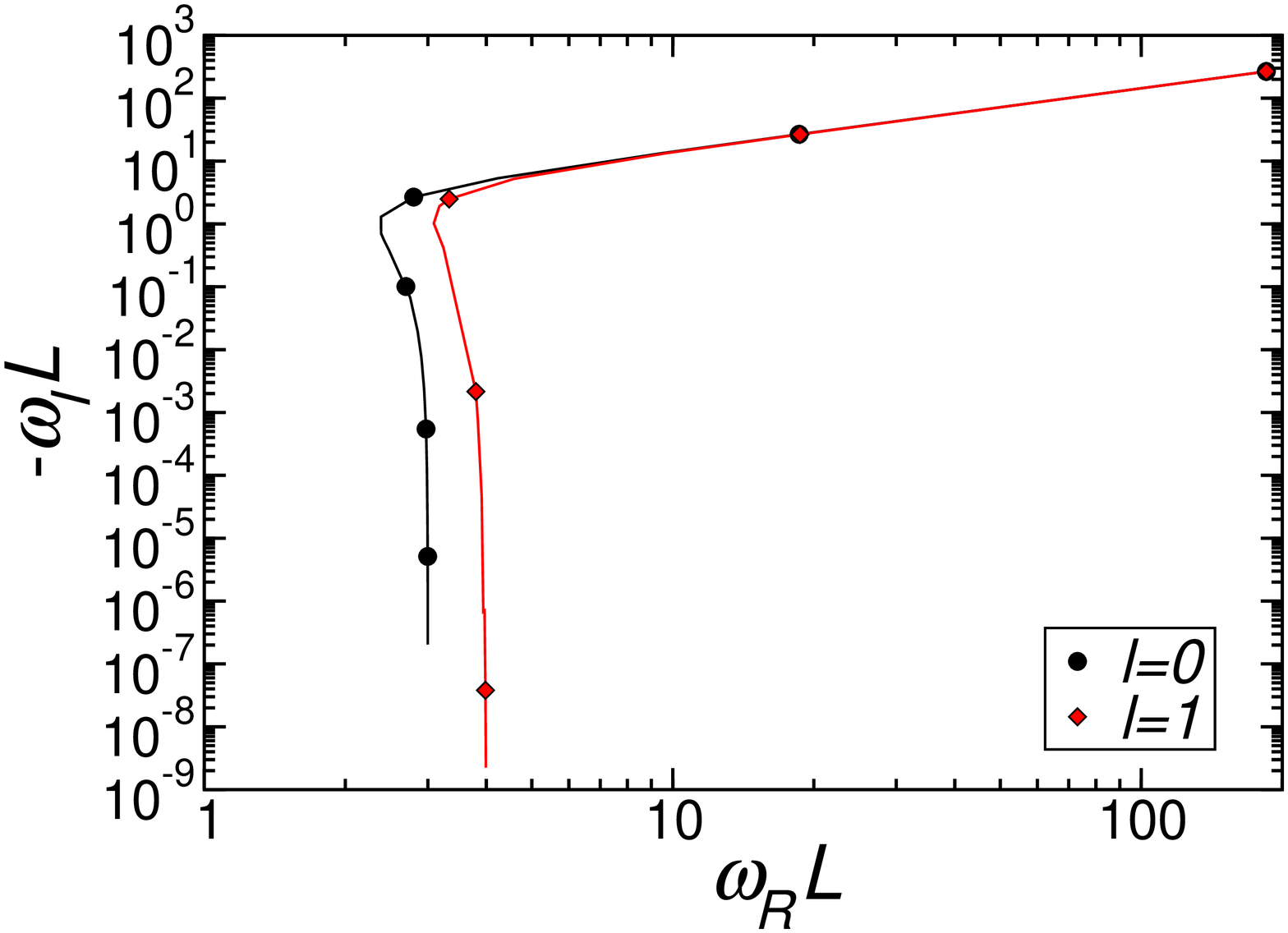,width=5.2cm,angle=-90}
\end{tabular}
\caption{Left: Potential for scalar field propagation of $l=0$ modes in a SAdS
  background, for different values of the BH size $r_+/L$. A local maximum
  (and a potential well) only exist for small BHs. Right: Track described by
  the fundamental scalar field QNMs with $l=0$ and $l=1$ as we vary the BH
  size $r_+/L$. Counterclockwise starting from the top-right of the figure we
  marked the points corresponding to different decades in $r_+/L$
  ($r_+/L=10^2,10^1,10^0,10^{-1},\dots$). Modes with different $l$ coalesce in
  the large BH regime, as long as $l\ll r_+/L$. \label{fig:sadssmall}}
\end{center}
\end{figure*}

The series solution method, which works so well for large BHs, converges very
slowly for small BHs \cite{Zhu:2001vi,Konoplya:2002zu,Cardoso:2003cj}. Recent
results make use of Breit-Wigner type resonances in the scattering
cross-section to study the very small BH regime \cite{Berti:2009wx}, described
briefly in Section \ref{sec:resonance}. Small and large BHs have a very
different behavior from a QNM perspective. This is related to the
qualitatively different behavior of the potential in the two regimes, which is
shown in the left panel of Fig.~\ref{fig:sadssmall} for $s=0$: for small BHs
($r_+/L<1$) the potential develops a well capable of sustaining trapped,
long-lived modes, corresponding to quasi-stationary states in quantum
mechanics \cite{Grain:2006dg,Festuccia:2008zx,Berti:2009wx}.
In the right panel of Fig.~\ref{fig:sadssmall} we plot the QNMs of scalar
fields for different BH sizes and $l=0,~1$.  In this limit (and under the
assumption that $M \omega_R \ll 1$) it is possible to prove that
\cite{Cardoso:2004hs}
\beq 
&&\omega_R\,L\simeq l+3+2n-k_{ln}\,r_+/L\,,\label{smallwr}\\
&&\omega_I\,L\simeq -\gamma_0(l,n) \left ( l+3+2n\right )\,
\left(r_+/L\right)^{\,2l+2}/\pi \,.\label{delta} \eeq
The constant $\gamma_0(l,n)$ ($n=0,\,1,\,2,\,\dots$) and selected values of
the constants $k_{l0}$ can be found in Ref.~\cite{Berti:2009wx}. The analytic
prediction is consistent with numerical results.
Likewise, for gravitational perturbations with $r_+/L\ll 1$ one finds
\beq
&&\omega_R\,L\simeq l+2+2n-c_{ln}\,r_+/L\,,\\
&&\omega_I\,L\simeq -\gamma_2(l,n) \left (l+2+2n\right )\,\left(r_+/L\right)^{\,2l+2}\,,\label{deltas2}
\eeq
where the constants $c_{ln}$ and $\gamma_2(l,n)$ can be found in
Refs.~\cite{Cardoso:2006wa,Berti:2009wx}.
In conclusion: for scalar field and gravitational perturbations of small SAdS
BHs the damping timescale $\omega_IL\sim (r_+/L)^{2l+2}$, and the oscillation
frequencies approach the pure AdS value in the limit $r_+/L \to 0$ (see
\cite{Konoplya:2002zu,Berti:2009wx} for more details). To our knowledge,
generalizations of these results to charged, higher-dimensional and rotating
AdS BHs are still lacking.

\subsubsection*{The highly-damped modes}

The large-damping regime ($n\gg 1$) can be studied using the monodromy method
for both small and large AdS BHs \cite{Natario:2004jd,Cardoso:2004up}. For
$n\gg 1$ and $r_+/L\gg 1$ (large BHs) the QNMs are given by
\be
\frac{\omega L^2}{r_+}=(d-1)\sin \left (\frac{\pi}{d-1}\right )\,e^{\frac{i\pi}{d-1}}\left
[n+\frac{d+1}{4}-i\frac{\log2}{2\pi}\right ]\label{highdampedAdS} \,.
\ee
Results for arbitrarily sized BHs can be found in
Refs.~\cite{Natario:2004jd,Cardoso:2004up}.  Eq.~(\ref{highdampedAdS})
describes scalar fields, gravitational tensor and gravitational vector-type
perturbations. It yields $\omega L^2/r_+\approx
(1.299+2.250i)n+(0.573+0.419i)$ for $d=4$ and $\omega L^2/r_+\approx
(2+2i)n+3.22064+2.77936$ for $d=5$, in good agreement with numerical results
for $d=4$ \cite{Cardoso:2003cj} and $d=5$ \cite{Friess:2006kw}.
For large SAdS BHs the asymptotic high-damping regime is approached very
quickly, so Eq.~(\ref{highdampedAdS}) describes fairly well even weakly-damped
modes.  There is some disagreement concerning the offset for scalar-type
gravitational perturbations \cite{Natario:2004jd}.

An asymptotic analysis based on the monodromy method was recently used to
predict the existence of a new family of modes \cite{Daghigh:2008jz}. So far,
this new part of the spectrum has not been confirmed by numerical studies.  In
Ref.~\cite{Amado:2008hw} the authors discuss the interesting possibility of
describing the highly damped QNM regime in terms of geodesics of the SAdS
spacetime.

\subsubsection*{The eikonal limit of large black holes}

In asymptotically AdS spacetimes the eikonal limit is especially interesting,
since large-$l$ modes can be very long-lived
\cite{Horowitz:1999jd,Festuccia:2008zx}. A WKB analysis
\cite{Festuccia:2008zx} and numerical investigations \cite{Morgan:2009vg} show
that for scalar field perturbations, $r_+/L\gg 1$ and $l\gg 1$
\beq
\omega\,L&=&
l+ 
\Pi_n 
\left (r_+/L\right )^
{\frac{2d-2}{d+1}} l^{-\frac{d-3}{d+1}}\,,
\label{eikonalads}\\
\Pi_n &\equiv & 
\f{1}{2}\left (\sqrt{\pi}(d-1)\left [\f{d+1}{2}+2n\right ]\,
\frac{\Gamma\left(3/2+1/(d-1)\right)}
{\Gamma\left(1/(d-1)\right)}\right )
^{\frac{2d-2}{d+1}}e^{-\frac{i2\pi}{d+1}}\,.
\nn
\eeq
So large-$l$ modes are very long-lived, and they could play a prominent role
in the BH's response to generic perturbations. This is at variance with the
asymptotically flat case, where the damping timescale is roughly constant as
$l$ varies. Notice also that the scaling with the BH size differs from that of
the weakly-damped and highly-damped modes. Other types of perturbations also
display a similar qualitative behavior \cite{Morgan:2009vg}.

\subsubsection*{The eikonal limit of small black holes}

For scalar and electromagnetic perturbations of small BHs the potential for
wave propagation for $l\gg 1$ develops a minimum, potentially supporting
quasi-stationary states, i.e. long-lived modes (see Fig.~\ref{fig:sadssmall}).
Define $r_b>r_c$ to be the two real zeros (turning points) of
$\omega_R^2-p^2f/r^2=0$. Then the real part of a class of long-lived modes in
four spacetime dimensions is given by the WKB condition
\be
2\int_{r_b}^{\infty}\frac{\sqrt{r^2\omega_R^2-p^2f}}{rf}\,dr=\pi\left (2n+1+\frac{3}{2}\right)\,,
\ee
where $p=l+d/2-3/2$. Their imaginary part is given by
\be
\omega_I=\frac{\gamma \Gamma}{8\omega_R}\,,\quad
\log \Gamma=2i\int_{r_b}^{r_c}\frac{\sqrt{r^2\omega_R^2-p^2f}}{rf}\,dr\,.
\ee
The prefactor $\gamma$, not shown in Ref.~\cite{Festuccia:2008zx}, can be
obtained by standard methods as shown in Ref.~\cite{Berti:2009wx}, where these
results are also supported by numerical calculations.

\subsection{Reissner-Nordstr\"om and Kerr anti-de Sitter black holes}

The analysis of QNMs of large RNAdS BHs was performed in
Ref.~\cite{Wang:2000gsa,Wang:2000dt} for weakly damped modes of a massless
scalar field, and later extended to charged scalar fields
\cite{Konoplya:2002ky}.  Scalar field, electromagnetic and gravitational
perturbations of RNAdS BHs were analyzed and compared in
Ref.~\cite{Berti:2003ud}. Half-integer spins were studied in
Refs.~\cite{Jing:2005uy,Zhang:2006bc}. Gravitational perturbations of
higher-dimensional charged solutions were considered in
\cite{Konoplya:2008rq}, and the highly-damped regime was explored analytically
by Nat\'ario and Schiappa \cite{Natario:2004jd}.

Ref.~\cite{Berti:2003ud} pointed out some interesting facts: (i) a
near-isospectrality of different classes of perturbations holds for large BHs,
but it breaks down as $r_+/L$ decreases, i.e. in the small BH limit; (ii) the
imaginary part of the purely damped modes found for electromagnetic and
vector-type gravitational perturbations tends to zero as the charge $Q$ tends
to the extremal value $Q_{\rm ext}$, possibly pointing to a marginal
instability of extremal RNAdS BHs; (iii) for all kinds of perturbations, the
real part of the fundamental QNM frequency, $L \omega_R$, has a minimum for
$Q/Q_{\rm ext}\simeq 0.366$, followed by a maximum at $L\omega_R\simeq 0.474$.
A reanalysis of massless scalar field perturbations found that the imaginary
part of scalar QNMs also tends to zero in the extremal limit
\cite{Wang:2004bv}. However, comparing time-evolutions of the field with
results from the Horowitz-Hubeny series solution, Wang {\it et al.}
\cite{Wang:2004bv} found that the BH response turns from a standard,
oscillatory QNM-type decay below some critical value of the charge ($Q<Q_{\rm
  crit}\simeq 0.3895 Q_{\rm ext}$) to a {\em non-oscillatory} behavior
characterized by a purely imaginary QNM frequency for $Q>Q_{\rm
  crit}$. Furthermore, these authors suggested that the potential marginal
instability proposed in Ref.~\cite{Berti:2003ud} does not pose a threat for
extremal RNAdS BHs, because (at least for scalar perturbations) the asymptotic
field in the near-extremal limit is dominated by a power-law tail of the type
first analyzed by Price \cite{price}. The implications of these findings in
the context of the AdS/CFT correspondence deserve further investigation.

There are few studies of QNMs of rotating BHs in AdS backgrounds. The
formalism to handle perturbations of these spacetimes was laid down by
Chambers and Moss \cite{Chambers:1994ap}.  Giammatteo and Moss
\cite{Giammatteo:2005vu} considered axially symmetric QNMs of large Kerr-AdS
BHs.  These spacetimes behave as a BH in a box. Superradiant amplification of
incident waves at the expense of the BH's rotational energy can then produce
instabilities in the non-axisymmetric modes of small Kerr AdS BHs
\cite{Cardoso:2004nk,Cardoso:2004hs,Cardoso:2006wa,Kodama:2007sf,Aliev:2008yk,Kodama:2009rq,Murata:2008xr}:
this is an interesting example of the ``black hole bomb'' first investigated
by Press and Teukolsky \cite{BHBomb}.

\subsection{Toroidal, cylindrical and plane-symmetric anti-de Sitter black holes}

The uniqueness theorems that apply for asymptotically flat BHs (see
e.g.~\cite{Heusler:1998ua}) can be evaded when the cosmological constant is
non-zero: solutions have been found with the topology of a cylinder, of a
plane or of a doughnut
\cite{Lemos:1994fn,Lemos:1994xp,Lemos:1995cm,Mann:1996gj,Vanzo:1997gw,
  Birmingham:1998nr}. Defining $f(r)=r^2/L^2-4M/r$, these spacetimes are
described by
\be
ds^{2}= f\,dt^{2}- f^{-1}dr^{2}-r^2\,dz^{2}-r^{2}d\phi^{2}.
\ee
The range of the coordinates $z$ and $\phi$ dictates the topology of the BH
spacetime. For a BH with toroidal topology the coordinate $z$ is compactified
such that $z/L$ ranges from $0$ to $2\pi$, and $\phi$ ranges from $0$ to
$2\pi$ as well. For the cylindrical BH, or black string, the coordinate $z$
has range $-\infty<z<\infty$, and $0\leq \phi <2\pi$. For the planar BH, or
black membrane, the coordinate $\phi$ is further decompactified
($-\infty<L\,\phi<\infty$).  For the torus, $M$ is related to the system's ADM
mass; for the cylinder, to the mass per unit length of a constant-$z$ line;
and for the plane, to the mass per unit area of the $(\phi, z)$ plane
\cite{Lemos:1994fn,Lemos:1994xp,Lemos:1995cm,Mann:1996gj,Vanzo:1997gw,
  Birmingham:1998nr}.

A formalism to handle electromagnetic and gravitational perturbations of these
spacetimes was developed and used to investigate numerically the QNMs of
scalar, electromagnetic and gravitational perturbations in
Ref.~\cite{Cardoso:2001vs}. A thorough analysis by Miranda and Zanchin
\cite{Miranda:2005qx,Miranda:2008vb} confirmed and extended the results of
Ref.~\cite{Cardoso:2001vs}.

\section{\label{sec:catalogue}The spectrum of asymptotically de Sitter and other black holes}

\subsection{Asymptotically de Sitter black holes}
BHs in a dS background have a rich structure. Consider an uncharged
spherically symmetric BH, the SdS$_d$ solution,
\be ds^{2}= fdt^{2}- f^{-1}dr^{2}-r^{2}d\Omega_{d-2}^2\,, \ee
where $f(r)=1-r^{2}/L^2-r_0^{d-3}/r^{d-3}$, $d\Omega_{d-2}^2$ is the metric of
the $(d-2)-$sphere, and $r_0$ is related to the mass $M$ of the spacetime by
$M=(d-2)A_{d-2}r_0^{d-3}/(16\pi)$. Depending on the value of $r_0/L^2$, in general this solution has two horizons, the event and the
cosmological horizon. When the cosmological and event horizons coalesce one
has a so-called extremal SdS$_d$ solution, which is related to a topologically
different solution, known as the Nariai spacetime
\cite{Nariai1,Nariai2,Caldarelli:2000wc,Cardoso:2004uz}.

\subsubsection*{Weakly damped modes} The first calculations of the fundamental
gravitational modes were done by Mellor and Moss \cite{Mellor:1989ac} for
Reissner-Nordstr\"om-de Sitter (RNdS) BHs, using numerical techniques. Brady
{\it et al.} \cite{Brady:1999wd} complemented this study through a numerical
time evolution of scalar fields in the SdS spacetime. Time evolutions were
also performed in Ref.~\cite{Abdalla:2003db}, where the results were compared
against WKB predictions (see also Ref.~\cite{Konoplya:2007jv}). Approximations
to the correct, lowest order QNMs were considered in Refs.~\cite{Moss:2001ga}
($s=2$, SdS), \cite{Suneeta:2003bj} ($s=0$, SdS), \cite{Jing:2003wq} ($s=1/2$,
RNdS) and \cite{Jing:2005bh} ($s=2$, RNdS), where the true potential was
approximated by a P\"oschl-Teller potential. Cardoso and Lemos
\cite{Cardoso:2003sw} showed that the true potential reduces to the
P\"oschl-Teller potential for near extremal SdS geometries, which explains why
previous results based on the P\"oschl-Teller approximation gave accurate
predictions for the QNMs. The results in \cite{Cardoso:2003sw} were later
generalized to $d$-dimensional RNdS BHs \cite{Molina:2003ff}. Other analytical
results used the WKB approximation for spins $s=0,1,2,1/2$ in the vicinity
of SdS BHs \cite{Zhidenko:2003wq}.

\subsubsection*{Highly-damped modes} The highly-damped limit, where the
imaginary part is much larger than the real part, was studied numerically by
Yoshida and Futamase \cite{Yoshida:2003zz} for near-extremal uncharged BHs
($s=1,2$) and in the general case in
Ref.~\cite{Choudhury:2003wd,Konoplya:2004uk}. The results were confirmed
analytically in Refs.~\cite{Cardoso:2004up} ($s=0,1,2$ SdS) and
\cite{Natario:2004jd} ($s=0,1,2$ RNdS).

\subsubsection*{Eikonal limit} In the large-$l$ limit, WKB techniques
\cite{barretozworski,Zhidenko:2003wq} yield
\be
3\sqrt{3}\,M\omega=\sqrt{1-27M^2/L^2}\left [l+1/2-i(n+1/2)\right ]\,.
\ee 
QNMs in the eikonal limit can be interpreted as perturbations of unstable
circular null geodesics \cite{Cardoso:2008bp}.  For studies on charged and
rotating BHs in dS backgrounds, see \cite{Konoplya:2007zx} and references
therein.

\subsection{Black holes in higher-derivative gravity}

Among all possible theories of gravity with higher derivative terms, theories
modified by the addition of a Gauss-Bonnet term
${\cal R}_{GB}=R_{\mu\nu\rho\sigma}R^{\mu\nu\rho\sigma}-4R_{\mu\nu}R^{\mu\nu}+R^2$,
are particularly attractive and have been considered by many authors
\cite{Moura:2006pz}. There are simple BH solutions in these theories
\cite{Wheeler:1985nh,Wheeler:1985qd,Boulware:1985wk,Kanti:1995vq,
  Torii:1996yi,Alexeev:1996vs,Pani:2009wy,Guo:2008hf,Guo:2008eq,Ohta:2009tb},
whose perturbations were studied in a series of works by Dotti and Gleiser
\cite{Dotti:2004sh,Dotti:2005sq,Gleiser:2005ra}, Moura and Schiappa
\cite{Moura:2006pz} and Takahashi and Soda \cite{Takahashi:2009dz}. In four dimensions the
Gauss-Bonnet term is a total divergence, and yields upon integration a
topological invariant, being therefore equivalent to Einstein's theory.
Therefore these theories are interesting in higher dimensions
only. Non-trivial four-dimensional scenarios can be accomplished by coupling
the Gauss-Bonnet term to a dilaton \cite{Pani:2009wy}: the resulting theory
also arises from the low-energy limit of certain string theories
\cite{Moura:2006pz}. Perturbations of BH spacetimes in these scenarios were
considered in Refs.~\cite{Pani:2009wy,Kanti:1997br,Torii:1996yi}.

QNMs of BHs in these theories were first investigated by Iyer {\it et al.}
\cite{Iyer:1989rd}, who used a WKB approach to study scalar
perturbations. QNMs of these spacetimes were also studied through a WKB
approach \cite{Abdalla:2005hu,Konoplya:2004xx,Chakrabarti:2006ei}, and
numerically \cite{Konoplya:2008ix,Abdalla:2005hu} for the low-lying modes.
The highly-damped regime was analytically explored in
Ref.~\cite{Daghigh:2006xg}. The eikonal limit was considered by Konoplya
\cite{Konoplya:2008ix} and interpreted in terms of circular null geodesics in
Ref.~\cite{Cardoso:2008bp}.

\subsection{Braneworlds}

Braneworld scenarios, where the standard model lives in a four-dimensional
brane embedded in a higher dimensional spacetime, have been a popular research
topic in the last decade. The extra dimensions can be compact
\cite{ArkaniHamed:1998rs,Antoniadis:1998ig} or even infinite, flat
\cite{Dvali:2000hr} or curved \cite{Randall:1999vf,Randall:1999ee}. BH
solutions in these theories are extremely difficult to find (see for instance
Ref.~\cite{Argyres:1998qn,Kaus:2009cg} for a discussion), some solutions are
known perturbatively in some regimes. For instance, in the case of flat
compact extra dimensions, Tangherlini BHs should be a good approximation to a
static BH solution as long as the horizon radius is much smaller than the size
of the extra dimension \cite{Argyres:1998qn}. In these scenarios the standard
model is localized on the brane and therefore BH oscillations are non-trivial,
especially for QNMs of standard-model fields
\cite{Clarkson:2005mg,Kanti:2005xa,Kanti:2006ua,alBinni:2007gk,Chen:2007jz,Nozawa:2008wf}.

\subsection{Black holes interacting with matter}

Astrophysical BHs are not expected to be in complete isolation, so it is
important to understand how QNMs change when BHs interact with the surrounding
environment. Leung {\it et al.} \cite{Leung:1999rh,Leung:1999iq} investigated
how the low-order QNMs of a BH are affected by a small amount of matter. For
such a BH, in the static, spherically-symmetric case, Medved {\it et al.}
\cite{Medved:2003rga,Medved:2003pr} proved that highly-damped QNMs depend only
on the surface gravity.
Specific models for BHs interacting with matter were constructed by several
authors. A popular cosmological scenario invokes the existence of dynamical
vacuum energy (``quintessence'', see e.g.~\cite{Carroll:1998zi,Zlatev:1998tr})
or phantom fields \cite{Caldwell:1999ew,McInnes:2001zw} to explain the
acceleration of the universe. The QNMs of BHs with quintessence or phantom
fields have been investigated in
Refs.~\cite{Chen:2005qh,Zhang:2006hh,Ma:2006by,Zhang:2006ij,
  Zhang:2007nu,Varghese:2008ky} and \cite{Chen:2008xb}, respectively.

\section{\label{sec:holography}Quasinormal modes and the gauge-gravity duality}

In this Section, we review a particular entry in the gauge-gravity duality
dictionary directly related to QNMs. It turns out that quasinormal spectra of
asymptotically AdS$_{d+1}$ and more general backgrounds correspond to poles of
the (retarded) thermal correlators of dual $d$-dimensional strongly
interacting quantum gauge theories.  The lowest quasinormal frequencies of
black branes have a direct interpretation as dispersion relations of
hydrodynamic excitations in the dual theory.
More information on near-equilibrium properties of BHs and black branes and
their holographic interpretation can be found in the reviews \cite{Son:2007vk,
  Damour:2008ji, Gubser:2009md, Hartnoll:2009sz, Herzog:2009xv}.

\subsection{\label{subsec:duality}The duality}

The discovery \cite{Polchinski:1995mt} and subsequent studies of $D$-branes in
string theory led to the concept of gauge-string duality. In the original
example of the duality, known as the AdS/CFT correspondence
\cite{Maldacena:1997re}, the full type IIB string theory on the background
AdS$_5\times S^5$ (five-dimensional AdS space times a five-sphere) was
conjectured\footnote{During the last decade, the AdS/CFT correspondence and,
  more generally, the gauge-string duality, survived numerous, often very
  non-trivial tests of validity. At the moment, there is very little doubt, if
  any, that the conjecture is valid.}  to be {\it equivalent} to the specific
supersymmetric gauge theory, ${\cal N}=4$ $SU(N_c)$ supersymmetric Yang-Mills
theory (SYM) in flat four-dimensional spacetime.  The equivalence is
understood as an equivalence of quantum partition functions.  In quantum field
theory, there is very strong evidence supporting the claim that ${\cal N}=4$
SYM is a conformal field theory (CFT), with the beta-function identically
equal to zero and coupling constant $g_{YM}$ being independent of the energy
scale. The two parameters characterizing ${\cal N}=4$ SYM are the 't Hooft
coupling $\lambda \equiv g_{YM}^2 N_c$ and the number of colors $N_c$. In the
AdS/CFT correspondence, these parameters are mapped into the string theory
parameters $L$ and $g_s$:
\begin{equation}
g^2_{YM}\, N_c \sim g_sN_c \sim L^4 / l_s^4\,, \qquad  N_c \sim L^4 / l_P^4\,,
\end{equation}
where $L$ is the parameter of AdS$_5$ and the radius of the five-sphere, $g_s$
is the string coupling and $l_s$, $l_P$ are the string and Planck lengths,
respectively. The full quantum string theory on AdS$_5\times S^5$ is poorly
understood.  However, its low-energy limit, type IIB supergravity, has been
extensively studied since the 1980s.  Restricting duality to the supergravity
limit of the full string theory restricts the values of the gauge theory
parameters to $g^2_{YM}\, N_c \gg 1$, $N_c \gg 1$.  Thus the gauge theory at
large values of coupling and large $N_c$ is effectively described by classical
gravity in the AdS background.

Following the original example of AdS/CFT correspondence, many more dual pairs
have been discovered, including those involving non-supersymmetric and
non-conformal theories. Gauge-string duality thus includes the original
AdS/CFT correspondence and all its ``non-conformal'' and ``non-AdS''
generalizations, often commonly referred to as ``AdS/CFT''.  The gauge-string
duality in the supergravity approximation is known as the gauge-gravity
duality. The duality provides a quantitative correspondence between classical
gravity in ten (or five) dimensions and a gauge theory (in the limit
$g^2_{YM}\, N_c \gg 1$, $N_c \gg 1$) in flat four-dimensional spacetime. Such
a correspondence between higher-dimensional gravity and lower-dimensional
non-gravitational theory is often referred to as ``holography''. The
gauge-gravity duality serves as a quantitative example of the ``holographic
principle'' proposed by 't Hooft and Susskind
\cite{'tHooft:1993gx,Susskind:1994vu}.

Since classical higher-dimensional gravity is holographically encoded into the
dual gauge theory's properties, one may wonder about the gauge theory
interpretation of the QNM spectrum. The short answer, conjectured in
\cite{Birmingham:2001pj}, established in \cite{Son:2002sd}, and further
generalized in \cite{Kovtun:2005ev} and many subsequent publications, is that
{\it the QNM spectrum of the fluctuation $\delta \phi$ of a higher-dimensional
  gravitational background coincides with the location of the poles of the
  retarded correlation function of the gauge theory operator ${\cal O}$ dual
  to the fluctuation $\delta \phi$.}  In the rest of this Section, we
elaborate on this statement and provide some explicit examples.

The main ingredient of the gauge-gravity duality is the ten-dimensional
(super)gravity background characterized by the values of the metric and other
supergravity fields such as the dilaton, the axion and various tensor
fields. The background fields must satisfy supergravity equations of
motion. (In most cases, only bosonic supergravity fields are considered, thus
eliminating the need for the prefix ``super''.)
For example, the near-horizon limit of the black three-brane background, which
is the basic ingredient of the AdS/CFT duality at finite temperature, consists
of the metric
\begin{equation}\label{near_horizon_metric}
  ds^2_{10} = \frac{r^2}{L^2} \Biggl[ -f dt^2 + dx^2 + dy^2 +dz^2 \Biggr]
+ \frac{L^2}{r^2} f^{-1}dr^2 + L^2 d\Omega_5^2\,,
\end{equation}
where $f(r) = 1- r_0^4/r^4$, and the Ramond-Ramond five-form field,
\begin{equation}
F_5 = - {4 r^3\over L^4} (1 + * )\,
 d t\wedge d x \wedge d y
 \wedge d z  \wedge d r\,,
\label{near_horizon_5form}
\end{equation}
with all other fields vanishing.
According to the gauge/gravity correspondence, the background
(\ref{near_horizon_metric})--(\ref{near_horizon_5form}) with non-extremality
parameter $r_0$ and Hawking temperature $T = r_0/\pi L^2$ is dual to ${\cal
  N}=4$ $SU(N_c)$ SYM at finite temperature $T$ in the limit of
$N_c\rightarrow \infty$, $g^2_{YM}N_c \rightarrow \infty$.
The ${\cal N}=4$ SYM is defined in Minkowski space with coordinates $t,x,y,z$.
The fifth (radial) coordinate $r$ of the dual metric
(\ref{near_horizon_metric}) plays the role of the energy scale in the gauge
theory (with the boundary at $r\rightarrow \infty$ corresponding to the ultraviolet in
the gauge theory), and the five-sphere describes internal degrees of freedom
associated with the $R$-symmetry group $SU(4)$ specific to theories with
${\cal N} = 4$ supersymmetries.

Quite often, the internal degrees of freedom are of less interest, and the
background can be dimensionally reduced from ten to five dimensions.  For the
metric (\ref{near_horizon_metric}), the result of such a reduction is the
five-dimensional Schwarzschild-AdS metric with translationally invariant
horizon
\begin{equation}\label{near_horizon_metric_sads}
  ds^2_{5} = \frac{r^2}{L^2} \Biggl[ -f dt^2 + dx^2 + dy^2 +dz^2 \Biggr]+
   \frac{L^2}{r^2} f^{-1}dr^2\,,
\end{equation}
obeying Einstein's equations $R_{\mu\nu} = 2 \Lambda/3\, g_{\mu\nu}$ in a
five-dimensional space with cosmological constant\footnote{The cosmological
  constant in five dimensions arises as a result of the dimensional reduction
  of the five-form (\ref{near_horizon_5form}) on $S^5$.}  $\Lambda = -6/L^2$.
Thus the gauge-gravity duality often appears as a correspondence involving
the five-dimensional (super)gravity bulk and the four-dimensional boundary
gauge theory.  One should always remember, however, that all five-dimensional
fields and their fluctuations have ten-dimensional origin.

As a remark, we note that the metric (\ref{near_horizon_metric}) with
translationally invariant horizon can be obtained from the Schwarzschild-AdS
BH metric
\begin{equation}
 ds^2_{5} =  - f\, dt^2 + \frac{dr^2}{f} + r^2 d\Omega_3^2\,, \qquad f = 1 +\frac{r^2}{L^2}
- \frac{r_0^4}{r^2 L^2}
\end{equation}
by rescaling $r\rightarrow \lambda^{1/4} r$, $r_0\rightarrow \lambda^{1/4}
r_0$, $t\rightarrow \lambda^{-1/4} t$, and taking the limit $\lambda
\rightarrow \infty$ while simultaneously blowing up the sphere
\begin{equation}
L^2\,  d\Omega_3^2 \rightarrow \lambda^{-1/2} \left( dx^2 + dy^2 +dz^2 \right)\,.
\end{equation}
Similar rescalings for more general metrics can be found in
\cite{Chamblin:1999tk, Son:2006em}.  This difference between black hole and
black brane metrics leads to the fact that black brane QNM spectra are
functions $\omega = \omega(q)$ of the {\it continuous} parameter $q$ (rather
than a discrete parameter $l$), where $q$ is the momentum along the
translationally invariant directions.

\subsection{\label{dualqnms}Dual quasinormal frequencies as poles of the retarded correlators}
Quantitatively, the gauge-string (gauge-gravity) duality is the equivalence of
the partition functions
\begin{equation}\label{gauge-grav}
Z_{YM} [J] = \langle e^{- \int J {\cal O} d^4x}\rangle_{YM} 
\equiv Z_{\rm string}[J] \simeq e^{-S_{\rm grav}[J]}\,,
\end{equation}
where the semiclassical approximation on the right-hand side corresponds to
passing from the gauge-string to the gauge-gravity duality in the appropriate
limit (e.g.  in the limit $N_c\rightarrow \infty$, $g^2_{YM}N_c \rightarrow
\infty$ for ${\cal N}=4$ SYM).  The equivalence (\ref{gauge-grav}) means that
the classical gravity action effectively serves as a generating functional for
correlation functions of gauge-invariant operators ${\cal O}$ in the dual
gauge theory.  On the gravity side, the role of $J$ for a given operator
${\cal O}$ is played by the boundary value $\delta \phi_0$ of the background
{\it fluctuation} $\delta \phi$ (for a moment, we are ignoring all indices the
field $\delta \phi$ might have).  For example, the boundary value of the
background metric fluctuation $h_{\mu\nu}$ plays the role of $J$ in computing
the correlators of the energy-momentum tensor $T_{\mu\nu}$ in a
four-dimensional gauge theory (see Table \ref{tab:gauge-grav-id}).

The recipe for applying the equivalence (\ref{gauge-grav}) is the following.
To compute the correlators of a gauge-invariant operator ${\cal O}$, one has
to
\begin{itemize}

\item
identify the dual fluctuation field $\delta \phi$ associated with ${\cal O}$

\item
solve the linearized bulk equations of motion satisfied by $\delta \phi$ with
the boundary condition\footnote{The second boundary condition on the
  fluctuation $\delta \phi$ is either a regularity condition (e.g. for
  zero-temperature global AdS space) or the incoming (outgoing) wave boundary
  condition. For metrics with horizons the incoming (outgoing) wave condition
  corresponds to computing the retarded (advanced) correlators in the boundary
  theory.}  $\delta \phi \rightarrow \delta \phi_0 \equiv J$

\item 
using this solution, compute the on-shell supergravity action $S_{\rm
  grav}[J]$ as a functional of $\delta \phi_0 \equiv J$

\item
compute the correlators in the usual field theory sense by taking functional
derivatives of $\exp{(-S_{\rm grav}[J])}$ with respect to $J$.

\end{itemize}
The recipe given above is sufficient for computing Euclidean correlation
functions from dual gravity. For Minkowski space correlators, there are
subtleties resolved in \cite{Son:2002sd, Herzog:2002pc} (recent work on the
Lorentzian AdS/CFT includes \cite{Marolf:2004fy, Skenderis:2008dg}).
Ref.~\cite{Son:2002sd} contains an exact prescription on how to compute the
Minkowski space two-point functions from fluctuations of a dual gravity
background. As a byproduct, the prescription establishes a one-to-one
correspondence between poles of the retarded quantum field theory correlators
and QNM spectra of the dual background.
Indeed, the dual gravity fluctuation field $\delta \phi (r,t,x,y,z)$
associated with the operator ${\cal O}$, whose retarded two-point function
$G^R$ we are interested in, satisfies an ordinary linear second-order
differential equation with respect to the radial coordinate $r$.  The
fluctuation's dependence on the ``usual'' four-dimensional space-time
coordinates $t,x,y,z$ in the bulk is typically trivial, allowing one to
Fourier transform with respect to them:
\begin{equation}
\delta \phi (r,t,x,y,z) = \int \frac{d\omega d \bf{q}}{(2\pi)^4} \; e^{-i \omega t + i\bf{q}\bf{x}}
\; \delta\phi (r,\omega, {\bf{q}}) \,.
\end{equation}
Note that at finite temperature the Lorentz invariance is broken, thus the
components $\omega$, ${\bf{q}}$ of the four-momentum are independent
variables.  For theories with unbroken rotation invariance, the fluctuation
will depend on the magnitude of the three-momentum $q = |{\bf{q}}|$, and one
can conveniently choose ${\bf{q}}$ along the direction $z$ of the brane, with
the four-momentum being given by $(\omega, 0,0, q)$.  The ODE satisfied by the
fluctuation $\delta\phi (r,\omega, q)$ typically has singular points at the
horizon $r=r_0$ and at the boundary $r\rightarrow \infty$.  The solution
$\delta\phi (r,\omega, q)$ satisfying the incoming wave boundary condition at
the horizon can be written near the boundary as
\begin{equation}
\delta\phi (r,\omega, q) = {\cal A} (\omega,q)\, r^{- \Delta_-} \left( 1 + \cdots \right)
+ {\cal B} (\omega,q)\, r^{- \Delta_+ } \left( 1 + \cdots \right)\,,
\end{equation}
where $\Delta_+$, $\Delta_-$ are the exponents of the ODE at $r=\infty$ (these
exponents are related to the conformal dimension of the operator ${\cal O}$),
and ellipses denote higher powers of $r$. In most cases, fields can be
redefined so that $\Delta_+>0$, $\Delta_-=0$. Applying the gauge-gravity
duality recipe for Minkowski correlators, for the retarded two-point function
of the operator ${\cal O}$ one finds
\begin{equation}
G^R (\omega,q) \sim \frac{{\cal B}}{{\cal A}} + \mbox{contact terms}\,.
\label{main_rat}
\end{equation}
Zeros of the coefficient ${\cal A}$ correspond to the poles of
$G^R(\omega,q)$.  On the other hand, from the general relativity point of
view, the condition ${\cal A}=0$ (for $\Delta_-=0$ this is just the Dirichlet
boundary condition) defines the QNM spectrum of the fluctuation $\delta \phi$.
Thus, {\it all the information about the poles of the retarded correlators of
  a quantum field theory with a gravity dual description is encoded in the QNM
  spectra of the dual gravity fluctuations.}
This statement is a useful entry in the gauge-gravity duality dictionary,
since the poles of thermal correlators carry important information about
transport properties and excitation spectra of the theory.  Consider, for
illustration, the relatively simple case of a two-dimensional CFT at finite
temperature dual to the $(2+1)$-dimensional BTZ BH background.  The retarded
two-point function of the operator of conformal dimension $\Delta =2$ at
finite temperature is given by \cite{Son:2002sd}
\be
G^R \sim \frac{\omega^2-q^2}{4\pi^2} \Biggl[ \psi \Biggl( 1 - i \frac{\omega - q}{4 \pi T} \Biggr)
+ \psi \Biggl( 1 - i \frac{\omega + q}{4 \pi T} \Biggr) \Biggr]\,,
\label{btz-correlator}
\ee
where $\psi(z) = \Gamma'(z)/\Gamma(z)$; we put $T_L=T_R$ and ignored the
constant prefactor for simplicity.  The correlator (\ref{btz-correlator}) has
infinitely many poles in the complex frequency plane, located at
\begin{equation}
\omega_n = \pm q - i \, 4 \pi T (n+1)\,, \qquad n=0,1,2,...\,.
\end{equation}
These poles are precisely the BTZ quasinormal frequencies \cite{Chan:1996yk,
  Cardoso:2001hn, Birmingham:2001hc}.

The role of QNM spectra as frequencies defining relaxation times in the dual
thermal field theory was realized and discussed in early publications
\cite{KalyanaRama:1999zj, Horowitz:1999jd, Danielsson:1999fa}.  Birmingham,
Sachs and Solodukhin \cite{Birmingham:2001pj} were the first to note
explicitly that the QNM spectrum of the BTZ BH coincides with the poles of the
retarded correlators of the dual $(1+1)$-dimensional CFT.
With the appearance of the Minkowski AdS/CFT recipe for the retarded
correlators \cite{Son:2002sd}, the relationship between QNMs and the poles of
the correlators has been established quantitatively first for the scalar
\cite{Son:2002sd}, \cite{Starinets:2002br} and later for general fluctuations
\cite{Nunez:2003eq, Kovtun:2005ev}.  In the case of non-scalar fluctuations,
considering gauge-invariant combinations of fluctuating fields is especially
useful \cite{Kovtun:2005ev}, although this is not the only possible approach
\cite{Policastro:2002tn, Amado:2009ts}.

A word of caution is necessary. As we have seen, asymptotically AdS spacetimes
offer a variety of choices for the boundary conditions at spatial infinity.
Not {\it all} such choices produce QNM spectra which have a meaningful
interpretation in the dual quantum field theory. In order to say that a
computed QNM corresponds to a pole of a dual theory correlator, one has to
analyse the bulk fluctuation along the lines leading to Eq.~(\ref{main_rat})
to establish the precise form of the boundary condition.


\begin{table}
\caption{\label{tab:gauge-grav-id} The correspondence between the boundary
  gauge theory operators and the dual five-dimensional gravity bulk fields.}
\begin{tabular}{|c||c|}  \hline
Gauge theory operator ${\cal O}$ &  Dual gravity fluctuation \\
\hline
energy-momentum tensor $T_{\mu\nu}$ & metric fluctuation $h_{\mu\nu}$\\
conserved current $J_{\mu}$ & Maxwell field $A_{\mu}$\\
Tr $F_{\mu\nu}^2$ & minimally coupled massless scalar $\varphi$\\
\hline
\end{tabular}
\end{table}

\subsection{\label{sec:hydlim}The hydrodynamic limit}

The most interesting results for QNM spectra with a dual field theory
interpretation are obtained for five-dimensional gravitational backgrounds
(note, however, the growing body of work on the AdS-Condensed Matter Theory
correspondence \cite{Hartnoll:2009sz, Herzog:2009xv}, where, for the purposes
of studying $(2+1)$-dimensional condensed matter systems, one is interested in
$(3+1)$-dimensional gravitational backgrounds).
For example, the poles of the retarded thermal two-point function
\be
G^R_{\mu\nu,\lambda\sigma} (\omega,q)  = -i \int d^4 x\, 
 e^{-i q\cdot x} \, \theta (t)\, 
 \langle \left[ T_{\mu\nu}(x),
T_{\lambda\sigma}(0)\right]\rangle_{\tiny T}
\ee
of the stress-energy tensor in four-dimensional ${\cal N}=4$ $SU(N_c)$ SYM in
the limit $N_c\rightarrow \infty$, $g^2_{YM}N_c \rightarrow \infty$ are given
by the QNM frequencies of the gravitational perturbation $h_{\mu\nu}$ of the
metric (\ref{near_horizon_metric_sads}). By symmetry, the perturbations are
divided into three groups \cite{Policastro:2002se, Kovtun:2005ev}. Indeed,
since the dual gauge theory is spatially isotropic, we are free to choose the
momentum of the perturbation along, say, the $z$ direction, leaving the $O(2)$
rotational symmetry of the $(x,y)$ plane intact.  The perturbations
$h_{\mu\nu}(r,t,z)$ are thus classified according to their transformation
properties under $O(2)$.  Following \cite{Policastro:2002se, Kovtun:2005ev},
we call them the scalar ($h_{xy}$), shear ($h_{tx}$, $h_{zx}$ or $h_{ty}$,
$h_{zy}$) and sound ($h_{tt}$, $h_{tz}$, $h_{zz}$, $h_{xx}+h_{yy}$) channels.
(Here we partially fixed the gauge by requiring $h_{\mu r}$=0.)  The $O(2)$
symmetry ensures that the equations of motion for perturbations of different
symmetry channels {\it decouple}.  Linear combinations of perturbations
invariant under the (residual) gauge transformations $h_{\mu\nu}\rightarrow
h_{\mu\nu} - \nabla_\mu\xi_\nu -\nabla_\nu\xi_\mu$ form the gauge-invariant
variables
\begin{eqnarray}
   &\,& 
    Z_1 = q H_{tx} + \omega H_{zx}\,,\label{l2}
\label{l1}\\
    &\,& 
    Z_2 = q^2 f H_{tt} + 2 \omega q H_{tz} +\omega^2 H_{zz} +
    \left[ q^2 \left( 2 - f\right) - \omega^2\right] H\,, \\
    &\,& 
    Z_3 = H_{xy}\,
\label{l3}
\end{eqnarray} 
in the shear, sound and scalar channels, respectively (here $H_{tt}=
L^2h_{tt}/r^2f$, $H_{ij}= L^2 h_{ij}/r^2$ ($i,j\neq t$),
$H=L^2(h_{xx}+h_{yy})/2r^2$). From the equations of motion satisfied by the
fluctuations one obtains three independent second-order ODEs for the
gauge-invariant variables $Z_1$, $Z_2$, $Z_3$.

\begin{figure*}[ht]
\begin{center}
\begin{tabular}{cc}
\epsfig{file=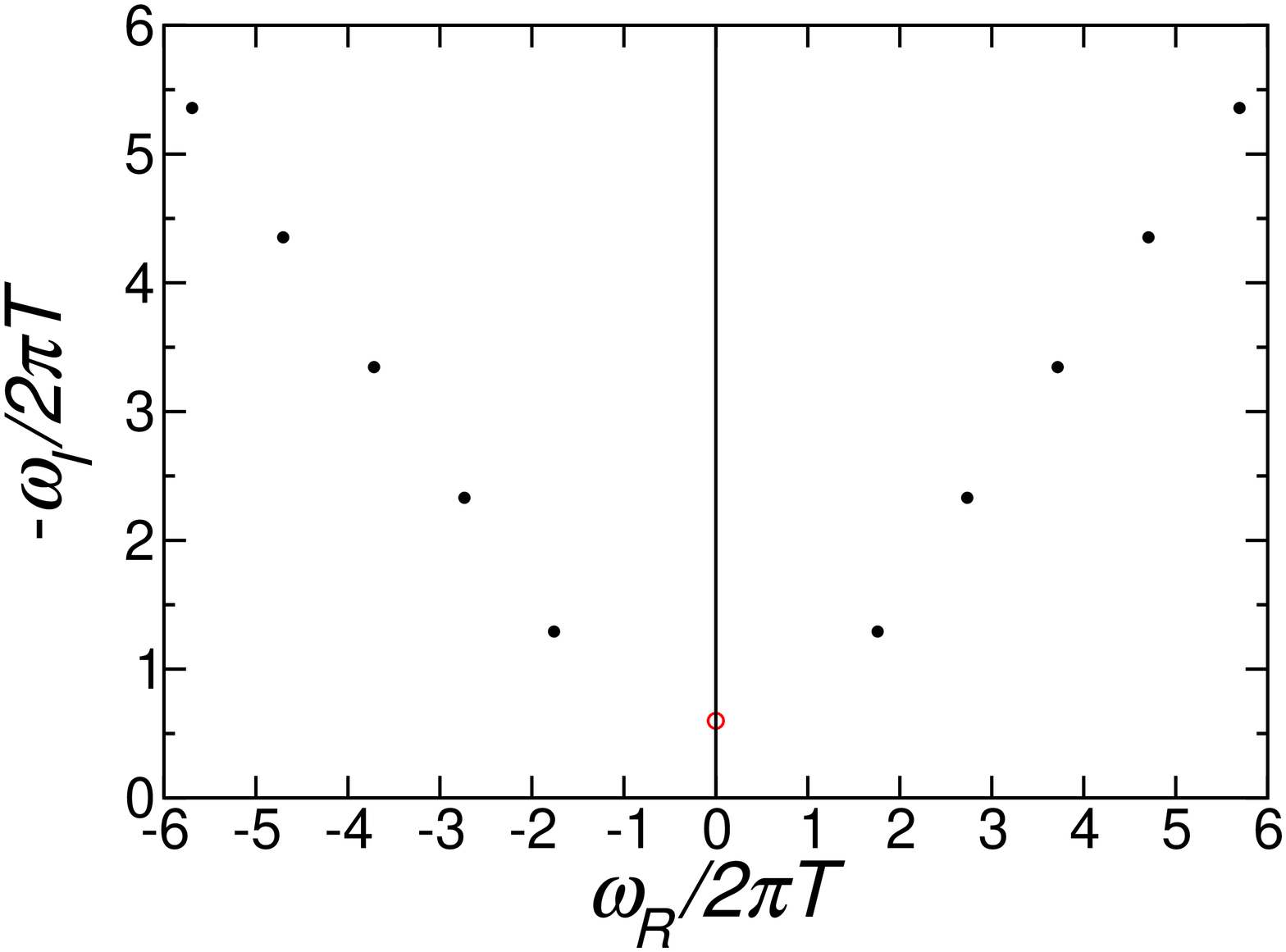,width=5.2cm,angle=-90} &
\epsfig{file=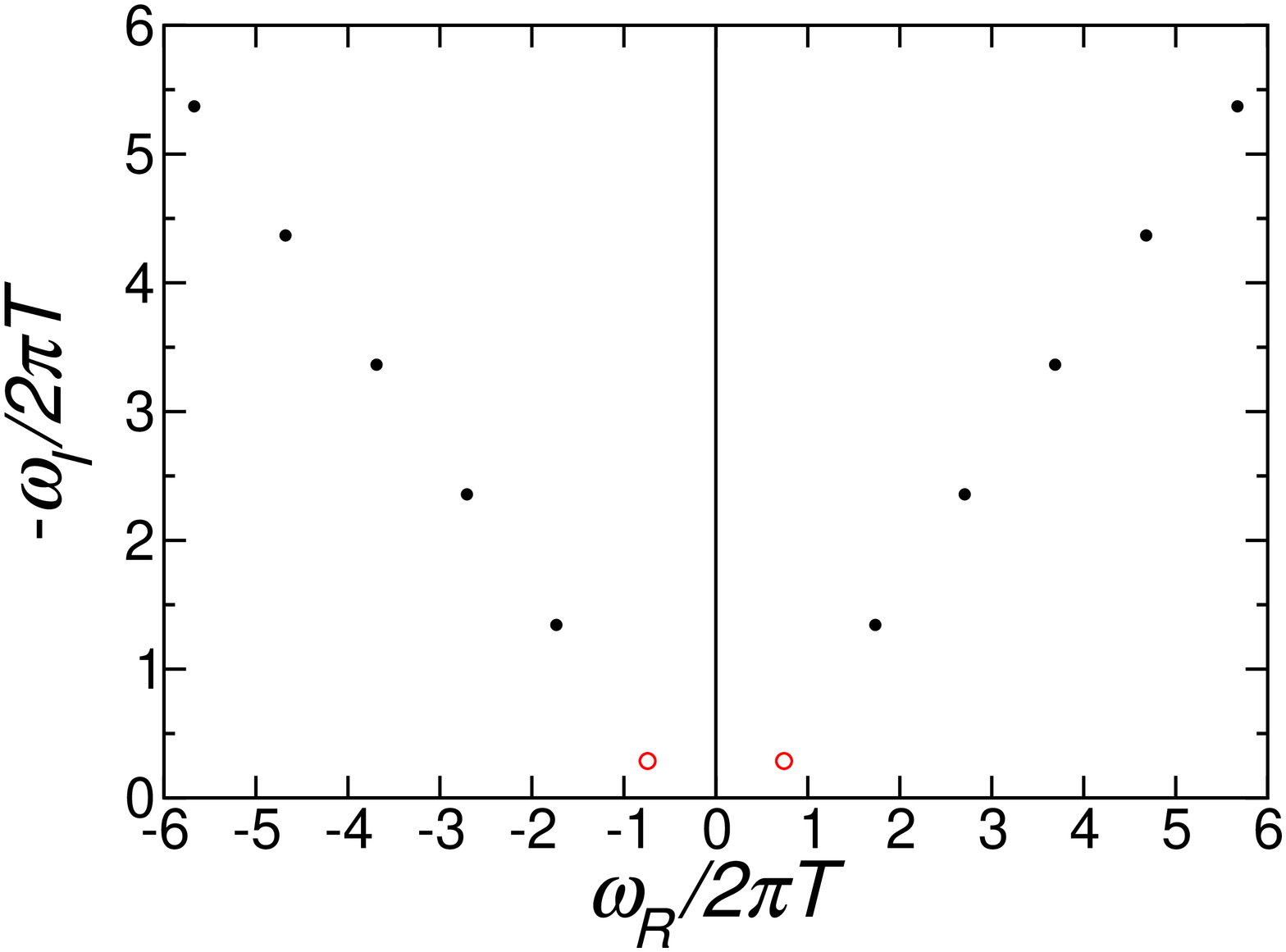,width=5.2cm,angle=-90} \\
\end{tabular}
\end{center}
\caption{Quasinormal spectrum of black three-brane gravitational fluctuations
  in the ``shear'' (left) and ``sound'' (right) channels, shown in the plane
  of complex frequency $\wn = \omega /2\pi T$, for fixed spatial momentum $\qn
  = q/2\pi T = 1$. Hydrodynamic frequencies are marked with hollow red dots
  (adapted from \cite{Kovtun:2005ev}).} \label{fig:complexplane}
\end{figure*}

The QNM spectra in all three channels share a characteristic feature: an
infinite sequence of (asymptotically) equidistant frequencies approximated
(for $q=0$) by a simple formula \cite{Starinets:2002br, Nunez:2003eq,
  Natario:2004jd}
\begin{equation}
\omega_n = 2\pi T n \left( \pm 1 - i\right) + \omega_0\,, \qquad  n\rightarrow \infty\,. 
\label{modeapprox}
\end{equation}
Each frequency has a non-trivial dependence on $q$ \cite{Starinets:2002br,
  Nunez:2003eq}.  Spectra in the shear and sound channels are shown in
Fig.~(\ref{fig:complexplane}).  In addition to the sequence mentioned above,
they contain the so called hydrodynamic frequencies shown in
Fig.~\ref{fig:complexplane} by hollow red dots.  The hydrodynamic frequencies
are remarkable, in that their existence and dependence on $q$ are predicted by
the hydrodynamics of the dual field theory. For example, low-frequency, small
momenta fluctuations of the stress-energy tensor $T_{\mu\nu}$ of any
$d$-dimensional theory are characterized by the dispersion relations (see
e.g. \cite{Policastro:2002se, Parnachev:2005hh, Baier:2007ix})
\begin{eqnarray}
\omega &=& - i \frac{\eta}{\epsilon +P}\, q^2 + O\left( q^4 \right)\,, \label{shmd} \\
\omega &=& \pm c_s q - \frac{i}{\epsilon +P}\, \left[  \frac{\zeta}{2}
 + \left( 1 - \frac{1}{d} \right) \eta \right]\, q^2 \, 
+  O\left( q^3 \right)\,, \label{smd}
\end{eqnarray}
where $\epsilon$, $P$, $\eta$, $\zeta$ are, respectively, energy density,
pressure, shear and bulk viscosities, and $c_s$ is the speed of sound.  For
{\it conformal} theories, some of the higher order terms are also known (see
e.g. \cite{Baier:2007ix}). In quantum field theory, the dispersion relations
(\ref{shmd}), (\ref{smd}) appear as poles of the retarded correlation
functions of the stress-energy tensor. Therefore, according to the holographic
dictionary discussed in Section~\ref{dualqnms}, Eqs.~(\ref{shmd}), (\ref{smd})
are precisely the lowest quasinormal frequencies in the spectrum of
gravitational perturbations.  For ${\cal N}=4$ SYM in $d=4$ with zero chemical
potential we have $\zeta=0$, $c_s=1/\sqrt{3}$, $\epsilon + P = s T$, where $s$
is the (volume) entropy density, and the dispersion relations (\ref{shmd}),
(\ref{smd}) become
\begin{eqnarray}
\omega &=& - i \frac{\eta}{s T}\, q^2 + O\left( q^4 \right)\,, \label{shmd1} \\
\omega &=& \pm \frac{q}{\sqrt{3}} - \frac{i \eta}{s T}\, q^2 \, 
+  O\left( q^3 \right)\,. \label{smd1}
\end{eqnarray}
On the other hand, the lowest QNMs of the fluctuations $Z_1$, $Z_2$ of the
dual black brane background (\ref{near_horizon_metric_sads}) can be computed
analytically \cite{Policastro:2002se, Policastro:2002tn}:
\begin{eqnarray}
\omega &=& - i \frac{1}{4 \pi T}\, q^2 + O\left( q^4 \right)\,, \label{shmd2} \\
\omega &=& \pm \frac{q}{\sqrt{3}} - \frac{i}{4 \pi T}\, q^2 \, 
+  O\left( q^3 \right)\,. \label{smd2}
\end{eqnarray}
Comparing Eqs.~(\ref{shmd1}) and (\ref{shmd2}), (\ref{smd1}) and (\ref{smd2}),
one finds that 1) the real part of the mode (\ref{smd2}) is correctly
predicted by hydrodynamics (yet another piece of evidence in favor of the
AdS/CFT conjecture) and 2) assuming the validity of AdS/CFT, the ratio
$\eta/s$ is equal to $1/4\pi$ in ${\cal N}=4$ SYM theory (in the limit of
infinite coupling and infinite $N_c$, where the dual gravity description is
valid).

This simple example illustrates a general method of extracting physical
quantities of strongly coupled quantum field theories from the QNM spectra of
their gravity duals.  Bulk fluctuations corresponding to operators of
conserved currents in the dual field theory in Minkowski spacetime are
guaranteed to have hydrodynamic frequencies in their QNM spectra. All such
frequencies have the generic property that $\omega \rightarrow 0$ for
$q\rightarrow 0$, characteristic of long wavelength, small frequency
fluctuations in flat space.
For ${\cal N}=4$ SYM, such operators are (in addition to $T_{\mu\nu}$) the
$R$-current and the supercurrent. Their corresponding bulk fluctuations are,
respectively, the $U(1)$ and the Rarita-Schwinger fluctuations in the
background (\ref{near_horizon_metric_sads}).
The hydrodynamic QNMs of these fluctuations were first computed in
Refs.~\cite{Policastro:2002se, Nunez:2003eq, Policastro:2008cx}.  The full QNM
spectrum of electromagnetic/gravitational fluctuations in the background
(\ref{near_horizon_metric_sads}) was computed in Refs.~\cite{Nunez:2003eq,
  Kovtun:2005ev}. The full QNM spectrum of the Rarita-Schwinger field has not
yet been determined.
Note that for theories in a finite volume (dual to black holes rather than
branes) hydrodynamic QNMs, strictly speaking, do not exist as the momentum $q$
is discrete.  However, for the large-radius asymptotically AdS BHs, the
emergence of the hydrodynamic behavior in the limit $r_+/L \rightarrow \infty$
is easily detected (see e.g.  Eq.~(\ref{longlived2})).

The interpretation of non-hydrodynamic QNMs is obscured by the fact that very
little is known about thermal correlation functions at strong coupling.
Typical singularities of such correlators at weak coupling appear to be cuts
rather than poles (this problem is discussed in Ref.~\cite{Hartnoll:2005ju}).
Poles associated with the QNM sequence (\ref{modeapprox}), for example, cannot
be interpreted as quasiparticles of ${\cal N}=4$ SYM, since their imaginary
part is large.

For backgrounds more complicated than the one of the black brane
(\ref{near_horizon_metric_sads}), the gauge-invariant variables $Z_i$ may
involve fields other than metric fluctuations.  Generically, these variables
form a system of coupled ODEs (simple examples can be found in
\cite{Parnachev:2005hh, Benincasa:2005iv, Son:2006em}).

A separate problem is the computation of the corrections to QNM spectra coming
from higher-derivative terms in the relevant supergravity actions. In the dual
field theory such corrections correspond, in particular, to coupling constant
corrections to transport coefficients. For instance, higher-derivative
corrections to the type IIB supergravity action result in 't Hooft coupling
corrections to the shear viscosity-entropy density ratio in ${\cal N}=4$ SYM
\cite{Buchel:2004di, Buchel:2008sh, Myers:2008yi}
\begin{equation}
\frac{\eta}{s} = \frac{1}{4\pi} \Biggl( 1 + \frac{15 \zeta (3)}{ \lambda^{3/2}} 
+ \cdots \Biggr)\,, \qquad \lambda \gg 1\,,
\label{visc_corr}
\end{equation}
where $\zeta(3) \approx 1.202$ is Ap\'{e}ry's constant.  Corrections to other
transport coefficients have also been considered \cite{Haack:2008xx,
  Ritz:2008kh, Myers:2009ij, Cremonini:2009sy}.

\subsection{\label{sec:visco}Universality of the shear mode and other developments}

The existence of ``hydrodynamic'' QNMs is a generic feature of black branes:
these QNMs must appear in the spectra of fluctuations dual to conserved
currents in translationally invariant backgrounds. In some cases, the lowest
QNMs can be computed analytically even for rather general metrics. For
example, for a black $p$-brane metric of the form
\begin{equation}
ds^2 = g_{tt}(r) dt^2 + g_{rr}(r) dr^2 + g_xx(r) \sum\limits_{i=1}^p \left( dx^i\right)^2\,,
\label{generic_m}
\end{equation}
with all other background fields vanishing, one finds that the component of an
electric field fluctuation parallel to the brane possesses a ``hydrodynamic''
frequency
\begin{equation}
\omega = - i D\, q^2 + O \left( q^4\right)\,,
\label{diff_w}
\end{equation}
where the coefficient $D$ is given in terms of the metric components:
\begin{equation}
D = \frac{\sqrt{-g(r_H)}}{g_{xx}(r_H) \sqrt{-g_{tt}(r_H)g_{rr}(r_H)}}\,
 \int\limits_{r_H}^\infty dr \frac{- g_{tt}(r)g_{rr}(r)}{\sqrt{-g(r)}}\,
\label{D_const}
\end{equation}
and $r_H$ is the position of the horizon.  If there exists a quantum field
theory (QFT) dual to this background, the electromagnetic fluctuation
naturally couples to a conserved $U(1)$ current in the theory, and $D$ is
interpreted as a diffusion constant of a corresponding $U(1)$ charge.

Similarly, the lowest gravitational QNM in the shear channel (\ref{l2}) for
the metric (\ref{generic_m}) can be computed analytically
\cite{Starinets:2008fb}.  The result is an
expression similar to Eqs.~(\ref{diff_w})-(\ref{D_const}), but it can be
further simplified using either Buchel-Liu's theorem \cite{Buchel:2003tz} or
the alternative proof given in \cite{Son:2007vk}.  Surprisingly, it turns out
that this mode is {\it universal}: its frequency is given by Eq.~(\ref{shmd2})
for {\it any} background with a metric of the form (\ref{generic_m}). (It is
important to note that the shear mode fluctuation decouples from the
fluctuations of other fields \cite{Kovtun:2004de}.)

The universality of the gravitational quasinormal shear mode has an important
consequence for a dual QFT: the ratio of shear viscosity to entropy density
has a universal value $\eta/s = 1/4\pi$ for all theories with gravity duals,
in the limit where the gravity dual description is valid (e.g., in the limit
of infinite $N_c$ and infinite 't Hooft coupling for ${\cal N}=4$ $SU(N_c)$
SYM in four dimensions). Earlier proofs of the shear viscosity universality
not involving QNMs were given in \cite{Kovtun:2004de, Buchel:2004qq}.  Charged
black brane backgrounds were considered in \cite{Son:2006em, Maeda:2006by,
  Mas:2008qs, Mas:2008ks,
  Ge:2008ak,Erdmenger:2008rm,Cremonini:2008tw,Matsuo:2009yu, Myers:2009ij}.

The viscosity bound conjecture \cite{Kovtun:2003wp} prompted further
investigation of the black brane QNM spectrum in higher-derivative gravity
\cite{Brigante:2007nu,Kats:2007mq,Brigante:2008gz,Buchel:2008vz,Buchel:2008ae,
  Cai:2009zv,Banerjee:2009wg,Ge:2009eh,Banerjee:2009fm}.  The conjecture
states that the ratio $\eta/s$ is bounded from below by $1/4\pi$ in all
physical systems. In the language of QNMs, this would mean that a correction
to the $q^2$-coefficient in Eq.~(\ref{shmd2}) coming from higher-derivative
terms in the action describing gravity duals of such systems is always
positive, as in Eq.~(\ref{visc_corr}).  In some gravity models with
higher-derivative terms this appears not to be the case
\cite{Brigante:2007nu,Kats:2007mq,Brigante:2008gz,Buchel:2008vz,Cremonini:2009sy}.
It would be highly desirable to have a comprehensive understanding of the
influence of higher-derivative terms on the shear mode and other QNMs.

QNMs of $Dp$-branes and more complicated backgrounds in the holographic
context were computed in \cite{Mas:2007ng, Maeda:2005cr, Benincasa:2005iv,
  Buchel:2005cv,Kanitscheider:2009as}.  Time-dependent backgrounds especially
relevant for modeling the behavior of the quark-gluon plasma in heavy ion
collisions were investigated in \cite{Janik:2006gp, Friess:2006kw,
  Michalogiorgakis:2006jc}.  QNMs in models with holographic mesons,
Sakai-Sugimoto model and other QCD-like models were studied in
\cite{Hoyos:2006gb,Myers:2007we,Erdmenger:2007ja,
  Myers:2008cj,Evans:2008tu,Evans:2008tv,Paredes:2008nf,
  Ammon:2009fe,Johnson:2009ev}.

The gauge-gravity duality is primarily used to investigate strongly 
coupled gauge theories with the help of a dual classical gravity theory. However, 
at least in principle, the holographic dictionary can be used to 
explore aspects of quantum gravity with the help of a weakly coupled gauge theory.
QNMs in this and similar contexts of duality have been studied in 
\cite{Fidkowski:2003nf,Festuccia:2005pi,Festuccia:2008zx,Amado:2008hw}.

Recently, holographic methods were extended to include gravity backgrounds
dual to non-relativistic theories as well as systems with spontaneous symmetry
breaking.  Exploration of QNM spectra in these models is an active area of
research \cite{Hartnoll:2009sz,Herzog:2009xv}.

\section{\label{sec:astro}Quasinormal modes of astrophysical black holes}

In the context of the Einstein-Maxwell equations, the most general solution
describing stationary axisymmetric BHs is the Kerr-Newman metric
~\cite{Heusler:1998ua}. For astrophysical BHs the electric charge $Q$ is
likely to be negligible, being shorted out by the surrounding plasma
\cite{Gibbons:1975kk,Blandford:1977ds}.  Astrophysical BHs are effectively the
simplest of all macroscopical objects, characterized only by their mass $M$
and angular momentum parameter $a$.  As a consequence, their structure and
their oscillation spectra are remarkably simple.

The complex frequency of a QNM yields two observables: the actual oscillation
frequency and the damping time of the oscillation. For each given mode, these
observables depend only on $M$ and $a$. Therefore a measurement of the
frequency and damping time of a QNM can be used to infer the mass and angular
momentum of the BH with potentially high accuracy
\cite{Echeverria:1989hg,Finn:1992wt,Berti:2005ys,Berti:2007zu}. Since the
whole QNM spectrum depends solely on $M$ and $a$, the measurement of {\em two
  or more} QNM frequencies provides a stringent observational test of the
no-hair theorem of general relativity
\cite{Dreyer:2003bv,Berti:2006cc,Berti:2007zu,Berti:2006qt}.  The prospects
for detecting the signature of BH oscillations in gravitational waves are the
main topic of this section.

There is strong and growing observational evidence for the existence of at
least two different classes of astrophysical BHs. Solar-mass BHs with $M\sim
5-20~M_\odot$ are usually found in X-ray binaries \cite{Remillard:2006fc}, and
SMBHs with $M\sim 10^6-10^{9.5}~M_\odot$ are believed to harbor most Active
Galactic Nuclei (AGN) \cite{Melia:2007vt}.  At the moment there is only
tentative evidence for intermediate-mass BHs (IMBHs) of mass $M\sim
10^2-10^5~M_\odot$ \cite{Miller:2003sc,Hurley:2007as,Berghea:2008rc}. Ringdown
detectability from these different classes of astrophysical BHs depends on
several factors, the first of which is the sensitivity of any given
gravitational wave detector.  Earth-based detectors, such as LIGO and Virgo,
have an optimal sensitivity band corresponding to stellar-mass BHs and IMBHs,
while the planned space-based detector LISA is most sensitive to ringdowns
from high-mass IMBHs and SMBHs (see Section \ref{sec:detectors} below).
Another important factor is the extent to which QNMs are excited in
astrophysical settings. The most promising scenarios to excite ringdown to a
detectable level are reviewed in Section \ref{sec:excitation}. Whatever the
source of the excitation, the frequency and amplitude of ringdown waves depend
on the mass and spin of the BH.  The evidence for astrophysical BHs and the
present understanding of their mass and spin distributions are reviewed in
Section \ref{sec:massspin}.  Measuring ringdown waves will allow us to extract
interesting information, ranging from accurate measurements of the mass and
spin of BHs to tests of the no-hair theorem of general relativity. These
applications of gravitational wave detection are reviewed in Sections
\ref{sec:inferringmassrd}-\ref{insptrans}.

\subsection{\label{sec:detectors}Physical parameters affecting ringdown detectability} 

Present and planned gravitational wave detectors are located at large distance
from astrophysical BHs. Therefore, for all practical purposes, a QNM as seen
by a detector is well approximated by the asymptotic behavior of the wave
equation at infinity, Eq.~(\ref{bcinfinityflat}). We can express the waveform
measured at the detector as a linear superposition of the gauge-invariant
polarization amplitudes $h_+,\,h_{\times}$, where, for a given mode
$(l,~m,~n)$,
\beq h_{+}&=& \frac{M}{r} {\rm Re}\left[ {\cal A}^{+}_{lmn}
e^{i(\om t+\ph^{+})}e^{-t/\ta} \Slm (\iota,\beta)
\right] \,,\\
h_{\times}&=&\frac{M}{r} {\rm Im}\left[ {\cal A}^{\times}_{lmn}
e^{i(\om t+\ph^{\times})} e^{-t/\ta} \Slm (\iota,\beta)
\right]\,.
\label{waveform0}
\eeq 
Here ${\cal A}^{+,\times}_{lmn}$ and $\ph^{+,\times}$ are the (real) amplitude
and phase of the wave, and $\Slm (\iota,\beta)$ denotes spin-weighted
spheroidal harmonics of spin-weight $-2$ \cite{Berti:2005ys}. The angles
$(\iota,\beta)$ are adapted to the source, so that the $z-$axis is aligned
with the spin of the BH. Interferometric detectors are sensitive to the
effective strain
\beq h = h_{+} F_+(\theta_S,\phi_S,\psi_S) + h_{\times}
F_\times(\theta_S,\phi_S,\psi_S) \,,
\label{detectwave}
\eeq
where $F_{+,\times}$ are pattern functions that depend on the orientation of
the detector and the direction of the source (specified by the polar angles
$\theta_S$, $\phi_S$) and on a polarization angle $\psi_S$
\cite{Berti:2005gp,KipThorneBook}.
A crucial quantity for gravitational wave detection is the signal-to-noise
ratio (SNR) $\rho$, defined as
\be\label{SNRdef}
\rho^2 = 4\int_0^\infty \f{\tilde h^*(f) \tilde h(f)}{S_h(f)}df\,,
\ee
where $\tilde h(f)$ is the Fourier transform of the waveform, and $S_h(f)$ is
the noise spectral density of the detector \cite{Finn:1992wt}. In discussing
the SNR we will usually average over source direction, detector and BH
orientations, making use of the sky averages: $\langle F_+^2\rangle=\langle
F_\times^2\rangle=1/5$, $\langle F_+F_\times\rangle=0$, and $\langle |\Slm|^2
\rangle=1/4\pi$. An analysis taking into account different sky locations and
orientations of the source probably requires Monte Carlo methods. At the
moment, such an analysis is still lacking.

Our chances of detecting and measuring ringdown waves are mainly determined by
the BH's mass $M$, by the spin parameter $a$, and by the ringdown efficiency
$\epsilon_{\rm rd}$. The latter quantity is defined as the fraction of the
total mass-energy of the system radiated in ringdown waves, and it is well
approximated by \cite{Berti:2005ys,Flanagan:1997sx}
\be\label{epsrd}
\epsilon_{\rm rd}\approx 
\frac{Q_{lmn}M\omega_{lmn}}{32\pi}
\left [\left ({\cal A}^{+}_{lmn}\right )^2 +
\left ({\cal A}^{\times}_{lmn}\right )^2 \right ]\,.
\ee
These three parameters (mass, spin and efficiency) affect the detectability
and measurability of the signal in different ways.

The BH {\em mass} sets the frequency scale and damping time of the emitted
radiation. For a Schwarzschild BH, the fundamental QNM with $l=2$ (that
dominates the radiation in most cases \cite{Flanagan:1997sx}) has frequency
$f$ and damping time $\tau$ given by
\beq\label{QNMestimate}
f&=&1.207\cdot 10^{-2}(10^6M_\odot/M)~{\rm Hz}\,,\\
\tau&=&55.37(M/10^6M_\odot)~{\rm s}\,.
\eeq

Earth-based detectors are limited at low frequency by a seismic cutoff $f_s$
(a plausible estimate for second-generation detectors being $f_s\sim 10$~Hz
for the Einstein Gravitational Wave Telescope, and $f_s\sim 20$~Hz for
Advanced LIGO).
Therefore they can detect the fundamental $l=m=2$ QNM as long as the BH mass
$M\lesssim 1.2\times 10^4 ({\rm Hz}/f_s)M_\odot$ if the BH is non rotating,
and $M\lesssim 2.7\times 10^4 ({\rm Hz}/f_s)M_\odot$ if the BH is rotating
near the Kerr limit (see Table I in Ref.~\cite{Berti:2007zu}). LISA is limited
at high masses (low frequencies) by acceleration noise, and at low masses by
the condition that the damping time $\tau$ should be longer than the
light-travel time $T_{\rm light}\simeq 16.7$~s corresponding to the planned
armlength ($L\simeq 5\cdot 10^9$~m). Thus LISA can detect ringdown waves from
BHs in the range $10^5M_\odot\lesssim M\lesssim 10^9M_\odot$
\cite{Berti:2005ys}. To summarize: Earth-based detectors are sensitive to the
ringdown of stellar-mass BHs and of relatively low-mass IMBHs, and LISA can
observe mergers of IMBHs and SMBHs throughout the whole universe. There is a
chance that the ongoing ringdown searches in data from Earth-based
gravitational wave detectors
\cite{Creighton:1999pm,Goggin:2006bs,Abbott:2009km,Nakano:2003ma,Nakano:2004ib,Tsunesada:2004ft,Tsunesada:2005fe}
may provide the first incontrovertible evidence of the existence of IMBHs.

As discussed in Section \ref{sec:qnmkerr}, the BH {\em spin} (for a given BH
mass) determines all frequencies of the Kerr QNM spectrum. For QNMs with $m>0$
the quality factor increases with spin (see Fig.~\ref{fig:fQKerr}).  Since the
detectability of a gravitational wave signal by matched filtering scales with
the square root of the number of cycles, highly spinning BHs could be the best
candidates for detection \cite{Flanagan:1997sx}.  However, exciting QNMs of
fast-rotating BHs seems to be harder, as the excitation factors tends to zero
as $a/M\to 1$ (see \cite{Ferrari:1984zz,Berti:2006wq} and Section
\ref{inversioncontour}).  In hindsight, this is not surprising: the build-up
of energy in a long-lived resonant mode usually takes place on a timescale
similar to the eventual mode damping, so it should be difficult to excite a
QNM with characteristic damping several times longer than the dynamical
timescale of the excitation process \cite{Glampedakis:2001js}. Numerical
simulations of the merger of comparable-mass BH binaries suggest that QNM
excitation is mildly dependent on the initial spin of the components (see
\cite{Berti:2007nw} and Section \ref{sec:excitation}), but further
investigation is required to clarify this issue.

There are different ways of quantifying the excitation of QNMs by generic
initial data
\cite{Leaver:1986gd,Sun:1988tz,Andersson:1995zk,Nollert:1998ys,Berti:2006wq}.
One can operationally define a {\em ringdown efficiency} $\epsilon_{\rm rd}$,
which is directly related to the gravitational wave amplitude as illustrated
by Eq.~(\ref{epsrd}), and hence to the SNR of the signal for a given detector
\cite{Flanagan:1997sx,Berti:2005ys}.  Different empirical notions for the
``ringdown starting time'' (which is intrinsically ill-defined
\cite{nollertthesis}) yield very different ringdown efficiencies
\cite{Dorband:2006gg,Berti:2007fi,Buonanno:2006ui}. From the point of view of
detection, a suitable definition of the ringdown starting time is the one
proposed by Nollert, using an ``energy-maximized orthogonal projection'' of a
given numerical waveform onto QNMs \cite{Berti:2007fi,Berti:2007zu}.  The {\em
  relative} excitation of different modes is even harder to determine than the
overall ringdown efficiency, but it is particularly relevant for tests of the
no-hair theorem using ringdown waves
\cite{Dreyer:2003bv,Berti:2005ys,Berti:2007zu}.  Ref.~\cite{Berti:2006hb}
discusses this issue in a general context, and Berti {\it et al.} discuss this
issue in a general context \cite{Berti:2006hb}, and give preliminary estimates
of the excitation of different multipoles in binary BH mergers
\cite{Berti:2007fi,Berti:2007nw}.

In Section \ref{sec:excitation} we review promising astrophysical scenarios
that could produce detectable ringdown signals (i.e., large efficiencies):
accretion, stellar collapse leading to BH formation and compact object
mergers. In Section \ref{sec:massspin} we summarize the current theoretical
and experimental understanding of BH masses and spins.

\subsection{\label{sec:excitation}Excitation of black hole ringdown in astrophysical settings} 

In principle, most dynamical processes involving BHs excite QNMs to some
degree. For the purpose of gravitational wave detection from astrophysical
BHs, the question is not whether QNMs are excited, but whether they are
excited {\em to a detectable level}. BH QNMs can be excited in a variety of
astrophysical settings, such as accretion, collapse and compact binary
mergers. As we will see, analytical estimates and numerical calculations show
that the most promising source of detectable ringdown waves is the merger of
two compact objects leading to BH formation.

\subsubsection*{Ringdown excitation by accretion} 

\begin{figure*}[ht]
\begin{center}
\begin{tabular}{cc}
\epsfig{file=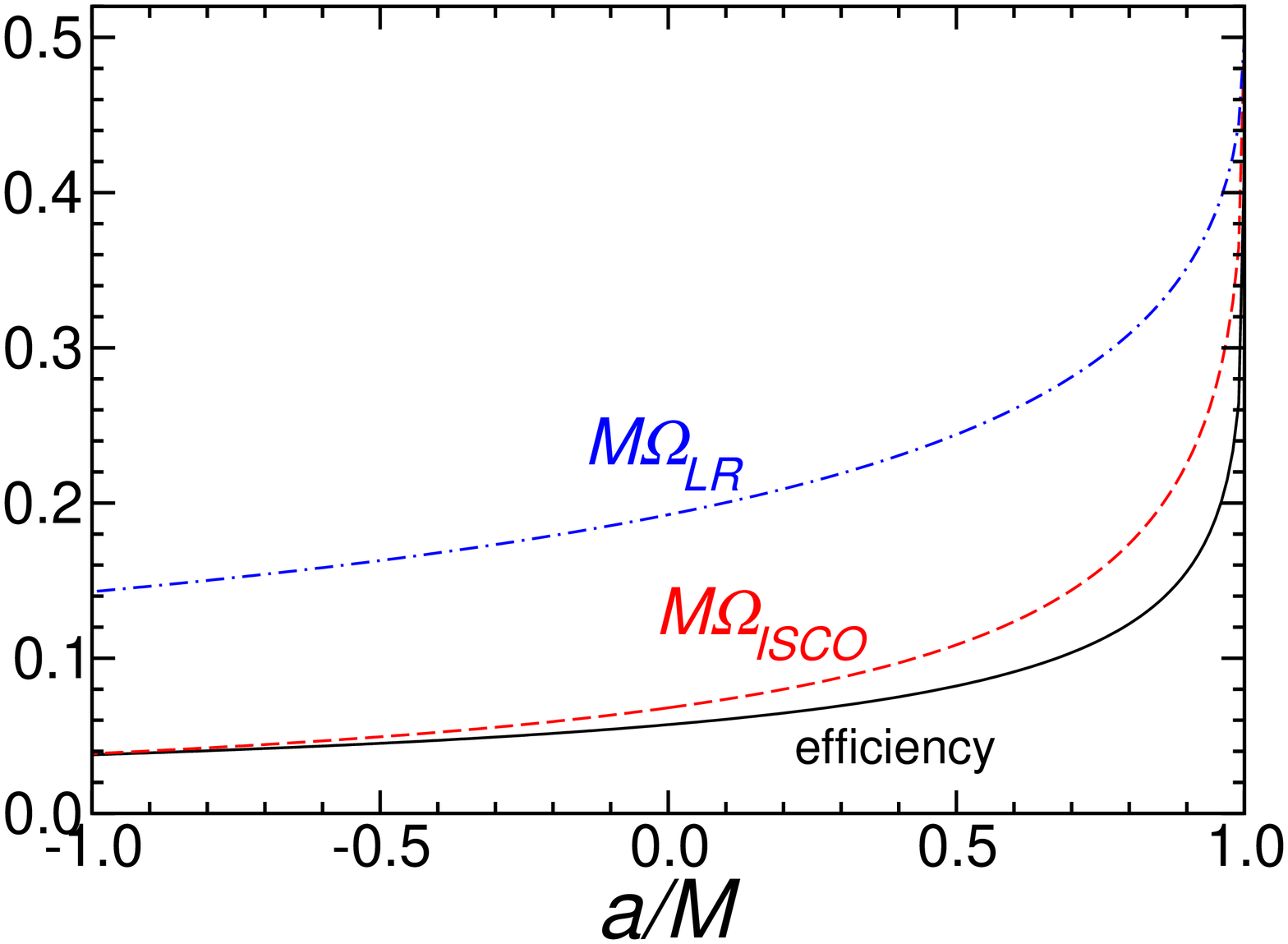,width=5.2cm,angle=-90} &
\epsfig{file=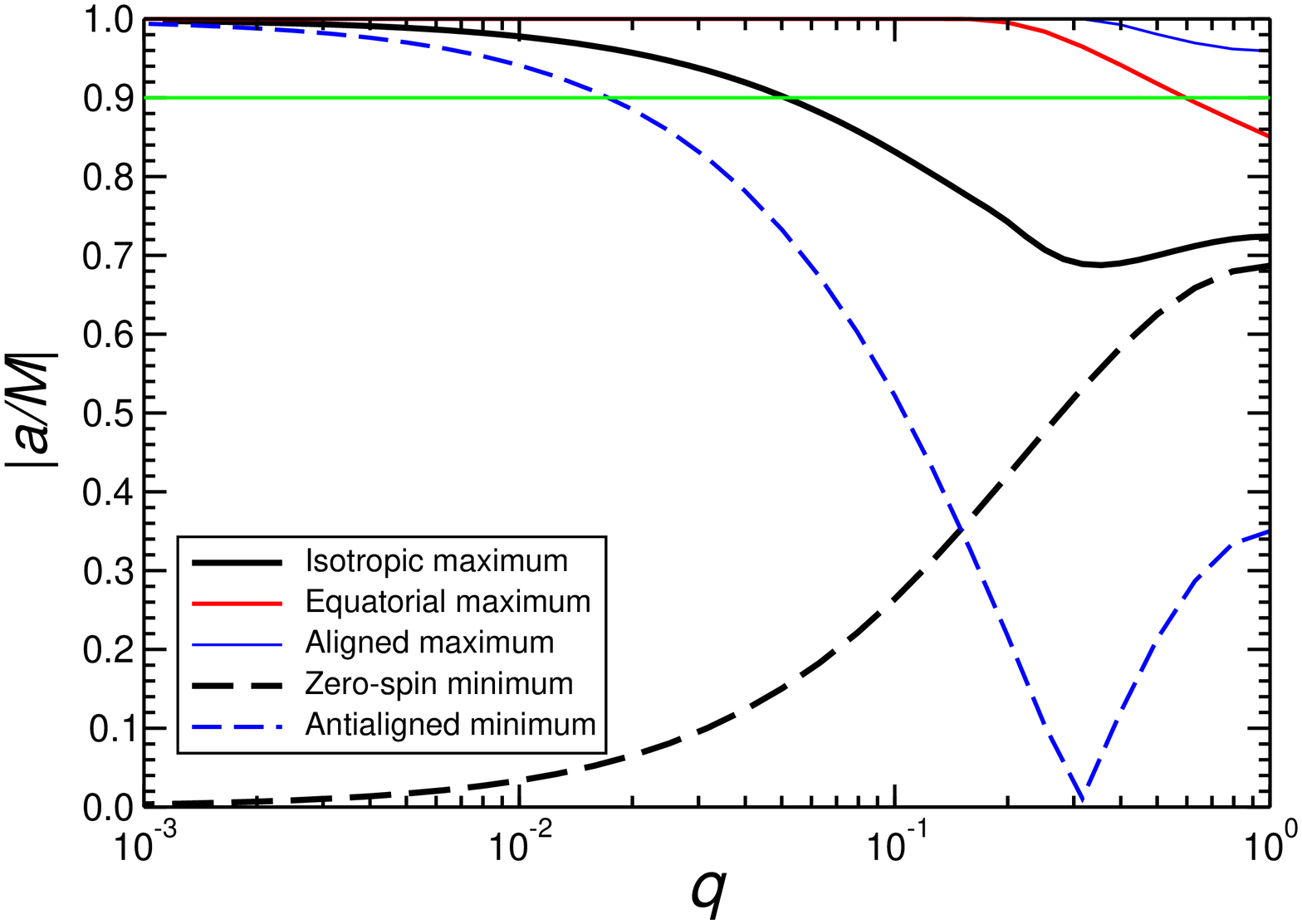,width=5.2cm,angle=-90} 
\end{tabular}
\end{center}
\caption{\label{fig:lightring} Left: ISCO frequency, light-ring frequency and
  radiation efficiency for Kerr BHs (positive values of $a/M$ refer to
  corotating orbits, negative to counterrotating orbits). Right: bounds on the
  final spin expected to result from a binary BH merger as a function of the
  binary's mass ratio $q$ (from Ref.~\cite{Berti:2008af}).}
\end{figure*}

An early study highlighting the importance of QNM ringing is the classic
analysis of the gravitational radiation emitted by particles falling radially
into a Schwarzschild BH \cite{Davis:1971gg}.
Unlike stellar oscillation modes (which play an important role in the orbital
dynamics of compact binaries \cite{Bildsten:1992my,
Gualtieri:2001cm,Pons:2001xs,
Flanagan:2007ix}) BH QNMs are hard to excite by matter orbiting around the
BH. The reason is that weakly damped QNMs are associated with unstable
geodesics at the light ring (see \cite{Cardoso:2008bp} and references
therein), and for Kerr BHs the light-ring frequency is always larger than the
frequency of the innermost stable circular orbit (ISCO), as illustrated in the
left panel of Fig.~\ref{fig:lightring}. Higher overtones may have lower
frequencies, but they are harder to excite because their quality factor is too
small.
According to this intuitive description, QNM excitation requires the accreting
matter to cross the light ring. Indeed, QNMs are always excited by particles
falling along generic geodesics into Kerr BHs (see \cite{Nakamura:1987zz} and
Appendix C of \cite{Berti:2007fi} for comprehensive lists of
references). Lumps of matter accreting onto a BH at appropriate rates could
potentially excite QNMs to a detectable amplitude. Unfortunately, even the
most optimistic estimates suggest that the wave amplitude is too small
\cite{Fryer:2001zw}. For generic accretion flows, simple analytical arguments
\cite{Berti:2006hb} show that {\em destructive} interference reduces the
ringdown amplitude \cite{Nakamura:1987zz,Lousto:1996sx,Nagar:2006eu}. This
conclusion is confirmed by numerical simulations of neutron star (NS)-BH
mergers: if the NS is tidally disrupted well before merger, accretion of the
NS material onto the BH proceeds incoherently, and the ringdown signal is
replaced by an abrupt cutoff in Fourier space at the tidal disruption
frequency (see Section \ref{sec:mergers} and Ref.~\cite{Shibata:2009cn}). If
it occurs in nature, a highly intermittent hyperaccretion scenario
\cite{ArayaGochez:2003vb} may excite QNM ringing in a nearly resonant fashion
for large spin parameters, and potentially lead to ringdown radiation strong
enough to be detectable by Advanced LIGO.

\subsubsection*{\label{ssec:collapse}Collapse to a black hole} 

There are several excellent reviews on gravitational wave signatures from
supernova core collapse and on the underlying physics
\cite{Nakamura:1987zz,Fryer:2003,Kotake:2005zn,Janka:2006fh,Ott:2008wt}.  When
the core of a massive star collapses, it produces a nonspherical protoneutron
star. 
Depending on details of the supernova explosion, the protoneutron star may
collapse to a BH, emitting a burst of gravitational waves due to the rapidly
shriking mass-quadrupole moment of the protoneutron star and to the QNM
ringing of the nascent BH.

Perturbative calculations of gravitational wave emission from rotating
gravitational collapse to a BH were first carried out in the 1970s by
Cunningham, Price and Moncrief \cite{Cunningham1,Cunningham2,Cunningham3},
improved upon by Seidel and collaborators
\cite{Seidel:1987in,Seidel:1988za,Seidel:1990xb,Seidel:1991my} and more
recently by Harada, Iguchi and Shibata \cite{Harada:2003bu}. These studies
suggest that gravitational waves are mainly generated in the region where the
Zerilli potential is large, and that the signal is usually dominated by QNM
ringing of the finally formed BH (see \cite{Harada:2003bu} for
exceptions). Simplified simulations based on a free-fall (Oppenheimer-Snyder)
collapse model yield a small energy output, with a typical core collapse
radiating up to $\simeq 10^{-7} M$ in gravitational waves. Most of the
radiation is quadrupolar ($l=2$), radiation in $l=3$ being typically two to
three orders of magnitude smaller (see Fig.~9 in \cite{Cunningham1}).  Recent
perturbative studies suggest that magnetic fields could increase the energy
output by several orders of magnitude \cite{Sotani:2007jv,Sotani:2009cm}.

The first numerical simulation of collapse in two dimensions was carried out
by Nakamura \cite{Nakamura:1981}, but numerical problems prevented
gravitational radiation extraction. For a long time the seminal 1985 work of
Stark and Piran \cite{Stark:1985da} has been the only nonperturbative,
axisymmetric calculation of gravitational wave emission from stellar
collapse. The waveform resembles that emitted by a point particle falling into
a BH, but with a reduced amplitude.  The total energy emitted increases with
the rotation rate, ranging from $\sim 10^{-8} M$ for $a/M=0$ to $\sim 7\times
10^{-4} M$ as $a/M\to 1$. Rotational effects halt the collapse for some
critical value of $(a/M)_{\rm crit}$ which is very close to unity, and depends
on the artificial pressure reduction used to trigger the collapse. Stark and
Piran find that the energy emitted $E/M\simeq 1.4\times 10^{-3}(a/M)^4$ for
$0<a/M<(a/M)_{\rm crit}$, and for larger spins it saturates to a maximum value
$\epsilon_{\rm max}\sim 10^{-4}$. Note however that the maximum energy
radiated is very sensitive to the amount of artificial pressure reduction used
to trigger the collapse: the 99\% pressure reduction used in the simulations
of Stark and Piran essentially produced a free-fall collapse, presumably {\it
  overestimating} the radiation efficiency.

This calculation has recently been improved using a three-dimensional code
\cite{Baiotti:2004wn,Baiotti:2005vi,Baiotti:2006wm,Baiotti:2007np}.
Ref.~\cite{Baiotti:2005vi} choose as the initial configuration the most
rapidly rotating, dynamically unstable model described by a polytropic
equation of state (EOS) with $\Gamma=2$ and $K=100$, having a dimensionless
rotation rate $\simeq 0.54$, and trigger collapse by reducing the pressure by
$\lesssim 2\%$. The ``$+$'' polarization is essentially a superposition of
modes with $l=2$ and $l=4$, and the ``$\times$'' polarization is a
superposition of modes with $l=3$ and $l=5$.  The energy lost to gravitational
waves according to these simulations is at most $\simeq 1.45\times 10^{-6}
(M/M_\odot)$, two orders of magnitude smaller than the estimate by Stark and
Piran for the same value of the angular momentum, but larger than the energy
losses found in recent calculations of rotating stellar core collapse to
protoneutron stars \cite{Mueller:2003fs}. Ref.~\cite{Baiotti:2007np} confirms
the basic scaling $E/M\approx (a/M)^4$ for $a/M\lesssim 0.54$ (the largest
rotation rates yielding equilibrium models in uniform rotation), but at
variance with Stark and Piran's simulations it finds that the efficiency has a
local maximum for large rotation rates.  If the collapse is triggered only by
pressure depletion, the overall efficiency for uniformly rotating models is
quite low ($E/M\approx 10^{-7}-10^{-6}$), the mass quadrupole does not change
very rapidly, and higher multipoles are not significantly excited.  However
the total energy radiated can increase by up to two orders of magnitude if
velocity perturbations are present in the collapsing star.  A rapidly rotating
polytropic star at 10~kpc can produce maximum characteristic amplitudes
$h_c\approx 5\times 10^{-22}(M/M_\odot)$ at characteristic frequencies ranging
between $f_c\approx 500$~Hz and $f_c\approx 900$~Hz, and could be detectable
with SNRs as large as about 30 by Earth-based interferometers.

In conclusion, the ringdown efficiency is very sensitive to poorly known
details of the mechanism triggering the collapse.  Simulations with more
realistic microphysics are required to study the complex signal preceding the
ringdown phase, as well as details of QNM excitation by matter accreting onto
the newly formed BH.

\subsubsection*{\label{sec:mergers}Mergers of compact objects leading to black hole formation}

Numerical relativity simulations of compact binary mergers made enormous
progress since 1999, when the two classic reviews on QNMs
\cite{Kokkotas:1999bd,Nollert:1999ji} were written. This section is an attempt
to summarize aspects of this progress of interest for the detection of
ringdown waves from astrophysical BHs.

Since NSs cannot have masses larger than about $3M_\odot$,
Eq.~(\ref{QNMscale}) implies that NS-NS mergers are potentially relevant only
for the detection of BH ringdowns by Earth-based interferometers, such as LIGO
and Virgo. On the other hand, NS-BH and BH-BH mergers are plausible targets
for ringdown detection by both Earth-based and future space-based detectors,
so we will discuss them in more detail.

\noindent
{\it (1) NS-NS mergers -} In 1999 Shibata and collaborators carried out the
first successful equal-mass NS-NS merger simulation for a polytropic index
$\Gamma=2$ \cite{Shibata:1999hn,Shibata:1999wm}.  The first reasonably
accurate calculation of gravitational waveforms was possible a few years
later, in 2002 \cite{Shibata:2002jb}. Unequal-mass binaries with a $\Gamma=2$
polytropic EOS were studied in \cite{Shibata:2003ga}, where it was found that
a BH forms when the total rest-mass of the system is larger than $\sim 1.7$
times the maximum allowed rest-mass of spherical NSs, irrespective of mass
ratio (which however affects significantly the waveforms and the mass of the
disk forming around the newly born BH).  More realistic EOSs and larger
parameter spaces were considered in Refs.~\cite{Shibata:2005ss,Shibata:2006nm}.
Ref.~\cite{Shibata:2006nm} uses stiff EOSs and models binary NSs of ADM mass
$M\gtrsim 2.6M_\odot$. For all mass ratios $0.65\lesssim q\leq 1$, a BH forms
whenever the mass $M>M_{\rm th}$. The threshold value $M_{\rm th}\simeq
1.3-1.35M_{\rm max}$ (where $M_{\rm max}$ is the maximum mass allowed by the
given EOS for cold, spherical NSs) depends on the EOS. If $M<M_{\rm th}$ the
merger results in a hypermassive NS of large ellipticity, emitting
quasiperiodic gravitational waves at frequencies $\sim 3-4~$kHz for $\lesssim
100~$ms. After this phase the NS may or may not collapse to a BH. For total
binary masses in the range $M\approx 2.7-2.9M_\odot$, the $l=m=2$ ringdown
frequency emitted in the collapse to a BH is $\approx 6.5-7$~kHz, with
amplitude $\sim 10^{-22}$ at a distance of 50~Mpc. Therefore BH ringdown from
NS-NS mergers is unlikely to be detected: the amplitude is too low and (most
importantly) the frequency is too high for present and planned Earth-based
gravitational wave detectors. It is interesting to note that NS-NS merger
simulations typically lead to final BH spins $a/M\approx 0.8$
\cite{Shibata:2006nm,Yamamoto:2008js}, not very different from the value
$a/M\simeq 0.69$ predicted by equal-mass BH merger simulations.  Various
groups have recently carried out NS-NS merger simulations with magnetic
fields, finding that aligned poloidal fields can delay the merger and strongly
affect the gravitational wave signal
\cite{Anderson:2008zp,Liu:2008xy,Baiotti:2009gk}.

\noindent
{\it (2) NS-BH mergers -} 
NS-BH binaries are potentially among the most interesting BH ringdown
sources. From Eq.~(\ref{QNMestimate}), the QNM frequency for non-spinning BHs
with $M\gtrsim 10M_\odot$ is within the sensitivity window of LIGO and
Virgo. Theory and observations
\cite{Belczynski:2007xg,Remillard:2006fc,Silverman:2008ss} suggest that
binaries containing BHs with masses in this range should be
common. Furthermore, high mass-ratio systems do not tidally strip the NS,
producing a merger with a clean ringdown signal. The simulations described in
the rest of this section suggest that NS-BH mergers are promising sources of
ringdown waves and (when the BH is spinning) excellent candidates as central
engines for gamma-ray bursts (GRBs).

NS-BH merger simulations became possible once the moving puncture approach
\cite{Campanelli:2005dd,Baker:2005vv} proved successful in evolving BH-BH
binaries.  Shibata and Uryu \cite{Shibata:2006bs,Shibata:2006ks} first evolved
two binaries consisting of (1) a non-rotating BH of mass $\approx 3.2M_\odot$
or $\approx 4M_\odot$, respectively, and (2) a NS of rest mass $\approx
1.4M_\odot$, modeled by a $\Gamma$-law EOS with $\Gamma=2$. A larger set of
simulations with non-spinning BHs of masses $M\sim 3.3-4.6M_\odot$ and NSs of
mass $M\sim 1.4M_\odot$ was performed in Ref.~\cite{Shibata:2007zm}. The NS is
tidally disrupted, and the system results in a BH with spin $a/M\approx
0.5-0.6$. Non-spinning NS-BH mergers can only produce massive tori and fuel
short-hard GRBs if the NS has compactness $M/R\lesssim 0.145$. Furthermore,
when tidal disruption occurs the QNM amplitude quickly decreases, because the
incoherent accretion of material is ineffective at resonantly exciting the
mode (see Fig.~6 of Ref.~\cite{Shibata:2007zm}).

The first code capable of handling NS-NS, NS-BH and BH-BH binary evolutions
comprising more than 4 orbits was recently described in
Ref.~\cite{Yamamoto:2008js}.  Gravitational waveforms from NS-BH mergers were
presented in Ref.~\cite{Shibata:2009cn} and classified into three classes as follows: 1)
For small mass ratio $q$ and small NS compactness (e.g., if $q\lesssim 3$ for
$M/R=0.145$ and $\Gamma=2$) tidal disruption occurs outside the ISCO, and
there is no QNM excitation; 2) For some systems, mass shedding occurs before
the binary reaches the ISCO. Most of the NS is swallowed by the BH before
tidal disruption is completed and QNMs are excited, but only to a low
amplitude; 3) If tidal effects do not play an important role the waveforms
shows significant QNM excitation, as they always do for BH-BH mergers. The
latter class of NS-BH mergers is clearly the most promising for BH ringdown
detection. Whenever ringdown is not significantly excited, kicks are also
suppressed: this confirms the crucial role played by the merger/ringdown phase
in determining the magnitude of the kick resulting from compact binary mergers
(see Section \ref{insptrans} below). The results of Ref.~\cite{Shibata:2009cn}
are in good agreement with NS-BH codes developed by other groups
\cite{Etienne:2007jg,Etienne:2008re,Duez:2008rb}. Ref.~\cite{Etienne:2007jg}
deals with systems where the NS is irrotational, the BH is non-rotating and
the mass ratio $q=1/3$. These simulations lead to the formation of BHs with
$a/M\approx 0.5-0.8$. Most of the NS material is prompty accreted, and no more
than 3\% of the NS mass is ejected into a gravitationally bound disk. This
disk mass, while larger than the typical values found in
Ref.~\cite{Shibata:2009cn}, is probably not enough to trigger short-hard GRBs.

Results from non-spinning NS-BH mergers show that the disk mass is typically
too low to fuel GRBs. For example, Fig.~7 of Ref.~\cite{Shibata:2009cn} shows
that, if the BH is non-spinning, the formation of a disk requires a ``fat'' NS
with radius $R\gtrsim 14$~km. The few simulations presently allowing for
(aligned or anti-aligned) BH spins have a drastically different outcome,
indicating that {\em spin plays a crucial role in fueling GRBs}
\cite{Etienne:2008re}. In these simulations the number of orbits before merger
increases as the spin is varied between $a/M=-0.5$ (anti-aligned), $0.0$ and
$0.75$ (for fixed mass ratio $q\simeq 1/3$). In the latter case the final BH
spin is $a/M\simeq 0.88$, and the tidal disruption of the NS leads to the
formation of a massive disk of about $0.2M_\odot$, potentially capable of
driving GRBs. The production of GRBs by NS-BH mergers and the possibility to
detect the final spin orientation using observations of the merger/ringdown
phase are promising areas of research for multi-messenger astronomy using a
network of ground-based detectors in conjunction with traditional
electromagnetic observations (see e.g.~\cite{Cutler:1994ys,Seto:2005xb,
  Kanner:2008zh,Bloom:2009vx,Nissanke:2009kt}).

\noindent
{\it (3) BH-BH mergers -} 
After 40 years of developments in numerical relativity, recent breakthroughs
\cite{Pretorius:2005gq,Campanelli:2005dd,Baker:2005vv} finally allowed
simulations of the merger and ringdown of BH binaries. Extensive collaborations
to use numerical merger waveforms in gravitational wave searches have just
started \cite{Aylott:2009ya}. A discussion of the accuracy of numerical
simulations and of their rapid progress in the last few years would take us
too far (see Refs.~\cite{Pretorius:2005gq,Hannam:2009rd} for reviews).
From the point of view of this review, the main result of these merger
simulations is that QNM ringing is observed in all binary BH merger
simulations. This is quite unlike NS-BH and NS-NS mergers, where tidal effects
can sometimes suppress ringdown excitation because of the incoherent accretion
of material onto the newly formed BH. Another important difference is that the
gravitational waveform from a BH-BH merger depends on the total mass of the
system via a trivial rescaling, so BH-BH systems are interesting for
ringdown detection by both Earth-based and space-based detectors.

The mass-ratio dependence of the ringdown efficiency can be estimated by
simple arguments \cite{Flanagan:1997sx,Berti:2007fi}.  The quadrupole moment
of a body of mass $M$ with a "bump" of mass $\mu\ll M$ is $Q\sim \mu M^2$. The
oscillation frequency of the system $f\sim 1/M$; hence, the radiated power
$dE/dt\sim \left(d^3Q/dt^3\right)^2\sim (f^3 Q)^2\sim \mu^2/M^2$. For a binary
with mass ratio $\mu/M$ the inspiral lasts $\sim (M/\mu)$ cycles times the
orbital time scale $T\sim M$, so the total energy loss during the inspiral is
$E_{\rm insp}\sim (M/\mu)(\mu^2/M^2)M\sim \mu$. By contrast, a typical
ringdown waveform lasts only a few cycles, so the ringdown energy loss $E_{\rm
  ringdown}\sim M(\mu^2/M^2)\sim \mu^2/M$ (compare with the classic result for
infalling particles in Ref.~\cite{Davis:1971gg}) and $E_{\rm ringdown}/E_{\rm
  insp}\sim \mu/M$: {\it ringdown is negligible with respect to inspiral for
  extreme-mass ratio binaries}. If we naively extrapolate these estimates to
bodies of comparable masses (interpreting $M$ as the total mass of the binary
and $\mu\to m_1 m_2/M$ as the reduced mass) we find that $E_{\rm
  ringdown}/M\sim \eta^2$, where $\eta\equiv \mu/M$ is the so-called {\em
  symmetric mass ratio} ($\eta\to 1/4$ in the comparable-mass limit).

Physical arguments to estimate the prefactors suggest that the merger/ringdown
waveform should actually dominate over inspiral for binaries with mass ratio
$q\gtrsim 1/10$ \cite{Flanagan:1997sx,Berti:2005ys}, and numerical simulations
of quasicircular inspirals of comparable-mass mergers have borne out this
expectation. For non-spinning binary BH mergers, the fraction of energy
radiated $(M-M_{\rm fin})/M$ (where $M$ denotes the total mass of two BHs in
isolation and $M_{\rm fin}$ is the mass of the final BH), as well as the final
spin $a/M$, have been extensively studied
\cite{Buonanno:2006ui,Berti:2007fi,Buonanno:2007pf,Baker:2008mj,Baker:2008mj}.
Ref.~\cite{Buonanno:2007pf} fits data from the simulations by the Goddard
group by a relation of the form
\beq
&&M_{\rm fin}/M=1+(\sqrt{8/9}-1)\eta-0.498(\pm 0.027)\eta^2\,,\\
&&a/M_{\rm fin}=\sqrt{12}\eta-2.900(\pm 0.065)\eta^2\,.
\eeq
This result is consistent with the fitting formula given in
\cite{Berti:2007fi} using a different normalization. Frequencies and damping
times of different multipolar components can be estimated using either a
standard least-squares algorithm \cite{Buonanno:2006ui} or Prony methods,
which are in many ways optimal to estimate the parameters of damped
exponentials in noise \cite{Berti:2007dg}. By monitoring the frequencies and
damping times after merger we can monitor inaccuracies in the higher
multipolar components of numerical simulations, and possibly explore
non-linear effects (see e.g. Section IV in Ref.~\cite{Berti:2007fi}).

Quantifying the fraction of energy radiated in ringdown is inherently
ambiguous. The reason for this difficulty can essentially be traced back to
the non-completeness of QNMs
\cite{nollertthesis,RevModPhys.70.1545,Nollert:1998ys,Berti:2007fi}.  An
operational viewpoint to isolate the ringdown contribution is given by
Nollert's energy-maximized orthogonal projection (EMOP) criterion
\cite{nollertthesis,Berti:2007fi}. The idea is to determine the starting time
of a ringdown waveform by assuming that the frequency of the ringdown waveform
is known, and performing matched-filtering (in white noise) of the numerical
waveform, using a damped sinusoid as the ``detection template''. Since
ringdown is essentially monochromatic, this should give a reasonably good,
frequency-independent and detector-independent estimate of the fraction of
energy that we can expect to detect by a ringdown search. For more details on
ringdown search techniques, see Section \ref{sec:detectability} below.

An application of the EMOP criterion to numerical simulations shows that
ringdown typically contributes $\sim 42\%$ of the energy radiated by the last
two cycles of BH-BH mergers with mass ratio $q\geq 1/4$ (see Tables VI and VII
of Ref.~\cite{Berti:2007fi}). By combining Prony methods for estimating
frequencies and the EMOP approach to estimate the ringdown starting time, we
find the following estimates for the energy emitted in ringdown as a result of
the merger of non-spinning, quasi-circular BH binaries \cite{Berti:2007fi}
\be
\f{E_{\rm ringdown}^{l=2}}{M}\simeq 0.271 \f{q^2}{(1+q)^4}\,,\qquad
\f{E_{\rm ringdown}^{l=3}}{M}\simeq 0.104 \f{q^2(q-1)^2}{(1+q)^6}\,.
\ee

For non-spinning, equal-mass binaries, the merger/ringdown signal as the
system's orbital angular momentum is reduced (so a quasicircular merger slowly
turns into an head-on collision) has been studied in
Refs.~\cite{Sperhake:2007gu,Hinder:2007qu}.  In the head-on limit one gets a
radiated energy $\sim 0.1\%M$. In the intermediate regime, where two
equal-mass BHs merge along orbits with large residual eccentricity, the
radiated energy decays (roughly) exponentially (see Table I and Fig.~7 in
Ref.~\cite{Sperhake:2007gu}) and the final BH spin has a local maximum $j_{\rm
  fin}\simeq 0.724\pm 0.13$ \cite{Sperhake:2007gu,Hinder:2007qu}. For
preliminary studies of the ringdown efficiency in the merger of quasicircular,
{\em spinning} binaries, see
Refs.~\cite{Berti:2007nw,Vaishnav:2007nm,Shoemaker:2008pe}.

Predictions of the spin of a BH resulting from a merger are very interesting
from the point of view of ringdown detection. For example, if we measure the
masses and spins during the inspiral we may be able to {\em predict} the final
spin and reduce the errors in parameter estimation
\cite{Luna:2006gw,Ajith:2009fz}. For a summary of semianalytic models and
fitting formulas to predict the final spin from generic mergers, we refer the
reader to
\cite{Buonanno:2007sv,Rezzolla:2007rz,Rezzolla:2008sd,Barausse:2009uz}. Some
insight into the general outcome of a spinning merger can be obtained by
looking at the right panel of Fig.~\ref{fig:lightring}. For simplicity,
consider three different merger scenarios: i) in the {\em isotropic} scenario,
both BH spins are distributed isotropically; ii) in the {\em aligned} spin
scenario, the individual BH spins in the binary are assumed to be aligned (for
example, in ``wet mergers'' the alignment could be caused by torques from
accreting gas, as suggested in \cite{Bogdanovic:2007hp}); iii) in the {\em
  equatorial} merger scenario, the smaller BH is supposed to orbit in the
equatorial plane of the larger hole (e.g. because of Newtonian dynamical
friction in a flattened system), but the spin orientation of the smaller BH is
distributed isotropically. The right panel of Fig.~\ref{fig:lightring} shows
the maximum and minimum spin resulting from a merger in these three
scenarios. These curves are obtained by (1) fixing some value of the mass
ratio $q$, (2) averaging over angles according to the three different
assumptions listed above, and (3) maximizing or minimizing the final {\em
  average} spin resulting from a merger (where the average is computed using
the fitting formulas of Ref.~\cite{Rezzolla:2007rz}) in the $(\hj{1},\hj{2})$
plane, where $\hj{i}$ is the spin magnitude of BH $i=1,~2$. Not surprisingly,
the minimum average final spin always corresponds to the case where both BHs
are non-spinning (dashed black line). The maximum average spin in the three
cases is shown by the continuous black (isotropic), red (equatorial) and blue
(aligned) lines. The dashed blue line shows the (modulus of) the minimum spin
that could be achieved if we allow for {\it antialignment} of both spins with
respect to the orbital angular momentum (a spin flip becomes possible when the
mass ratio $q\approx 1/3$). The most interesting prediction of this plot is
the existence of a narrow funnel between the solid black and dashed black
lines: on average, isotropic major mergers (with $q\gtrsim 0.2$ or so) always
produce a final spin which is very close to the spin resulting from equal-mass
non-spinning BH binaries, i.e. $a/M\sim 0.69$. Furthermore, in all three
scenarios the most likely spin resulting from ``major'' mergers with $q\gtrsim
0.1$ is quite close to $\hjf \simeq 0.69$.

A critical assessment of the available predictions on the final spin and on
the final kick is outside the scope of this review; the interested reader is
referred to Ref.~\cite{Kesden}.

\subsection{\label{sec:massspin}Astrophysical black holes: mass and spin
  estimates}

From the discussion in the previous section it should be clear that ringdown
detectability depends crucially on the physical parameters of astrophysical
BHs.  What are the most promising scenarios leading to the formation of
massive stellar-mass BHs that would predominantly be seen in merger/ringdown
by Earth-based interferometers? What are the event rates for mergers to be
observed by LISA?  Given that ringdown detectability scales strongly with mass
ratio, what is the most likely mass ratio for binaries whose merger is
detectable by LIGO and LISA?  What is the most likely spin magnitude of BHs in
binaries, and what are the odds that we can unveil connections between
merger/ringdown waveforms and the central engine of GRBs? In this section we
briefly survey theoretical expectations and state-of-the-art measurements of
the mass and spin distribution of BHs.  We summarize some of the most relevant
information that astronomers have collected by working as busy bees over the
past thirty years,
paying particular attention to the implications for the detection of
gravitational waves from the merger/ringdown of BH binaries.

Our focus here is on testing the BH nature of astrophysical objects by
gravitational wave observations, but there are excellent reviews on measuring
BH masses, spins and (possibly) providing evidence of an event horizon by
``traditional'' electromagnetic astronomy. Narayan reviews the status of BH
astrophysics, focusing on observational progress in measuring mass and spin
and on (circumstantial) observational evidence for the defining property of a
BH: the event horizon \cite{Narayan:2005ie}. Psaltis discusses how
electromagnetic observations of BHs and neutron stars can be used to probe
strong-field gravity; in the process he describes various ways of identifying
BHs and measuring their properties, including continuum spectroscopy, line
spectroscopy, and attempts at imaging the vicinity of BHs to constrain their
angular size \cite{Psaltis:2008bb}.

\subsubsection*{Stellar-mass black hole candidates} 

The most accurate mass measurements for stellar-mass BH candidates are made
via dynamical methods, that is, by looking at how the unseen BH affects the
orbit of a companion star. Consider a test particle in circular orbit around
the BH. If the orbit is wide enough for Newtonian physics to apply, then the
mass $M=\omega^2 r^3=v^2r=v^3/\omega$, where $r$ is the orbital separation,
$v$ is the orbital velocity and $\omega=2\pi/T$ with $T$ the orbital period
(simple modifications can account for orbital eccentricity).  By measuring any
two of $v$, $r$ and $\omega$, we may estimate the BH mass $M$. In the case of
BH X-ray binaries it is relatively easy to measure $\omega$ and the maximum
line-of-sight Doppler velocity $K_c=v\sin \iota$ of the companion star. From
these quantities one can compute the ``mass function''
\be
f(M)\equiv\f{K_c}{\omega}=\f{M\sin^3\iota}{(1+M_c/M)^2}\,,
\ee
where $M,M_c$ are the masses of the BH candidate and of the companion,
respectively.  The inclination angle $\iota$ of the orbit can be estimated
from the light curve of the binary, and sometimes it's even possible to
estimate $M_c$. By combining measurements of $\omega$ and $K_c$ with estimates
of $\iota$ and $M_c$, one can in principle determine the masses of both binary
members. However, to identify BH candidates the essential point is to note
that the mass function $f(M)$, which depends only on $\omega$ and $K_c$,
provides a strict {\em lower bound} on $M$. Since NSs cannot be more massive
than about $3M_\odot$ \cite{Rhoades:1974fn}, all X-ray binaries for which
$f(M)\gtrsim 3M_\odot$ should contain a BH.  Remillard and McClintock
\cite{Remillard:2006fc} review the phenomenology of 20 X-ray binaries with
dynamically confirmed BHs, presenting a census of BH candidates and a critique
of different methods for measuring spins. Their Table I provides a list of
(lower bounds on) the mass of about 20 BH candidates. The most massive
stellar-mass BH candidate to date is IC 10 X-1, with a minimum mass $M=23.1\pm
2.1M_\odot$ if the companion's mass $M_c=17M_\odot$ ($M=32.7\pm 2.6M_\odot$ if
one trusts an estimate of $M_c=35M_\odot$ for the companion)
\cite{Silverman:2008ss}. This system is particularly interesting for
gravitational wave detection, because it should become a close double BH
binary with coalescence time of $\sim 2-3$~Gyr \cite{Bulik:2008ab}.

The relative importance of ringdown with respect to inspiral waves decreases
for extreme mass ratios. For the stellar-mass binaries of interest for
Earth-based interferometers, population synthesis codes suggest that $q$
should always be very close to unity \cite{Belczynski:2007xg}. A rather
speculative kind of source (in the absence of solid evidence for IMBHs)
consists of the intermediate mass ratio inspiral (IMRI) of a compact object,
such as a NS or BH, into an IMBH. The relatively low energy content in
ringdown waves is compensated, in this case, by the fact that the ringdown
frequency is close to the minimum of the Advanced LIGO noise power spectral
density (see Appendix B of Ref.~\cite{Mandel:2007hi}). Another promising
ringdown source for advanced Earth-based interferometers (albeit with highly
uncertain event rates) are IMBH-IMBH inspirals. These binaries, if they are
numerous enough to be detectable, present an interesting data analysis
challenge: the initial inspiral phase could be detected by LISA, while the
ringdown phase is in the optimal bandwidth for Advanced LIGO, that could
therefore be used for ``follow-up'' ringdown searches \cite{Fregeau:2006yz}.

\subsubsection*{Supermassive black hole candidates} 

A good review of SMBH observations from a historical perspective can be found
in Ref.~\cite{Melia:2007vt}. The first quasar was identified in 1963
\cite{Schmidt:1963}, when the Kerr solution had just been discovered
\cite{Kerr:1963ud} and its astrophysical relevance was unclear. In the
intervening years astronomers gathered strong observational evidence for the
presence of SMBHs in the bulges of nearly all local, massive galaxies
\cite{Kormendy:1995er,Magorrian:1998,McLure:2002}. Reliable mass estimates are
available for many of these systems.
The most precise measurement comes from observations of stellar proper motion
at the center of our own galaxy, indicating the presence of a ``dark object''
of mass $M\simeq (4.1\pm 0.6)\times 10^6M_\odot$
\cite{Schodel:2002py,Ghez:2008ms} and size smaller than about one astronomical
unit \cite{Shen:2005cw}. A Schwarzschild BH of the given mass has radius
$R\simeq 0.081$ astronomical units, compatible with the observations. Any
distribution of individual objects within such a small region (with the
possible exceptions of dark matter particles or asteroids, which however
should be kicked out by three-body interactions with stars) would be
gravitationally unstable \cite{Maoz:1998,ColemanMiller:2005tg}.  Theoretical
alternatives to SMBHs (e.g., boson stars and gravastars) have been proposed by
various authors, but the formation process of these hypothetical objects is
unclear, and many of these exotic alternatives can be shown to be unstable
\cite{Cardoso:2008kj,Pani:2009fd}.
Another accurate mass measurement comes from the motion of the gas disk at the
centre of the nearby galaxy NGC 4258 \cite{Herrnstein:1999cw}, as monitored by
radio interferometry of the waves emitted from water molecules via maser
emission. The observations imply the presence of an object of mass $3.5\times
10^7M_\odot$ within $\sim 4\times 10^{15}$~m.
Other techniques include applications of the virial theorem to the velocity
dispersion of stars near the galactic center \cite{Kormendy:1995er} and
reverberation mapping to obtain more crude estimates for distant, variable
AGNs \cite{Gebhardt:2000sb}. The reader interested in a SMBH mass census can
consult Graham's survey \cite{Graham:2008uh}, listing 76 galaxies with direct
SMBH mass measurements and (when available) their host bulge's central
velocity dispersions. Graham also lists 8 stellar systems that could
potentially host intermediate-mass BHs.
For our purposes, it suffices to note that SMBHs have masses in the range
$M\sim 10^5-10^9M_\odot$, approximately proportional to the mass of the host
galaxies, $M\sim 10^{-3}~M_{\rm galaxy}$ \cite{Merritt:2000mi}.
There is an almost-linear relation between the mass of a SMBH and the mass of
the galactic bulge hosting it
\cite{Kormendy:1995er,Magorrian:1998,McLure:2002}. The BH mass is also tightly
correlated with other properties of the galactic bulge, such as the central
stellar velocity dispersion $\sigma$, the bulge light concentration and the
near-infrared bulge luminosity
\cite{Ferrarese:2000se,Marconi:2003hj,Graham:2008uh}.  These correlations
clearly indicate that SMBHs are causally linked to the surrounding galactic
environment. The growth of galaxies and SMBHs must be intertwined, and
observations of the merger and ringdown of SMBHs with LISA holds great promise
to clarify their formation history.

\subsubsection*{Supermassive black hole binary candidates} 

The SMBHs harboring nearby galactic cores are expected to grow via a
combination of mergers and accretion, and hierarchical merger models of galaxy
formation predict that binary SMBHs should be common in the Universe
\cite{Merritt:2004gc,Hopkins:2005fb,Hopkins:2005ms}. The merger of SMBH
binaries is one of the most luminous gravitational wave events in the
universe, and it is the strongest conceivable astrophysical source of ringdown
waves. Observational evidence for close SMBH binaries started emerging very
recently, and it is one of the most exciting frontiers of relativistic
astrophysics.

The formation of SMBHs during galaxy mergers is a challenging problem in
theoretical astrophysics. The general scenario was outlined in a pioneering
analysis by Begelman, Blandford and Rees \cite{Begelman:1980vb} (see
Ref.~\cite{Merritt:2004gc} for a more recent review). The evolution of a SMBH
binary can be roughly divided in three phases: i) as the galaxies merge, SMBHs
sink to the center via dynamical friction; ii) the binary's binding energy
increases because of gravitational slingshot interactions, i.e. the ejection
of stars on orbits intersecting the binary (these stars' angular momentum must
be in a region of phase space called the ``loss cone''); iii) if the binary
separation becomes small enough, gravitational radiation carries away the
remaining angular momentum. Notice that the gravitational wave coalescence
time is shorter for more eccentric binaries \cite{Peters:1963ux}, so
high-eccentricity binaries are slightly more likely to coalesce within a
Hubble time (see e.g.~\cite{Berti:2006ew}).
The transition from phase ii) to phase iii) is a field of active research,
that has been referred to as the ``final parsec problem''
\cite{Merritt:2004gc}. Since the binary will quickly eject all stars through
gravitational slingshot interaction, the problem is to find some mechanism
(such as gas accretion, star-star encounters and triaxial distortions of
galactic nuclei) to refill the loss cone. It is generally believed, based on
both theoretical and observational arguments, that efficient coalescence
should be the norm \cite{Berti:2006ew}.  The main point here is that only SMBH
binaries with separations $\lesssim 1$~pc can merge within a Hubble time under
the sole influence of gravitational radiation.

The mass ratio distribution is an important variable from the point of view of
ringdown detection.  The impact of different SMBH assembly models on the mass
and mass ratio distribution of detectable binaries has been discussed by
various authors. The general consensus is that mass ratios $q\lesssim 1/10$
(and down to $q\approx 10^{-3}-10^{-4}$) should be common
\cite{Volonteri:2002vz,Volonteri:2004cf,Sesana:2007sh,Koushiappas:2005qz,
  Gergely:2008mt,Gergely:2007ny}. This may not be true for mergers between BHs
in separate dark matter halos, because the smaller halo could get tidally
stripped and it would not be able to sink efficiently toward the center of the
main halo \cite{Taffoni:2003sr}.
In any case, SMBH mergers will be strong LISA ringdown sources even for modest
mass ratios.

Possible observational smoking guns of SMBH binaries (such as X-shaped radio
galaxies, double-double radio galaxies, helical radio jet patterns,
semi-periodic signals in lightcurves, double-peaked emission-line profiles and
galaxies which lack central cusps) are reviewed by Komossa
\cite{Komossa:2003wz}.
At the time of her review, the most spectacular example of a SMBH binary was
the ultraluminous infrared galaxy NGC 6240 \cite{Komossa:2002tn}, containing
two active SMBHs separated by a relatively short projected distance
$\sim 1$~kpc.
Since then, more quasars have been identified as promising hosts of SMBH
binaries\footnote{A binary AGN with separation $\sim 4.6$~kpc has been claimed
  in Arp 299 \cite{Ballo:2003ww}, and a system with projected distance $\sim
  10.5$~kpc has been found in the galactic pair ESO 509-IG066
  \cite{Guainazzi:2004ky}. Evans {\it et al.} \cite{Evans:2007nq} reveal the
  AGN nature of the companion of the FRII radio source 3C 321. Bianchi {\it et
    al.}  identify a binary AGN in Mrk 463 with projected separation $\sim
  3.8$~kpc \cite{Bianchi:2008un}.
Finally, using multifrequency observations with the Very Long Baseline Array
(VLBA), Rodriguez {\it et al.} \cite{Rodriguez:2006th,Rodriguez:2009ax} report
the discovery of a SMBH binary in the radio galaxy 0402+379 with a total
estimated mass of $1.5\times 10^8M_\odot$ and a projected orbital separation
of just 7.3~pc. This is the smallest orbital separation by more than two
orders of magnitude, but even for this relatively close binary the emitted
gravitational waves have frequency $\sim 2\times 10^{-13}$~Hz, way too low to
be observed by LISA, and a merger time $\sim 10^{18}$~yr (much longer than the
age of the Universe) {\it if} gravitational radiation is the only dissipative
mechanism.}.
Even more interestingly, in the last year three systems have been proposed to
host binary SMBHs at separations smaller than one parsec
\cite{Valtonen:2008tx,Boroson:2009va}. These observations could be extremely
important for our understanding of SMBH binary mergers and their rates
\cite{Volonteri:2009nh}.  Since they are still controversial and they would
take us too far from the main topic of this review, we briefly summarize them
for the interested reader in \ref{app:SMBHB}.

\subsubsection*{Spins: theoretical expectations} 

Estimating BH spins is an exciting observational frontier for the next decade
of observational astronomy
\cite{Brenneman:2009hs,McClintock:2009as,Miller:2009hh}. Present spin
estimates based on electromagnetic observations are all, to some extent,
model-dependent. Few dependable and accurate measurements are available, and
there are different opinions on the expected spin magnitude of both
stellar-mass BHs and SMBHs. The punchline of the theoretical work summarized
in this section is as follows: 1) stellar-mass BHs should typically retain
their natal spin, and so they can be used to infer information about the
mechanism triggering the collapse; 2) SMBH spins encode the history of the
hole, and particularly the relative importance of mergers and accretion in the
hierarchical formation process responsible for growing the holes.  The rest of
this section gives arguments in support of these conclusions.

\noindent {\em 1) Stellar-mass black holes:} Theoretical arguments suggest
that stellar-mass BHs in binaries retain the spin they had at birth: neither
accretion nor angular momentum extraction are likely to change significantly
their mass or spin.  A BH must accrete an appreciable fraction of its original
mass in order to significantly change its spin. For BHs with low-mass
companions, even the accretion of the entire companion star will only change
the spin by a small fraction; for BHs with high-mass companions, even
Eddington-limited accretion will only grow the BH spin by a small amount
before the high-mass companion explodes \cite{King:1999aq}. Therefore the spin
of stellar-mass BHs should depend mainly on their formation process. 

Detailed studies of spin evolution in compact binaries have been carried out
by the Northwestern group, focusing mostly on NS-BH binary systems and using
progressive improvements of the STARTRACK stellar evolution code (see
\cite{Belczynski:2007xg} and references therein).  The evolution prior to the
supernova explosion involves mass-transfer phases, which are expected to align
the spins of both the BH and the NS progenitor, but a significant natal kick
of the NS at birth is required to form a coalescing NS-BH binary
\cite{Kalogera:1999tq}. In general, the plane of the post-supernova orbit is
tilted with respect to the pre-supernova plane, and hence tilted with respect
to the BH spin axis by some angle $\iota$, inducing precession of the binary's
orbital plane. Preliminary results suggest that precession only marginally
impacts the detection of gravitational waves from the inspiral waves, but it
should be significant for parameter estimation
\cite{O'Shaughnessy:2005qc,Belczynski:2007xg,Grandclement:2003ck}.
Belczynski {\it et al.} \cite{Belczynski:2007xg} consider both BH-BH and NS-BH
binaries. For NS-BH binaries they confirm the qualitative predictions of
\cite{King:1999aq}: BHs cannot be significantly spun up by accretion in the
common envelope phase. For example, only 20\% of initially non-spinning BHs
spin up to $a/M>0.1$, and no BHs spin up to $a/M\gtrsim 0.5$. The spin-up is
even smaller for BH-BH binaries, with the highest attainable spins being very
close to the initial spins of the individual BHs (see Fig.~5 and Fig.~9 in
\cite{Belczynski:2007xg}). Furthermore, the kick-induced tilt angle
$\iota<45^\circ$ for $\sim 50\%$ of NS-BH systems. The fraction of events that
can potentially produce short-hard GRBs, and which therefore is relevant for
merger/ringdown searches in association with gamma-ray bursts, is significant
(of order $\sim 40\%$) {\it only} if the initial BH spin $a/M\gtrsim 0.6$
\cite{Belczynski:2007xg}.

These studies suggest that Advanced-LIGO measurements of BH spins in the
inspiral of binaries containing stellar-mass BHs should be an excellent probe
of the collapse mechanism that produced the BH in the first place, because the
BH essentially retains the spin it had at birth. Furthermore, in view of the
NS-BH simulations discussed in Section \ref{sec:mergers}, NS-BH binaries where
the BH is rapidly spinning are good candidates as central engines of GRBs, and
the resulting gravitational wave signal {\em may} have a significant ringdown
component.  This could potentially allow a ringdown-based measurement of the
final BH spin magnitude and direction, with interesting implications for
coincident electromagnetic/gravitational observations of GRB events.

\noindent
{\em 2) Supermassive black holes:} Since SMBHs are expected to grow by a
combination of mergers and accretion, their spin will depend on three main
ingredients \cite{Gammie:2003qi}: i) the spin of ``seed'' BHs at birth, that
in some sense defines the initial conditions for the problem; ii) the spin
resulting from a binary BH merger; iii) the maximum spin attainable by
accretion. We discuss these points, in turn, below.

\noindent
{\em i) Natal spins -} Little is known observationally about the formation of
the first BHs in the universe.  A popular formation scenario involves the
collapse of primordial, massive ($M\sim 30-300~M_\odot$), metal-free
Population III stars at cosmological redshift $z\sim 20$ to form primordial
BHs with $M\sim 10^2~M_\odot$, clustering in the cores of massive dark-matter
halos \cite{Madau:2001sc}, but details of the collapse are uncertain
\cite{Ohkubo:2005pu,Suwa:2007nq}.
Shibata and Shapiro simulated the collapse of uniformly rotating stars
supported by radiation pressure and spinning at the mass-shedding limit,
finding numerically \cite{Shibata:2002br} that the final BH spin
(independently of the progenitor mass) is $a/M\approx 0.75$, and supporting
this result by analytical arguments \cite{Shapiro:2002kk}. Alternative
scenarios suggests that BH seeds would form at $z\gappreq 12$ from low-angular
momentum material in protogalactic discs
\cite{Koushiappas:2003zn,Begelman:2006db}; these seeds would have larger mass
$M\sim 10^5~M_\odot$, but their angular momenta depend on the dynamics of the
collapsing material.
Whether this ``initial value problem'' is relevant for the overall spin
distribution of SMBHs is a matter of debate. Merger tree simulations where the
accretion disk orientation is chosen randomly show that the spin distribution
does indeed retain memory of the initial conditions
\cite{Volonteri:2004cf,Berti:2008af}.  However, a recent model where the BH
spin directions are ``linked with the galaxies'' suggests that the spin
distribution could be largely independent of the initial conditions
\cite{Lagos:2009xr}.

\noindent
{\em ii) Spin from mergers -} A pioneering attempt to study massive BH spin
evolution by repeated mergers, predating the 2005 numerical relativity
breakthrough, is due to Hughes and Blandford
\cite{Hughes:2002ei}. Extrapolating results from BH perturbation theory they
found that ``minor mergers'' ($q\lesssim 1/10$) of a large BH with an
isotropic distribution of small objects tend to spin down the hole.  An
implementation of the results shown in the right panel of
Fig.~\ref{fig:lightring} within merger tree scenarios confirms these
results. Furthermore, it shows that SMBH spins cluster around $a/M\sim 0.7$ if
alignment and accretion are inefficient, so that the BH spin growth is
dominated by mergers \cite{Berti:2008af}. However, Volonteri {\it et al.}
\cite{Volonteri:2004cf,Berti:2008af} argued that on average {\em accretion
  should dominate over mergers} in determining the spin evolution in
hierarchical SMBH formation scenarios.

\noindent
{\em iii) Spin from accretion -} The details of SMBH growth by accretion are
very uncertain. It is usually believed that prolonged accretion should lead to
large spins \cite{Volonteri:2004cf}. However, King {\it et al.}
\cite{King:2005mv,King:2006uu,King:2008au,Wang:2009ws} suggested that gas
accretion may occur through a series of chaotically oriented episodes, leading
to moderate spins $a/M\sim 0.1-0.3$. Since numerical relativity suggests that
comparable-mass mergers should not be efficient at spinning up BHs, and such
comparable-mass mergers are expected to be common \cite{Bogdanovic:2007hp}, a
few measurements of spins $a/M\gtrsim 0.9$ (such as the value of
$a/M=0.989^{+0.009}_{-0.002}$ claimed by Brenneman and Reynolds
\cite{Brenneman:2006hw} for the Seyfert 1.2 galaxy MCG-06-30-15) would imply
that chaotic accretion is not the norm.
\begin{figure*}[ht]
\begin{center}
\begin{tabular}{ccc}
\epsfig{file=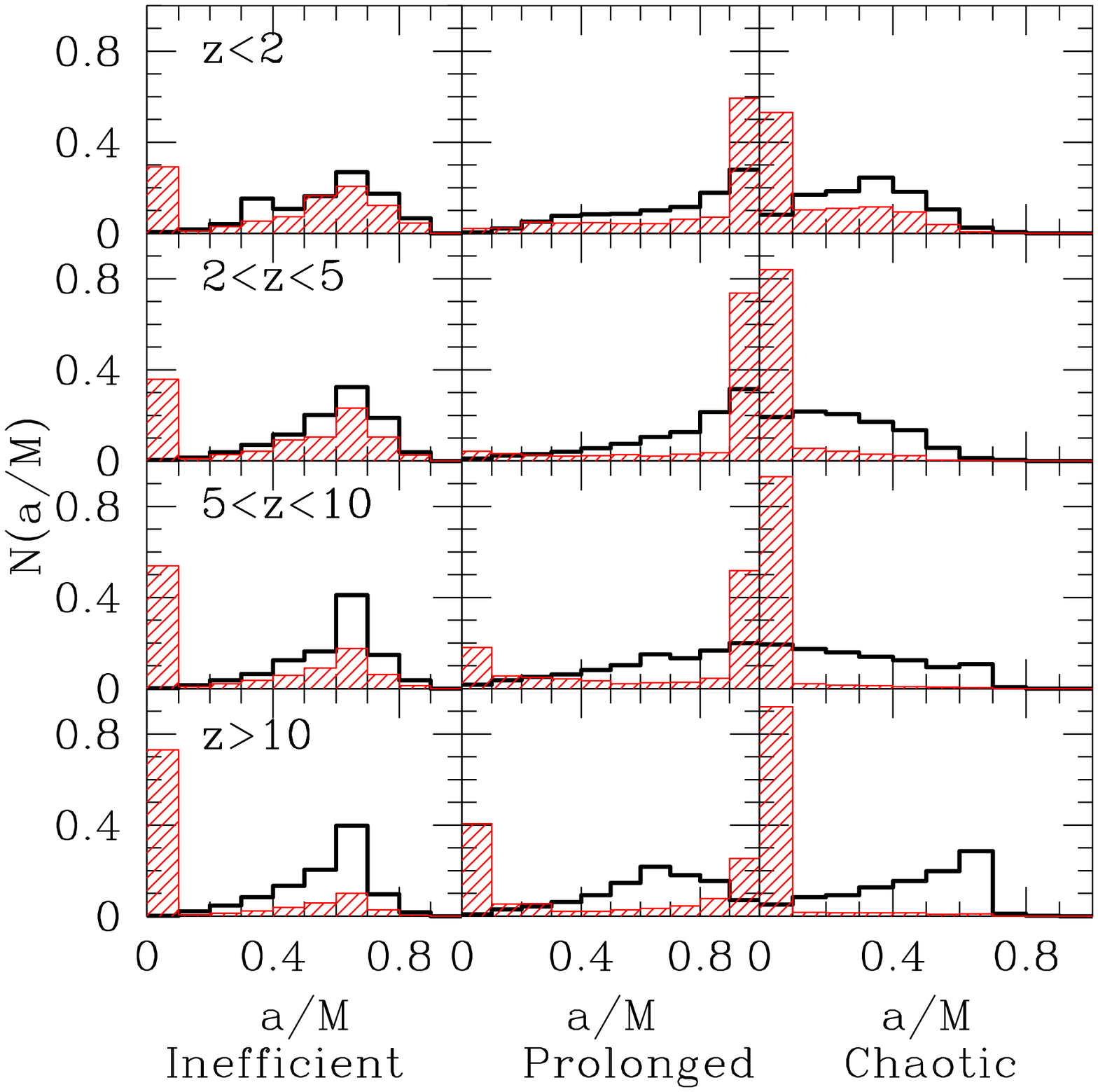,width=6cm,angle=0} &
\epsfig{file=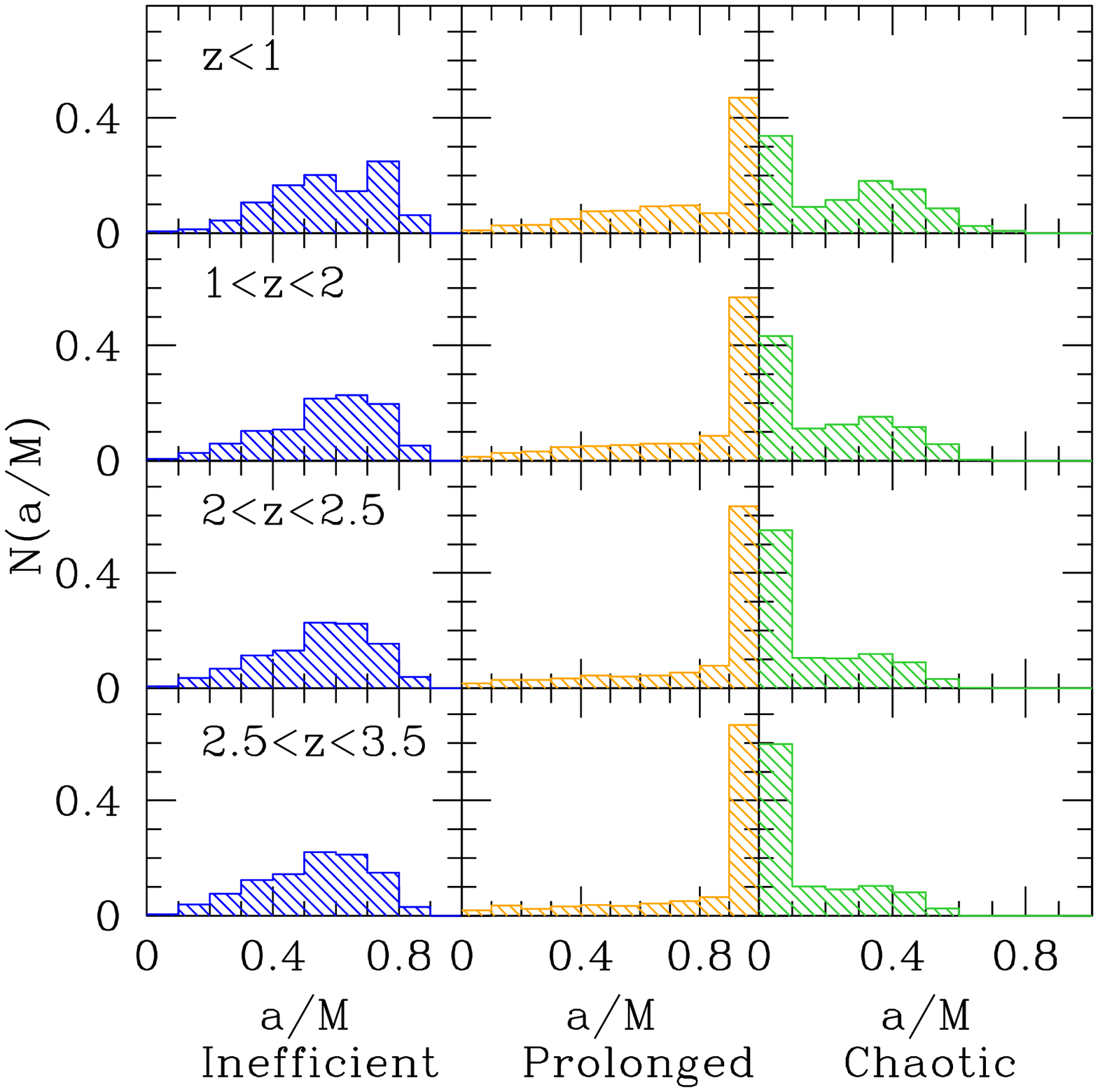,width=6cm,angle=0} \\
\end{tabular}
\end{center}
\caption{Left panel: spin distributions for merging BHs. Red dashed histograms
  show the spin distribution of inspiralling SMBHs in binaries, while black
  histograms show the spin distribution of SMBHs post-merger. Right panel:
  spin distribution of the whole MBH population in the mass range
  $10^5M_\odot\leq M \leq 3\times 10^7M_\odot$. For simplicity, we consider on
  ly the isotropic merger scenario. Left to right in each panel, accretion is
  assumed to be inefficient, prolonged or chaotic, respectively (see
  Ref.~\cite{Berti:2008af} and the text for details).}
\label{fig:spinBV}
\end{figure*}

Given the present uncertainty on the physical agents responsible for SMBH
growth, it would be extremely valuable to find observational signatures of
different formation scenarios.  Figure~\ref{fig:spinBV} (adapted from
Ref.~\cite{Berti:2008af}) shows that electromagnetic spin measurements and
gravitational wave measurements from the inspiral and ringdown may provide an
excellent way of discriminating between different mechanisms of BH growth. The
spins are assumed to be isotropically distributed (but efficient alignment, as
suggested in Ref.~\cite{Bogdanovic:2007hp}, would only marginally alter the
picture: see Ref.~\cite{Berti:2008af}) and the seed BHs are assumed to be
non-spinning.  In the left panel, the red dashed histograms show theoretical
estimates of the spin distribution that could be measured by observing
gravitational radiation from inspiralling SMBH binaries; the black histograms
show the distribution of SMBH spins post-merger, as would be measured by
observations of ringdown waves. Finally, the histograms in the right panel
show the distribution of spins of the whole BH population, that can be probed
by electromagnetic observations of the kind described in \ref{app:spins}. In
each panel time runs upward (the histograms correspond to different redshift
cuts, as indicated in the inset) and the three ``columns'', from left to
right, correspond to: (1) spin growth being determined only by mergers
(inefficient accretion), (2) spin growth being driven by mergers and prolonged
accretion \cite{Volonteri:2004cf}, (3) spin growth being driven by mergers and
chaotic accretion \cite{King:2005mv,King:2006uu,King:2008au}. The spin
distributions are obviously very different. According to the chaotic accretion
scenario, ringdown measurements would never observe BH spins larger than $\sim
0.7$ or so, and most spins would be very low. If prolonged accretion
dominates, then most BHs should be rapidly spinning, and if accretion is
inefficient the spin distribution should have an attractor around $j\approx
0.7$. The spin distribution clearly encodes information on the SMBH merger
history.

It is important to keep in mind that the expected spins of SMBHs may well
depend strongly on their masses. It has been suggested that SMBHs with
$M\lesssim 2\times 10^6 M_\odot$ may grow primarily by disruption of stars
(see e.g. Fig.~9 of \cite{Wang:2003ny}), which would then lead to low
spins. These BHs would be in the optimal sensitivity window for LISA, but they
are more difficult to observe electromagnetically.  In contrast, the Soltan
\cite{Soltan:1982vf} argument that SMBHs grow mainly be accretion (see
\ref{app:spins}) really applies only to $M\gtrsim 10^7 M_\odot$, because only
a small fraction of SMBH mass is in BHs with $M\lesssim 10^7 M_\odot$ .  As a
result, higher-mass SMBHs could grow mainly by accretion, but lower-mass BHs
would grow mainly by stellar disruption, mergers with stellar-mass compact
objects and comparable-mass mergers.  Therefore, the observation of spin {\em
  as a function of mass} could be a powerful diagnostic of SMBH evolution.

\begin{table}[ht]
  \caption{\label{tab:spins} A list of spin estimates available in the
    literature, along with the method used for the estimate (see
    \ref{app:spins}) and the relevant references. Tables 1 and 2 of
    Ref.~\cite{Daly:2009ih} list spin estimates for 19 powerful FRII radio
    sources (FRIIb) and for 29 central dominant galaxies (CDGs). For
    MS0735.6+7421 Ref.~\cite{Daly:2009ih} estimates a spin of $0.83 \pm 0.39$,
    consistent with Ref.~\cite{McNamara:2008dy}.}
\begin{tabular}{llll}  
\hline
\hline
System & Estimated spin & Method &Reference\\
\hline
\multicolumn{4}{c}{Stellar-mass BHs}\\ 
Cygnus X-1         & $0.05\pm 0.01$ & Line spectroscopy & \cite{Miller:2009cw}\\
LMC X-3            & $\approx 0.2-0.4$ & Continuum& \cite{Davis:2006cm} \\
4U 1543-475        & $0.3\pm 0.1$  & Line spectroscopy & \cite{Miller:2009cw} \\
                   & $0.75-0.85$   & Continuum & \cite{Shafee:2005ef} \\
SAX J1711.6-3808   & $0.6^{+0.2}_{-0.4}$ & Line spectroscopy & \cite{Miller:2009cw}\\
XTE J1550-564      & $\approx 0.1-0.8$     & Continuum & \cite{Davis:2006cm} \\
                   & $0.76\pm 0.01$ & Line spectroscopy & \cite{Miller:2009cw} \\
SWIFT J1753.5-0127 &$0.76^{+0.11}_{-0.15}$ & Line spectroscopy &\cite{Reis:2009jk}\\
M33 X-7            & $0.77\pm 0.05$& Continuum & \cite{Liu:2008tk} \\
XTE J1908+094      & $0.75\pm 0.09$ & Line spectroscopy & \cite{Miller:2009cw}\\
XTE J1650-500      & $0.79\pm 0.01$& Line spectroscopy & \cite{Miller:2009cw}\\
GRS 1915+105       & $0.7-0.8$     & Continuum & \cite{Middleton:2006kj} \\
                   & $0.98-1$      & Continuum & \cite{Zhang:1997dy,McClintock:2006xd} \\
LMC X-1            & $0.90^{+0.04}_{-0.09}$ & Continuum & \cite{Gou:2009ks} \\
GX 339-4           & $0.94\pm 0.02$& Line spectroscopy & \cite{Miller:2008vc,Miller:2009cw}\\
GRO J1655-40       & $\geq 0.25$   & QPOs & \cite{Strohmayer:2001yn} \\
                   & $0.65-0.75$   & Continuum & \cite{Zhang:1997dy,Shafee:2005ef} \\
                   & $\approx 0.1-0.8$     & Continuum & \cite{Davis:2006cm} \\
                   & $0.98\pm 0.01$   & Line spectroscopy &\cite{Miller:2009cw}\\
XTE J1655-40       & $\approx 1$   & Line spectroscopy & \cite{Miller:2004cg}\\
XTE J1550-564      & $\approx 1$   & Line spectroscopy & \cite{Miller:2004cg}\\
\hline
\hline
\multicolumn{4}{c}{SMBHs}\\ 
$29$ CDGs  & $0.1-0.8$ & Energetics & \cite{Daly:2009ih}\\
$19$ FRIIb & $0.7-1$ & Energetics & \cite{Daly:2009ih}\\
SWIFT J2127.4+5654 & $0.6\pm 0.2$  & Line spectroscopy & \cite{Miniutti:2009dg}\\
MCG-06-30-15  & $0.989^{+0.009}_{-0.002}$ & Line spectroscopy & \cite{Brenneman:2006hw}\\
1H0419-577    & $\approx 1$   & Line spectroscopy & \cite{Fabian:2005qj} \\
MS0735.6+7421 & $\approx 1$   & Energetics & \cite{McNamara:2008dy}\\
              & Large & Average efficiency & \cite{Elvis:2001bn,Yu:2002sq,Wang:2006bz,Wang:2008ms,Cao:2008pd}\\
              & Small & Average efficiency & \cite{Shankar:2007zg,Merloni:2008hx,Daly:2008zk,Wang:2009ws}\\
\hline
\hline
\end{tabular}
\end{table}

\subsubsection*{Spins: observational estimates} 

Spin estimates based on electromagnetic observations made enormous progress in
the last three years. A summary of estimates available in the literature is
provided in Table \ref{tab:spins}. Notice that in some cases (most notably for
4U 1543-475 and GRO J1655-40) different methods yield sensibly different spin
estimates (see e.g.~\cite{Miller:2009cw}).  The main methods used so far to
estimate spins are continuum spectroscopy of accretion disks, spectroscopy of
relativistically broadened Fe K$\alpha$ fluorescence lines, and energetic
arguments based on the radiative efficiency of quasars. A discussion of these
topics would take us too far, but it is essential to appreciate the
statistical and systematic errors involved in the numbers listed in Table
\ref{tab:spins}. To improve readability, we review recent literature in this
field in \ref{app:spins}.

A glance at the Table shows that the estimated spin magnitudes of stellar mass
BHs cover the whole range from zero to one. This seems to confirm the doctrine
that stellar-mass BHs in X-ray binaries essentially retain the spin they had
at birth \cite{King:1999aq}. The situation is even more unclear for SMBHs,
where (as shown in the last two lines of Table \ref{tab:spins}) uncertainties
in observational data on the mean efficiency of quasars lead to very different
conclusions on the average values of SMBH spins. Gravitational wave detection
could be instrumental in resolving this controversy.

\subsection{\label{sec:detectability}Detection range for Earth-based and space-based detectors}

We usually say that a gravitational wave signal is detectable when the SNR, as
defined in Eq.~(\ref{SNRdef}), is larger than some threshold, typically
$\rho>10$. Since the gravitational wave amplitude decreases linearly with the
(luminosity) distance $D_L$ from the source, the distance corresponding to
$\rho=10$ is sometimes called the detector {\em range}.
\begin{figure*}[ht]
\begin{center}
\begin{tabular}{cc}
\epsfig{file=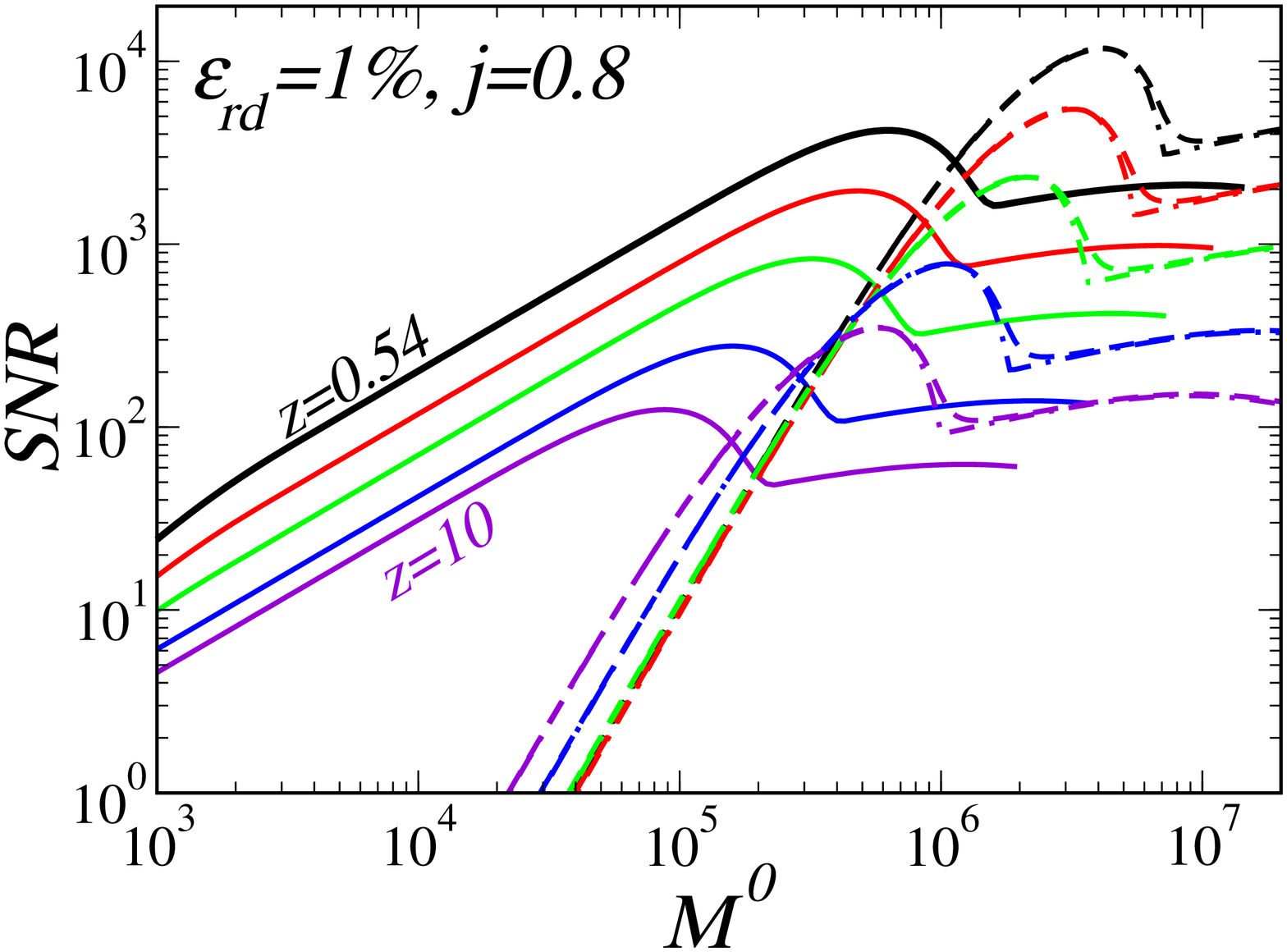,width=5.2cm,angle=-90} &
\epsfig{file=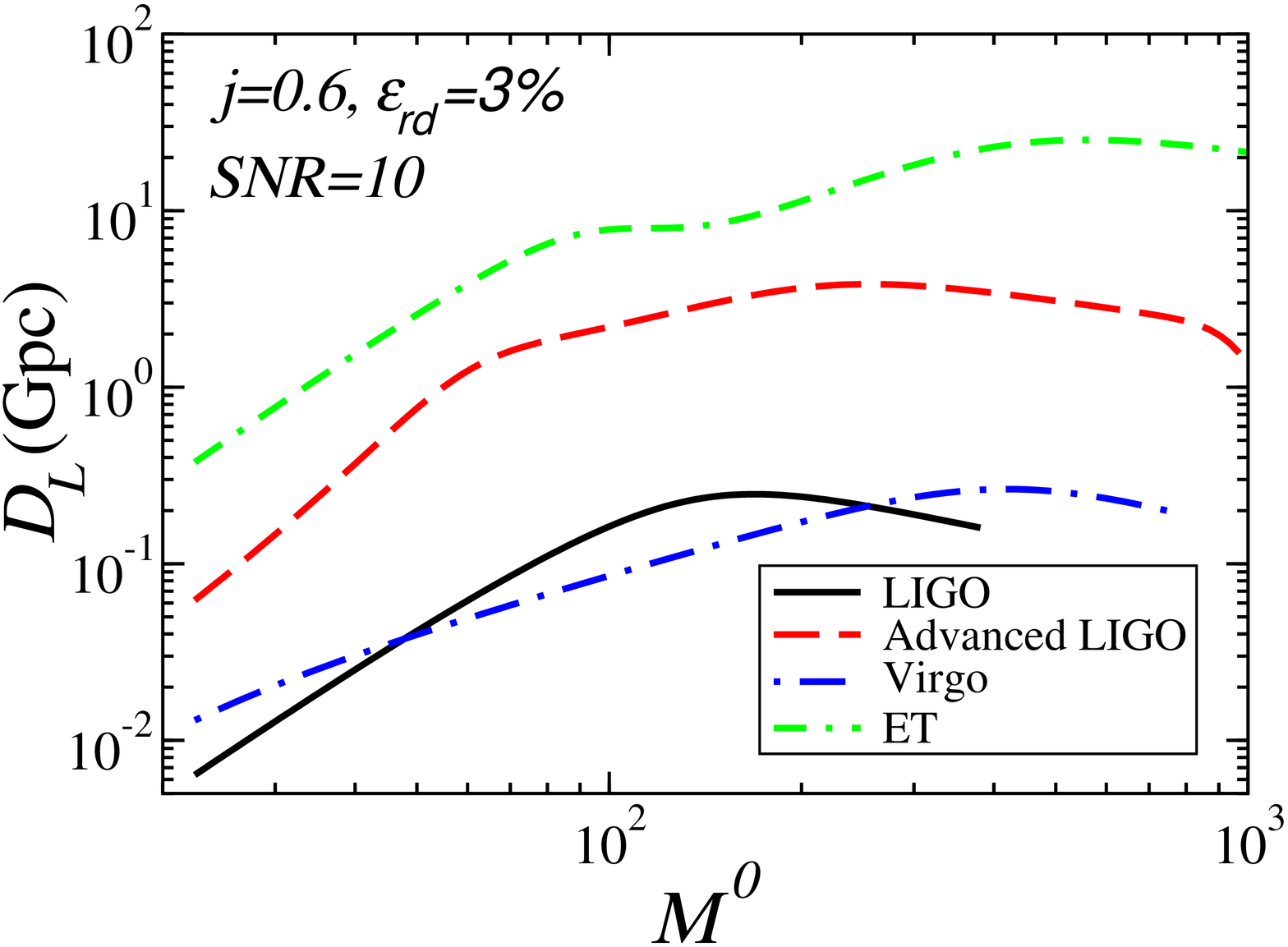,width=5.2cm,angle=-90} \\
\end{tabular}
\end{center}
\caption{Left: typical SNR for equal-mass inspirals and ringdown signals
  detected by LISA, assuming a final spin $j=0.8$ and a conservative ringdown
  efficiency $\epsilon_{\rm rd}=1\%$ (from \cite{Berti:2005ys}). Top to
  bottom, the lines correspond to sources at redshift
  $z=0.54,~1,~2,~5,~10$. Right: reach of Earth-based interferometers for a
  ringdown producing a BH with final spin $j=0.6$ and $\epsilon_{\rm rd}=3\%$,
  as a function of the mass of the final hole (from
  \cite{Berti:2007zu}).} \label{fig:reach}
\end{figure*}
In the left panel of Fig.~\ref{fig:reach} we show LISA's SNR for equal-mass
inspirals (solid lines) and the ensuing ringdown (dashed lines) as a function
of the total mass of the system $M^0$, where the superscript ``0'' means that
the mass is measured in the source frame (see Ref.~\cite{Berti:2005ys} for
details of the assumptions going into this calculation). Each line corresponds
to a different luminosity distance or, which is the same, to a different
cosmological redshift $z$: for example, a redshift $z\simeq 0.5$ corresponds
to a luminosity distance of about 3~Gpc in a standard $\Lambda$CDM
cosmology. The plot illustrates a number of important points: (1) LISA can
detect the last year of the inspiral of equal-mass binaries with total mass
$10^3M_\odot<M^0<10^6M_\odot$ out to cosmological distances ($z\gtrsim 10$);
(2) the ringdown phase has (under reasonable assumptions, which are confirmed
by numerical simulations: see e.g. Fig.~14 in Ref.~\cite{Ajith:2007kx}) a
larger maximum SNR than the inspiral phase, and this maximum is achieved at
larger values of the SMBH's mass. This is important, because it implies that
ringdown searches are better suited for the detection of binaries with
$10^5M_\odot<M^0<10^7M_\odot$, which is closer to the typical mass range
estimated for SMBHs at galactic centres (see Section \ref{sec:massspin}).

In the right panel of Fig.~\ref{fig:reach} we show the ringdown detection
range corresponding to $\rho=10$ for Earth-based detectors. The plot shows
that Advanced LIGO and the Einstein Gravitational Wave Telescope (ET) have the
potential to detect ringdown from IMBH-IMBH systems of mass up to $\sim
10^3M_\odot$ out to a luminosity distance of a few Gpc
\cite{Fregeau:2006yz}. In fact, third-generation Earth-based interferometers
could probe the first generation of ``light'' seed BHs of $M\sim
10^2-10^3M_\odot$, providing information complementary to LISA on the earliest
BHs in the universe \cite{Sesana:2009wg}. A more speculative source is the
intermediate-mass ratio inspiral (IMRI) of stellar-mass BHs into IMBHs. For
these systems, ringdown would be suppressed relative to inspiral because of
the small mass ratio. However, what is lost in terms of number of cycles is
gained in terms of detector sensitivity: IMRI ringdowns would happen in the
optimal frequency band of second- and third-generation detectors, and
therefore they could be detectable by a network of Advanced LIGOs
\cite{Mandel:2007hi}.

\subsection{Event rates}

Gravitational wave interferometers (unlike traditional electromagnetic
observatories) respond to the waves' amplitude, not to their energy. Since the
wave amplitude decays linearly with distance, a modest increase of (say) a
factor two in sensitivity means that the detectable volume increases by a
factor eight. Given that merging compact binaries are the most promising
ringdown source, the relevant question for ringdown detection is then: how
many merging compact binary events can we expect in a given volume? The issue
of estimating event rates is one of the most pressing in gravitational wave
detection. The uncertainties involved are so large, and progress in the field
is so rapid, that any estimates we quote are likely to become rapidly
obsolete. For this reason we dedicate little space to event rate estimates,
providing a few references as a guide for the interested reader.

\subsubsection*{Stellar-mass and intermediate-mass black hole ringdowns}

In the simplest models, compact binary coalescence rates should be
proportional to the stellar birth rate in nearby spiral galaxies, which can be
estimated from their blue luminosity. Therefore the coalescence rates are
usually given in units of ${\rm Myr}^{-1}L_{10}^{-1}$, where $L_{10}$ is
$10^{10}$ times the blue solar luminosity. To convert these numbers into
detection rates, one must take into account the fact that detection ranges for
ringdowns are different from those for inspirals.  If the distance to an event
is above $\sim 50$~Mpc the local over-density of galaxies can be ignored, and
the number of galaxies containing possible sources is $N=[(4\pi)/3](D/{\rm
  Mpc})^3 (2.26)^{-3} 0.0117$ Milky-way equivalent galaxies, where 2.26 is a
correction factor to include averaging over all sky locations/orientations,
and $0.0117$Mpc$^{-3}$ is the density of Milky-Way equivalent galaxies.  For
shorter distances one should use a sky catalog, such as
Ref.~\cite{Kopparapu:2007ib}.

For NS-NS binaries, early and conservative estimates were made by Phinney
\cite{Phinney:1991ei}. At present, the most reliable NS-NS merging rate
estimates are obtained by extrapolating from observed binary pulsars
\cite{Kalogera:2003tn,Kalogera:2003tnErratum}. Expected rates are $\approx
50{\rm Myr}^{-1}L_{10}^{-1}$, but they could be an order of magnitude lower or
larger.  For NS-BH and BH-BH rates we have to rely mostly on
population-synthesis models \cite{O'Shaughnessy:2005qs,O'Shaughnessy:2006wh}.
Plausible rate estimates are $\approx 2{\rm Myr}^{-1}L_{10}^{-1}$ for NS-BH
binaries and $\approx 0.4{\rm Myr}^{-1}L_{10}^{-1}$ for BH-BH binaries, but
they could be roughly two orders of magnitude larger or lower. These rates
translate into tens to thousands of inspiral events per year in Advanced
LIGO. The typical end-product of these mergers are BHs of mass $\sim 10
M_\odot$, and the range for Advanced LIGO detection of these BH ringdowns is
more than an order of magnitude less than the inspiral range, so ringdown
rates should be $\sim 10^2-10^3$ times smaller than inspiral rates.
Rates for ringdowns involving IMBHs are even more uncertain. In optimistic
scenarios, Advanced LIGO could see $\sim 10$ IMBH binary mergers per year
\cite{Fregeau:2006yz} and perhaps $\sim 20$ ringdowns from the merger of
stellar-mass BHs into IMBHs \cite{Mandel:2007hi} (see also
\cite{Brown:2006pj,Mandel:2008bc}).

These predictions rapidly change as our understanding of the underlying
physics and compact binary observations improve. For example,
Ref.~\cite{Belczynski:2006zi} argues that potential BH-BH binary progenitors
may undergo a common envelope phase while the donor is evolving through the
Hertzsprung gap. This would probably lead to a merger, thus shutting off a
channel for BH-BH production and sensibly reducing BH-BH merger rates. 
On the other hand, based on observations of a very massive BH binary,
Ref.~\cite{Bulik:2008ab} estimates an {\em initial LIGO} rate of order
$0.5/$yr for relatively massive BH binaries (that therefore would be observed
mostly in the merger/ringdown phase).

Particularly interesting for ringdown detection are compact binary mergers
from dense star clusters. Ref.~\cite{Miller:2008yw} points out that a
``high-mass'' selection occurs because, in nuclear star clusters at the
centers of low-mass galaxies, three-body interactions usually ``pair up'' the
two heaviest members of a triple system. This typically produces binary
mergers with $M\gtrsim 20M_\odot$, which means that Earth-based
interferometers would usually observe the merger/ringdown phase. These
findings are consistent with work by other authors
\cite{O'Leary:2005tb,O'Leary:2007qa,O'Leary:2008xt}.

\subsubsection*{Supermassive black hole ringdowns}

The LISA noise curve determines the optimal mass and redshift range where
binary inspiral and ringdown events have large SNR, allowing a precise
measurement of the source parameters. Reliable estimates of the number of
events detectable during the mission's lifetime, and of their mass spectrum as
a function of redshift, will play a key role in the planning of LISA data
analysis. For this reason, over the last few years the calculation of SMBH
merger event rates and of their mass spectrum has become an active field of
research.

\begin{table}[hbt]
\centering
\caption{SMBH binary rates (events/year) predicted by different models
  (adapted and updated from Ref.~\cite{Berti:2006ew}).}
\vskip 12pt
\begin{tabular}{@{}lcc@{}}
\hline
\hline
Reference                                       & Rate             & Redshift range\\
\hline
\hline
Haehnelt 2003 \cite{Haehnelt:2004}                  & 0.1-1            & $0<z<5$ (gas collapse only) \\
                                                & 10-100           & $z>5$ (hierarchical buildup) \\
\hline
Enoki {\it et al.} 2004 \cite{Enoki:2004ew}             & 1                & $z>2$\\
\hline
Menou {\it et al.} 2001 \cite{Menou:2001hb}     & 10               & $z<5$ \\
\hline
Rhook and Wyithe 2005 \cite{Rhook:2005pt,Wyithe:2002ep}   & 15               & $z\sim 3-4$\\
\hline
Volonteri, Haardt and Madau \cite{Volonteri:2002vz,Sesana:2007sh,Arun:2008zn} & $\sim 30$ & $z\sim 4-16$\\
\hline
Begelman, Volonteri and Rees \cite{Begelman:2006db,Sesana:2007sh,Arun:2008zn} & $\sim 20$ & $z\sim 3-10$\\

\hline
Koushiappas and Zentner 2005 \cite{Koushiappas:2005qz}          & $\gtrsim 10^3$ & mostly $z\sim 10$, down to $z\sim 1$\\
\hline
Islam {\it et al.} 2003 \cite{Islam:2003nt}              & $10^4-10^5$      & $z\sim 4-6$\\
\hline	     	   	       	            	 	     
\hline								     
\end{tabular}
\label{tab:rates}
\end{table}

A discussion of rate estimates is out of the scope of this review (see
e.g.~\cite{Berti:2006ew,Arun:2008zn,Tanaka:2008bv}), but the large
uncertainties in SMBH binary formation models and in the predicted event rates
are quite evident from Table~\ref{tab:rates}. The numbers we list should be
interpreted with caution. Each prediction depends on a large number of poorly
known physical processes, and the notion of ``detectability'' of a merger
event is defined in different ways: some authors define detectability setting
a threshold on the SNR, others set a threshold on the gravitational wave
effective amplitude. Furthermore, different authors use different LISA noise
curves.
A tentative bottom line is that we could face one of the following two
scenarios. According to a class of models, we should observe $\approx 10$
events/year at redshifts (say) $2\lappreq z\lappreq 6$. However, we cannot
exclude the possibility that hundreds or thousands of SMBH binaries will
produce a large (and perhaps even stochastic) background in the LISA
data. Clearly, the detection strategy to use strongly depends on which of the
two scenarios actually occurs in nature.

Besides being able to observe SMBH mergers throughout the universe, LISA
should also be able to detect IMBH-IMBH binary mergers (that is, binaries
containing a $10-100~M_\odot$ BH orbiting a $100-1000~M_\odot$ BH). Rates for
IMBH binary detections were first estimated by Miller \cite{Miller:2002vg} and
then revised by Will \cite{Will:2004fj}. The revised estimates are very
pessimistic, predicting $\sim 10^{-6}$ events/year for typical values of the
parameters. A more promising scenario involves 
an IMBH spiralling into a SMBH \cite{Matsubayashi:2004bd}. For these
systems, Ref.~\cite{Miller:2004va} estimates a detection rate of a few
events/year, suggesting that mergers of a $10^3~M_\odot$ IMBH into a
$10^6~M_\odot$ SMBH could be observed out to $z\sim 20$ with an SNR of 10 in a
one-year integration. Ref.~\cite{PortegiesZwart:2005zp} predicts an even more
optimistic rate of $\sim 10^2$ events/year throughout the universe. These
estimates are very preliminary and even more uncertain than the corresponding
estimates for SMBH binaries, but they should be taken into account in design
choices concerning (for example) the optimal armlength of LISA.

\subsection{\label{sec:inferringmassrd}Inferring black hole mass and spin from ringdown measurements}

We have seen in Section \ref{sec:detectability} that the prospects for
detection of ringdown radiation by LISA and advanced Earth-based detectors are
quite encouraging.  Interesting physics can be extracted from the observation
of BH ringdowns \cite{Berti:2005ys,Finn:1992wt,Dreyer:2003bv}.  Since
astrophysical BHs in general relativity are fully characterized by their mass
and angular momentum, the detection of a single mode is in principle
sufficient to estimate the mass and spin parameter of the hole, by inverting
the experimentally determined $\omega_{lmn}(M,a/M), Q_{lmn}(a/M)$.  Indeed,
one finds that accurate measurements of SMBH mass and angular momentum can be
made.

\begin{figure*}[ht]
\begin{center}
\begin{tabular}{cc}
\epsfig{file=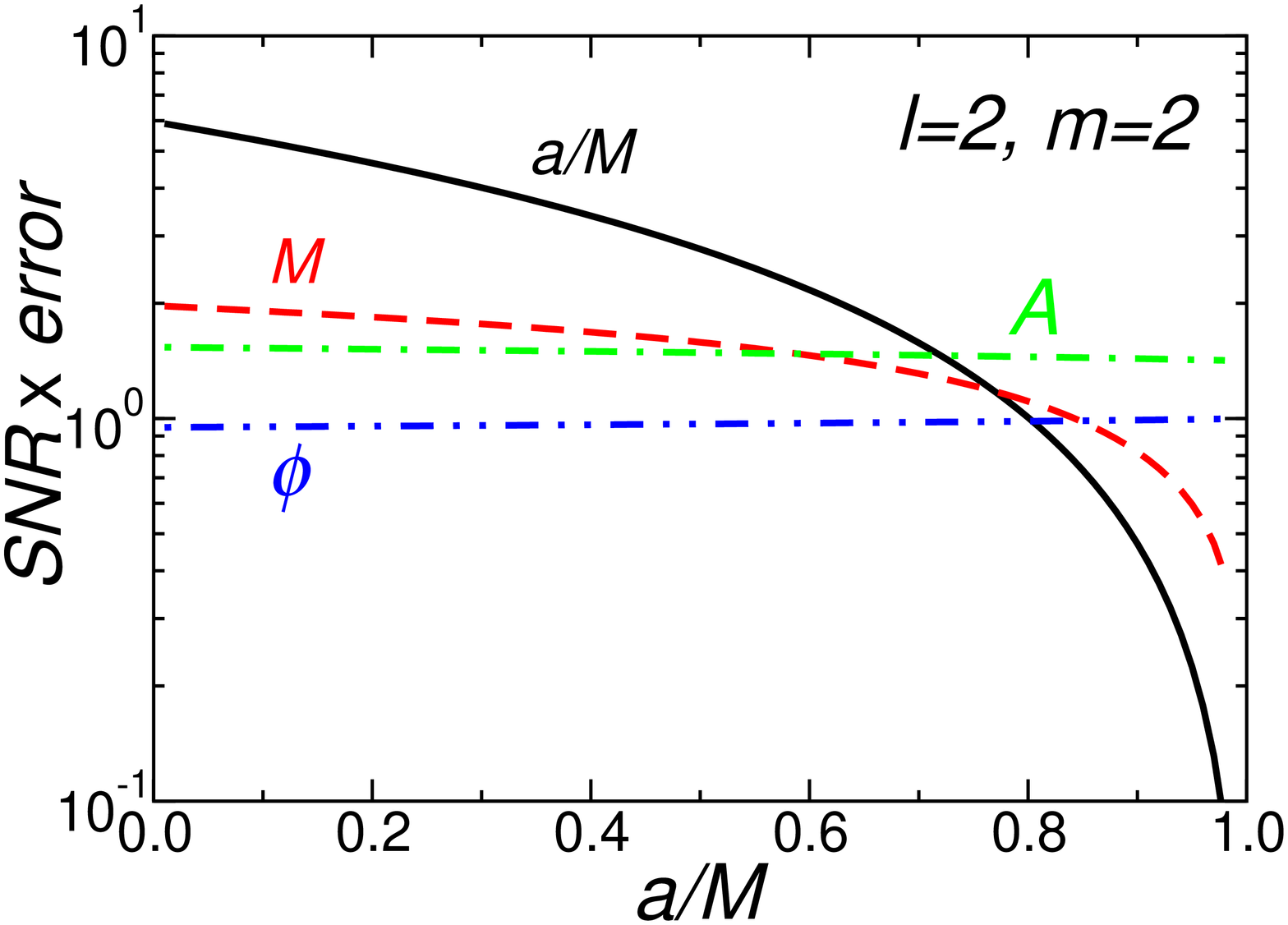,width=5.2cm,angle=-90}&
\epsfig{file=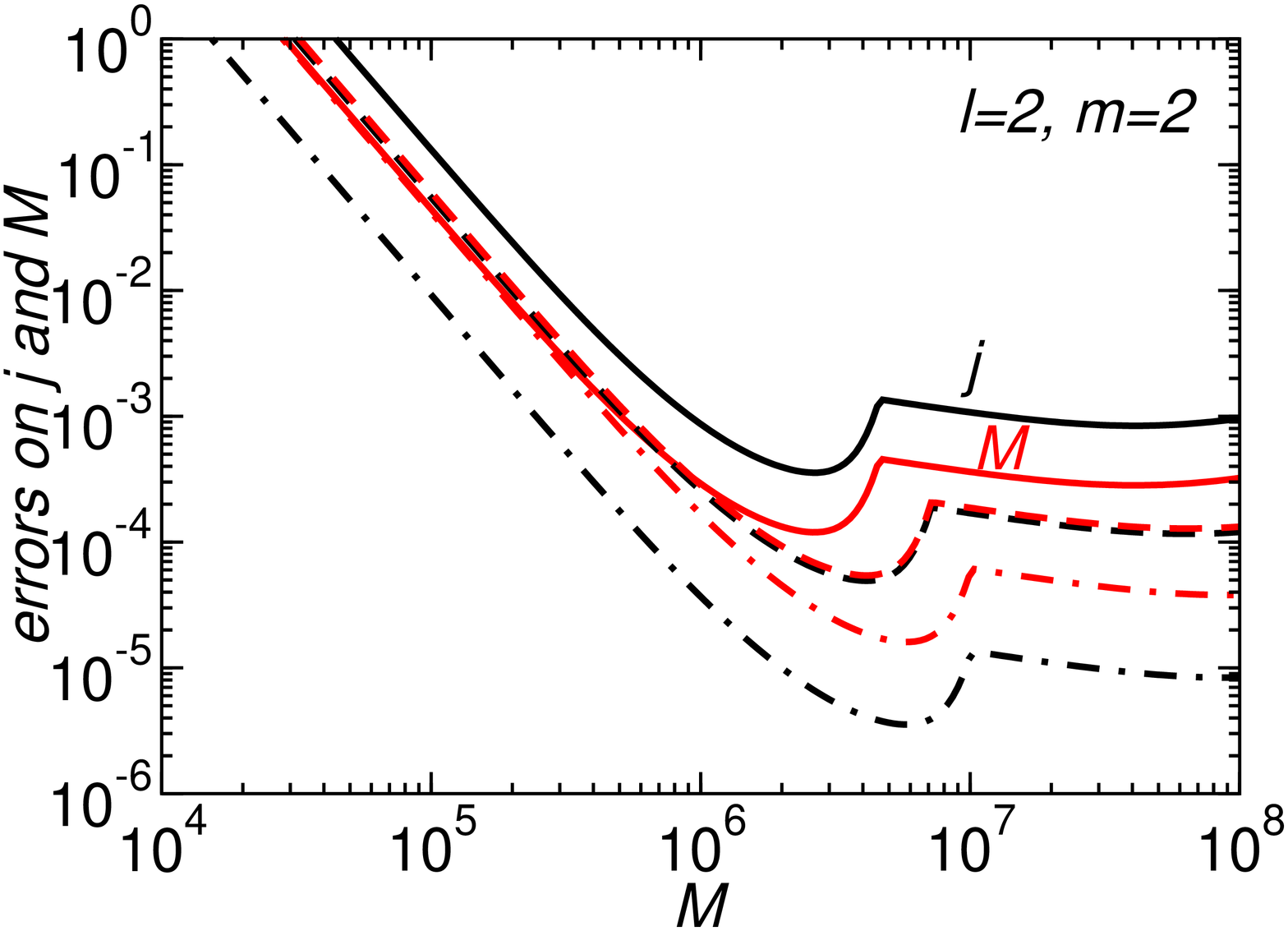,width=5.2cm,angle=-90}\\
\end{tabular}
\end{center}
\caption{Left: errors (multiplied by the signal-to-noise ratio $\rho$) in
  measurements of different parameters for the fundamental $l = m = 2$ mode as
  function of the rotation parameter $a/M$. Solid (black) lines give $\rho
  \sigma_{a/M}$ , dashed (red) lines $\rho\sigma_{M}/M$, dot-dashed (green)
  lines $\rho \sigma_{A}/A$, dot-dot-dashed (blue) lines $\rho \sigma_{\phi}$,
  where $\sigma_{k}$ denotes the estimated rms error for variable $k$, $M$
  denotes the mass of the BH, and $A$ and $\phi$ denote the amplitude and
  phase of the wave (adapted from \cite{Berti:2005ys}.) Right: actual errors
  in a LISA measurement for a source located ad $D_L=3~$Gpc, with ringdown
  efficiency $\epsilon_{\rm rd}=3\%$.} \label{fig:accuracyBHparameters}
\end{figure*}
For example, the left panel of Fig.~\ref{fig:accuracyBHparameters} shows the
estimated error (multiplied by the SNR $\rho$) in measuring the SMBH mass $M$,
angular momentum parameter $a/M$, QNM amplitude $A$, and phase $\phi$ for
circularly polarized radiation from the fundamental $l=m=2$ bar mode
(cf. Eq.~(\ref{waveform0}) for definitions). If an energy $\sim 10^{-4}M$ is
radiated into the fundamental mode of a $10^6 M_{\odot}$ SMBH with $a/M = 0.8$
at 3 Gpc ($\rho \sim 200$), $M$ and $a/M$ could be measured to 1$\%$; if the
energy deposition is only $10^{-6}$, they could still be measured to
10$\%$. These numbers were computed for LISA in Ref.~\cite{Berti:2005ys}, but
they carry over to other detectors through a simple rescaling by
$\rho$. Generalizing to multi-mode detection (and specifically to the
detection of two modes with a range of relative amplitudes) one finds similar
results \cite{Berti:2005ys}.  Gravitational wave detectors will be able to
determine the mass and spin of BHs with excellent precision from observations
of the ringdown phase.

\subsubsection*{Event loss and bias in parameter estimation using single-mode templates}

Current ringdown searches are performed using matched filtering and
single-mode templates, consisting of a single exponentially damped
sinusoid. These are the simplest possible templates, and they are expected to
capture the relevant physics of the problem when one ringdown mode dominates
over the others.  Unfortunately, this notion of dominance must be precisely
formulated. Consider for illustration the case of non-spinning binary BH
mergers, and suppose for simplicity that there are only two modes in the
signal, say $l=m=2$ and $l=m=3$.  In particular, assume that the strain $h$ as
seen by a detector, Eq.~(\ref{detectwave}), is of the form
\be
h=
{\cal A}_1e^{-\pi\,f_1t/Q_1}\sin(2\pi f_1\,t-\phi_1)+
{\cal A}_2e^{-\pi\,f_2t/Q_2}\sin(2\pi f_2 t-\phi_2)\,.
\label{htwomode}
\ee
Estimates of the multipolar energy distribution give a relative amplitude
${\cal A}_2/{\cal A}_1=h_{33}/h_{22}\sim 0.3-0.4(1-1/q)$, where $q$ is the
mass ratio and $h_{22}$ ($h_{33}$) are the amplitudes of the $l=m=2~(3)$ modes
of the radiation, respectively \cite{Berti:2007fi}. For $q>3$, which includes
most likely merger scenarios, we get $h_{33}/h_{22}\geq 0.2-0.3$.  It is now
natural to ask: given a relative amplitude of this order, how many events
would we miss in a search with single-mode ringdown templates?

\begin{figure}[ht]
\begin{center}
\begin{tabular}{ll}
\epsfig{file=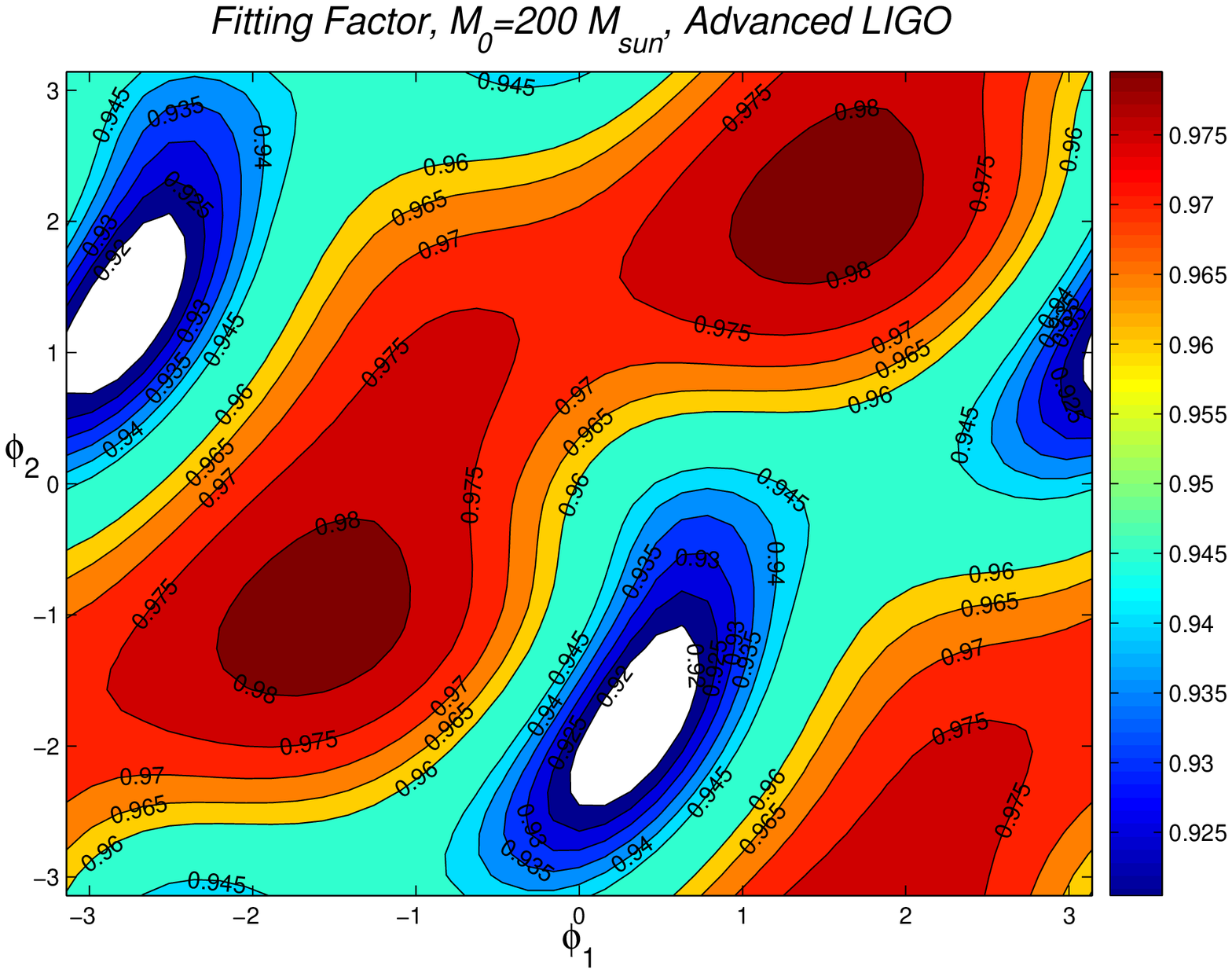,width=6.5cm,angle=0}
\epsfig{file=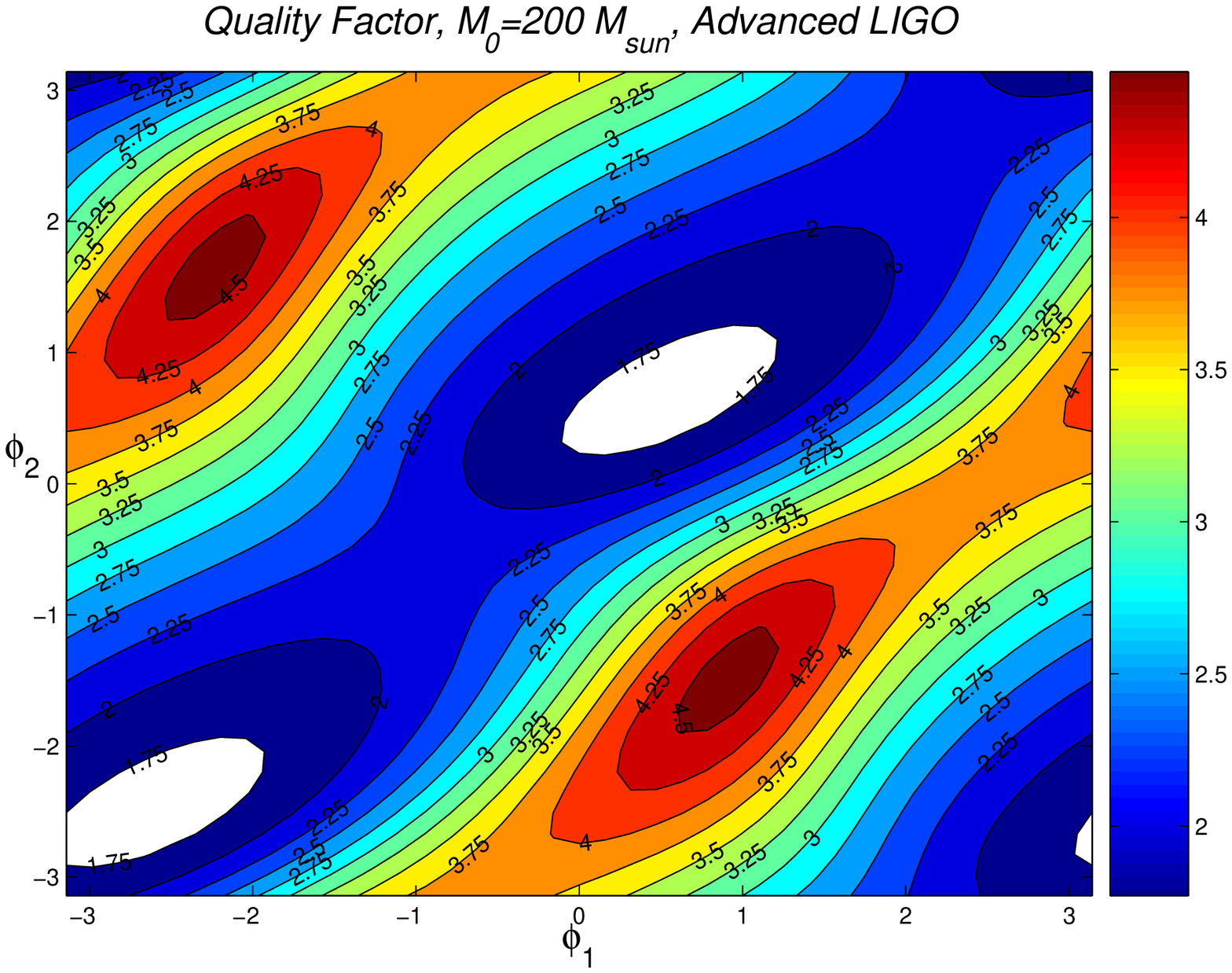,width=6.5cm,angle=0}\\
\end{tabular}
\end{center}
\caption{Left: contour plots of the fitting factor as a function of the phase
  angles of the signal. For this illustrative calculation we assume that the
  Kerr parameter of the final BH is $j=0.6$, and that the relative amplitude
  of the second mode is ${\cal A}=0.3$. Right: quality factor estimated by a
  single-mode filter. The ``true'' frequencies and quality factors for
  $a/M=0.6$ are $M\omega_{R\,1}=0.4940$, $Q_1=2.9490$,
  $M\omega_{R\,2}=0.7862$, $Q_2=4.5507$. See \cite{Berti:2007zu} for more
  details.}
\label{fig:FFcontours}
\end{figure}
The answer is quantified in the left panel of Fig.~\ref{fig:FFcontours}, where
we show Apostolatos' fitting factor (FF) \cite{Apostolatos:1995pj,Owen:1995tm}
resulting from searching a two-mode signal with single-mode templates from a
$200 M_{\odot},\,a/M=0.6$ BH, with relative amplitude $A_2/A_1=0.3$, in
Advanced LIGO data. Contour plots of the FF are shown as a function of the
(unknown) phases in Eq.~(\ref{htwomode}). The FF is essentially the ratio
$\rho/\rho_{\rm max}$, where $\rho$ is the actual SNR achieved by matched
filtering and $\rho_{\rm max}$ is the maximum possible SNR, attained when the
template and waveform coincide. FFs larger than 0.97, leading to a loss of
less than about $10\%$ in events, are typically considered acceptable. The FF
achievable by a detection is usually assumed to correspond to the minimum of
the FF in the space of the unknown parameters $(\phi_1,~\phi_2)$ (this is
known as the ``minimax'' criterion). Therefore, according to this simplified
calculation, single-mode templates may well lead to unacceptable SNR
degradation in a search. The situation can be even worse for larger masses and
other detectors \cite{Berti:2007zu}; similar conclusions apply also to LISA,
albeit in a completely different mass range.

Besides leading to a loss in the number of events, single-mode templates lead
to a large bias in the estimation of the BH's mass and spin.  The estimated
frequency has relatively small bias, and it always corresponds to the dominant
(least-damped) mode in the pair. Results are more interesting for the
estimated quality factor, shown in the right panel of
Fig.~\ref{fig:FFcontours} as a function of the phase angles. For our chosen
value of the Kerr parameter the quality factor of the $l=m=2$ and $l=m=3$
modes are $Q_1 = 2.9490$ and $Q_2 = 4.5507$, respectively. Comparing with the
left panel, we see that relative minima in the FF (white ``islands'' in the
left panel) occur, roughly speaking, when the quality factor ``best fits'' the
subdominant, $l=m=3$ mode. This is a rather remarkable result: the minimax
filter ``best fits'' the subdominant mode in the pair, leading to significant
bias in the estimation of the quality factor (and hence of the
spin). Unfortunately, maxima in the FF do not correspond to the filter being
optimally adapted to the $l=m=2$ mode. As the filter tries to maximize the
SNR, the estimated value of the quality factor becomes significantly biased,
and it deviates quite sensibly from the value expected for the dominant
($l=m=2$) mode. The bottom line is, again, that single-mode filters may be
useful for detection, but a multi-mode post-processing will be necessary for
accurate spin measurements.

There is, of course, a price to pay when using multi-mode templates.  About
${\cal N}\sim 6Q_{\rm max} \log{f_{\rm max}/f_{\rm min}}\sim 500$ single-mode
templates are enough to cover the parameter space for Earth-based detector
searches, if we assume an event loss of no more than $10\%$ (i.e., a minimal
match larger than 0.97 \cite{Owen:1995tm}). For two-mode templates, rough
estimates suggest that this number may increase up to ${\cal N} \sim b\times
10^6$, with $b\sim 1$ a detector-dependent constant \cite{Berti:2007zu}. A
more detailed data analysis study (e.g. using better template placing
techniques, along the lines of
\cite{Nakano:2003ma,Nakano:2004ib,Tsunesada:2004ft,Tsunesada:2005fe}) will be
needed to reduce computational requirements.

\subsection{\label{sec:nohair}Tests of the no-hair theorem}
The fact that all information is radiated away in the process leading to BH
formation, so that astrophysical BHs in Einstein's theory are characterized
completely by their mass and angular momentum, is known as the ``no-hair
theorem''. To test this theorem, it is necessary (but not sufficient) to
resolve two QNMs \cite{Berti:2005ys,Dreyer:2003bv}. Roughly speaking, one mode
is used to measure $M$ and $a$, and the other to test consistency with the
Kerr solution.  

Can we tell if there really are two or more modes in the signal, and can we
resolve their parameters? Physical intuition suggests that if the noise is
large and the amplitude of the weaker signal is very low (or if the two
signals have almost identical frequencies) the two modes could be difficult to
resolve. A crude lower limit on the SNR required to resolve frequencies and
damping times was presented in Refs.~\cite{Berti:2005ys,Berti:2007zu}. The
analysis uses the statistical uncertainty in the determination of each
frequency and damping time, which a standard Fisher-matrix calculation
estimates to be \cite{Berti:2005ys}
\beq
\rho \sigma_{f_1}&=&
\f{\pi}{\sqrt{2}}
\left\{
\f{f_1^3\left(3+16Q_1^4\right)}{{\cal A}_1^2 Q_1^7}
\left[
\f{{\cal A}_1^2 Q_1^3}{f_1\left(1+4Q_1^2\right)}+
\f{{\cal A}_2^2 Q_2^3}{f_2\left(1+4Q_2^2\right)}
\right] \right\}^{\f{1}{2}},\label{sigmaffh}\\
\rho \sigma_{\tau_1}&=& \f{2}{\pi} \left\{ \f{\left(3+4Q_1^2\right)}{{\cal A}_1^2 f_1 Q_1} \left[ \f{{\cal
A}_1^2 Q_1^3}{f_1\left(1+4Q_1^2\right)}+ \f{{\cal A}_2^2 Q_2^3}{f_2\left(1+4Q_2^2\right)} \right]
\right\}^{\f{1}{2}}. \eeq
These errors refer to mode ``1'' in a pair; errors on $f_2$ and $\tau_2$ are
simply obtained by exchanging indices ($1\leftrightarrow 2$)
\cite{Berti:2005ys,Berti:2007zu}. The expressions above further assume
white noise for the detector, and that modes ``1'' and ``2'' correspond to
different values of $l$ or $m$.

A natural criterion ({\it \'a la} Rayleigh) to resolve frequencies and damping
times is that $|f_1-f_2|>{\rm max}(\sigma_{f_1},\sigma_{f_2})$,
$|\tau_1-\tau_2|>{\rm max}(\sigma_{\tau_1},\sigma_{\tau_2})$.  
In interferometry this would mean that two objects are (barely) resolvable if
the maximum of the diffraction pattern of object 1 is located at the minimum
of the diffraction pattern of object 2. We can introduce two ``critical'' SNRs
required to resolve frequencies and damping times,
$\rho_{\rm crit}^f =
{\rm max}(\rho \sigma_{f_1},\rho \sigma_{f_2})/|f_1-f_2|\,,
\rho_{\rm crit}^\tau = 
{\rm max}(\rho \sigma_{\tau_1},\rho \sigma_{\tau_2})/|\tau_1-\tau_2|$,
and recast our resolvability conditions as
\beq \label{minimal}
\rho&>&
\rho_{\rm crit}
={\rm min}(\rho_{\rm crit}^f,\rho_{\rm crit}^\tau)\,,\\
\label{both}
\rho&>&
\rho_{\rm both}
={\rm max}(\rho_{\rm crit}^f,\rho_{\rm crit}^\tau)\,. \eeq
The first condition implies resolvability of either the frequency or the
damping time, and the second implies resolvability of both.

Now let us consider how to resolve {\it amplitudes}, i.e. the minimum SNR
needed to determine whether two or more modes are present in a given ringdown
signal.  Suppose again, for simplicity, that the true signal is a two-mode
superposition. Then we expect the weaker signal to be hard to resolve if its
amplitude is low and/or if the detector's noise is large. Appendix B of
Ref.~\cite{Berti:2007zu} quantifies this statement by deriving a critical SNR
for amplitude resolvability $\rho_{\rm GLRT}$ based on the generalized
likelihood ratio test. The derivation of this critical SNR is based on the
following simplifying assumptions: (i) using other criteria one has already
decided for the presence of at least one damped exponential in the signal, and
(ii) the parameters of the ringdown signal (frequencies and damping times), as
well as the amplitude of the dominant mode, are known. In practice the latter
assumption is not valid, so these estimates of the minimum SNR should be
considered optimistic.
\begin{figure*}[ht]
\begin{center}
\epsfig{file=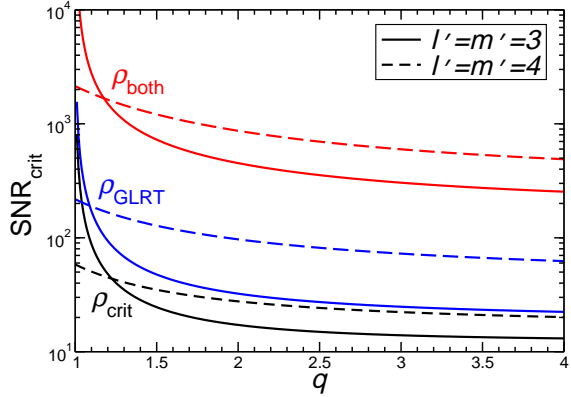,width=6.5cm,angle=-90} 
\end{center}
\caption{Minimum SNR required to resolve two QNMs from a BH resulting from the
  inspiralling of a BH binary with mass ratio $q$.  The dominant mode (mode
  ``1'') is assumed to be the fundamental QNM with $l=m=2$; mode ``2'' is
  either the fundamental $l=m=3$ QNM (solid lines) or the fundamental $l=m=4$
  QNM (dashed lines). The relative mode amplitude ${\cal A}(q)$ is estimated
  from numerical simulations \cite{Berti:2007fi}. If $\rho>\rho_{\rm GLRT}$ we
  can tell the presence of a second mode in the waveform, if $\rho>\rho_{\rm
    crit}$ we can resolve either the frequency or the damping time, and if
  $\rho>\rho_{\rm both}$ we can resolve both.}
\label{fig:minimumSNR}
\end{figure*}
Fig.~\ref{fig:minimumSNR} compares the critical SNR $\rho_{\rm GLRT}$
(Eq.~(B12) of Ref.~\cite{Berti:2007zu}) and the two different criteria for
frequency resolvability, Eqs.~(\ref{minimal}) and (\ref{both}). The plot shows
that $\rho_{\rm crit}<\rho_{\rm GLRT}<\rho_{\rm both}$ for all values of
$q$. Therefore, given a detection, the most important criterion to determine
whether we can carry out no-hair tests is the GLRT criterion. If
$\rho>\rho_{\rm GLRT}$ we can decide for the presence of a second mode in the
signal. Whenever the second mode is present, we also have $\rho>\rho_{\rm
  crit}$: that is, we can resolve at least the frequencies (if not also the
damping times) of the two modes. A SNR $\rho \sim 30-40$ is typically enough
to perform the GLRT test on the $l=m=3$ mode, as long as $q\gtrsim 1.5$. From
Figs.~\ref{fig:reach}-\ref{fig:minimumSNR} we conclude that not only LISA,
but also advanced Earth-based detectors (Advanced LIGO and ET) have the
potential to identify Kerr BHs as the vacuum solutions of Einstein's general
relativity.

In conclusion, ringdown radiation can be used to distinguish BHs from exotic
alternatives, such as boson stars \cite{Berti:2006qt} or gravastars
\cite{Chirenti:2007mk}. Ringdown tests of the Kerr nature of astrophysical BHs
are independent from (and complementary to) proposed tests using a multipolar
mapping of the Kerr spacetime, as encoded in EMRI signals according to
``Ryan's theorem'' and its generalizations
\cite{Ryan:1995wh,Ryan:1997hg,Li:2007qu}.

\subsection{\label{insptrans}Matching inspiral and ringdown: problems and applications}

Models of gravitational waveforms from inspiralling compact binaries usually
rely on the post-Newtonian approximation to general relativity
\cite{Blanchet:2002av}. Ideally, for matched-filtering detection we would like
to have detection template banks with ``complete'' waveforms encompassing the
inspiral, merger and ringdown. Phenomenological template families based on
physically motivated fits of numerical merger waveforms or on
effective-one-body models are now available
\cite{Ajith:2007qp,Ajith:2007kx,Pan:2007nw,Boyle:2009dg}. However, it would be
desirable to have a full understanding of the merger process, connecting the
post-Newtonian approximation during the inspiral to a description of the
ringdown as a superposition of QNMs.

QNM fits of numerical relativity waveforms are routinely used to check that
the total angular momentum of the system is conserved during BH merger
simulations. This is usually achieved by computing three independent
quantities: the angular momentum radiated at infinity, the angular momentum
obtained from isolated horizon calculations, and the angular momentum obtained
by inverting the frequencies and damping times resulting from a QNM fit of the
ringdown of the final Kerr BH (see
e.g.~\cite{Buonanno:2006ui,Berti:2007fi,Dain:2008ck}).  Buonanno, Cook and
Pretorius noticed that, as more and more overtones are included, a QNM
expansion gets in better agreement with numerical waveforms for equal-mass
binaries \cite{Buonanno:2006ui}. 

So far, efforts to ``stitch'' the post-Newtonian approximation (or one of its
effective-one-body variants) to the ringdown have used rather crude models for
ringdown excitation.
The original effective-one-body approach simply attached the plunge waveform
to the fundamental $l=m=2$ QNM of a Kerr BH \cite{Buonanno:2000ef}.  In their
study of recoils, Damour and Gopakumar attached the plunge waveform to the
fundamental Schwarzschild QNMs with $l=2$ and $l=3$ \cite{Damour:2006tr}.  A
comprehensive analysis of Pretorius' numerical waveforms for equal-mass BH
binaries \cite{Buonanno:2006ui} clearly illustrated the importance of
higher-order overtones.  Damour and Nagar obtained a good match to numerical
relativity waveforms by requiring continuity of the plunge waveform with a
(Schwarzschild) ringdown waveform including five overtones on a grid of
points, or ``comb'' \cite{Damour:2007xr}. A ``hybrid comb'' procedure,
imposing the continuity of the waveform {\it and its derivatives}, was later
introduced in Ref.~\cite{Buonanno:2009qa}.  These stitching methods have great
phenomenological interest, and they can do remarkably well at reproducing
numerical waveforms.  For example, the authors of Ref.~\cite{Buonanno:2009qa}
extend an effective-one-body waveform through merger by stitching the inspiral
waveform to QNMs, finding striking agreement (at the $0.1\%$ level) between
the numerical QNM frequency and the perturbative prediction for the same
frequency computed from the final mass and spin of the numerical simulation. A
slightly different approach was adopted in Ref.~\cite{Pan:2007nw}, where the
frequency-domain waveform was truncated abruptly at some cutoff frequency
close to the QNM frequency. The truncated waveforms display a Gibbs phenomenon
in the time domain, and they are (surprisingly) quite effective at detecting
the full signal. The reason is that a cutoff in the frequency domain around
the fundamental QNM frequency corresponds, in the time domain, to a damped
sinusoid with frequency close to the cutoff frequency.

These phenomenological ``stitching recipes'' may well be enough for the
purpose of gravitational wave detection, but it is desirable to have a
complete understanding of QNM excitation in a compact object merger within
perturbation theory, based on the concept of excitation
coefficients. Ref.~\cite{Berti:2006wq} carried out a first step in this
direction by computing the excitation factors for Kerr BHs (see Section
\ref{inversioncontour}), but more work is required to compute the excitation
{\it coefficients} for generic initial data, and to understand the
validity of the linear approximation in a BH merger. A correct matching of
effective-one-body waveforms with the ringdown signal may have astrophysical
implications because QNMs play an important role in the recoil of merging BH
binaries, through a process sometimes called ``ringdown braking'' (see
\cite{Blanchet:2005rj,Damour:2006tr,Schnittman:2007ij} and references
therein). A complete analytical description of the merger of spinning,
precessing binaries would also prove useful for statistical studies of
astrophysical relevance ~\cite{Schnittman:2007sn}.

Even if phenomenological waveforms may be good enough at detecting
gravitational wave signals, a complete description of the waveform (including
both inspiral and ringdown) has been shown to improve parameter estimation by
effectively decorrelating the source parameters (see \cite{Luna:2006gw} for a
study predating the numerical relativity breakthrough, and
\cite{Babak:2008bu,Thorpe:2008wh,Ajith:2007qp,Ajith:2007kx,Ajith:2009fz} for
recent efforts in this direction). Large-scale efforts (dubbed the NINJA and
Samurai projects, respectively) are now attempting to use complete numerical
waveforms in LIGO data analysis, and to cross-validate numerical waveforms
produced by different codes \cite{Aylott:2009ya,Hannam:2009hh}.

An important point to keep in mind is that inspiral and merger/ringdown really
probe different BH populations, so they provide complementary
information. This should be quite clear by inspecting the red and black
histograms in Fig.~\ref{fig:spinBV}, which show that the spin distribution of
SMBHs in binaries is usually different from the spin distribution of the
remnant BHs formed as a result of the merger. This complementarity can be used
for interesting physical applications. For example, if we can determine
accurately enough the masses and spins of BHs {\em before and after merger}
for the same system, we could use this information to test Hawking's area
theorem in astrophysical settings \cite{Hughes:2004vw}.

\section{\label{sec:otherdev}Other recent developments}

\subsection{Black hole area quantization: in search of a $\log$}
One of the main driving forces behind the development of new tools to study
QNMs is the possibility that classical BH oscillations could yield insights
into their quantum behavior. First suggested by York \cite{York:1983zb} and
Hod \cite{Hod:1998vk}, this idea was further explored by Dreyer
\cite{Dreyer:2002vy} in the context of Loop Quantum Gravity, and subsequently
revisited by many authors (see
e.g.~\cite{Corichi:2002ty,Kunstatter:2002pj,Oppenheim:2003sq,Ling:2003iz,
  Setare:2003bd,Kiefer:2004ma,Setare:2004uu,Maggiore:2007nq,Vagenas:2008yi,
  Medved:2008iq,Wei:2009yj}).

The idea can be traced back to arguments by Bekenstein and collaborators
\cite{Bekenstein:1974jk,Bekenstein:1995ju}. A semi-classical reasoning
suggests that the BH area spectrum is quantized according to
\be
A_{\,N}=\gamma\, l_{P}^2\, \,,\quad N=1,2,...\quad\,,
\label{areaspectrum}
\ee
where $l_{P}$ is the Planck length (but see \cite{Khriplovich:2004fd} for
criticism).  Statistical physics arguments impose a constraint on $\gamma$
\cite{Bekenstein:1995ju,Bekenstein:1973ur}:
\be
\gamma=4 \log k\,,
\label{areaspectrum1}
\ee
where $k$ is an integer, which in general is left undetermined (but see
\cite{Bekenstein:1998aw}). Inspired by Bekenstein's ideas, Hod
\cite{Hod:1998vk} proposed to determine $k$ via a version of Bohr's
correspondence principle in which the highly-damped QNM frequencies play a
fundamental role.
At the time, the only available exploration of highly-damped QNMs was the
numerical study by Nollert \cite{Nollert:1993}, which indicated that
gravitational highly-damped QNMs in the Schwarzschild geometry asymptote to
\be
M\omega =0.0437123-i(2n+1)/8\,.
\label{asymptQNnollert}
\ee
While looking for classical oscillation frequencies proportional to the
logarithm of an integer, Hod realized that $0.0437123 \sim \ln3/(8\pi)$, and
went on to suggest that the emission of a quantum with frequency
$\omega_R=\ln3/(8\pi)$ corresponds to the smallest energy a BH can emit.  The
corresponding change in surface area would then be
\be
\Delta A= 32\pi MdM=32\pi M{\hbar} \omega= 4{\hbar}\ln 3 \,.
\label{surfacearea}
\ee
and by comparing with Eq.~(\ref{areaspectrum}) we get $k=3$, thus fixing the
area spectrum to
\be
A_{\rm n}=4l_{P}^2\ln{3}\,  n\,;\,\,n=1,2,...
\label{areaspectrumfinal}
\ee
A few years later Dreyer \cite{Dreyer:2002vy} used similar arguments to fix
the Barbero-Immirzi parameter of Loop Quantum Gravity. Shortly afterwards,
Motl \cite{Motl:2002hd} and Motl and Neitzke \cite{Motl:2003cd} showed
analytically that the highly damped QNM frequencies of a Schwarzschild BH are
indeed given by Eq.~(\ref{Mresult}), containing the desired logarithm of an
integer.

Motl and Neitzke's work lent some support to Hod's ideas. It was followed by a
flurry of activity to explore the highly-damped QNMs of several BH spacetimes,
whose outcome has been described and summarized in previous
sections. Unfortunately the conjectured relation between QNMs and area
quantization stumbled before the charged and rotating four-dimensional
geometries, for which the highly-damped regime is not as simple as suggested
by Hod's original argument (see Sections \ref{sec:qnmkerr}-\ref{sec:qnmrn}).
Recently Maggiore \cite{Maggiore:2007nq} observed that Hod's prescription was
based on experience with ``normal'' quantum systems, for which the relevant
frequency is $\omega_R$. For highly-damped systems, Maggiore noted that one
should rather consider the imaginary part of the QNMs $\omega_I$, solving a
number of puzzles and obtaining a different expression for the quantum of
area. A similar prescription has been extended also to Kerr BHs
\cite{Vagenas:2008yi,Medved:2008iq} and other geometries \cite{Wei:2009yj}.
It is not clear whether Maggiore's suggestion can be extended consistently to
all geometries. Whether a relation between QNMs and the quantum behavior of
BHs exists or not, Hod's suggestion was at the very least an important thrust
to complete our understanding of classical BH oscillation spectra.

\subsection{Thermodynamics and phase transitions in black hole systems}

In the last few years, remarkable relations between classical and
thermodynamical properties of black objects have been uncovered. For instance,
a correspondence between classical and thermodynamical instabilities of a
large class of black branes conjectured by Gubser and Mitra
\cite{Gubser:2000ec,Gubser:2000mm} was proved by Reall \cite{Reall:2001ag}
(see \cite{Harmark:2007md} for a review). Manifestations of this duality are
expected to appear in the QNM spectra. Indeed, some indications that phase
transitions correspond to changes in the QNM spectrum were provided in
specific cases by various authors
\cite{Koutsoumbas:2006xj,Shen:2007xk,Rao:2007zzb,
  Myung:2008ze,Koutsoumbas:2008pw,Koutsoumbas:2008yq}. However, at present
there seems to be no obvious correspondence between thermodynamical phase
transitions of the kind suggested by Davies \cite{Davies:1978mf,Davies:1989ey}
and QNM spectra (see \cite{Berti:2008xu} and references therein).

Another connection between QNMs and BH thermodynamics may follow from Hod's
proposed ``universal relaxation bound''
\cite{Hod:2006jw,Hod:2007tb,Hod:2008se}.  Hod's proposal asserts that the
relaxation time $\tau$ of {\it any} thermodynamic system is bounded by $\tau
\geq \hbar/(\pi T)$, where $T$ is the temperature of the system. For BHs this
implies the existence of (at least one) QNM frequency with imaginary part
$\omega_I \leq \pi T$, where $T$ is now the Hawking temperature of the BH. The
bound seems to be valid for various kinds of BHs, and it may be saturated by
extremal BHs. The significance of Hod's bound is not completely clear to
us. The bound is trivially satisfied by any physical system exhibiting
hydrodynamic behavior, since such a system always possesses sufficiently
long-lived modes (and, correspondingly, QNMs with imaginary part sufficiently
close or even infinitely close to zero: these hydrodynamic frequencies are
discussed in Section \ref{sec:hydlim}).  Finite-volume systems might be more
interesting, but then the concept of the "relaxation time" used in the bound
needs a proper definition.  In any case, this is an interesting idea which
might require better understanding. The relation between Hod's bound and the
viscosity-entropy bound (see Section \ref{sec:visco}) was discussed in
Ref.~\cite{Hod:2009rt}.

\subsection{Non-linear quasinormal modes}

QNMs are usually defined and studied by considering only first-order
perturbations. Being an intrinsically non-linear theory, general relativity is
expected to display non-linear effects, which might conceivably be captured by
going to higher orders in perturbation theory. The second-order formalism laid
down several years ago by various authors
\cite{Gleiser:1995gx,Gleiser:1998rw,Garat:2000gp} has recently been used to
compute corrections to the QNM frequencies, their detectability and their
influence on the late-time behavior of the system
\cite{Ioka:2007ak,Nakano:2007cj,Okuzumi:2008ej}. The encouraging outcome of
these studies is that non-linear effects may well be observable by future
gravitational wave interferometers. Favata \cite{Favata:2008yd} explores the
interesting possibility of detecting another important non-linear effect of general
relativity (the so-called gravitational wave memory) through gravitational
wave observations of merging binaries.

Higher-order perturbations of BH spacetimes have been explored systematically
by Brizuela {\it et al.} \cite{Brizuela:2006ne,Brizuela:2007zza}, using the
Gerlach-Sengupta formalism and the computer algebra methods described in
\cite{Brizuela:2008ra} (see also Kol \cite{Kol:2007mw}). A complete gauge-invariant formalism for second-order perturbations of
Schwarzschild BHs was recently reported \cite{Brizuela:2009qd}.

\subsection{\label{ref:acoustic}Quasinormal modes and analogue black holes}

Strong-field effects of general relativity are very small in Earth-bound
experiments. For this reason BH physics is most easily studied via
observations of astrophysical phenomena such as accretion, X-ray spectra, and
hopefully gravitational wave emission. The possibility to devise gravitational
experiments probing strong-field general relativity in the lab, as appealing
as it sounds, may seem out of reach. However, Unruh, Visser and others
\cite{Unruh:1980cg,Barcelo:2005fc} ingeniously showed that some defining
properties of BHs can be reproduced and studied by ``analogue BHs''. These
systems display at least a subset of the properties traditionally associated
with BHs and event horizons. Unruh's analogue BHs do not carry information
about the dynamics of Einstein's equations, but share many kinematical
features with true general-relativistic BHs.

The basic idea behind these analogue BHs is quite simple. Let us focus on a
particular analogue BH, the acoustic or ``dumb'' hole
\cite{Unruh:1980cg,Visser:1997ux}.  Consider a fluid moving with a
space-dependent velocity $v_0^i(x^i)$, for example water flowing through a
variable-section tube. Suppose the fluid velocity increases downstream, and
that there is a point where the fluid velocity exceeds the local sound
velocity $c(x^i)\equiv \sqrt{\partial p/\partial \rho}$, in a certain
frame. At this point, in that frame, we get the equivalent of an apparent
horizon for sound waves. In fact, no (sonic) information generated downstream
of this point can ever be communicated upstream (for the velocity of any
perturbation is always directed downstream, by simple velocity addition). This
is the acoustic analogue of a BH, sometimes referred to as a {\it dumb
  hole}. These are not true BHs, because the acoustic metric satisfies the
equations of fluid dynamics and not Einstein's equations.  However, sound
waves propagate according to the usual curved space Klein-Gordon equation with
the effective metric \cite{Unruh:1980cg,Visser:1997ux,Barcelo:2005fc},
\be
g^{u \nu}\equiv \frac{1}{\rho _0 c} \left[
\begin{array}{ccc}
-1 &\vdots&-v_0^{j}\\
\ldots \ldots &.&\ldots \ldots \ldots\\
-v_0^{i}&\vdots& (c^2 \delta _{ij}-v_0^iv_0^j)
\end{array}
\right]\,. \label{metricvisserinv} \ee
Analogue BHs should Hawking-radiate, though an experimental verification of
Hawking radiation in the lab is not an easy feat
\cite{novellovisser,Barcelo:2005fc}. Furthermore, sound wave propagation in
these metrics should reproduce many features of wave propagation in curved
spacetimes. Most importantly for this review, acoustic BHs have a (this time
literally!) ``characteristic sound'' \cite{Nollert:1999ji} encoded in their
QNM spectrum. QNMs of acoustic BHs, which may be important in experimental
realizations of the idea, were computed in
Refs.~\cite{Berti:2004ju,Cardoso:2004fi,Lepe:2004kv,Chen:2006zy} for a simple
$(2+1)-$dimensional acoustic hole, the ``draining bathtub''
\cite{Visser:1997ux}, for which ${\vec v}_0=(-A {\vec r}+B {\vec\phi})/r$. The
modes of $(3+1)$-dimensional acoustic holes with a ``sink'' at the origin were
computed in Refs.~\cite{Berti:2004ju,Xi:2007yb,Saavedra:2005ug}.  Scattering
from these holes is discussed in Ref.~\cite{Dolan:2009hd}. 

The experimental confirmation of these predictions is an interesting topic for
future research.
In practice, one may need a device to accelerate the fluid up to supersonic
velocities, such as a Laval nozzle \cite{Cardoso:2005ij,Cadoni:2005nh} (see
also Unruh's discussion in \cite{novellovisser}). Laval nozzles were first
used in steam turbines, but they find applications in other contexts,
including rocket engines and nozzles in supersonic wind tunnels. They consist
of a converging pipe where the fluid is accelerated, followed by a throat (the
narrowest part of the tube) where the flow undergoes a sonic transition, and
finally a diverging pipe where the fluid continues to accelerate.  QNMs of
flows in de Laval nozzles were discussed and computed in
Refs.~\cite{Okuzumi:2007hf,Abdalla:2007dz}.

Following on Unruh's ``dumb hole'' proposal, many different kinds of analogue
BHs have been devised, based on condensed matter physics, slow light
etcetera. We refer the reader to Refs.~\cite{novellovisser,Barcelo:2005fc} for
thorough reviews on the subject.  QNMs of condensed-matter analogue BHs were
recently computed in Refs.~\cite{Nakano:2004ha,Barcelo:2007ru}.  Finally, we
should mention that accretion of material onto astrophysical BHs can give rise
to supersonic walls, i.e., acoustic horizons outside the event horizon.  This
process is particularly interesting: astrophysical BHs may provide a
nature-given setting for producing analogue BHs \cite{Das:2006an,Das:2007ix}.

\section{Outlook}

The investigations of the last decade show that quasinormal spectra encode a
wealth of information on the classical and quantum properties of BHs and black
branes. From a holographic, high-energy perspective, BH QNMs yield important
information on the quasiparticle spectra and transport coefficients of the
dual theory, and have an intriguing hydrodynamic description. From an
astrophysical viewpoint, gravitational wave measurements of QNMs may allow us
to accurately measure BH masses and spins with unprecedented accuracy and to
test the no-hair theorem of general relativity. The suggestion that BH
oscillation frequencies may be related to their quantum properties is
controversial, but at the very least it has stimulated tremendous technical
progress in the calculation of previously unexplored regimes of the QNM
spectrum. In this review we tried to summarize these recent
developments. There is no doubt that further interesting connections between
QNM research, fundamental physics and astrophysics will be unveiled in the
future. In closing, we think it might be useful to list some of the
outstanding, unresolved issues.

\noindent
{\it General mathematical problems, asymptotic expansions and related issues --}
 It would probably be fair to say that the level of mathematical rigor in
QNM studies deserves an improvement. Whether or not boundary value problems for 
dissipative systems can be treated in the framework of a sufficiently general 
mathematical theory appears to be an open question. Less ambitiously, a 
unified treatment of resonances (QNMs included) at the level of mathematical 
physics would be desirable. Reliable regular and asymptotic expansions based on systematic 
procedures rather than {\it ad hoc} recipes have not yet been developed in many cases.
Improving numerical algorithms, in particular for QNMs with large imaginary parts, also 
remains an avenue of research. 

\noindent
{\it Black hole perturbation theory in asymptotically-flat spacetimes --}
Despite a fifty-year long history of investigation, BH perturbation theory in
asymptotically flat spacetimes has a number of unresolved issues.  The problem
of metric reconstruction from the Teukolsky formalism is still open. QNM excitation for generic perturbations of Kerr BHs
have not been fully explored yet. The eikonal limit of Kerr QNMs is still
poorly understood, as is the relation of QNMs with generic $(l,m)$ to unstable
circular geodesics in the Kerr metric. Finally, an outstanding problem
concerns the decoupling of linear perturbations of the Kerr-Newman metric.

\noindent
{\it Gauge-gravity duality --} At the moment, computing QNMs in the context of
the gauge-gravity duality is a thriving industry, naturally expanding to
include models with higher-derivative gravity, holographic models with
spontaneous symmetry breaking, or backgrounds dual to non-relativistic
theories.  As discussed in Section \ref{sec:holography}, one distinguishes
hydrodynamic-type modes from the rest of the QNM spectrum. On a technical
level, these modes are characterized by small parameters and thus can often be
determined analytically.  A full analysis of hydrodynamic QNMs for generic
backgrounds is currently lacking.  In particular, it would be helpful to
obtain an expression for the parameters of the sound mode similar to
Eq.~(\ref{D_const}) for the diffusion constant. The spectra of fermionic
fluctuations (e.g. of the Rarita-Schwinger field, see
\cite{Policastro:2008cx}) in black brane backgrounds have not yet been fully
computed. It would be very interesting to explore further the connection
between holography and the black hole membrane paradigm
\cite{Damour:2008ji,Parikh:1997ma,Starinets:2008fb,Iqbal:2008by}.

\noindent
{\it Gravitational wave astronomy --} An important problem in gravitational
wave physics concerns the matching of inspiral waveforms (as computed by
post-Newtonian theory) to the ringdown phase. This is necessary to produce
complete phenomenological templates to be used in gravitational wave
detection. Most attempts at solving this problem have adopted a purely
phenomenological approach, but a study of the excitation coefficients induced
by generic initial data (possibly considering non-linear corrections) could
provide a more consistent and systematic solution to the matching problem. A
study of the QNM contribution to the Green function has recently been shown to
hold promise as a computational approach to solve the self-force problem
\cite{Casals:2009zh}, which is of fundamental importance to model extreme- and
intermediate-mass ratio inspirals for LISA data analysis. Other open issues
concern ringdown data analysis. Simple calculations show that single-mode
templates only produce moderate losses in SNR when detecting multi-mode
signals. However, the loss in terms of parameter estimation accuracy is much
more significant. The problem of optimal template placing for detection by
multi-mode templates needs to be addressed. The potential of higher multipoles
in estimating the source parameters without angle-averaging should also be
explored: the different angular dependence of subdominant multipoles may
provide information on the spin direction of the final BH. If BH spins are
linked to jets, this information could be used for coincident searches of
gamma-ray bursts or other electromagnetic counterparts to compact binary
mergers \cite{Kanner:2008zh,Bloom:2009vx}.

In the last few years, unexpected connections between QNMs and seemingly
unrelated phenomena (such as analogue BHs or thermodynamical BH phase
transitions) have been uncovered or proposed. More intriguing connections will
surely emerge in the coming decades.

\ack

We thank Alessandra Buonanno, Manuela Campanelli, Marc Casals, Ruth Daly, Sam
Dolan, Guido Festuccia, Martin Gaskell, Leonardo Gualtieri, Shahar Hod, Roman Konoplya, Cole
Miller, Christian Ott, Luciano Rezzolla, Andrzej Rostworowski, Ana Sousa, Uli
Sperhake and Vilson Zanchin for very useful discussions and for sharing with
us some unpublished material.  We also thank Ilya Mandel for clarifications on
rate calculations; Luca Baiotti, Bruno Giacomazzo and Luciano Rezzolla for
providing numerical data from Ref.~\cite{Baiotti:2008ra}; Marta Volonteri for
preparing Figure~\ref{fig:spinBV}; Lubos Motl and Andrew Neitzke for
permission to reproduce Figure~\ref{fig:monodromy2} from
Ref.~\cite{Motl:2003cd}; Shijun Yoshida for providing numerical data from
Ref.~\cite{Cardoso:2003vt}, and Alex Miranda for numerous suggestions.  V.C.'s
work was partially funded by Funda\c c\~ao para a Ci\^encia e Tecnologia (FCT)
- Portugal through projects PTDC/FIS/64175/2006, by a Fulbright Scholarship,
and by the National Science Foundation through LIGO Research Support grant NSF
PHY-0757937.



\appendix

\section{\label{app:relpotentials}} 
{\bf Isospectrality, algebraically special modes and naked singularities}
\vskip 2mm

\noindent
A number of striking relations among gravitational perturbations of the Kerr
geometry were revealed by Chandrasekhar and colleagues
\cite{MTB,chandrarelation,Chandrasekhar:1975qn,chandraspecial,Wald1973}. Here
we review these relations specializing most of the discussion to the
Schwarzschild geometry. The reader is referred to the original works for more
details and for extensions to the Kerr spacetime
\cite{MTB,chandrarelation,Chandrasekhar:1975qn,chandraspecial}.
 
By direct substitution, it can be checked that the gravitational potentials
$V_{s=2}^{\pm}$ in Eqs.~(\ref{vodd}) and (\ref{veven}) can be rewritten in the
form
\begin{equation}
V_{s=2}^\pm=W^2\mp\frac{dW}{dr_*}+\beta\,, \qquad \beta=-\frac{\lambda^2(\lambda+1)^2}{9M^2}\,,\label{V2}
\end{equation}
where
\be
W=\frac{3M\left (6 Mr^2+2\lambda L^2(2M-r) \right
)}{\lambda r^2\left (6M+2\lambda r\right)L^2}-
\frac{\lambda (\lambda+1)}{3M}-\frac{3M}{\lambda L^2}\,.
\ee
(these equations correct some typos in
Ref.~\cite{Cardoso:2001bb}). Eq.~(\ref{V2}) emerged from Chandrasekhar's
investigations \cite{Chandrasekhar:1975qn,chandrarelation,chandraspecial} of
the nature of the gravitational potentials. Potentials of this form are called
superpartner potentials \cite{Cooper:1994eh}, and they imply the following
relation between the corresponding wavefunctions $\Psi^-$ and $\Psi^+$
\cite{MTB,Cardoso:2001bb}:
\be
\Psi^{\pm}=\frac{1}{\beta-\omega^2}\left (\mp W\,\Psi^{\mp}+\frac{d\Psi^{\mp}}{dr_*}\right )\,.\label{superpartner1}
\ee
Eq.~(\ref{V2}) seems to be unique to four-dimensional spacetimes, and does not
generalize to higher dimensions \cite{Kodama:2003jz,Ishibashi:2003ap}.  The
potential for electromagnetic-type and gravitational-type perturbations of
{\it extremal} RN BHs can also be expressed in the form (\ref{V2})
\cite{Onozawa:1996ba,Okamura:1997ic}. This justifies the fact that
electromagnetic perturbations with angular index $l$ are isospectral to
gravitational with index $l-1$, as discussed in Section \ref{sec:qnmrn}. The
isospectrality is a manifestation of supersymmetry between electromagnetic and
gravitational perturbations for extremal charged BHs
\cite{Onozawa:1996ba,Okamura:1997ic,Kallosh:1997ug}.

\noindent
{\it Isospectrality and asymptotic flatness} 
Suppose that $\omega$ is a QNM frequency for $\Psi^+$, i.e. that $\Psi^+
\rightarrow A^{+}e^{-i\omega r_*}$ when $r \rightarrow r_+$ and $\Psi^+
\rightarrow\,a^{+}_{\rm out}e^{i\omega r_*} $ when $r\rightarrow \infty$.
Here $a^{+}_{\rm out}=1$ for the Schwarzschild metric and, if we impose
Dirichlet boundary conditions, $a^{+}_{\rm out}=0$ for the SAdS metric.
Substituting into Eq.~(\ref{superpartner1}) we see that
\beq
\Psi^-&\rightarrow& \frac{A^{+}}{\beta-\omega^2}\left (W(r_+)-i\omega \right )e^{-i\omega r_*}\,,\qquad r \rightarrow r_+\,,\\
&\rightarrow & \,\frac{1}{\beta-\omega^2}\left (-a^{+}_{\rm out}\,W(\infty)e^{i\omega r_*}+\frac{d \Psi^{-}}{dr_*}\right ) \,,\qquad r\rightarrow \infty\,.
\eeq
The key point is that $d \Psi^{-}/dr_* \sim e^{i\omega r_*} $ at infinity for
the Schwarzschild geometry. Therefore, if $\omega$ is a QNM frequency for
$\Psi^{+}$ it is also a QNM frequency for $\Psi^{-}$, and the two types of
gravitational perturbations are isospectral
\cite{Chandrasekhar:1975qn,chandrarelation}. In general $d\Psi^{-}/dr_* \neq
0$ for SAdS, so the isospectrality is broken.  The above relations are valid
also for dS backgrounds: since the outer boundary conditions are imposed at
the cosmological horizon, it is easy to see that gravitational perturbations
of both parities are again isospectral.  A specialized analysis is needed at
the points where $\beta-\omega^2=0$: this condition defines the so-called
algebraically special modes, discussed below.

\noindent
{\it Algebraically special modes}
A class of special modes can be found analytically using Eq.~(\ref{V2}). In
the Petrov classification, this condition corresponds to a change in the
algebraic character of the spacetime. For this reason, Chandrasekhar called
these modes ``algebraically special'' (AS) \cite{chandraspecial,MTB}.
Defining $\pm W=\frac{d}{dr_*}\log \chi_\pm$, the wave equation can be written
as
\begin{equation}
\frac{1}{\Psi_{\pm}}\frac{d^2\Psi_{\pm}}{dr_*^2}+\left
(\omega^2-\beta \right )=
\frac{1}{\chi_\pm}\frac{d^2\chi_\pm}{dr_*^2}\,,
\label{equationgeneral}
\end{equation}
where we have used the identity $
\frac{1}{\chi_\pm}\frac{d^2\chi_\pm}{dr_*^2}=W^2\pm\frac{dW}{dr_*}$. AS modes
have frequencies $\omega=\tilde \Omega_l$ such that $\beta-\tilde \Omega_l^2=
0$. In this case Eq.~(\ref{equationgeneral}) can be easily integrated:
\be
\Psi_{\pm}=\chi_\pm \int dr_*/\chi_{\pm}^2=\chi_\pm\left
(C_{\pm 1}+C_{\pm 2}\int_{0}^r dr_*/\chi_{\pm}^2\right )\,, \label{solalge}
\ee
where $\chi_\pm=\exp\left[\pm\int W dr_*\right]$.  The special relation
between the two gravitational potentials extends to the Kerr geometry, where
AS modes correspond to a vanishing Teukolsky-Starobinsky constant \cite{MTB}.
The nature of the boundary conditions at the Schwarzschild AS frequency is
extremely subtle, and has been studied in detail by Maassen van den Brink
\cite{MaassenvandenBrink:2000ru}. Let us introduce some terminology
\cite{MaassenvandenBrink:2000ru}:

\noindent 1) ``standard'' QNMs have outgoing-wave boundary conditions at both
sides (that is, they are outgoing at infinity and ``outgoing into the
horizon'');

\noindent 2) total transmission modes from the left (TTM$_L$'s) are incoming
from the left (the BH horizon) and outgoing to the right (spatial infinity);

\noindent 3) total transmission modes from the right (TTM$_R$'s) are incoming
from the right and outgoing to the left.

The Regge--Wheeler equation and the Zerilli equation must be treated on
different footing at the AS frequency, since the supersymmetry transformation
leading to the proof of isospectrality is singular there. It turns out that
the Regge-Wheeler equation has {\it no modes at all}, while the Zerilli
equation has {\it both a QNM and a TTM$_L$ at the AS frequency}
\cite{MaassenvandenBrink:2000ru}.

Numerical calculations of AS modes have yielded some puzzling
results. Studying the Regge-Wheeler equation, Leaver \cite{Leaver:1985ax}
found a QNM very close (but not exactly {\it at}) the AS frequencies of
Eq.~(\ref{AlgSp}). Namely, he found QNMs at frequencies $2M\tilde
\Omega'_2=0.000000-3.998000i$ and $2M\tilde \Omega'_3=-0.000259-20.015653i$.
Similarly, in the extremal RN case one finds a QNM frequency very close to,
but {\it not exactly equal to}, the AS frequency \cite{Berti:2004md}. Maassen
van den Brink \cite{MaassenvandenBrink:2000ru} speculated that the numerical
calculations may be inaccurate and that no conclusion can be drawn on the
coincidence of $\tilde \Omega_l$ and $\tilde \Omega'_l$, {\it ``if the latter
  does exist at all''}.

An independent calculation was carried out by Andersson
\cite{Andersson:1994tt}, who found that the Regge--Wheeler equation has pure
imaginary TTM$_R$'s very close to $\tilde \Omega_l$ for $2\leq l\leq 6$. He
therefore suggested that the modes he found coincide with $\tilde \Omega_l$,
which would then be a TTM.  Maassen van den Brink
\cite{MaassenvandenBrink:2000ru} observed again that, if all figures in the
computed modes are significant, the coincidence of TTM's and QNMs is not
confirmed by this calculation, since $\tilde \Omega'_l$ and $\tilde \Omega_l$
are numerically (slightly) different.

Onozawa \cite{Onozawa:1996ux} calculated the (TTM) AS mode for Kerr BHs,
improving upon the accuracy of the Kerr AS frequencies computed in
Ref.~\cite{chandraspecial}. He showed that the Kerr QNM with overtone $n=9$
tends to the AS frequency $\tilde \Omega_l$ (as defined by the
Teukolsky-Starobinsky identity) when $a\to 0$ and suggested that QNMs
approaching $\tilde \Omega_l$ from the left and from the right may cancel at
$a=0$, leaving only a special (TTM) mode.
The situation concerning Kerr QNMs branching from the AS Schwarzchild mode is
still far from clear. Using slow-rotation expansions of the perturbation
equations, Maassen van den Brink \cite{MaassenvandenBrink:2000ru} drew two
basic conclusions on these modes. The first is that, for $a>0$, the so--called
Kerr special modes are all TTM's (left or right, depending on the sign of
$s$). The TTM$_R$'s should not survive as $a\to 0$, since they do not exist in
the Schwarzschild limit. In particular, in this limit, the special Kerr mode
becomes a TTM$_L$ for $s=-2$; furthermore, the special mode and the TTM$_R$
cancel each other for $s=2$. Studying the limit $a\to 0$ in detail, Maassen
van den Brink reached a second conclusion: the Schwarzschild special frequency
$\tilde \Omega_l$ should be a limit point for a multiplet of ``standard'' Kerr
QNMs, which for small $a$ are well approximated by
\be\label{VDBsmalla}
2M\omega=-4i -{33078176\over 700009}\frac{ma}{2M}+{3492608\over 41177}i  \frac{a^2}{4M^2}
+{\cal O}(ma^2)
+{\cal O}(a^4)
\ee
when $l=2$, and by his Eq.~(7.33) when $l>2$. Numerical studies found QNMs
close to the AS frequency, but {\it not} in agreement with this analytical
prediction \cite{Berti:2003jh}.
It was suggested (note [46] in Ref.~\cite{MaassenvandenBrink:2000ru}) that
QNMs corresponding to the AS frequency with $m>0$ may have one of the
following three behaviors in the Schwarzschild limit: they may merge with
those having $m<0$ at a frequency $\tilde \Omega'_l$ such that $|\tilde
\Omega'_l|<|\tilde \Omega_l|$ (but $|\tilde \Omega'_l|>|\tilde \Omega_l|$ for
$l\geq 4$) and disappear, as suggested by Onozawa \cite{Onozawa:1996ux}; they
may terminate at some (finite) small $a$; or, finally, they may disappear
towards $\omega=-i \infty$.  Recently another alternative was suggested
\cite{Leung:2003eq}, that in the Schwarzschild geometry a pair of
``unconventional damped modes'' should exist.  For $l=2$ these modes were
identified by a fitting procedure to be located on the unphysical sheet lying
behind the branch cut (hence the name ``unconventional'') at
$2M\omega_\pm=\mp0.027+(0.0033-4)i$. An approximate analytical calculation
confirmed the presence of these modes, yielding
$2M\omega_+\simeq-0.03248+(0.003436-4)i$, in reasonable agreement with the
above prediction.  If the prediction is true, an {\it additional} QNM
multiplet should emerge near $\tilde \Omega_l$ as $a$ increases.  This
multiplet {\it ``may well be due to $\omega_\pm$ splitting (since spherical
  symmetry is broken) and moving through the negative imaginary axis as $a$ is
  tuned''} \cite{Leung:2003eq}. The emergence of such multiplets was shown in
Ref.~\cite{Berti:2003jh}, but these do not seem to behave exactly as predicted
\cite{Leung:2003eq}.

\noindent
{\it Instability of naked singularities}
AS modes play an important role in the stability analysis of certain
spacetimes containing naked singularities.  An example of such a spacetime is
the negative-mass Schwarzschild geometry, Eq.~(\ref{lineelementads}) with
$L\to \infty$ and $M<0$.
From the general solution (\ref{solalge}) with $C_{-2}=0$ it follows that
\be
\Psi_{-}=C_{-1}\chi_{-}=C_{-1}r\left (6M+(l-1)(l+2)r\right )^{-1}e^{-\frac{\lambda(\lambda+1)}{3M}r_*}\,,
\ee
where $r_*=(r+2M\log[-2 M + r])$, is a regular Zerilli or scalar-type
gravitational mode in the entire spacetime with frequency
$\omega=i\lambda(\lambda+1)/(3M)$. Since $\omega_I>0$, the spacetime is
unstable.  This instability was first found numerically by Gleiser and Dotti
\cite{Gleiser:2006yz} and recognized to be an ``algebraically special
instability'' in Ref.~\cite{Cardoso:2006bv}. In fact, the calculation above
can be extended to (negative-mass) charged BHs in de Sitter spacetimes
\cite{Cardoso:2006bv}. The AS mode is not unstable for negative-mass Kerr
geometries, but numerical results show that both negative-mass and
overspinning ``Kerr'' geometries are unstable
\cite{Dotti:2006gc,Dotti:2008yr,Cardoso:2008kj,Pani:2009fd}.

\clearpage
\section{\label{app:highdampinglimitKerr}}
{\bf Highly damped modes of Kerr black holes}
\vskip 2mm

\noindent
Here we reproduce the formulas from Kao and Tomino \cite{Kao:2008sv}, that
allow one to compute analytically the highly damped modes of the Kerr
spacetime, and which have been used to compare against numerical results in
Fig. \ref{fig:fig13}.
Define 
$u_1=a(\lambda^{1/3}- \lambda^{-1/3})/2\sqrt{3}$,
$v_1=a(\lambda^{1/3}+ \lambda^{-1/3})/2$
and
$\lambda=3\sqrt{3} M/a +\sqrt{1+27M^2/a^2}$. Define also
\beq
f_0(r_0) &=& i r_0 \sqrt{3u_1^2 + v_1^2 + 2i u_1 v_1} E\left[\Upsilon_0\right]-\frac{i (r_0-u_1)(9u_1^2+v_1^2)}{\sqrt{3u_1^2+v_1^2+2i u_1 v_1}} K\left[\Upsilon_0\right] \nonumber \\
&-& \frac{i2r_0(3u_1 + iv_1) (r_0-u_1+iv_1)}{\sqrt{3u_1^2+v_1^2+2i u_1 v_1}} \Pi\left[\Upsilon_{-1},\Upsilon_0\right] \nonumber \\
&-& \frac{i(3u_1+iv_1)(3u_1^2-v_1^2-2r_0^2)}{\sqrt{3u_1^2+v_1^2+2i u_1 v_1}} \Pi\left[\frac{-2iv_1}{3u_1-iv_1}, \Upsilon_0\right]\,, \label{f0D4}
\eeq
with $\Upsilon_{-1}=\frac{-2i
  v_1(r_0+2u_1)}{(r_0-u_1-iv_1)(3u_1-iv_1)},\,\Upsilon_0=\frac{4i u_1
  v_1}{3u_1^2+v_1^2+2i u_1 v_1}$ and
\beq
f_m(r_0) &=& \frac{2}{r_0 \sqrt{3u_1^2 + v_1^2 - 2i u_1 v_1}} \times 
\left\{ F\left[\sin^{-1}\Upsilon_1,\Upsilon_2 \right] - K\left[\Upsilon_2
  \right]  
\right\} \nonumber \\
&-& \frac{4u_1}{r_0(r_0+2u_1) \sqrt{3u_1^2 + v_1^2 - 2i u_1 v_1}} \nonumber \\
&&\times \left\{ \Pi\left[\Upsilon_3, \sin^{-1}
\Upsilon_1, \Upsilon_2 \right]- \Pi\left[\Upsilon_3,\Upsilon_2 \right] \right\}. \label{fmD4}
\eeq
Here 
$\Upsilon_1\equiv 
\sqrt{\frac{3u_1^2 + v_1^2 - 2i u_1 v_1}
{3u_1^2 + v_1^2 + 2i u_1 v_1}}$, 
$\Upsilon_2=
\frac{3u_1^2 + v_1^2 + 2i u_1 v_1}
{3u_1^2 + v_1^2 - 2i u_1 v_1}$ 
and 
$\Upsilon_3=\frac{r_0(3u_1-i v_1)}{(r_0+2u_1)(u_1-iv_1)}$.
The functions $E(m), E(\varphi,m), K(m), F(\varphi,m), \Pi(n,m)$ and
$\Pi(n,\varphi,m)$ are the elliptical integrals: 
\beq
E(m) &=& \int_0^{\frac{\pi}{2}} \sqrt{1-m\sin^2\theta} \,d\theta, \, E(\varphi,m) = \int_0^{\varphi} \sqrt{1-m\sin^2\theta} \,d\theta,\nonumber \\
K(m)&=& \int_0^{\frac{\pi}{2}} \frac{1}{\sqrt{1-m\sin^2\theta}} \,d\theta, F(\varphi,m) = \int_0^{\varphi} \frac{1}{\sqrt{1-m\sin^2\theta}} \,d\theta, \nonumber\\
\Pi(n,m)&=& \int_0^{\frac{\pi}{2}} [1-n\sin^2\theta]^{-1}[1-m\sin^2\theta]^{-1/2} \,d\theta, \nonumber\\
\Pi(n,\varphi,m)&=& \int_0^{\varphi}
[1-n\sin^2\theta]^{-1}[1-m\sin^2\theta]^{-1/2} \,d\theta\,.\nonumber \eeq
Because of the pole at $r=r_0$ and branch cuts, there is an ambiguity in
Eqs.~(\ref{f0D4}) and (\ref{fmD4}). The correct analytic expression is
\beq
\delta_0 = -2i\left[
\frac{f_0(r_+) - f_0(r_-)}{r_+ - r_-} + \frac{i4\pi M r_+}{r_+ - r_-} 
\Theta(3u_1 r_+ - 3 u_1^2 - v_1^2) \right],
\eeq
where $\Theta(x)$ is the step function.  The term with a step function is
introduced to compensate the discontinuity caused by the term
$\Pi\left[\frac{-2i v_1(r_+ + 2u_1)}{(r_+ - u_1 - iv_1)(3u_1-iv_1)},\frac{4i
    u_1 v_1}{3u_1^2+v_1^2+2i u_1 v_1}\right]$. Similarly,
\beq 
\delta_m = 4iMa \left[r_+ f_m(r_+,u_1,v_1) - r_- f_m(r_-,u_1,v_1) \right]
/(r_+ - r_-).
\eeq 
Note that here there is no discontinuity and we do not need to introduce any
term with step function.  The quantities $\varpi,\,\delta \omega_I$, defined
in Eqs.~(\ref{varpikerr}) and (\ref{deltakerr}), are related to
$\delta_0,\,\delta_m$ via $\varpi=\delta_m/\delta_0\,,\, \delta
\omega_I=2\pi/\delta_0$.

\section{\label{app:SMBHB}}
{\bf Supermassive black hole binary candidates}
\vskip 2mm

\noindent
At the moment of writing there are three plausible SMBH binary candidates at
close separation. Due to the relevance of these observations for SMBH mergers
and their rates (see e.g.~\cite{Volonteri:2009nh}), here we briefly summarize
the observations and their current interpretations.

\noindent {\em SDSSJ092712.65+294344.0 --} 
This quasar shows two sets of narrow emission lines, only one of which has
associated broad lines, separated in velocity by 2650 km\,s$^{-1}$, as well as
additional emission and absorption lines at intermediate redshift. Komossa and
collaborators \cite{Komossa:2008qd} interpret the velocity separation between
the two sets of lines as evidence for a recoiling BH with mass $M\approx
10^{8.8}M_\odot$. This large recoil speed can only be achieved if the final BH
formed from the merger of two large-spin BHs in the so-called ``superkick''
configuration \cite{Gonzalez:2006md,Campanelli:2007ew,Campanelli:2007cga}. The
recoiling BH interpretation has been criticized by various authors. Bogdanovic
{\it et al.}  \cite{Bogdanovic:2008uz} propose a model where the blueshifted
narrow lines originate from an accretion stream within the inner rim of the
circumbinary disk of a massive BH binary with mass ratio $q\approx 0.1$ and
mass $M_2\approx 10^8M_\odot$ for the secondary.  Dotti {\it et al.}
\cite{Dotti:2008yb} also propose a model with a massive BH binary embedded in
a circumbinary disk, where the blueshifted lines originate from gas swirling
around the secondary BH. The BH binary has mass ratio $q\approx 0.3$,
$M_1\approx 2\times 10^9M_\odot$, a semi-major axis $a\approx 0.34$~pc and an
orbital period $P\approx 370$~yrs.  More detailed observations seem to favor
the idea that the system represents the superposition of two AGN, rather than
a recoiling SMBH \cite{Shields:2008kn}.  Other interpretations suggest that
the system is a more distant analog of NGC1275 \cite{Heckman:2008en}, a large
and a small galaxy interacting near the center of a rich cluster
\cite{Conselice:2001gp}.

\noindent {\em SDSSJ153636.22+044127.0 --}
This quasar shows two broad-line emission systems, separated in velocity by
3500~km$/$s, and unresolved absorption lines with intermediate
velocity. Boroson and Lauer \cite{Boroson:2009va} interpret this quasar 
as a binary system of two SMBHs with masses $M_1\approx
10^{8.9}M_\odot$ and $M_2\approx 10^{7.3}M_\odot$ (hence $q\approx 40$)
separated by $\sim 0.1$~pc, with an orbital period of $\sim 100$~yrs. 
Depending on unknown geometrical factors of the
orbit, this system could coalesce either in $3\times 10^{11}$~yrs or in
$7\times 10^9$~yrs. 
%
Several alternative interpretations of the observations have been proposed.
Gaskell \cite{Gaskell:2009av} points out that the blueshift/redshift of the
broad emission lines in this system can be interpreted in terms of normal line
emission from a disk in an AGN.  Radio observations by Wrobel and Laor
\cite{Wrobel:2009na} identify two faint, spatially distinct radio sources
within the quasar's optical localization region, and suggest that the system
could consist of a quasar pair separated by $\sim 5$~kpc.  Chornock {\it et
  al.}  \cite{ChornockB,Chornock:2009sb} find a {\it third} broad emission
component in the system, note a lack in velocity shift between the blueshifted
and redshifted components, and argue that the system may be an unusual member
of a class of AGNs known as ``double-peaked emitters,'' rather than a SMBH
binary or a quasar pair.  Lauer and Boroson \cite{Lauer:2009us} remark that
the lack of amplitude variations is unusual for a double-peaked emitter, and
that longer temporal baselines are required to rule out the binary hypothesis.

\noindent {\em OJ287 --}  
This BL Lac object shows quasi-periodical optical outbursts at 12-year
intervals, with two outburst peaks (separated by one to two years) per
interval. Optical observations of this source date back to 1890. Valtonen and
collaborators interpret the two outbursts as happening when a smaller BH
pierces the accretion disk of the primary BH, producing two impact flashes per
period (see \cite{Valtonen:2007} for details of the model).  A model with
non-spinning BHs of mass $M_1=1.8\times 10^{10}M_\odot$, $M_2=1\times
10^8M_\odot$, semi-major axis
$a\simeq 0.045$~pc and eccentricity $e\simeq 0.66$ was able to predict the
date of the latest outburst. Remarkably, this model only reproduces the
observations if gravitational radiation reaction is included
\cite{Valtonen:2008tx}. Indeed, if the interpretation is correct, the system
would inspiral very quickly and merge in $\sim 10^4$~years.  Alternative
models attribute the observed behavior to (1) oscillations in the accretion
disk or jet of a single BH, or (2) variations of the accretion rate in a disk
or a wobble of a jet in a binary BH (see \cite{Komossa:2002tn,Valtonen:2008tx}
for more details).

If the binary BH interpretation is correct, the orbital parameters of the two
candidate BH binaries from Refs.~\cite{Valtonen:2008tx,Boroson:2009va} may not
be too dissimilar: in gravitational wave lingo, both systems would be IMRIs
(intermediate mass ratio inspirals).

\section{\label{app:spins}}
{\bf Black hole spin estimates}
\vskip 2mm

\noindent
Estimating BH masses is relatively easy because mass has a measurable effect
even at large radii, where Newtonian gravity applies. Spin, on the contrary,
is only measurable by looking at orbits close to the BH, where relativistic
effects are important. The reason is that the gravitational potential around a
rotating object can be expanded as
\be
\Phi(r,\theta)=
-(M/r)
-q (M/r)^3
P_2(\cos\theta)
+{\cal O}(r^{-4})\,.
\ee
The parameter $q=Q/M^3$ is a measure of the mass quadrupole moment, and to
lowest order it scales quadratically with the spins
\cite{Laarakkers:1997hb,Berti:2003nb}; for BHs, $Q=-Ma^2$.  Since the
leading-order spin contribution scales like $(M/r)^3$, spin measurements must
rely on observations of test particles on orbits with very small
radii. 

Luckily, we do have a chance to observe such orbits by looking at accreting
gas close to the ISCO allowed by general relativity.  All methods to measure
spin using electromagnetic observations are based on variants of this
idea, which is illustrated in the left panel of Fig.~\ref{fig:lightring}.
There we show the variation of the ISCO frequency $M\Omega_{\rm ISCO}$ and of
the light-ring frequency $M\Omega_{\rm LR}$ with the dimensionless spin
parameter $a/M$ \cite{Bardeen:1972fi}. Positive values of $a/M$ correspond to
corotating orbits, and negative values correspond to counterrotating
orbits. The orbital frequency at the ISCO has a maximum when $a/M=1$ ($R_{\rm
  ISCO}=M$) and a minimum when $a/M=-1$ ($R_{\rm ISCO}=9M$); for a
Schwarzschild BH, $R_{\rm ISCO}=6M$ and $M\Omega_{\rm ISCO}=6^{-3/2}$. The gas
in an accretion disk spirals in through a sequence of quasi-circular orbits as
it viscously loses angular momentum; when it reaches the ISCO, it accelerates
radially and falls into the BH, so the ISCO can be thought of as the inner
edge of the accretion disk. The radiation efficiency $\eta$, also plotted in
Fig.~\ref{fig:lightring}, yields the energy radiated by the accretion disk per
unit accreted mass. In principle $\eta$ is determined by the binding energy of
gas at the ISCO, which depends only on $a/M$.

Most spin estimates are based on the observation of electromagnetic emission
from accreting gas close to the ISCO. Unfortunately, all of these estimates
are to some degree model-dependent. Some corrections to the highly idealized
scenario described above are discussed in a number of papers
\cite{Zhang:1997dy,Reynolds:2007rx,Beckwith:2008pu}. For example, Beckwith and
Hawley \cite{Beckwith:2008pu} claim that an accurate modeling of magnetic
fields would significantly displace the ``inner edge'' of the accretion disk
from the standard general relativistic ISCO, and that this would lower the
spin estimates inferred from Fe K$\alpha$ observations. More in general,
systematic errors originate from our limited understanding of i) the physical
processes producing the X-ray spectra 
\footnote{The most pessimistic viewpoint on electromagnetic spin measurements
  is perhaps that of Miller and Turner: according to them, ``with the
  uncertainty and debate on the physical processes that dominate the X-ray
  spectrum in this energy range [...] it is not currently possible to
  constrain the BH spin in a model-independent way'' \cite{Turner:2009mv} (see
  also \cite{Miller:2008zq}).}
, and ii) accretion physics in strong-field gravity regions. Because of these
uncertainties, different models are sometimes capable of fitting the
observations while predicting different values of the spin (see
e.g. Ref.~\cite{Psaltis:2008bb} for examples). A discussion of accretion disk
physics as applied to a study of the strong-field gravity region around BHs is
outside the scope of this paper, and we refer the reader to the many excellent
review articles on this topic
\footnote{Liedahl and Torres \cite{Liedahl:2005zb} focus on the basic general
  relativistic equations, scattering processes and atomic
  transitions. Reynolds and Nowak \cite{Reynolds:2002np} discuss theoretical
  problems in the modeling of relativistic disk lines. Fabian and Miniutti
  \cite{Fabian:2005hr} and Miller \cite{Miller:2007tj} review relativistic
  X-ray emission lines from the inner accretion disks around rotating BHs and
  observational prospects for constraining the spin history of SMBHs (see also
  \cite{Miller:2006te} for a similar review focusing on stellar-mass BHs).}.

Schematically, we can classify the main techniques used so far to estimate BH
spins as belonging to four groups
\cite{Narayan:2005ie,Remillard:2006fc,Psaltis:2008bb}:

\noindent
(a) {\it Continuum spectroscopy of accretion disks} has been applied to
various stellar-mass BH candidates. This method was pioneered by Zhang {\it et
  al.} \cite{Zhang:1997dy} to suggest that two galactic superluminal jet
sources, GRO J1655-40 and GRS 1915+105, should harbor rapidly spinning
BHs. Narayan, McClintock and collaborators have embarked in a program to
estimate spins for about a dozen BH candidates
\cite{Shafee:2005ef,McClintock:2006xd,Narayan:2007ks,
  Shafee:2007sa,McClintock:2007we,Liu:2008tk,Gou:2009ks} (but other groups are
also very active
\cite{Middleton:2006kj,Davis:2006bk,Davis:2006cm,Brown:2006cj}).
They also proposed that accretion can be used to provide {\it hints} of the
presence of an event horizon for galactic stellar-mass BH binaries
\cite{Psaltis:2008bb,Narayan:2005ie} (but see \cite{Abramowicz:2002vt} for
criticism).

The basic idea is that when BHs have a large mass accretion rate the accreting
gas tends to be optically thick, radiating approximately as a blackbody. In
this spectral state (known as the ``high soft state'') the flux of radiation
$F(R)$ emitted by the accretion disk can be computed \cite{Page:1974he} and
used to obtain an effective temperature profile
$T(r)=\left[F(R)/\sigma\right]^{1/4}$, where $\sigma$ is the Stefan-Boltzmann
constant. By comparing the blackbody radiation with the spectral flux received
at Earth one can estimate the quantity $R_{\rm in}\cos\iota/D$, where $R_{\rm
  in}$ is the inner edge of the accretion disk, $D$ is the distance to the
source and $\iota$ the inclination angle. Unfortunately the method relies on
independent estimates of $\iota$ and $D$. Perhaps the major weakness of the
method consists in the fact that, in practice, the observed spectrum deviates
significantly from a blackbody. These deviations are usually modeled by
artificially increasing the temperature of the emitted radiation by a poorly
known ``spectral hardening factor'', which is usually (and roughly)
approximated by a constant $f_{\rm col}\simeq 1.7$.
Perhaps the most precise spin measurements to date using this technique regard
the eclipsing X-ray binary M33 X-7, with a claimed value $a/M=0.77\pm 0.05$
\cite{Liu:2008tk}, and the first extragalactic X-ray binary LMC X-1, for which
$a/M=0.90^{+0.04}_{-0.09}$ \cite{Gou:2009ks}.

\noindent
(b) {\it Spectroscopy of relativistically broadened Fe K$\alpha$ fluorescence
  lines} has been proposed as a promising alternative to continuum
fitting. This method originated from the discovery of a strong, broad spectral
line in the X-ray spectrum of the Seyfert 1.2 galaxy MCG-6-30-15
\cite{Tanaka:1995en}. Brenneman and Reynolds \cite{Brenneman:2006hw} applied
the method to {\it XMM-Newton} observations of this system, estimating the
spin to be very near the maximal limit: $a/M=0.989^{+0.009}_{-0.002}$. In this
case, the largest source of error comes from the unknown dependence of the
line emissivity on the disk radius $R$, which is usually modeled as a power
law \cite{Narayan:2005ie}. An analysis of systematic errors involved in the Fe
K$\alpha$ measurements was carried out by Reynolds and Fabian
\cite{Reynolds:2007rx}.  Their main finding is that systematic errors can be
significant for modest values of the spin, but they decrease for large spins,
so large-spin measurements (such as the one of Ref.~\cite{Brenneman:2006hw})
should be more reliable. The method has been applied to other AGN
\cite{Nandra:1996vv,Nandra:2006mf,Nandra:2007rp,Reeves:2006ne} and even to
stellar mass BH candidates \cite{Miller:2006te}. A very accurate measurement
of $a/M=0.935\pm 0.01$(statistical)$\pm 0.01$(systematic) has recently been
claimed by Miller {\it et al.}  \cite{Miller:2008vc} and Reis {\it et al.}
\cite{Reis:2008ja} for the stellar-mass BH GX 339-4 in
outburst. Ref.~\cite{Miller:2009cw} discusses BH spin estimates for several
different systems and their correlations with other physical properties of
X-ray binaries.

\noindent
(c) {\it Quasi-periodic oscillations} (QPOs) are likely to offer the most
reliable spin measurements for accreting systems that harbor stellar-mass BH
candidates, once the correct model is known. Unfortunately, at present there
are several competing models to explain QPOs (see
\cite{Remillard:2006fc,Psaltis:2008bb,Rezzolla:2003zx} for a discussion and
further references), and models based on different physical assumptions
typically yield very different spin estimates. For example, Figure 12 in
Psaltis's review \cite{Psaltis:2008bb} shows that the diskoseismology model of
Ref.~\cite{Wagoner:2001uj} and the parametric resonance model of
Ref.~\cite{Abramowicz:2001bi} give very different BH spin estimates in the
case of GRO J1655-40. At the very least, an indication that QPOs are related
to dynamical frequencies near the ISCO comes from the fact that the QPO
frequencies roughly scale with the inverse of the BH masses, as they should in
general relativity. A relatively model-independent lower limit on the spin can
then be obtained for GRO J1655-40: for this system one gets a lower limit of
$a/M\geq 0.25$ by requiring {\it only} that the observed 450~Hz oscillation
frequency must be limited by the azimuthal frequency at the ISCO, a rather
solid assumption \cite{Strohmayer:2001yn}.

\noindent
(d) {\it Statistical methods based on the radiative efficiency of AGN}.  A
rather general argument constrains, in principle, the average properties of
the spin distribution in AGN. The general idea is suggested by a glance at the
efficiency $\eta=L_{\rm acc}/\dot Mc^2$ (i.e., the energy radiated $L_{\rm
  acc}$ per unit accreted mass) in the right panel of Figure
\ref{fig:lightring}. For a non-rotating BH $\eta=1-(8/9)^{1/2}\simeq 0.057$,
while for a maximally rotating Kerr BH $\eta\simeq 0.42$ is much larger. In a
typical accretion system one can measure $L_{\rm acc}$ (provided the distance
is known), but not $\dot M$, so there is no way of knowing $\eta$ accurately
enough to estimate $a/M$. However, by observing high-redshift AGN one can
estimate the {\em mean} energy radiated by SMBHs per unit volume. By
considering SMBHs in nearby galaxies we can estimate the mean mass in SMBHs
per unit volume of the current universe. If we assume that SMBHs grow mostly
by accretion, by dividing these two quantities we can get the average
radiative efficiency of AGN, hence their average spin (this is a variant of
the famous ``Soltan argument'' \cite{Soltan:1982vf}).

Elvis {\it et al.} \cite{Elvis:2001bn} and Yu and Tremaine \cite{Yu:2002sq}
use observational data to infer that the mean efficiency $\eta\sim 0.10-0.15$
on average, and $\eta\sim 0.2$ (or larger) for the most massive systems. This
is possible only if SMBHs have significant rotation. Their arguments have been
supported by several later studies, some of which claim an average efficiency
$\eta\sim 0.3-0.35$ for quasars at moderate redshift
\cite{Wang:2006bz,Wang:2008ms,Cao:2008pd}. However there is no general
consensus on the interpretation of the data. Shankar {\it et al.}
\cite{Shankar:2007zg}, Merloni and Heinz \cite{Merloni:2008hx} and Daly
\cite{Daly:2008zk} present arguments in favor of low average radiative
efficiencies, hence low spins. Wang {\it et al.}  \cite{Wang:2009ws} argue
that the radiative efficiency has a strong cosmological evolution, decreasing
from $\eta\approx 0.3$ at $z\approx 2$ to $\eta\approx 0.03$ at $z\approx
0$. The uncertainties in disentangling radiative efficiencies from quasar
lifetimes were pointed out in Ref.~\cite{Rafiee:2008rz}. Daly
\cite{Daly:2009ih} provides spin estimates for 19 powerful FRII radio sources
and for 29 central dominant galaxies (CDGs). For the first class of sources
the spins seem to decrease from near-extremal values at $z=2$ to $\approx 0.7$
at $z=0$, while for the (lower power) CDGs the estimated spins are in the
range $0.1-0.8$. In conclusion, the jury is still out on the mean radiative
efficiency (and mean spin) of quasars. The outcome of this debate is
fundamental to constrain SMBH evolution models
\cite{Berti:2008af,Lagos:2009xr,Sijacki:2009mn}.

\noindent
(e) {\it Other methods}. Besides the methods listed above, there are other
avenues for measuring BH spin that hold promise for the future. One of these
is polarimetry. The idea is to exploit the fact that, in general relativity,
the plane of polarization of BH disk radiation changes with energy. This is a
purely relativistic effect, absent in Newtonian gravity, and the magnitude of
the change of the plane of polarization can give a direct measure of $a/M$
\cite{Connors:1980,Remillard:2006fc}. A second idea is based on Very Long
Baseline Interferometry (VLBI) imaging of the silhouette of the SMBH in
M87. This system is at a distance of 16~Mpc, much farther away than the SMBH
at the galactic center; however, because of its larger mass of $\sim 3.4\times
10^9M_\odot$, the apparent diameter of M87's horizon is 22$\,\mu$as, about
half as large as Sgr~A$^*$. Unlike Sgr~A$^*$, M87 exhibits a powerful radio
jet, hence it holds promise for exploring the relation between the BH spin and
the jet generation mechanism \cite{Broderick:2008qf}. A third, indirect way to
constrain BH spins is based on energetic considerations. The AGN outburst in
the MS0735.6+7421 cluster's central galaxy implies that its putative SMBH grew
by about $1/3$ of its mass in the past 100~Myr, accreting matter at $\sim
3-5M_\odot$/yr, inconsistent with the Bondi mechanism. The energetics of the
system could be explained instead by angular momentum released from a rapidly
spinning SMBH with $M>10^{10}M_\odot$ \cite{McNamara:2008dy}.

\vskip 1cm

\bibliographystyle{h-physrev4}

\bibliography{qnmcqg4}

\end{document}